\documentclass[11pt,oneside,english,french,intoc,bibliography=totoc,index=totoc,BCOR10mm,captions=tableheading,titlepage,fleqn]{scrbook}

\usepackage[T1]{fontenc}
\usepackage[latin9]{inputenc}
\usepackage[a4paper]{geometry}
\usepackage{fancyhdr}
\usepackage{babel}
\makeatletter
\addto\extrasfrench{%
   \providecommand{\og}{\leavevmode\flqq~}%
   \providecommand{\fg}{\ifdim\lastskip>\z@\unskip\fi~\frqq}%
}

\makeatother
\usepackage{float}
\usepackage{calc}
\usepackage{framed}
\usepackage{textcomp}
\usepackage{bm}
\usepackage{pdfpages}
\usepackage{amsthm}
\usepackage{amsmath}
\usepackage{amssymb}
\usepackage{graphicx}
\usepackage{setspace}
\usepackage{esint}
\usepackage{nomencl}
\providecommand{\printnomenclature}{\printglossary}
\providecommand{\makenomenclature}{\makeglossary}
\makenomenclature
\usepackage[unicode=true,
 bookmarks=true,bookmarksnumbered=true,bookmarksopen=true,bookmarksopenlevel=1,
 breaklinks=true,pdfborder={0 0 0},backref=false,colorlinks=false]
 {hyperref}
\hypersetup{pdftitle={Une théorie de la fonctionnelle de la densité moléculaire pour la solvatation dans l'eau},
 pdfauthor={Guillaume Jeanmairet},
 pdfpagelayout=OneColumn, pdfnewwindow=true, pdfstartview=XYZ, plainpages=false}

\makeatletter

\newcommand{\lyxmathsym}[1]{\ifmmode\begingroup\def\b@ld{bold}
  \text{\ifx\math@version\b@ld\bfseries\fi#1}\endgroup\else#1\fi}

\providecommand{\tabularnewline}{\\}

\numberwithin{figure}{section}
\numberwithin{equation}{section}

\@ifundefined{date}{}{\date{}}
\AtBeginDocument{
\renewcommand{\ref}[1]{\mbox{\autoref{#1}}}}
\addto\extrasfrench{%
 \renewcommand*{\equationautorefname}[1]{éq.\thinspace}

}

\usepackage[figure]{hypcap}

\let\myTOC\tableofcontents
\renewcommand\tableofcontents{%
  \myTOC
 }
\frontmatter

\setkomafont{captionlabel}{\bfseries}
\setcapindent{1em}

\usepackage{calc}

\let\mySection\section\renewcommand{\section}{\suppressfloats[t]\mySection}

\makeatother

\begin{document}
\includepdf[pages=-]{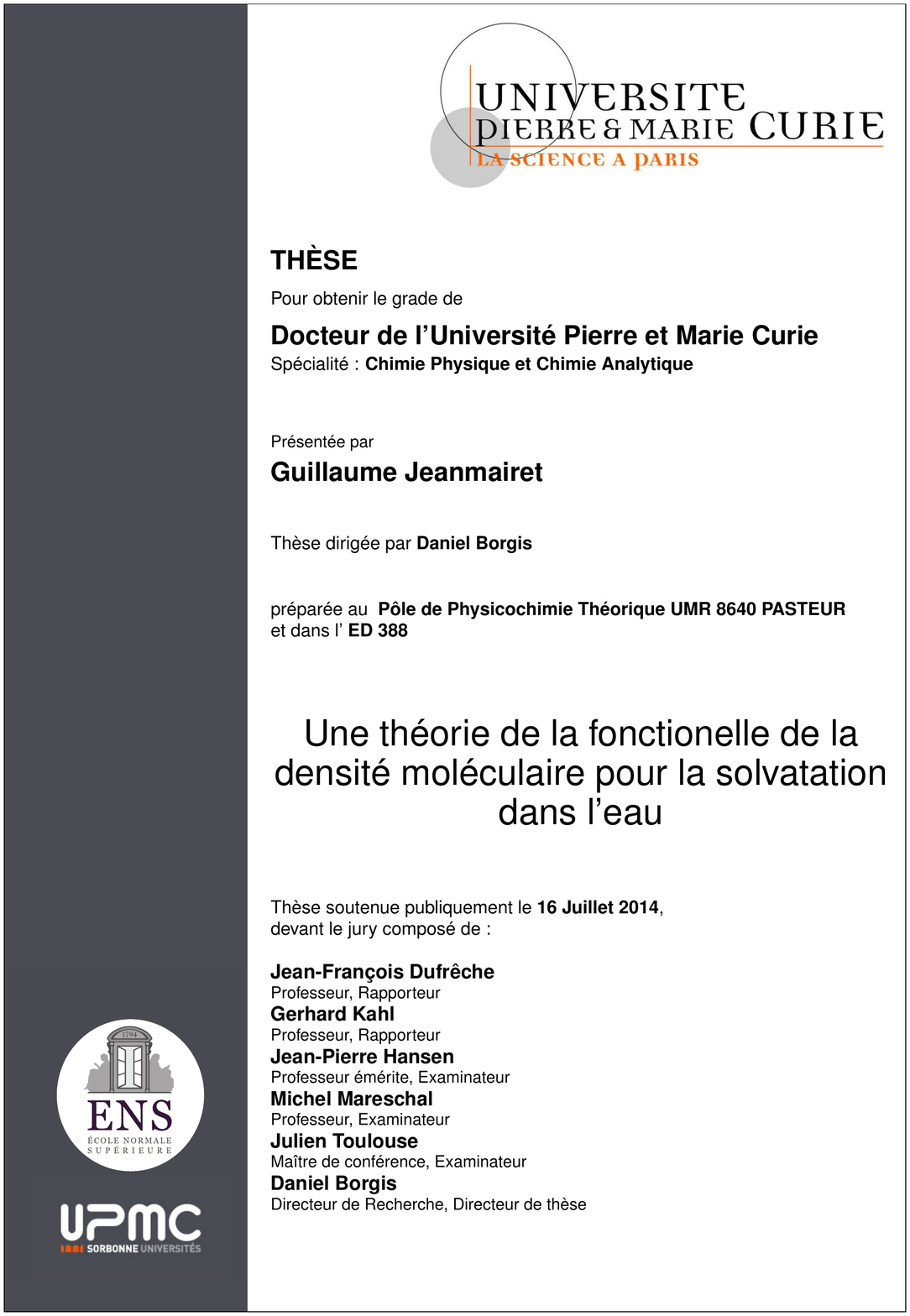}

\chapter*{Remerciements}

En premier lieu, je souhaite remercier les personnes ayant accepté
d'évaluer ces travaux de thèse. Merci à Jean-François Dufrêche et
Gerhard Kahl qui m'ont fait l'honneur de rapporter le présent manuscrit.
Merci également à Jean-Pierre Hansen, Michel Mareschal et Julien Toulouse
d'avoir pris part au jury de cette thèse. 

Je remercie également Ludovic Jullien, directeur de l'UMR 8640 PASTEUR,
qui m'a permis d'effectuer ce travail dans un environnement de grande
qualité.

Ce travail a été réalisé sous la direction de Daniel Borgis à qui
j'adresse mes sincères remerciements pour sa pédagogie, sa rigueur
scientifique, son enthousiasme et la confiance qu'il m'a accordé.
Toutes ces qualités ont permis de faire de ces trois années une aventure
exaltante scientifiquement et humainement.

Merci également à Maximilien Levesque qui a participé à mon encadrement
et qui a été à mon écoute tout au long de cette thèse. Sa sympathie,
sa gentillesse et nos discussions scientifiques resteront un souvenir
précieux.

Je souhaite également remercier les différents membres du pôle que
j'ai eu le plaisir de côtoyer pendant ces trois ans. Parfois, les
discussions autour d'un café sont plus fructueuses que la lecture
d'un article.

Ces trois années ont également été pour moi l'occasion d'enseigner
en tant qu'agrégé préparateur. Je remercie donc mes collègues de l'ENS
et de la préparation à l'agrégation, en particulier Jérôme Quérard,
mon partenaire de galère avec la \og vieille polarographie \fg{}.

J'aimerais aussi remercier mes copains de promo à l'ENS pour nos débats,
scientifiques ou non, où le respect n'avait d'égal que la mauvaise
foi.

Enfin, je souhaite terminer par une pensée pour ceux qui me soutiennent
depuis longtemps, ma famille mais aussi Fanny pour sa patience, ses
encouragements, ses relectures, ses talents de coiffeuse et tout le
reste.

\newpage{}\tableofcontents{}

\pagestyle{plain}

\mainmatter\addcontentsline{toc}{chapter}{Résumé}

\section*{Une théorie de la fonctionnelle de la densité moléculaire pour la
solvatation dans l'eau}

La théorie de la fonctionnelle de la densité classique est utilisée
pour étudier la solvatation de solutés quelconques dans le solvant
eau. Une forme approchée de la fonctionnelle d'excès pour l'eau est
proposée. Cette fonctionnelle nécessite l'utilisation de fonctions
de corrélation du solvant pur. Celles-ci peuvent être calculées par
simulations numériques, dynamique moléculaire ou Monte Carlo ou obtenues
expérimentalement. La minimisation de cette fonctionnelle donne accès
à l'énergie libre de solvatation ainsi qu'à la densité d'équilibre
du solvant. Différentes corrections de cette fonctionnelle approchée
sont proposées. Une correction permet de renforcer l'ordre tétraédrique
du solvant eau autour des solutés chargés, une autre permet de reproduire
le comportement hydrophobe à longue distance de solutés apolaires.
Pour réaliser la minimisation numérique de la fonctionnelle, la théorie
a été implémentée sur une double grille tridimensionnelle pour les
coordonnées angulaires et spatiales, dans un code de minimisation
fonctionnelle écrit en Fortran moderne, mdft. Ce programme a été utilisé
pour étudier la solvatation en milieu aqueux de petits solutés atomiques
neutres et chargés et de petites molécules polaires et apolaires ainsi
que de solutés plus complexes, une argile hydrophobe et une petite
protéine. Dans chacun des cas la théorie de la fonctionnelle de la
densité classique permet d'obtenir des résultats similaires à ceux
théoriquement exacts obtenus par dynamique moléculaire, avec des temps
de calculs inférieurs d'au moins trois ordres de grandeurs.

\rule[0.5ex]{1\columnwidth}{1pt}

\selectlanguage{english}%

\section*{A molecular density functional theory to study solvation in water}

A classical density functional theory is applied to study solvation
of solutes in water. An approximate form of the excess functional
is proposed for water. This functional requires the knowledge of pure
solvent direct correlation functions. Those functions can be computed
by using molecular simulations such as molecular dynamic or Monte
Carlo. It is also possible to use functions that have been determined
experimentally. The functional minimization gives access to the solvation
free energy and to the equilibrium solvent density. Some correction
to the functional are also proposed to get the proper tetrahedral
order of solvent molecules around a charged solute and to reproduce
the correct long range hydrophobic behavior of big apolar solutes.
To proceed the numerical minimization of the functional, the theory
has been discretized on two tridimensional grids, one for the space
coordinates, the other for the angular coordinates, in a functional
minimization code written in modern Fortran, mdft. This program is
used to study the solvation in water of small solutes of several kind,
atomic and molecular, charged or neutral. More complex solutes, a
neutral clay and a small protein have also been studied by functional
minimization. In each case the classical density functional theory
is able to reproduce the exact results predicted by MD. The computational
cost is at least three order of magnitude less than in explicit methods.\selectlanguage{french}%

\cleardoublepage{}

\lhead[\chaptername~\thechapter]{\rightmark}

\rhead[\leftmark]{}

\lfoot[\thepage]{}

\cfoot{}

\rfoot[]{\thepage}

\chapter*{Notations}

Les notations suivantes sont utilisées dans ce manuscrit.\\

\begin{itemize}
\item $(\theta,\phi)$ désigne les deux angles d'Euler des cordonnées sphériques,
$\theta$ est la colatitude qui varie entre $0$ et $\pi$, et $\phi$
la latitude qui varie entre $0$ et $2\pi$.
\item $\psi$ est le troisième angle d'Euler désignant la rotation propre
autour du vecteur radial.
\item $\bm{\Omega}$ est une notation compacte des trois angles d'Euler
$(\theta,\phi,\psi)$.
\item $(x,y,z)$ désigne les coordonnées cartésiennes.
\item $\bm{r}$ est le vecteur position, c'est donc une notation compacte
des cordonnées cartésiennes.
\item $\star$ désigne le produit de convolution, par exemple $\left[f*g\right](x)=\int_{-\infty}^{\infty}f(x-u)g(u)\mathrm{d}u$.
\item $\hat{f}(k)=\iiint_{R^{3}}f(\bm{r})\exp(-2i\pi\bm{k}\cdot\bm{r})d\bm{r}$
désigne la transformée de Fourier de la fonction $f$.
\item $\bm{k}$ est le vecteur réciproque.
\item $\beta=(\mathrm{k_{B}T})^{-1}$ est l'inverse de l'énergie thermique.
\item $\tilde{\bm{u}}$ désigne le vecteur unitaire $\bm{u}/\left\Vert \bm{u}\right\Vert $.
\item $\bar{f}$ désigne la fonction gros grains obtenue à partir de la
fonction $f$.
\item ${\cal F}\left[n(\bm{r})\right]$ désigne une fonctionnelle de la
fonction $n$, de variable $\bm{r}$. La notation en crochets est
réservée aux fonctionnelles, qui sont généralement écrites en caractères
calligraphiés. 
\item $\frac{\delta{\cal F}\left[n(\bm{r})\right]}{\delta n(\bm{r})}$ désigne
la dérivée fonctionnelle par rapport à la fonction $n$.
\item $\delta(x)$ désigne la fonction de Dirac.
\item $\left\Vert .\right\Vert $ désigne la norme euclidienne.
\item $\left|.\right|$ est la fonction valeur absolue.\end{itemize}

\cleardoublepage{}

\pagestyle{fancy}

\lhead[\chaptername~\thechapter]{\rightmark}

\lhead[\chaptername~\thechapter]{\rightmark}

\rhead[\leftmark]{}

\lfoot[\thepage]{}

\cfoot{}

\rfoot[]{\thepage}

\part{État de l'art }

\chapter{Modélisation des propriétés de solvatation\label{chap:1}}

\section{Solvatation}

La solvatation est définie par l'Union Internationale de Chimie Pure
et Appliquée (IUPAC)\cite{nic_iupac_2009} comme:

\textit{}%
\fbox{\begin{minipage}[t]{1\columnwidth}%
\textit{Toute interaction stabilisante d'un soluté et d'un solvant
{[}...{]}. De telles interactions mettent en jeu généralement des
forces électrostatiques et de Van der Waals, ainsi que des effets
plus spécifiques chimiquement, tels que la formation de liaisons hydrogènes.}%
\end{minipage}}

Une solution est un mélange homogène dans lequel une entité chimique
est présente en grande quantité (le solvant), et où une ou plusieurs
autres entités chimiques (les solutés) sont présentes en plus faible
quantité. Bien que les solutions solides existent, elles ne seront
pas du tout abordées dans ce manuscrit où on se limitera à l'étude
des solutions liquides.

Dans le cadre du traitement complet des interactions entre noyaux
et électrons réalisé en chimie quantique, les forces agissant entre
solutés et molécules de solvant ont toutes pour origine l'interaction
électrostatique. On peut s'interroger dès lors sur la signification
des termes \textit{forces de Van-der-Waals} et \textit{liaisons hydrogènes}
qui, dans la définition, semblent venir en plus des interactions électrostatiques.
Un traitement complet et précis des interactions électrostatiques
est complexe. Historiquement, les chimistes ont été amenés à les représenter
par des modèles plus simples d'interactions comme les forces de Van-der-Waals
et les liaisons hydrogènes. 

Il y a formation d'une phase homogène si les interactions qui se développent
entre solutés et solvant sont globalement plus favorables que la somme
des interactions solutés-solutés et solvant-solvant perdues lors de
la mise en solution. Ces interactions entre solutés et solvant jouent
un rôle clé dans la mise en solution mais aussi dans la mise en œuvre
des réactions chimiques qui se produisent en solution. Comme nous
allons le voir dans la suite de ce chapitre, l'estimation de ces interactions
a depuis longtemps intéressé les physico-chimistes. Le but de cette
thèse est le développement d'une méthode théorique et son implémentation
numérique, pour étudier les questions liées à la solvatation en milieu
aqueux. Il nous faut donc définir plus précisément les grandeurs qui
vont être utilisées pour décrire la solvatation. On s'intéressera
d'abord à l'aspect énergétique mentionné auparavant puis à la structure
de solvatation, qui décrit l'arrangement spatial des molécules de
solvant autour des solutés.

\subsection{Aspect énergétique de la solvatation\label{sec:The-next-section}}

Nous avons vu que la solvatation est la conséquence d'interactions
(ou de forces) intermoléculaires se développant entre toutes les molécules
de la solution, solvant et solutés.

Au premier abord, on pourrait penser qu'une bonne estimation de l'effet
de la solvatation est de considérer la différence entre la somme de
toutes ces interactions dans le cas de chacun des constituants pris
dans des phases pures et la somme de ces mêmes interactions dans la
solution. Le signe de cette différence renseigne directement sur la
stabilisation ou la déstabilisation causée par la mise en solution.
Cependant, cette énergie microscopique dépend de la position de chacune
des molécules. Ainsi, à chaque configuration différente (description
détaillée des positions de toutes les molécules) est associée une
valeur différente de l'énergie. De plus, il est évident qu'en procédant
de cette façon, c'est-à-dire en ne considérant que les termes énergétiques,
on manque un terme important qui est le terme entropique. Les observables
appropriées sont des grandeurs macroscopiques qui sont des moyennes
des configurations microscopiques et qui ne dépendent plus des positions
instantanées.

La bonne façon de quantifier la solvatation est de considérer plutôt
les différentes énergies libres mises en jeu lors de ce processus.
Pour le physico-chimiste la grandeur macroscopique la plus couramment
considérée est l'énergie libre de Gibbs, aussi appelée enthalpie libre.
C'est la mesure pertinente du travail d'un système à nombre de particules,
température et pression fixés.

\begin{figure}[h]
\noindent \begin{centering}
\includegraphics[width=0.6\textwidth]{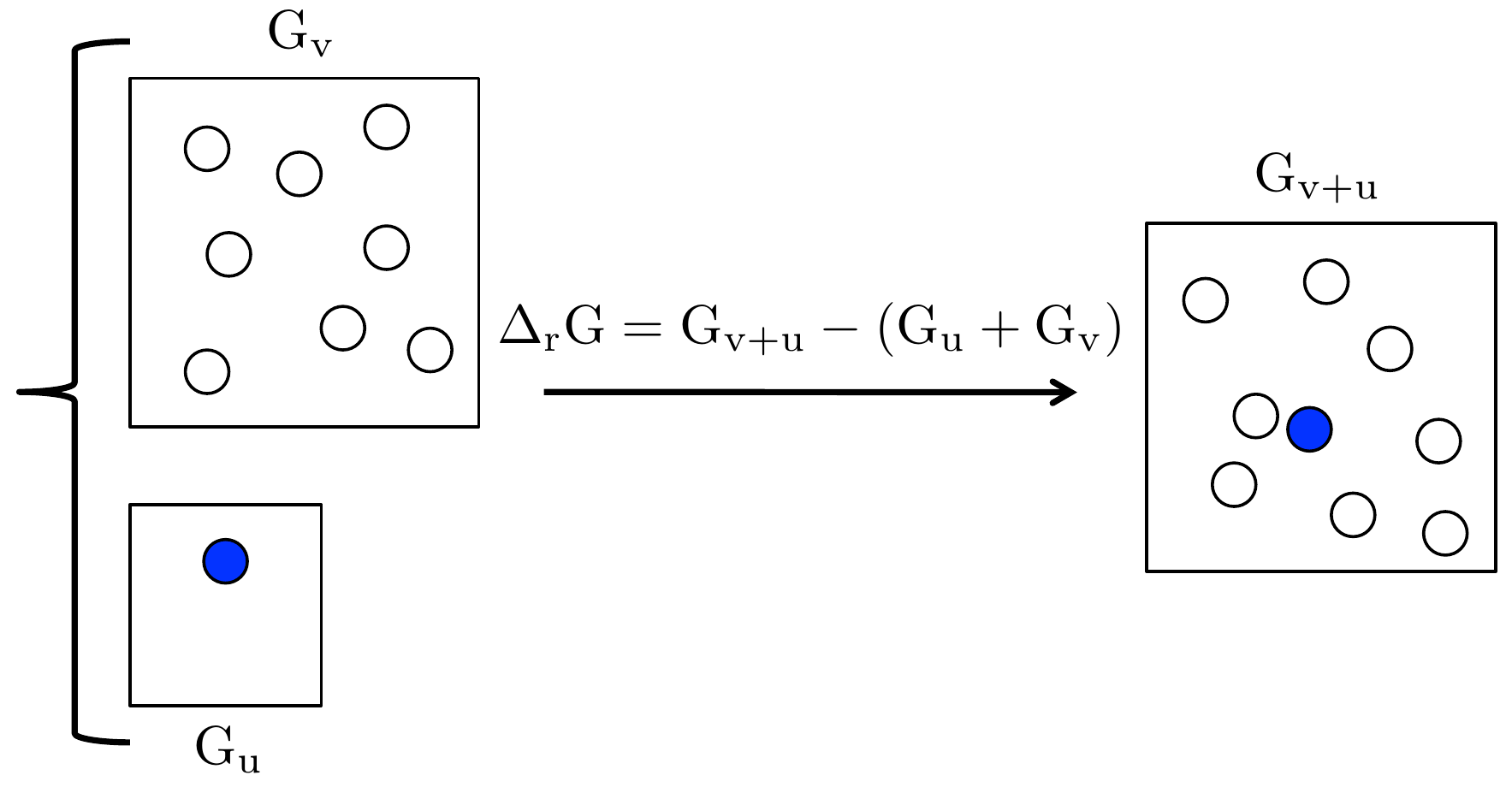}
\par\end{centering}

\protect\caption{Vue schématique de l'enthalpie libre de solvatation, définie comme
la différence entre l'enthalpie libre de la solution ($\mathrm{G_{v+u}}$)
et la somme des enthalpies libres des molécules de solvants ($\mathrm{G_{\text{v}}}$)
et solutés ($\mathrm{G_{\text{u}}}$) purs.\label{fig:ShemaDeltar}}
\end{figure}

On considérera des variations (d'énergie, d'entropie, etc) entre l'état
final de la réaction (ici les espèces mises en solution) et son état
initial (les espèces prises dans leurs phases pures respectives),
ces deux états étant pris à l'équilibre thermodynamique, voir \ref{fig:ShemaDeltar}.
On utilise généralement des grandeurs intensives, les grandeurs molaires.
Toutes ces grandeurs peuvent être mesurées expérimentalement afin
de conclure sur la stabilisation relative obtenue par la mise en solution. 

L'énergie libre de solvatation (dimensionnée comme une énergie divisée
par une quantité de matière) est parfaite pour mesurer à quel point
la mise en solution est favorable, mais ne donne en revanche aucune
information sur l'arrangement spatial des molécules. L'organisation
des molécules joue pourtant un rôle important dans l'explication des
propriétés physiques et chimiques des réactions en solution. Il est
donc judicieux de s'intéresser à des grandeurs informant sur l'organisation
des molécules dans la solution pour connaître l'influence du soluté
sur la structure spatiale du solvant.

\subsection{La structure de solvatation}

\begin{figure}[h]
\noindent \begin{centering}
\includegraphics[width=0.6\textwidth]{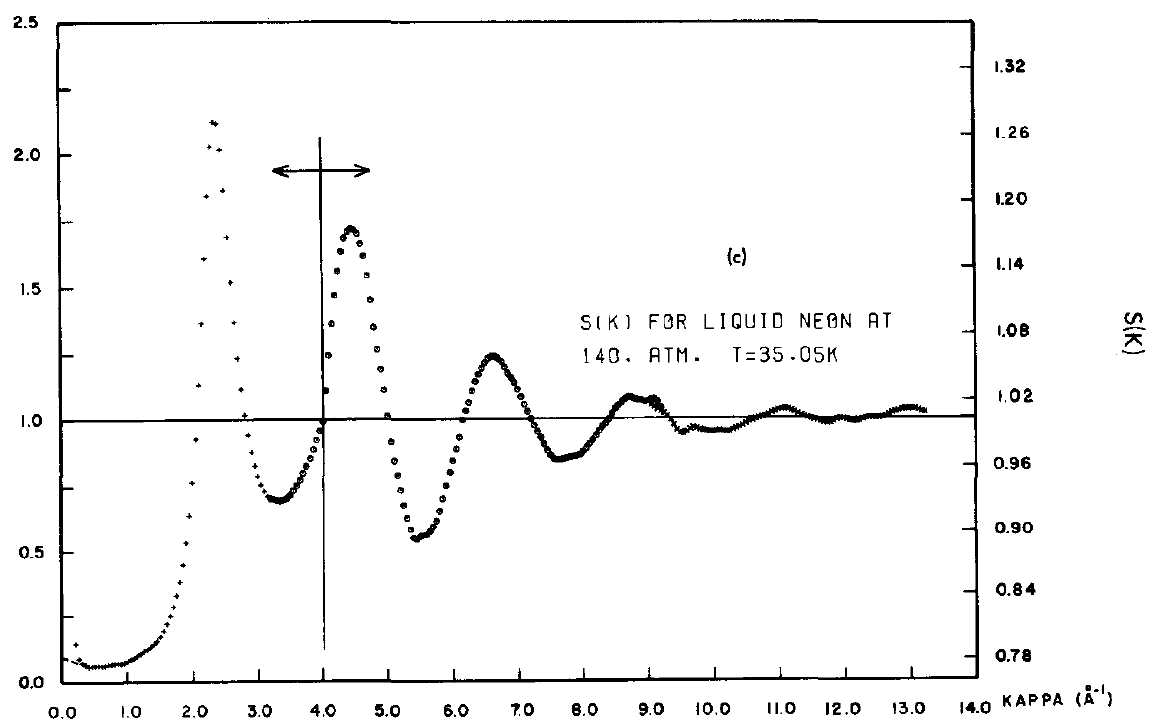}
\par\end{centering}

\protect\caption{Facteur de structure du néon liquide à 35.05K à une densité $\rho=$0.03469
$\textrm{\AA}^{-3}$\label{fig:Facteur-de-structureNEDegraaf} obtenue
par diffraction de neutrons \cite{graaf_structure_2003} en fonction
de la norme du vecteur de diffraction.}

\end{figure}

En ce qui concerne l'arrangement tridimensionnel des molécules dans
la solution, le saut microscopique-macroscopique effectué dans la
section précédente est encore plus pertinent. En effet, la connaissance
exacte de la position des molécules à chaque instant est bien sûr
la donnée contenant le plus d'information sur la structure. Cependant,
les positions des molécules sont sans cesse changeantes ce qui rend
l'information de leurs positions instantanées très peu lisible et
surtout impossible à mesurer expérimentalement. Il est donc indispensable
d'introduire des grandeurs permettant de mesurer l'ordre moyen relatif
de la solution. Expérimentalement ceci peut être fait en déterminant
le facteur de structure $S$ qui est lié à l'intensité du signal lumineux
diffracté par l'échantillon. Comme pour un solide celui-ci est mesuré
par des expériences de diffraction de neutrons ou de rayons X\cite{fischer_neutron_2006}.
Contrairement à un cristal, le liquide est isotrope. De ce fait ce
facteur de structure ne dépend que de la norme du vecteur de diffraction
$\left\Vert \bm{k}\right\Vert $ et non de son orientation. Si l'on
représente le facteur de structure en fonction de la norme du vecteur
de diffraction, un exemple est donné en \ref{fig:Facteur-de-structureNEDegraaf}
pour le néon, on remarque que celui ci comporte des pics larges témoins
de l'existence d'un ordre à courte distance. 

Dans le cadre de la physique des liquides, on utilise plus souvent
la fonction de distribution radiale, notée $g$, qui est liée au facteur
de structure par la relation\cite{hansen_theory_2006}:
\begin{equation}
S(k)=1+\rho\iiint_{\mathbb{R}^{3}}e^{-i\bm{kr}}g(r)\mathrm{d}\bm{r},
\end{equation}
où $\bm{k}$ est le nombre d'onde du vecteur diffraction et $\rho$
la densité du liquide homogène. Cette fonction de distribution radiale
mesure le nombre moyen de molécules dans une coquille de rayon $r$
et d'épaisseur $\mathrm{d}r$, centrée sur une molécule donnée, normalisée
par la quantité de molécule dans cette même coquille. Cette définition
justifie donc que la fonction de distribution radiale tende vers 1
pour les grandes valeurs de $r$.

\fbox{\begin{minipage}[t]{1\columnwidth}%
Une façon intuitive de voir cette fonction est de constater qu'elle
représente la probabilité à l'équilibre thermodynamique, de trouver
une molécule à une distance $r$ d'une molécule donnée.%
\end{minipage}}

\begin{figure}[h]

\noindent \centering{}\includegraphics[width=0.6\textwidth]{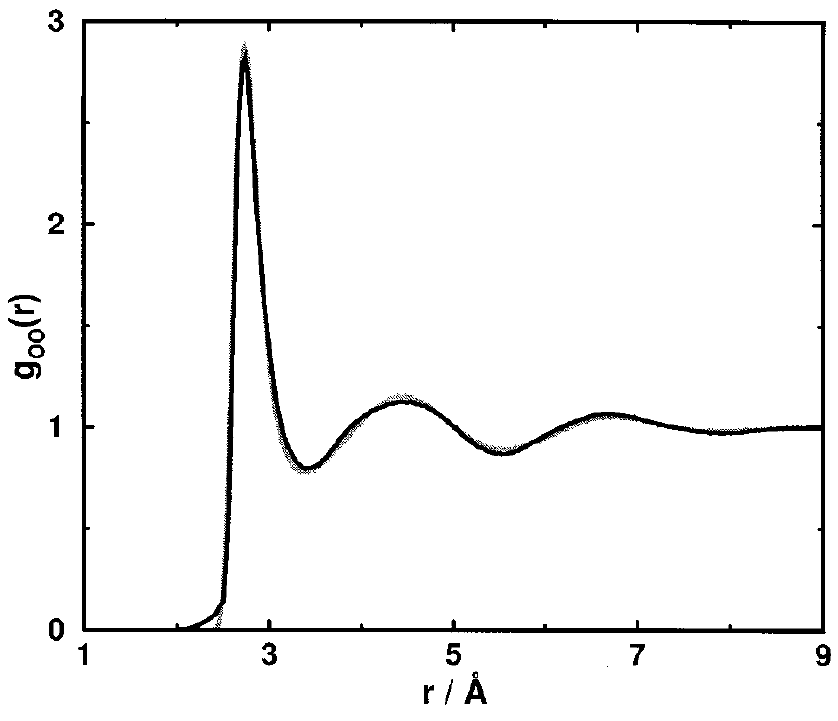}\protect\caption{Fonction de distribution radiale entre les oxygènes des molécules
d'eau. La courbe noire est obtenue expérimentalement par diffraction
de rayons X. La courbe grise est obtenue par dynamique moléculaire
avec le modèle d'eau TIP5P\cite{sorenson_what_2000}.\label{fig:Fonction-de-distributionwater}}
\end{figure}

Un exemple de fonction de distribution radiale entre les oxygènes
des molécules d'eau dans le solvant pur est donné en \ref{fig:Fonction-de-distributionwater}.
On remarque que cette fonction présente un premier pic relativement
fin autour de 3 $\textrm{\AA}$. Ce pic correspond à une probabilité
plus importante que dans un liquide parfaitement homogène de rencontrer
deux atomes d' oxygènes séparés de cette distance. Ce premier pic
correspond à une première sphère de solvatation bien définie. L'eau
voisine de la molécule d'eau que l'on regarde est très ordonnée. La
déplétion observée juste après ce premier pic s'explique par des considérations
stériques, il est en effet moins probable de trouver des atomes d'oxygène
près de la région de haute densité précédente. On observe ensuite
des oscillations avec des extrema de moins en moins importants jusqu'à
atteindre la valeur asymptotique 1. On peut en conclure que la fonction
de distribution radiale est une grandeur pertinente lorsqu'on s'intéresse
à la structure de solvatation. La courbe grise qui apparait sur la
\ref{fig:Fonction-de-distributionwater} est obtenue par dynamique
moléculaire en utilisant le modèle d'eau non-polarisable TIP5P, un
des modèles les plus récents, parmi les centaines de modèles qui ont
été proposés pour l'eau. Cette courbe reproduit parfaitement les résultats
expérimentaux. 

On constate que l'accès aux structures et aux grandeurs énergétiques
de solvatation par des méthodes numériques est donc possible. Comme
mis en évidence sur la \ref{fig:Fonction-de-distributionwater} il
faut pour cela développer et utiliser des méthodes ou théories qui
permettent, pour un modèle de molécule donné, d'obtenir ces propriétés
physiques à l'équilibre thermodynamique.

\fbox{\begin{minipage}[t]{1\columnwidth}%
L'étude des propriétés de solvatation, c'est-à-dire la bonne représentation
des interactions entre molécules de soluté et de solvant est un problème
physico-chimique important. Le but étant de reproduire ou prédire
des résultats expérimentaux et notamment les propriétés énergétiques
et structurales de la solvatation.%
\end{minipage}}

Dans les deux prochaines parties de ce chapitre sont présentées différentes
méthodes permettant de calculer les propriétés de solvatation. Les
simulations moléculaires, dites explicites, décrivent le solvant par
ces configurations microscopiques. On peut également utiliser des
méthodes de solvant implicite où le solvant est traité à un niveau
théorique où certains détails moléculaires sont omis.

\section{Modèles explicites}

Pour calculer les propriétés de solvatation on peut envisager d'utiliser
une représentation microscopique réaliste du système étudié, c'est-à-dire
représenter explicitement les molécules de solvant et de soluté. En
choisissant un modèle des interactions entre molécules, on peut placer
un ensemble de molécules à des positions diverses, et calculer l'énergie
de cette configuration microscopique. La théorie physique microscopique
pertinente pour déduire les propriétés macroscopiques à l'équilibre
à partir de la connaissance des configurations microscopiques est
la physique statistique. On se propose d'en rappeler quelques point
clés dans le prochain paragraphe

\subsection{Méthodes de solvant explicite}

Dans les techniques de simulations de solvant explicite, on représente
la solution par une collection de molécules. Celles-ci interagissent
au travers de champ de force, ou de manière équivalente, d'une énergie
potentielle ${\cal U}(\bm{r}_{1},\bm{r}_{2},...,\bm{r}_{N})$ dont
dérive ces forces, modélisant les interactions entre molécules. 

On peut décomposer cette énergie potentielle en la somme de deux composantes
: une partie intramoléculaire, décrivant les interactions entre atomes
d'une même molécule, et un partie intermoléculaire décrivant les interactions
entre atomes de molécules différentes. Il existe de nombreux types
de champs de forces pouvant décrire ces deux composantes. Dans le
cadre de cette thèse, on ne considérera que des modèles de solvants
rigides, c'est-à-dire que les longueurs et angles de liaison des molécules
sont fixes. Une telle approximation revient à considérer que l'énergie
intramoléculaire n'est jamais modifiée, ce qui permet de ne pas avoir
à la décrire. Pour ce qui concerne les interactions intermoléculaires,
on se limite à deux types d'énergie potentielle. 

L'interaction entre deux sites chargés séparés d'une distance $r$
est décrite par l'interaction coulombienne, d'énergie potentielle
$\mathrm{V_{C}}$,
\begin{equation}
\mathrm{V_{C}}(r)=\frac{\mathrm{z_{1}z_{2}e^{2}}}{4\pi\epsilon_{0}r},
\end{equation}
où les charges des atomes sont respectivement $\mathrm{z_{1}e}$ et
$\mathrm{z_{2}e}$, avec $e\approx1.602.10^{-19}\,\mathrm{C}$ la
charge élémentaire, et $\mathrm{z_{1}}$, $\mathrm{z_{2}}$ les charges
relatives, $\epsilon_{0}\simeq8.854.10^{-12}\,\mathrm{F.m^{-1}}$
la permittivité du vide. 

\begin{figure}[h]
\noindent \centering{}\includegraphics[width=0.6\textwidth]{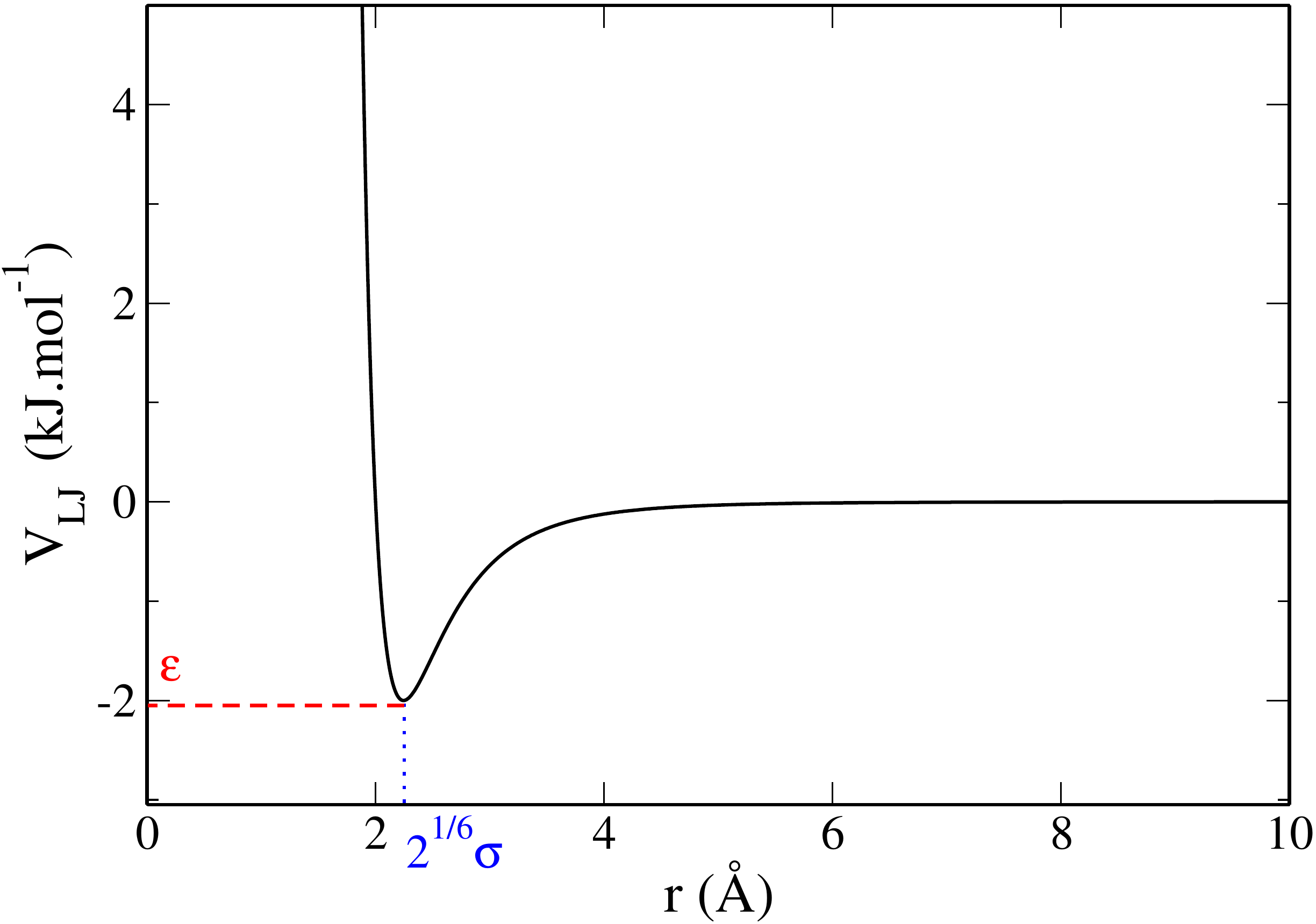}\protect\caption{Potentiel de Lennard-Jones avec $\epsilon=2.0$ kJ.mol$^{-1}$, et
$\sigma=2\,\textrm{\AA}$\label{fig:Potentiel-de-Lennard}}
\end{figure}

L'autre potentiel utilisé est le potentiel de Lennard-Jones, qui modélise
l'ensemble des interactions de Van der Waals, à savoir les forces
de Keesom, Debye et London, ainsi que la répulsion de Pauli à courte
distance. Ce potentiel d'interaction entre deux atomes séparés d'une
distance $r$ a pour expression,
\begin{equation}
\mathrm{V_{LJ}}(r)=4\epsilon\left[\left(\frac{\sigma}{r}\right)^{12}-\left(\frac{\sigma}{r}\right)^{6}\right]
\end{equation}
où $\epsilon$ est la profondeur du puit de potentiel, et $\sigma$
est relié à la position du minimum de ce puit de potentiel. Une représentation
de ce potentiel d'interaction est donnée en \ref{fig:Potentiel-de-Lennard}.

\subsection{Bases de thermodynamique statistique}

Supposons que l'on ait une collection de N particules dans un volume
V à une température T donnée, interagissant au travers d'une d'énergie
potentielle ${\cal U}$. Si l'on décrit la position de chaque particule
$i$ par leur vecteur position $\bm{r}_{i}$, l'énergie totale associée
à cette configuration est la somme d'une partie cinétique et d'une
partie potentielle due aux forces entre molécules. Elle peut être
décrite par la fonction de Hamilton, appelée par la suite hamiltonien,
\begin{equation}
{\cal H}(\bm{r}_{1},\bm{r}_{2},...,\bm{r}_{\mathrm{N}},\bm{p}_{1},\bm{p}_{2},...,\bm{p}_{\mathrm{N}})=\sum_{i=1}^{\mathrm{N}}\frac{\bm{p}_{i}^{2}}{2\mathrm{m}_{i}}+{\cal U}(\bm{r}_{1},\bm{r}_{2},...,\bm{r}_{\mathrm{N}}).
\end{equation}
Où $\bm{p}_{i}$ est la quantité de mouvement de la molécule $i$
de masse $\mathrm{m}_{i}$ et ${\cal U}(\bm{r}_{1},\bm{r}_{2},...,\bm{r}_{N})$
est l'énergie potentielle totale de la configuration étudiée.

Cet hamiltonien dépend de $2\mathrm{N}$ variables à valeurs dans
$\mathbb{R}^{3}$. Il existe un espace à $6\mathrm{N}$ dimensions,
appelé espace des phases, tel que chaque point (de cordonnées $\bm{r}_{1},\bm{r}_{2},...,\bm{r}_{N},\bm{p}_{1},\bm{p}_{2},...,\bm{p}_{N}$)
de cet espace représente une configuration de la collection de molécules.

Si le nombre de degrés de liberté, (le nombre de particules) est grand,
il est alors impossible de connaître exactement la configuration microscopique.
On utilise donc une description statistique des configurations microscopiques.
La distribution de probabilité des configurations de l'espace des
phases obéit à une statistique de Maxwell-Boltzmann\cite{frenkel_understanding_2002},
\begin{equation}
\hspace{-1cm}P(\bm{r}_{1},...,\bm{r}_{\mathrm{N}},\bm{p}_{1},...,\bm{p}_{\mathrm{N}})=\frac{\frac{1}{\mathrm{N!h^{3N}}}e^{-\beta{\cal H}(\bm{r}_{1},...,\bm{r}_{\mathrm{N}},\bm{p}_{1},...,\bm{p}_{\mathrm{N}})}}{\frac{1}{\mathrm{N!h^{3N}}}\iiint_{\mathbb{R}^{3}}\cdot\cdot\cdot\iiint_{\mathbb{R}^{3}}e^{-\beta{\cal H}(\bm{r}_{1},...,\bm{r}_{\mathrm{N}},\bm{p}_{1},...,\bm{p}_{\mathrm{N}})}\mathrm{d}\bm{r}_{1}...\mathrm{d}\bm{r}_{N}\mathrm{d}\bm{p}_{1}...\mathrm{d}\bm{p}_{\mathrm{N}}},
\end{equation}
où $\beta=1\mathrm{/k_{B}T}$, avec T la température absolue, $\mathrm{k_{B}}=1.38066\ 10^{-23}\ \mathrm{J.K^{-1}}$
la constante de Boltzmann et $\mathrm{h}=6.62606957.10^{\text{\textminus}34}\ \mathrm{J.s}$
la constante de Planck. Le terme au dénominateur est d'une importance
particulière, on le nomme fonction de partition et on le note usuellement
$\mathrm{Q}$. On peut remarquer que cette distribution de probabilité
est normalisée. La démonstration de ce résultat n'est pas présentée
ici mais on pourra se reporter à \cite{Diu_1995} pour l'y trouver.

L'énergie libre de Helmholtz, F, (aussi appelée énergie libre) est
reliée à la fonction de partition $\mathrm{Q}$,
\begin{equation}
\mathrm{F=-k_{B}T\ln Q.}\label{eq:F=00003DkbTlnQ}
\end{equation}
On remarque que F est définie à une constante près. Si l'on connait
toutes les configurations accessibles au système, c'est-à-dire si
on est capable de proposer une méthode permettant de visiter entièrement
l'espace des phases, on peut, en utilisant l'\ref{eq:F=00003DkbTlnQ},
calculer l'énergie libre du système. 

L'énergie libre est une grandeur particulièrement difficile à calculer
grâce à des simulations moléculaires car elle ne se réduit pas à un
calcul de valeur moyenne. Aussi on calcule toujours des différences
d'énergie libre entre deux systèmes. Ceci revient, pour une transformation
comme celle décrite sur la \ref{fig:ShemaDeltar}, à calculer l'énergie
libre de solvatation.

En ce qui concerne les propriétés structurales, telles que les fonctions
de distribution radiale, celles-ci sont facilement calculables en
utilisant des méthodes d'histogrammes.

La physique statistique est un outil puissant pour calculer les grandeurs
d'intérêt si on connait l'espace des phases. Pour échantillonner cet
espace on utilise principalement deux grandes familles de simulations
moléculaires\cite{frenkel_understanding_2002}.

La première est la dynamique moléculaire qui consiste, à partir d'une
configuration initiale de molécules à visiter l'espace des phases
en résolvant les équations de la mécanique classique. Pour simplifier,
à partir d'une configuration à un instant $t$, on calcule l'ensemble
des forces agissant sur les molécules grâce aux potentiels d'interaction
choisis. À partir du calcul de ces forces on utilise une version discrète
des lois de Newton pour déterminer la position des molécules au pas
suivant, c'est-à-dire à l'instant $t+\Delta t$, avec $\Delta t$
le pas de temps de la simulation. On refait alors un nouveau calcul
de forces et on itére le procédé.

L'espace des phases peut également être échantillonné par la méthode
de Monte-Carlo, qui est une méthode stochastique. Dans cette méthode,
l'espace des phases est visité par des déplacements aléatoires qui
sont acceptés ou refusés en fonction de la probabilité d'existence
des états atteints par ces déplacements. 

Une représentation microscopique réaliste du solvant nécessite d'utiliser
un grand nombre de molécules, jusqu'à plusieurs centaines de milliers,
pour étudier la mise en solution d'une molécule biologique. Une grande
partie de la puissance de calcul utilisée est principalement consacrée
à l'évaluation des interactions solvant-solvant, plutôt qu'à celles
mises en jeu entre les solutés et le solvant. De plus pour que l'espace
des phases soit correctement échantillonné il faut générer un grand
nombre de configurations ce qui rallonge d'autant plus le temps de
calcul. Pour ces raisons l'utilisation de simulations numériques pour
étudier les propriétés de solvatation est numériquement coûteux. Le
calcul de l'énergie libre de solvatation est encore plus problématique
puisqu'il nécessite l'utilisation de méthodes d'intégrations thermodynamiques
telles que l'Umbrella Sampling\cite{torrie_nonphysical_1977}, qui
impose la réalisation de plusieurs simulations moléculaires sur des
systèmes intermédiaires entre l'état de départ (solvant bulk) et l'état
d'arrivée (soluté en solution).

Il peut être intéressant de développer des méthodes numériquement
plus efficaces, notamment pour étudier des systèmes de taille importante.
Les méthodes de solvants implicites sont une alternative intéressante
puisque le solvant y est décrit avec moins de détails, ce qui diminue
le coût numérique de l'évaluation des interactions entre molécules
de solvant.

\section{Méthodes de solvant implicite}

Conceptuellement, on peut décomposer l'enthalpie libre de solvatation
en une somme de trois termes, 
\begin{equation}
\mathrm{\Delta G_{\mathrm{solv}}=\Delta G_{\mathrm{elec}}+\Delta G_{\mathrm{VdW}}+\Delta G_{\mathrm{cav}},}
\end{equation}
où $\Delta\mathrm{G_{\mathrm{elec}}}$ est la contribution due à l'électrostatique,
qui provient du travail nécessaire pour créer la distribution de charge
du soluté dans la solution et pour polariser cette distribution de
charge sous l'influence du solvant, $\Delta G_{\mathrm{VdW}}$ est
la contribution due aux forces de Van-der-Waals (dispersion et répulsion),
$\Delta G_{\mathrm{cav}}$ correspond au coût nécessaire à la création
de la cavité contenant le soluté. Ce dernier terme contient le coût
entropique de réorganisation des molécules de solvant et la résistance
à la pression du solvant pour former la cavité. Ces deux derniers
termes sont généralement réunis en une contribution non polaire,
\begin{equation}
\mathrm{\Delta G_{\mathrm{solv,np}}=\Delta G_{\mathrm{cav}}+\Delta G_{\mathrm{VdW}}.}
\end{equation}
Il existe plusieurs théories permettant d'estimer cette contribution
à l'énergie libre \textit{(Scaled Particle Theory\cite{pierotti_scaled_1976,reiss_scaled_1965,reiss_statistical_1959,stillinger_structure_1973}}
et \textit{Solvent-Exposed Surface Area}\cite{tanford_interfacial_1979})
en la décomposant en plusieurs termes dont le terme principal est
essentiellement un terme de tension de surface,
\begin{equation}
\Delta G_{\mathrm{solv,np}}\approx\gamma_{v}A,
\end{equation}

où A est la surface accessible au solvant et $\gamma_{v}$ la tension
de surface liquide-gaz du solvant. Cette expression est quasi-exacte
pour un soluté de sphères dures de grandes tailles. En pratique, pour
des solutés plus complexes, les paramètres sont ajustés pour reproduire
les énergies de solvatation expérimentales. La partie électrostatique
peut être évaluée à partir de nombreux modèles de solvants implicites,
qui diffèrent essentiellement par l'échelle de description du solvant.

\subsection{Milieu diélectrique continu}

\subsubsection{Poisson-Boltzmann}

Cette méthode suppose que le soluté est fixe et qu'il est défini comme
un milieu de faible permittivité diélectrique, $\epsilon\mathrm{^{u}}$
(variant typiquement de 1 à 8). Il forme une cavité dans un continuum
diélectrique polarisable caractérisé par une permittivité diélectrique
élevée $\epsilon\mathrm{^{v}}$ (80 pour l'eau). Les charges du soluté
sont traitées explicitement. Le problème se ramène à un problème d'électrostatique
courant qui peut être formulé grâce à une équation de Poisson,
\begin{equation}
\nabla\cdot\bm{D}(\bm{r})=\frac{\sigma(\bm{r})}{\epsilon_{0}},\label{eq:poisson-with_D}
\end{equation}
 où $\sigma$ est la densité de charge, $\epsilon_{0}$ la permittivité
du vide et $\bm{D}$ le déplacement diélectrique défini comme,
\begin{equation}
\bm{D}(\bm{r})=-\epsilon(\bm{r})\nabla\Phi(\bm{r}),
\end{equation}
 où $\epsilon(\bm{r})$ est la permittivité relative et $\Phi$ le
potentiel électrostatique. Si on connait $\Phi$, on peut calculer
l'énergie libre de solvation du soluté,
\begin{equation}
\Delta G_{\mathrm{elec}}=\frac{1}{2}\iiint_{\mathbb{R}^{3}}\sigma(\bm{r})\left(\Phi(\bm{r})-\Phi_{0}(\bm{r})\right)\mathrm{d}\bm{r},
\end{equation}
 avec $\Phi_{0}(\bm{r})$ le potentiel électrostatique créé par le
soluté dans le vide.

Si le système à modéliser est un soluté entouré d'un solvant contenant
des ions en solution ($\mathrm{N_{I}}$ types d'ions), la densité
de charge totale peut s'écrire,
\begin{equation}
\sigma(\bm{r})=\sigma^{\text{IA}}(\bm{r})+\sigma\mathrm{^{u}}(\bm{r}),
\end{equation}
où $\sigma\mathrm{^{u}}(\bm{r})$ désigne la densité de charge du
soluté et $\sigma^{\text{IA}}(\bm{r})$ celle de l'atmosphère ionique.
Il est possible de faire l'hypothèse d'une distribution de Boltzmann
pour les ions en solution à l'équilibre (d'ou le nom de la méthode)
sous l'effet du champ électrique,
\begin{equation}
\sigma^{\text{IA}}(\bm{r})=\sum_{i=1}^{\mathrm{N_{I}}}\mathrm{z}_{i}\mathrm{e}\mathrm{n}_{i}^{0}\exp\left(\frac{\mathrm{z}_{i}\mathrm{e}\Phi(\bm{r})}{\mathrm{k_{B}}\mathrm{T}}\right),
\end{equation}
où $\mathrm{n}_{i}^{0}$ est la densité ionique de l'ion $i$ en l'absence
d'interaction et $\mathrm{z}_{i}$ la valence de l'ion $i$. L'\ref{eq:poisson-with_D}
se réécrit alors,
\begin{equation}
\nabla\cdot\left[\epsilon(\bm{r})\nabla\Phi(\bm{r})\right]=\sum_{i=1}^{\mathrm{N_{I}}}\frac{\mathrm{z}_{i}\mathrm{e}}{\epsilon_{0}}n_{i}^{0}\exp\left(\frac{z_{i}\mathrm{e}\Phi(\bm{r})}{\mathrm{k_{B}}\mathrm{T}}\right)+\frac{\sigma^{\mathrm{u}}(\bm{r})}{\epsilon_{0}},
\end{equation}
qui est l'équation de Poisson-Boltzmann. 

Cette équation est exacte et constitue la référence pour les modèles
de solvants implicites. Malheureusement, elle n'est pas résolvable
analytiquement avec des systèmes d'intérêt pour lesquels la densité
de charge a une forme non-triviale. Il existe néanmoins de nombreux
développements permettant de résoudre cette équation de manière numérique\cite{roux_implicit_1999},
basés par exemple sur la méthode des éléments finis\cite{baker_electrostatics_2001}
ou des différences finies\cite{warwicker_calculation_1982,honig_classical_1995,honig_macroscopic_1993,bruccoleri_finite_1997,ray_luo_accelerated_2002}.
Il existe de plus des codes mis à la disposition de la communauté
par exemple le code DelPhi\cite{gilson_calculating_1988,nicholls_rapid_1991}
ou le code APBS\cite{baker_electrostatics_2001}

\subsubsection{Méthode de Born généralisée}

La méthode de Born généralisée est une approximation de la méthode
de Poisson-Boltzmann qui vise à donner des résultats comparables avec
un coût numérique inférieur. On souhaite évaluer l'énergie libre électrostatique
d'un ensemble de $\mathbb{\mathrm{N_{tot}}}$ particules, de rayon
de Born $\alpha_{i}$ et de charge $q_{i}$ dans un milieu de constante
diélectrique $\epsilon$. Celle-ci est définie comme la somme de l'énergie
de Coulomb dans un milieu diélectrique et d'une correction de Born,
\begin{equation}
\mathrm{\Delta G_{elec}}=-\frac{1}{8\pi}\left(1-\frac{1}{\epsilon}\right)\sum_{i=1}^{\mathrm{N_{tot}}}\frac{q_{i}^{2}}{\alpha_{i}}
\end{equation}

Cette expression est vraie si les rayons de Born des particules ne
se recouvrent pas. Malheureusement, pour un soluté réel on ne connait
pas les rayons de Born des différentes particules (ou sites chargés).

Une généralisation empirique pour des particules dont les rayons de
Born sont autorisés à se recouvrir a été proposée \cite{still_semianalytical_1990},
\begin{equation}
\mathrm{\Delta G_{elec}}=\frac{1}{8\pi\epsilon}\sum_{i=1}^{\mathrm{N_{tot}}}\sum_{j>i}^{\mathrm{N_{tot}}}\frac{q_{i}q_{j}}{\sqrt{r_{ij}^{2}+\alpha_{i}\alpha_{j}\exp(\frac{-r_{ij}^{2}}{4\alpha_{i}\alpha_{j}})}}\left(1-\frac{1}{\epsilon}\right)
\end{equation}

La détermination des rayons de Born $\alpha_{i}$ des atomes $i$
se fait en calculant l'énergie libre électrostatique de l'atome i,
avec les charges de toutes les autres particules égales à 0, $\mathrm{\Delta G_{elec,0}^{i}}$.
Notons cependant que le diélectrique est toujours égal à celui du
soluté pour les particules de charges nulles. Par exemple, ceci peut
être fait en utilisant l'équation de Poisson-Boltzmann. Le rayon de
Born de cette particule est $\alpha_{i}$ telle que l'énergie libre
électrostatique calculée avec la formule de Born simple d'une particule
sphérique calculée dans un continuum, $\mathrm{\Delta G_{Born}^{\mathit{i}}}$
\begin{equation}
\mathrm{\Delta G_{Born}^{\mathit{i}}}=-\frac{q^{2}}{8\pi\epsilon_{0}\alpha_{i}}\left(1-\frac{1}{\epsilon}\right)
\end{equation}

soit la même que celle obtenue, $\mathrm{\Delta G_{elec,0}^{\mathit{i}}}$.
On trouve donc,
\begin{equation}
\alpha_{i}=-\frac{1}{8\pi\epsilon_{0}}\left(1-\frac{1}{\epsilon}\right)\frac{1}{\mathrm{\Delta G_{elec,0}^{\mathit{i}}}}.
\end{equation}

Le coût numérique de cette méthode réside dans la détermination de
ces rayons de Born $\alpha_{i}$, qui requiert pour chaque particule
le calcul d'une énergie libre électrostatique pour laquelle on a éteint
les charges de toutes les autres particules. De plus, si on désire
utiliser cette méthode dans une simulation dynamique il faut recommencer
le processus à chaque pas. D'autres déterminations du rayon de Born,
plus efficaces numériquement, ont donc été proposées. La principale
est l'approximation CFA (Coulomb Field Approximation) qui consiste
à supposer que le déplacement diélectrique est égal au champ électrique
coulombien dans le vide, ceci revient à négliger la réponse due au
solvant puisque l'on néglige l'effet du continuum diélectrique ce
qui tend à sous-estimer les rayons de Born et les énergies libres
de solvatation. Les rayons de Born peuvent désormais être déterminés
numériquement ce qui rend la méthode extrêmement rapide et donc utilisable
dans des simulations de dynamique moléculaire. Cependant cette théorie
décrit le solvant par un continuum, et ne tient donc pas compte de
la nature moléculaire de celui-ci. D'autres théories conservent ce
niveau de description, on en décrit rapidement trois dans les paragraphes
suivants.

\subsection{Équations intégrales}

Les théories des équations intégrales sont basées sur l'équation de
Ornstein-Zernike, qui pour un fluide uniforme et isotrope s'écrit,
\begin{equation}
h(r)=c(r)+\rho\iiint_{\mathbb{R}^{3}}c(\left\Vert \bm{r}-\bm{r}^{\prime}\right\Vert )h(r^{\prime})\mathrm{d}\bm{r}^{\prime}\label{eq:OZ}
\end{equation}

avec $h$ la fonction de corrélation totale, $c$ la fonction de corrélation
directe et $\rho$ le densité du fluide. La fonction $h$ décrit la
variation de la densité locale de particule à une distance $r$ d'une
particule située à l'origine. Elle est reliée à la fonction de distribution
radiale par,
\begin{equation}
h(r)=g(r)-1.
\end{equation}

L'\ref{eq:OZ} a une interprétation physique forte, la fonction de
corrélation totale $h$ entre deux particules situées à une distance
$r$ est due à une interaction directe entre ces deux particules $c$,
ainsi qu'à la somme de toutes les contributions indirectes propagées
par l'ensemble de toutes les autres particules (le terme intégrale).

La fonction $h$ caractérise la structure du liquide à l'équilibre
thermodynamique et permet d'en calculer les propriétés d'équilibre.
Pour résoudre l'\ref{eq:OZ} et calculer les fonctions de corrélation
il est néanmoins nécessaire d'utiliser une autre équation qui relie
$h$ et $c$ appelée relation de fermeture.

On connait un certain nombre de relations de fermeture relativement
adaptées à l'étude de fluide atomique. On peut citer, par exemple,
la relation de fermeture HNC, pour hypernetted-chain,
\begin{equation}
g(r)=\exp\left[-\beta v(r)+h(r)-c(r)\right].\label{eq:HNC}
\end{equation}
Cette relation relie les fonctions $g,h$ et $c$ à une perturbation
extérieure modélisée par un potentiel de paire $v$. L'\ref{eq:HNC}
est une approximation, elle néglige en particulier des corrélations
à courte distance. Avec cette relation de fermeture il est désormais
possible de résoudre l'\ref{eq:OZ}.

Lorsque le fluide est composé de molécules, les interactions de paires
entre elles ne dépendent plus uniquement de la position relative des
centres de masse des deux molécules, mais aussi de leurs orientations
respectives.  Pour décrire un tel fluide il est donc nécessaire de
réécrire une version moléculaire de l'équation de Ornstein-Zernike
(MOZ).
\begin{equation}
h(r_{12},\boldsymbol{\Omega}_{1},\boldsymbol{\Omega}_{2})=c(r_{12},\boldsymbol{\Omega}_{1},\boldsymbol{\Omega}_{2})+\frac{1}{8\pi^{2}}\int_{\theta=0}^{\pi}\int_{\phi=0}^{2\pi}\int_{\psi=0}^{2\pi}\iiint_{\mathbb{R}^{3}}c(r_{13},\boldsymbol{\Omega}_{1},\boldsymbol{\Omega}_{3})\rho h(r_{32},\boldsymbol{\Omega}_{3},\boldsymbol{\Omega}_{2})\mathrm{d}\bm{r}_{3}\mathrm{d}\boldsymbol{\Omega}_{3}\label{eq:MOZ}
\end{equation}

On peut alors utiliser une voie similaire à l'étude des fluides simples
pour essayer de résoudre MOZ. Il faut à nouveau utiliser une relation
de fermeture, par exemple l'\ref{eq:HNC} précédente. Le problème
est désormais bien plus complexe puisqu'on étudie un fluide moléculaire.
En effet, les fonctions $g,h,v\text{ et }c$ intervenant dans HNC
sont également des fonctions moléculaires, elles dépendent donc de
l'orientation relative et de la distance séparant les deux molécules
considérées. La résolution de MOZ avec ces fonctions est numériquement
très complexe. Des développements ont néanmoins été réalisés pour
contourner ce problème. L'idée générale est de réaliser une projection
sur une base d'invariants rotationnels des fonctions mises en jeu\cite{blum_invariant_1972,blum_invariant2_1972}
$g,h,c,v$. Par exemple,
\begin{equation}
c\left(\bm{r}_{12},\bm{\Omega}_{1},\bm{\Omega}_{2}\right)=\sum_{m,n,l,\mu,\nu}c_{\mu\nu}^{mnl}\left(r\right)\Phi_{\mu\nu}^{mnl}\left(\bm{\Omega}_{1},\bm{\Omega}_{2},\tilde{\bm{r}}\right),
\end{equation}

avec les invariants rotationnels, $\Phi_{\mu\nu}^{mnl}$:\foreignlanguage{english}{
\begin{eqnarray}
\Phi_{\mu\nu}^{mnl}\left(\bm{\Omega}_{1},\bm{\Omega}_{2},\tilde{r}\right) & = & \sqrt{\left(2m+1\right)\left(2n+1\right)}\sum_{\mu^{\prime},\nu^{\prime},\lambda^{\prime}}\begin{pmatrix}m & n & l\\
\mu^{\prime} & \nu^{\prime} & \lambda^{\prime}
\end{pmatrix}\nonumber \\
 &  & \times R_{\mu^{\prime}\mu}^{m}\left(\bm{\Omega}_{1}\right)R_{\nu^{\prime}\nu}^{n}\left(\bm{\Omega}_{2}\right)R_{\lambda^{\prime}0}^{l}\left(\tilde{\bm{r}}\right).
\end{eqnarray}
}Les coefficients $\begin{pmatrix}m & n & l\\
\mu^{\prime} & \nu^{\prime} & \lambda^{\prime}
\end{pmatrix}$ sont les symboles $3-j$ de Wigner\cite{messiah_quantum_1970}, et
les $R_{\mu^{\prime}\mu}^{m}\left(\bm{\Omega}\right)$ sont les harmoniques
sphériques généralisées de Wigner, $\tilde{\bm{r}}=(\bm{r}_{1}-\bm{r}_{2})/\left\Vert \bm{r}_{1}-\bm{r}_{2}\right\Vert $
désigne le vecteur unitaire séparant les deux particules considérées.
Des projections similaires sont réalisées pour les autres fonctions
$h,v,g$.

Les invariants rotationnels sont indépendants du référentiel (d'où
leur intérêt) et forment une base orthogonale. Ils dépendent de l'orientation
relative des deux molécules, donc de 5 angles d'Euler, caractérisés
par les cinq indices $m$, $n$, $l$, $\mu$, $\nu$. En principe
le développement sur la base des invariants rotationnels est infini
mais pour des molécules linéaires (e.g., le fluide Stockmayer) trois
angles d'Euler sont suffisants : $m+n+l$ est pair et $\mu=\nu=0$.
On peut donc simplifier la notation dans ce cas précis de molécules
linéaires. De plus, il est supposé et vérifié que l'on peut limiter
ce développement à un nombre maximal de projections, caractérisées
par un doublet $(m,n)$ pour décrire quantitativement la dépendance
angulaire des corrélations. En utilisant ces projections on peut alors
résoudre MOZ avec la relation de fermeture HNC. Des développements
ont été réalisés dans ce formalisme qui permettent une description
précise des corrélations des liquides moléculaires\cite{puibasset_bridge_2012}.

\subsection{RISM (Reference Interaction Site Model)}

La théorie RISM postule que dans une molécule, constituée de sites
atomiques, le potentiel d'interaction entre les molécules peut être
décrit par une somme de potentiels de paire, qui ne dépendent que
de la distance entre sites atomiques. Ceci est une approximation extrêmement
courante en simulation moléculaire. On peut réécrire l'\ref{eq:MOZ}
en faisant intervenir des fonctions de corrélation de paire entre
sites des molécules $h_{\alpha\beta}(r)$, au lieu des fonctions moléculaires.
On les calcule en fixant la distance entre sites d'interaction et
en faisant une moyenne angulaire. Moyennant une série d'approximations
pour pouvoir faire cette moyenne angulaire dans l'\ref{eq:MOZ}, on
obtient l'équation matricielle fondamentale de RISM,
\begin{equation}
\bm{h}(r)=\bm{\omega}\star\bm{c}\star\bm{\omega}(r)+\rho\bm{\omega}\star\bm{c}\star\bm{h}(r),\label{eq:MOZ-RISM}
\end{equation}
$\star$ désigne un produit matriciel et un produit de convolution
et $\bm{h},\bm{c},\bm{\omega}$ sont des matrices dont les éléments
sont des fonctions décrivant des paires de sites moléculaires, qu'on
labélise $\alpha\gamma$. Ainsi, $h_{\alpha\gamma}$ est la fonction
de corrélation de paire entre les sites $\alpha$ et $\gamma$. $\omega_{\alpha\gamma}$
est définie par
\begin{equation}
\omega_{\alpha\gamma}=\rho\delta_{\alpha\gamma}(r)+\rho s_{\alpha\gamma}(r),
\end{equation}
où $s_{\alpha\gamma}$ est la fonction de distribution de paire site-site
intramoléculaire. Elle décrit la probabilité de trouver un site $\gamma$
à une position $\bm{r}$ sachant qu'un site $\alpha$ se trouve à
l'origine. Cette équation a été introduite pour des fluides composés
de sphères dures mais son utilisation a été généralisée à des fluides
composés de sphères Lennard-Jones et portant des charges. Le modèle
de molécule se choisit au travers de la relation de fermeture utilisée
conjointement à MOZ pour la résolution du problème, c'est-à-dire la
détermination des fonctions de corrélation de paire. Le formalisme
tel que décrit ici s'applique à un fluide pur.

Si on veut étudier un système soluté-solvant, on peut généraliser
l'\ref{eq:MOZ-RISM} pour un mélange de fluide,
\begin{equation}
\bm{\rho}\bm{h}\bm{\rho}(r)=\bm{\omega}\star\bm{c}\star\bm{\omega}(r)+\bm{\omega}\star\bm{c}\star\bm{\rho}\bm{h}\bm{\rho}(r),
\end{equation}
où les matrices $\bm{h},\bm{c},\bm{\omega}$ sont désormais formées
d'éléments $h_{\alpha_{M}\gamma_{M^{\prime}}}$, $c_{\alpha_{M}\gamma_{M^{\prime}}}$
, $\omega_{\alpha_{M}\gamma_{M^{\prime}}}$. L'indice $\alpha_{M}$
désigne le $\alpha^{i\grave{e}me}$ site de la molécule M et $\gamma_{M^{\prime}}$
le $\gamma^{i\grave{e}me}$ site de la molécule M$^{\prime}$. La
matrice $\bm{\rho}$ est la matrice diagonale des densités en nombre
des espèces, $\rho_{\alpha_{M}\gamma_{M^{\prime}}}=\delta_{\alpha\gamma}\delta_{MM^{\prime}}\rho_{M}$.
On peut réorganiser cette équation matricielle en séparant les termes
liés au soluté, désigné par l'indice $u$, et ceux liés au solvant,
désignés par l'indice $v$. Dans la limite où la dilution du soluté
est infinie, c'est à dire quand $\rho_{u}\rightarrow0$, on obtient
le jeu d'équations suivants,
\begin{gather}
\bm{h}_{vv}=\bm{w}_{v}\star\bm{c}_{vv}\star\bm{w}_{v}+\bm{w}_{v}\star\bm{c}_{vv}\star\bm{\rho}_{v}\bm{h}_{vv}\label{eq:RISMsolvSolv}\\
\bm{h}_{uv}=\bm{w}_{u}\star\bm{c}_{uv}\star\bm{w}_{v}+\bm{w}_{u}\star\bm{c}_{uv}\star\bm{\rho}_{v}\bm{h}_{uv}\\
\bm{h}_{uu}=\bm{w}_{u}\star\bm{c}_{uu}\star\bm{w}_{u}+\bm{w}_{u}\star\bm{c}_{uu}\star\bm{\rho}_{v}\bm{h}_{uv},\label{eq:RISM_solv_solv}
\end{gather}
où $\bm{w}_{v}=\bm{\omega}_{v}/\bm{\rho}_{v}$ et $\bm{w}_{u}=\bm{\omega}_{u}/\bm{\rho}_{u}$.
En utilisant cette théorie avec une relation de fermeture adaptée
il devient possible de calculer les fonctions de corrélation du système
et d'en déduire des grandeurs physiques d'intérêt, dont l'énergie
libre de solvatation, via une intégration thermodynamique nécessitant
la connaissance des fonctions de corrélation de paires à différents
états thermodynamiques entre le solvant pur et le soluté en solution.

Une des limitations de ce modèle est qu'il ne permet pas une description
de solutés de géométries complexes car les fonctions de corrélation
site-site sont moyennées sur les orientations des molécules. 

Pour ces raisons, une généralisation tridimensionnelle de RISM (3D-RISM)
autour d'un soluté non sphérique a été proposée.

\subsection{3D-RISM}

Cette théorie utilise des fonctions de corrélation tridimensionnelle
pour les sites du solvant proches du soluté. On écrit d'abord l'équation
moléculaire de Ornstein-Zernike dépendante des angles. 
\begin{flalign}
h^{uv}(r_{12},\boldsymbol{\Omega}_{1},\boldsymbol{\Omega}_{2})= & c^{uv}(r_{12},\boldsymbol{\Omega}_{1},\boldsymbol{\Omega}_{2})\nonumber \\
 & +\frac{1}{8\pi^{2}}\int_{\theta=0}^{\pi}\int_{\phi=0}^{2\pi}\int_{\psi=0}^{2\pi}\iiint_{\mathbb{R}^{3}}c^{uv}(r_{13},\boldsymbol{\Omega}_{1},\boldsymbol{\Omega}_{3})\rho^{v}h^{vv}(r_{32},\boldsymbol{\Omega}_{3},\boldsymbol{\Omega}_{2})\mathrm{d}\bm{r}_{3}\mathrm{d}\boldsymbol{\Omega}_{3}
\end{flalign}
 Contrairement à RISM où cette équation était complètement moyennée
sur l'ensemble des orientations du soluté et du solvant, la méthode
3D-RISM conserve la description angulaire du soluté. Les fonctions
de corrélation de paire des sites de solvants autour des sites $\gamma$
du soluté ont donc une dépendance angulaire en plus de leur dépendance
radiale,
\begin{equation}
h_{\gamma}^{uv}(\bm{r}_{1\gamma})\equiv h_{\gamma}^{uv}(r_{1\gamma},\boldsymbol{\Omega})=\frac{1}{\Omega}\iiint_{\pi^{2}}h^{uv}(\left\Vert \bm{r}_{1\gamma}-\bm{r}_{2\gamma}\right\Vert ,\boldsymbol{\Omega}_{1},\boldsymbol{\Omega}_{2})\mathrm{d}\boldsymbol{\Omega}_{1},
\end{equation}
où $\bm{r}_{1\gamma}$ est le vecteur intermoléculaire entre une molécule
de soluté dont le centre de masse est situé en $\bm{r}_{1}$ et un
site de solvant situé en $\bm{r}_{\gamma}$, et $\bm{r}_{2\gamma}$
le vecteur intramoléculaire entre le centre de masse du solvant situé
en 2 et le site $\gamma$ de la molécule de solvant ayant l'orientation
$\boldsymbol{\Omega}_{2}$. Pour pouvoir effectuer la moyenne angulaire
de l'équation de OZ avec dépendance angulaire, l'hypothèse de base
de 3D-RISM est que la fonction de corrélation directe 6D $c^{uv}(r_{12},\boldsymbol{\Omega}_{1},\boldsymbol{\Omega}_{2})$
peut se décomposer en la somme des contributions partielles des sites
de la molécule de solvant 2,
\begin{equation}
c^{uv}(r_{12},\boldsymbol{\Omega}_{1},\boldsymbol{\Omega}_{2})=\sum_{\alpha}c_{\alpha}^{uv}(\bm{r}_{1\alpha}).
\end{equation}

On transforme l'équation moléculaire de OZ dans l'espace de Fourier
et on la moyenne sur $\boldsymbol{\Omega}_{2}$, on obtient alors
l'équation fondamentale de 3D-RISM pour le soluté écrite dans l'espace
réciproque,
\begin{equation}
\hat{h}_{\gamma}^{uv}(\bm{k})=\hat{c}_{\alpha}^{uv}(\bm{k})\hat{\omega}_{\alpha\gamma}^{vv}(\bm{k})+\hat{c}_{\alpha}^{uv}(\bm{k})\rho^{v}\hat{h}_{\alpha\gamma}^{vv}(\bm{k})
\end{equation}

Cette équation peut être calculée numériquement pour obtenir les fonctions
de corrélation totale $\hat{h}_{\alpha\gamma}^{vv}(\bm{k})$ et directes
$\hat{c}_{\alpha}^{uv}(\bm{k})$ par discrétisation sur une grille
3D uniforme et utilisation de transformées de Fourier rapides (FFT)
pour calculer les produits de convolution. L'\ref{eq:RISMsolvSolv}
décrivant les corrélations entre molécules de solvant est inchangée.
Il faut bien entendu toujours se munir de relations de fermetures
pour pouvoir résoudre cette équation.

La théorie 3D-RISM a d'abord été développée par Roux et collaborateurs
avec la relation de fermeture HNC\cite{3D_RISM_beglov_integral_1997}
pour étudier la solvatation de solutés polaires, eau et N-méthyacétamide,
dans le solvant eau. Cette théorie permet de reproduire les propriétés
structurales de solvatation.

Malgré ce succès, la relation de fermeture HNC converge mal, notamment
pour l'étude d'un liquide associé près d'une interface cristalline.
Pour contourner ce problème, une version alternative de 3D-RISM a
été proposée par Kovalenko et Hirata\cite{kovalenko_self-consistent_1999}
basée sur une version linéarisée de HNC, la fermeture KH. Cette fermeture
permet l'étude d'une interface métal-eau, le métal étant décrit par
DFT électronique.

La théorie 3D-RISM est une manière rapide d'accéder aux propriétés
de solvatation tout en conservant un niveau de description moléculaire,
c'est pourquoi elle est de plus en plus utilisée. Parmi les utilisations
notables de cette théorie on peut signaler: le calcul de potentiels
de force moyenne\cite{kovalenko_potentials_2000,kovalenko_potential_1999}
et l'étude de molécules biologiques\cite{imai_water_2005,imai_locating_2007,imai_solvation_2004}.
Le couplage entre DFT électronique et 3D-RISM mentionné précédemment
a aussi été utilisé pour étudier l'effet du solvant sur des équilibres
de tautomérisation en solution\cite{casanova_evaluation_2007}. Récemment
des corrections de volume molaire partiel aux calculs 3D-RISM ont
été proposé pour la prédiction d'énergies libres de solvatation\cite{palmer_accurate_2010,palmer_towards_2010,sergiievskyi_3drism_2012}.

\lhead[\chaptername~\thechapter]{\rightmark}

\rhead[\leftmark]{}

\lfoot[\thepage]{}

\cfoot{}

\rfoot[]{\thepage}

\part{Théorie}

\chapter{La théorie de la fonctionnelle de la densité classique\label{chap:cDFT}}

Dans ce chapitre est fait un bref rappel de la théorie de la fonctionnelle
de la densité électronique pour souligner les points communs entre
cette théorie très connue des chimistes théoriciens avec son analogue
classique. On présente ensuite les bases théoriques et certaines des
équations fondamentales ainsi que diverses approximations de la théorie
de la fonctionnelle de la densité classique. À la fin du chapitre
est proposé un tableau récapitulatif des définitions et équations
clés pour la lecture de la suite de ce manuscrit.

\section{La théorie de la fonctionnelle de la densité électronique}

En chimie quantique, un des objectifs est de connaître la structure
électronique et l'énergie d'un édifice moléculaire, molécule, ion,
etc. Pour cela, on peut chercher à résoudre directement l'équation
de Schrödinger indépendante du temps. Résoudre cette équation revient
à trouver la fonction d'onde électronique $\Psi$ et l'énergie du
système $E$, solutions de cette équation aux valeurs propres, en
utilisant par exemple la méthode de Hartree-Fock\cite{Szabo_Ostlund}.
On peut remarquer que la fonction d'onde électronique est généralement
un vecteur de haute dimension puisque c'est une fonction qui dépend
des coordonnées de chacun des électrons, soit un vecteur à $3\text{N}$
dimensions (si on oublie le spin), où $\text{N}$ est le nombre d'électrons.
Du fait de ce nombre important de degrés de liberté, les méthodes
de détermination de la fonction d'onde sont numériquement coûteuses.

Trouver la fonction d'onde, qui apparait explicitement comme une variable,
semble être le moyen le plus naturel pour résoudre cette équation.
Celle-ci n'est pas directement une observable physique mais un objet
mathématique qui contient toutes les informations du système. La densité
de probabilité électronique, qui est égale à la norme au carré de
la fonction d'onde est, en revanche, une observable physique. Du fait
de l'indiscernabilité des électrons, on peut laisser de côté cette
probabilité qui dépend de $3\text{N}$ variables. On introduit alors
une densité de probabilité de présence des électrons à une position
$\bm{r}$. Celle ci peut être définie comme
\begin{equation}
n(\bm{r})=\sum_{i=1}^{\text{N}}\psi_{i}^{*}(\bm{r})\psi_{i}(\bm{r}),\label{eq:densit=0000E9_=0000E9lec}
\end{equation}
où $\psi_{i}^{*}$ désigne l'adjoint de la fonction d'onde $\psi_{i}.$
Bien que cette fonction que l'on nomme densité électronique, soit
beaucoup plus simple que la fonction d'onde puisqu'elle ne dépend
plus que de 3 variables spatiales, elle contient encore la plupart
de l'information physique.

La théorie de la fonctionnelle de la densité électronique (eDFT, une
définition du terme fonctionnelle est donnée en \ref{chap:D=0000E9riv=0000E9-fonctionelle}),
introduite par Kohn et Hohenberg\cite{hohenberg_inhomogeneous_1964},
et développée par Kohn et Sham\cite{kohn_self-consistent_1965} au
milieu des années 1960, est basée sur deux principaux théorèmes. 

Le premier stipule que \textit{l''énergie d'un système polyélectronique
est une fonctionnelle unique de la densité électronique,} introduite
dans l'\ref{eq:densit=0000E9_=0000E9lec}. Ce théorème montre qu'il
existe une bijection entre la fonction d'onde décrivant l'état fondamental
et la densité électronique de ce même état. Toutes les propriétés
de l'état fondamental du système peuvent être connues via la connaissance
de la densité électronique au lieu de celle de la fonction d'onde.
Ceci est d'une grande importance puisque l'on ramène la résolution
d'un problème à $3\text{N}$ variables à celle d'un problème à 3 variables,
ce qui explique l'efficacité numérique de la eDFT et sa large utilisation
en chimie quantique et en physique du solide. Soulignons tout de même
qu'à cause d'approximations techniques qu'il n'est pas pertinent de
développer ici, les algorithmes de calculs de eDFT ont en réalité
une complexité qui augmente comme ${\cal O}(\text{N})$\cite{blum_ab_2009}.

Actuellement, on ne connait pas d'expression exacte de la fonctionnelle.
Le second théorème de Kohn et Hohenberg donne cependant une information
importante sur celle-ci, \textit{la densité d'électron qui minimise
la fonctionnelle est la densité électronique de l'état fondamental.
}Ce théorème stipule que si on connait une expression exacte de la
fonctionnelle, on pourrait, en minimisant celle-ci, trouver la densité
électronique de l'état fondamental. De nombreux développements ont
été réalisés durant les dernières décennies pour proposer une expression
approchée de la fonctionnelle\cite{ziegler_approximate_1991}, dont
la forme exacte reste inconnue malgré la preuve de son existence.
La diffusion de programmes de calcul de eDFT a permis de ne pas limiter
l'utilisation de cette méthode aux seuls chimistes théoriciens mais
de démocratiser son utilisation auprès des expérimentateurs. Le développement
d'approximations efficaces de la fonctionnelle permet le calcul de
propriétés structurales et énergétiques d'un grand nombre de système
à un coût numérique raisonnable. Ceci explique les succès de cette
méthode et la place qu'elle occupe aujourd'hui dans la production
scientifique et la littérature\cite{koch_chemists_2000}.

Un an après l'article fondateur de Kohn et Hohenberg, Mermin étend
la théorie de la fonctionnelle de la densité au cas du gaz inhomogène
d'électron à température non-nulle\cite{mermin_thermal_1965}. Ceci
revient en fait à réécrire la DFT dans le cadre de la physique statistique.
Bien que toujours appliquée au cas d'un hamiltonien quantique dans
cet article, ceci ouvre la porte à l'utilisation de la DFT avec un
hamiltonien classique. 

\begin{figure}[h]

\noindent \begin{centering}
\includegraphics[width=0.6\textwidth]{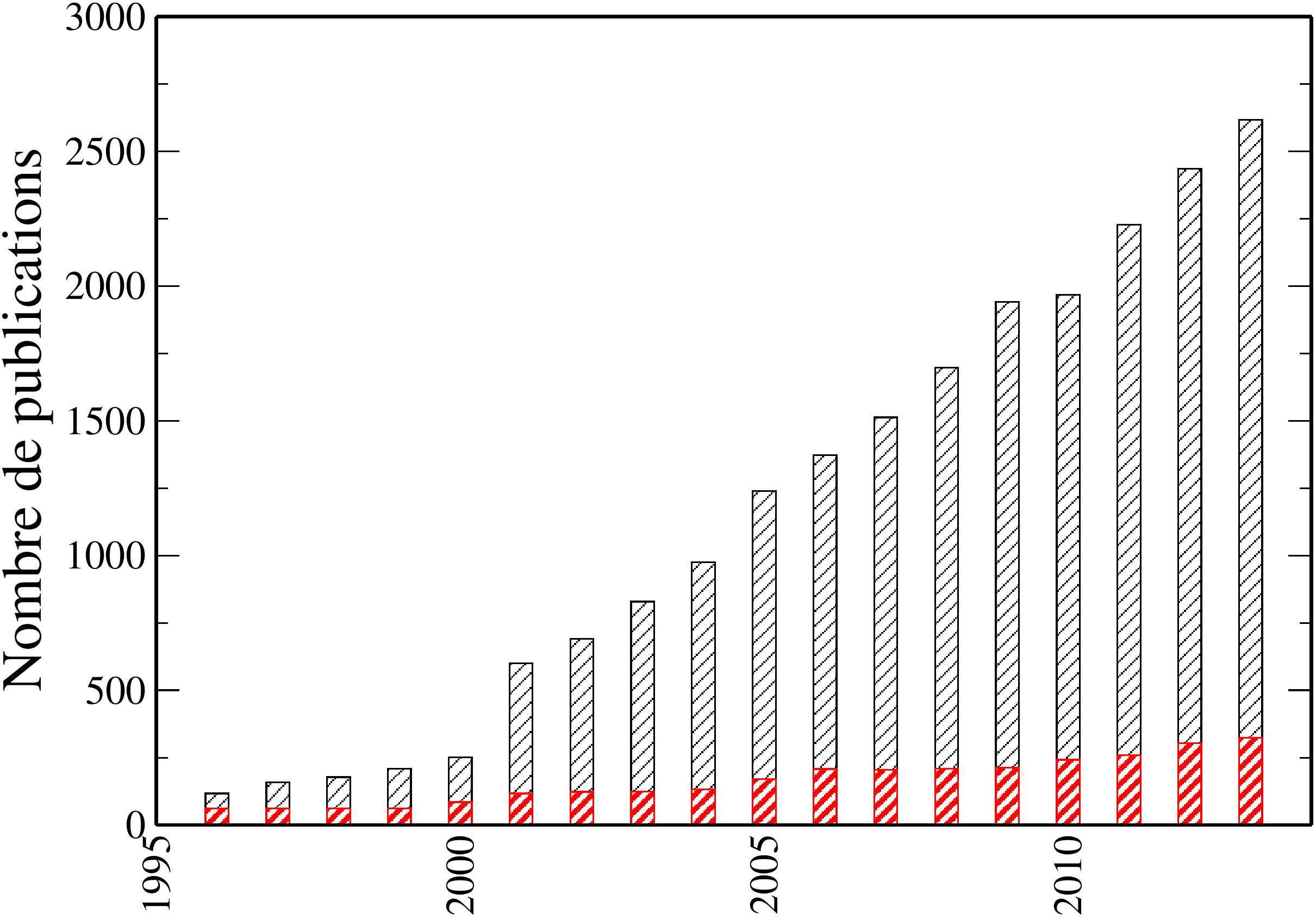}\protect\caption{Nombre de publications par an ayant pour sujet, \og Classical Density
Functional Theory \fg{} en rouge et \og Density Functional Theory \fg{}
en noir, selon le site Web of Science. Le nombre de publications ayant
pour sujet \og Density Functional Theory \fg{} a été divisé par
10.\label{fig:Nombre-de-publications}}

\par\end{centering}

\end{figure}

On peut alors imaginer son utilisation pour l'étude de la solvatation
dans un fluide décrit par un champ de force classique, ce qui est
l'objectif de ces travaux de thèse. Il peut apparaître surprenant
que, malgré leurs découvertes quasi-simultanées, la théorie fonctionnelle
de la densité classique (cDFT), au-delà de modèles atomiques simples,
n'ait pas connu le même succès que son analogue quantique. On peut
expliquer ce constat par le fait qu'il existait déjà des méthodes
permettant l'étude de problèmes à N corps classiques avant la formulation
de la DFT, comme les simulations Monte-Carlo ou de dynamique moléculaire.
Ces méthodes peuvent tout à fait être mises en oeuvre pour étudier
des systèmes complexes (de taille importante). Elles ont un coût numérique
élevé mais acceptable et sont théoriquement exactes (elles permettent
en principe, après un échantillonnage complet de l'espace des phases,
d'obtenir le résultat prévu par la physique statistique). C'est pourquoi
ces méthodes sont largement utilisées dans la communauté. Ceci peut
expliquer l'intérêt relativement limité de la communauté au développement
de la cDFT en chimie puisque les physicochimistes possèdent déjà des
méthodes ayant fait leurs preuves pour étudier des systèmes classiques.
En conséquence, pour l'instant, la cDFT n'a pas encore atteint le
même niveau de précision dans la prédiction de propriétés physicochimiques
d'intérêt pour les biologistes et les chimistes que la eDFT pour l'étude
des systèmes quantiques. De plus, contrairement à la eDFT, il n'existe
pas de programmes de cDFT largement répandus et accessibles à tous.
Néanmoins, certains systèmes de grande taille restent inaccessibles
aux simulations numériques, et il se peut que la cDFT, moins coûteuse
numériquement, soit un bon moyen d'étude de ces systèmes. Ceci peut
expliquer le regain d'intérêt dans la littérature ces dix dernières
années pour le développement de la cDFT. Sur la \ref{fig:Nombre-de-publications}
on donne le nombre de publications par an ayant pour sujet \og Classical
Density Functional Theory \fg{} qui reste néanmoins deux ordre de
grandeurs plus faible que celui ayant pour sujet \og Density Functional
Theory \fg{}.

Dans les parties suivantes, nous exposons les principes de la théorie
fonctionnelle de la densité classique dans le cas du fluide idéal.
Nous présenterons ensuite la construction de la fonctionnelle dans
le cas de l'approximation du fluide homogène de référence qui sera
utilisée dans toute la suite de ce manuscrit.

\section{La théorie fonctionnelle de la densité classique (cDFT)}

\subsection{Exemple introductif, le cas du fluide idéal}

Les deux prochaines sous-parties présentent brièvement la DFT classique.
Dans le cas du fluide idéal et des ensembles canonique et grand-canonique,
on va montrer que l'on peut écrire des fonctionnelles de la densité
de solvant qui permettent, par minimisation fonctionnelle, de retrouver
les résultats exacts prévus par la thermodynamique statistique.

\subsubsection{L'ensemble canonique $\text{(N,V,T)}$\label{sub:L'ensemble-Canonique}}

On considère ici le cas de l'ensemble canonique, c'est-à-dire un ensemble
où le volume $\text{V}$, la température $\text{T}$ et le nombre
de particule N sont fixés. L'énergie du système est décrite par la
donnée de l'hamiltonien décrivant un système à N particules de masse
m, à la température T,
\begin{equation}
{\cal H}_{\text{N}}={\cal H}(\bm{r}^{\mathbf{\mathrm{N}}},\bm{p}^{\mathrm{N}})={\cal K}(\bm{p}^{\mathrm{N}})+{\cal U}(\bm{r}^{\mathrm{N}})+v(\bm{r}^{\mathrm{N}}),\label{H=00003DT+U+V}
\end{equation}

avec ${\cal K}$ l'énergie cinétique, ${\cal U}$ l'énergie potentielle
d'interaction et $v$ un potentiel extérieur, qui dépendent des positions
et des impulsions $(\bm{r}^{\mathrm{N}},\bm{p}^{\mathrm{N}})=(\bm{r}_{1},...,\bm{r}_{\mathrm{N}},\bm{p}_{1},...,\bm{p}_{\mathrm{N}})$
des N particules. Le potentiel extérieur est une énergie qui a pour
origine une perturbation qui ne provient pas des particules du système,
par exemple si un champ électrique non nul est imposé, celui ci interagit
avec les particules si celles-ci sont chargées.

La densité de probabilité pour ce système à $N$ particules est%
\footnote{On rappelle ici l'expression de la fonction de partition\cite{Diu_1995}:
\begin{equation}
\text{Q}=\frac{1}{\text{h}^{3\text{N}}\text{N}!}\idotsint_{\mathbb{R}^{3\text{N}}}\idotsint_{\mathbb{R}^{3\text{N}}}e^{-\beta{\cal H}_{\text{N}}}\text{d}\bm{r}_{1}...\text{d}\bm{r}_{\text{N}}\text{d}\bm{p}_{1}...\text{d}\bm{p}_{\text{N}}.
\end{equation}
}
\begin{equation}
f_{0}=Q^{-1}\exp\left[-\beta\left({\cal H}_{\text{N}}\right)\right].
\end{equation}
Dans le cas où le système étudié est le fluide idéal isolé, c'est-à-dire
que les particules n'interagissent pas entre elles $\left({\cal U}=0\right)$
et que le potentiel extérieur est nul $\left(v=0\right)$, on peut
calculer analytiquement la fonction de partition, on trouve,
\begin{equation}
\text{Q}_{\text{id}}=\left(\frac{\mathrm{V}}{\Lambda^{3}}\right)^{\mathrm{N}}\frac{1}{{\cal \text{N}}!}.
\end{equation}
Ce résultat a une interprétation immédiate en analyse combinatoire.
En effet, la longueur d'onde de de Broglie $\Lambda=\left(\frac{2\pi\text{mk}\mathrm{k_{B}T}}{\text{\ensuremath{h^{2}}}}\right)^{3/2}$,
avec $\text{h}=6.62606957.10^{-34}\ \text{m\ensuremath{{}^{2}}.kg.s\ensuremath{{}^{-1}}}$
la constante de Planck, représente l'étalement spatial de la particule.
$\Lambda^{3}$ est donc le plus petit volume élémentaire que l'on
puisse considérer dans l'espace des phases (en raison du principe
d'incertitude d'Heisenberg).

On peut alors \og découper \fg{} le volume V en $\text{V}/\Lambda^{3}$
éléments et, comme les particules n'interagissent pas, il y a pour
chaque particule $\text{V}/\Lambda^{3}$ possibilités pour placer
la particule. Le facteur $1/\mathrm{N}!$ provient de l'indiscernabilité
des particules.

Si on réinjecte l'expression de la fonction de partition dans l'\ref{eq:F=00003DkbTlnQ}
et que l'on utilise la formule de Stirling%
\footnote{$n!\sim\sqrt{2\pi n}\left(\frac{n}{e}\right)^{n}$%
}, on obtient pour l'énergie libre de Helmholtz, 
\begin{equation}
\mathrm{F_{id}}=\mathrm{k_{B}T}\left[\text{N}\ln\left(\rho\Lambda^{3}\right)-\text{N}\right],
\end{equation}
où on a introduit $\rho=\mathrm{\frac{N}{V}}$, la densité du fluide
homogène.

On pose une fonctionnelle de la densité inhomogène,
\begin{equation}
{\cal F}_{\text{id}}\left[\rho(\bm{r})\right]=\mathrm{k_{B}}\text{T}\iiint_{\mathbb{R}^{3}}\rho(\bm{r})\left[\ln\left(\rho(\bm{r})\Lambda^{3}\right)-1\right]\text{d}\bm{r}.\label{eq:Fid}
\end{equation}
En dérivant analytiquement cette fonctionnelle (voir \ref{chap:D=0000E9riv=0000E9-fonctionelle}
pour la définition d'une dérivée fonctionnelle) et en cherchant les
racines de la dérivée obtenue, on trouve que la fonctionnelle est
minimale pour une densité d'équilibre homogène $\rho_{\mathrm{eq}}=\text{N/V}$. 

\fbox{\begin{minipage}[t]{1\columnwidth}%
On a montré ici que pour le fluide idéal isolé, étudié dans l'ensemble
canonique, il est possible d'introduire une fonctionnelle de la densité
de ce fluide idéal. Cette fonctionnelle est égale à l'énergie libre
de Helmholtz du système homogène lorsqu'elle est minimale. De plus,
ce minimum est atteint pour une densité du fluide qui est celle du
fluide homogène. En minimisant cette fonctionnelle on retrouve donc
les propriétés énergétiques et structurales du système à l'équilibre
thermodynamique.%
\end{minipage}}

\subsubsection{L'ensemble grand-canonique $\text{(\ensuremath{\mu},V,T)}$\label{sub:L'ensemble-Grand-Canonique}}

On se place maintenant dans l'ensemble grand-canonique, c'est-à-dire
un ensemble où le volume V, la température T et le potentiel chimique
$\mu$ sont fixés.

Dans ce cas, la densité de probabilité d'un système à N particules
à l'équilibre thermodynamique s'exprime,
\begin{equation}
f_{\text{eq}}=\Xi^{-1}\exp\left[-\beta\left({\cal H}_{\text{N}}-\mu\text{N}\right)\right].\label{proba_GC}
\end{equation}
La fonction de grand-partition, $\Xi$, est donnée par,
\begin{equation}
\Xi=\text{Tr}_{\text{cl}}\exp\left[-\beta\left({\cal H}_{\text{N}}-\mu\text{N}\right)\right],
\end{equation}
où $\text{Tr}_{\text{cl}}$ désigne la trace classique, c'est-à-dire
\begin{equation}
\text{Tr}_{\text{cl}}=\sum_{\text{N}=0}^{\infty}\frac{1}{\mathrm{h}\text{\ensuremath{^{3\mathrm{N}}}N}!}\idotsint_{\mathbb{R}^{3\text{N}}}\idotsint_{\mathbb{R}^{3\text{N}}}\text{d}\bm{r}...\text{d}\bm{r}_{\text{N}}\text{d}\bm{p}_{1}...\text{d}\bm{p}_{\text{N}},
\end{equation}

Dans le cas du fluide idéal en l'absence de potentiel extérieur, on
a,
\begin{equation}
\Xi_{\text{id}}=\sum_{\text{N}=0}^{\infty}\frac{\text{\text{e}\ensuremath{{}^{\beta\mu\mathrm{N}}}}}{\text{N!\ensuremath{\Lambda^{3\mathrm{N}}}}}\idotsint_{\mathbb{R}^{3\text{N}}}\text{d}\bm{r}^{\text{N}}=\text{e}^{\text{z\text{V}}},
\end{equation}
en posant $\text{z=\ensuremath{\exp\left(\beta\mu\right)}/\ensuremath{\Lambda^{3}}}$
l'activité.

L'expression du grand potentiel dans cet ensemble est 
\begin{equation}
\Omega=\text{-}\mathrm{k_{B}T}\ln\Xi,
\end{equation}
ce qui dans le cas du gaz idéal donne,
\begin{equation}
\Omega_{\text{id}}=-\mathrm{k_{B}T}\mathrm{zV}.\label{Omegaid}
\end{equation}
Notons que le grand potentiel est également lié à l'énergie libre
de Helmholtz par la transformée de Legendre,
\begin{equation}
\Omega=\mathrm{F}-\mu\mathrm{N}\label{Omega=00003DF-muN}
\end{equation}
On peut introduire, comme en \ref{sub:L'ensemble-Canonique} et dans
l'\ref{eq:Fid}, une fonctionnelle qui dépend de la densité non-homogène,
\begin{equation}
\Omega\left[\rho(\bm{r})\right]={\cal F}_{\text{id}}[\rho(\bm{r})]-\mu\iiint_{\mathbb{R}^{3}}\rho(\bm{r})\text{d}\bm{r},\label{eq:Omega=00005Brho=00005D_GC_id}
\end{equation}
En minimisant cette fonctionnelle, c'est-à-dire en cherchant les racines
de la dérivé fonctionnelle de l'\ref{eq:Omega=00005Brho=00005D_GC_id},
on trouve $\rho_{\text{eq}}=\exp\left(\beta\mu\right)/\Lambda^{3}$.
En réinjectant cette expression de la densité à l'équilibre, on retrouve
de façon cohérente l'expression du grand potentiel pour le fluide
idéal donnée en \ref{Omegaid}.

On retrouve donc ici un résultat similaire à celui obtenu dans l'ensemble
canonique pour le fluide idéal.

On peut donc généraliser les principes de la théorie de la fonctionnelle
de la densité classique dans l'ensemble grand-canonique au cas du
fluide quelconque. La suite de notre étude sera menée exclusivement
dans l'ensemble grand-canonique.

\fbox{\begin{minipage}[t]{1\columnwidth}%
\begin{itemize}
\item Dans l'ensemble grand-canonique, pour un fluide donné, éventuellement
soumis à un potentiel extérieur, on peut écrire une fonctionnelle
unique de la densité de solvant.
\item Cette fonctionnelle est minimale pour la densité de solvant correspondant
à l'équilibre thermodynamique, elle est alors égale au grand-potentiel
du système.\end{itemize}
\end{minipage}}

Ce résultat est intéressant puisqu'il permet, si on connait une expression
de la fonctionnelle, de déterminer par principe variationnel la densité
à l'équilibre thermodynamique et le grand-potentiel, c'est-à-dire
les propriétés énergétiques et structurales d'équilibres du système.
Réaliser une minimisation fonctionnelle pourrait s'avérer beaucoup
moins coûteux numériquement que de réaliser l'échantillonnage de l'espace
de phases. Dans la partie suivante on présente la démonstration générale
de ces deux points, qui a été proposée par Evans\cite{evans79}.

\subsection{Principe variationnel pour le grand potentiel\label{sub:Principe-Variationel-pour}}

On a introduit la probabilité d'équilibre en \ref{proba_GC}. Supposons
qu'il existe une autre distribution de probabilité $f$ avec la seule
condition que celle-ci soit normée, c'est-à-dire $\text{Tr}_{\text{cl}}f=1$.
On définit alors une fonctionnelle de cette probabilité comme,
\begin{equation}
\Omega\left[f\right]=\mathrm{Tr_{cl}}\left[f\Bigl(\left({\cal H}_{\text{N}}-\mu\text{N}\right)+\text{k}{}_{\text{B}}\text{T}\ln f\Bigr)\right],\label{eq:Omega=00005Bf=00005D}
\end{equation}
avec ${\cal H}_{\mathrm{N}}$ le hamiltonien défini en \ref{H=00003DT+U+V}.
Il est évident que pour $f=f_{\text{eq}}$, on a: 
\begin{equation}
\Omega[f_{\text{eq}}]=\Omega=-\text{k}\text{\ensuremath{_{B}}T}\ln\Xi.\label{eq:Omega=00003D-kTlnPsi}
\end{equation}

Supposons que $f\neq f_{\text{eq}}$, comme ces deux distributions
de probabilité sont normalisées, on a: 
\begin{eqnarray}
\text{Tr}_{\text{cl}}f & = & \text{Tr}_{\text{cl}}f_{\text{eq}}.\label{eq:Trf=00003DTrf0}
\end{eqnarray}
On peut multiplier chaque membre de cette égalité par $\Omega$ qui
est un scalaire,
\begin{eqnarray}
\text{Tr}_{\text{cl}}\left(\Omega f\right) & = & \text{Tr}_{\text{cl}}\left(\Omega f_{\mathrm{eq}}\right).
\end{eqnarray}
Soit, en utilisant l'\ref{proba_GC} et l'\ref{eq:Omega=00003D-kTlnPsi},
\begin{equation}
\text{Tr}_{\text{cl}}\left[f\left(v+{\cal K}+{\cal U}-\mu\text{N}+\text{k}{}_{\text{B}}\text{T}\ln f_{\text{eq}}\right)\right]=\text{Tr}_{\text{cl}}\left[f_{\text{eq}}\left(v+K+U-\mu\text{N}+\text{k}{}_{\text{B}}\text{T}\ln f_{\text{eq}}\right)\right],
\end{equation}
 et donc avec l'\ref{eq:Omega=00005Bf=00005D}, 
\begin{equation}
\Omega\left[f\right]=\Omega\left[f_{\text{eq}}\right]+\text{k}{}_{\text{B}}\text{T}\text{Tr}_{\text{cl}}\left(f\ln\frac{f}{f_{\text{eq}}}\right)\label{eq:toto}
\end{equation}
On peut ajouter au membre de droite $\text{Tr}_{\text{cl}}f_{0}$
et retirer $\text{Tr}_{\text{cl}}f$ en conservant l'égalité grâce
à l'\ref{eq:Trf=00003DTrf0}. L'\ref{eq:toto} devient
\begin{eqnarray}
\Omega\left[f\right] & = & \Omega\left[f_{\text{eq}}\right]+\text{k}{}_{\text{B}}\text{T}\text{Tr}_{\text{cl}}\left[f_{\text{eq}}\Bigl(\frac{f}{f_{\text{eq}}}\ln\left(\frac{f}{f_{\text{eq}}}\right)-\frac{f}{f_{\text{eq}}}+1\Bigr)\right].
\end{eqnarray}
Il suffit de remarquer que la fonction entre parenthèses, $x\mapsto x\ln x-x+1$,
définie sur $\mathbb{R}_{+}$ est positive et ne s'annule qu'en $x=1$
pour montrer que la fonctionnelle $\Omega\left[f\right]$ est toujours
strictement supérieure à $\Omega$, sauf en $f=f_{\text{eq}}$ où
elle vaut $\Omega$.

On vient de démontrer un théorème clé de la DFT classique : la fonctionnelle
$\Omega\left[f\right]$ introduite est égale au grand potentiel $\Omega$
à son minimum qui est atteint pour la distribution de probabilité
d'équilibre $f_{\text{eq}}$.

On peut montrer que cette distribution de probabilité d'équilibre
$f_{\text{eq}}$ est une fonctionnelle unique de la densité d'équilibre
$\rho_{\mathrm{eq}}(\bm{r})$. Pour cela, il suffit de remarquer que
la densité de probabilité d'équilibre est, de manière évidente, une
fonctionnelle du champ extérieur $v$. La densité d'équilibre, en
tant que fonctionnelle de la densité de probabilité d'équilibre est
donc une fonctionnelle du champ extérieur $v$. On peut démontrer
(ceci est un cas simplifié du calcul présenté en \ref{sec:L'=0000E9nergie-libre-intrins=0000E8que_unique_fonctioelle})
que la densité de particule à l'équilibre est associée de manière
unique à un potentiel extérieur. En conséquence, la distribution de
probabilité d'équilibre est une fonctionnelle unique de la densité
d'équilibre. De ce fait, le potentiel extérieur est une fonctionnelle
unique de la densité d'équilibre.

On introduit alors la fonctionnelle intrinsèque
\begin{equation}
{\cal F}[\rho(\bm{r})]=\mathrm{Tr_{cl}}\Bigl[f\left[\left({\cal K}+{\cal U}\right)+\text{k}\text{\ensuremath{_{B}}T}\ln f\right]\Bigr]
\end{equation}
qui est elle aussi une unique fonctionnelle de la densité, même si
on n'en connait pas de forme qui dépende explicitement de la densité.
On peut ainsi définir une nouvelle fonctionnelle:
\begin{eqnarray}
\Omega_{v}\left[\rho(\bm{r})\right] & = & {\cal F}[\rho(\bm{r})]+\iiint_{\mathbb{R}^{3}}\rho(\bm{r})v(\bm{r})\text{d}\bm{r}-\mu\iiint_{\mathbb{R}^{3}}\rho(\bm{r})\text{d}\bm{r}\\
 & = & {\cal F}[\rho(\bm{r})]-\iiint_{\mathbb{R}^{3}}\rho(\bm{r})\psi(\bm{r})\text{d}\bm{r}\label{eq:Omega=00003DF-int_rho_psi}
\end{eqnarray}
où l'on a introduit $\psi(\bm{r})=\mu-v(\bm{r})$ le potentiel chimique
intrinsèque.

Il est évident que pour une densité $\rho(\bm{r})$ correspondant
à une distribution de probabilité $f$, on a l'égalité suivante:
\begin{equation}
\Omega\left[f\right]=\Omega_{v}\left[\rho(\bm{r})\right].
\end{equation}
En se servant des propriétés démontrées pour $\Omega\left[f\right]$,
on prouve directement que $\Omega_{v}\left[\rho(\bm{r})\right]$ est
égal au grand potentiel $\Omega$ à son minimum, atteint pour la densité
d'équilibre à une particule $\rho_{\mathrm{eq}}(\bm{r})$. Cependant,
nous ne connaissons pas d'expression de la fonctionnelle intrinsèque
${\cal F}$. Remarquons néanmoins que pour un fluide sans interactions
(i.e., ${\cal U}=0$) cette fonctionnelle est égale à la fonctionnelle
${\cal F}_{\text{id}}$ introduite en \ref{sub:L'ensemble-Canonique}.
Dès lors, on peut réécrire cette fonctionnelle intrinsèque comme la
somme d'une partie idéale, qui décrit l'entropie d'un fluide sans
interaction à la même densité, et d'un terme d'excès, défini comme
la différence entre la partie intrinsèque et la partie idéale.
\begin{equation}
{\cal F}\left[\rho(\bm{r})\right]={\cal F}_{\text{id}}\left[\rho(\bm{r})\right]+{\cal F}_{\text{exc}}\left[\rho(\bm{r})\right].\label{eq:F=00003DFid+Fexc}
\end{equation}
En dérivant l'\ref{eq:Omega=00003DF-int_rho_psi} par rapport à $\rho(\bm{r})$
on trouve immédiatement l'égalité suivante qui lie dérivée de la fonctionnelle
intrinsèque et potentiel chimique intrinsèque
\begin{equation}
\frac{\delta{\cal F}\left[\rho(\bm{r})\right]}{\delta\rho(\bm{r})}=\psi(\bm{r}).\label{eq:dF/drho=00003Dpsi}
\end{equation}
On a également, en dérivant l'\ref{eq:Omega=00003DF-int_rho_psi}
par rapport au potentiel chimique intrinsèque:
\begin{equation}
\frac{\delta\Omega_{v}\left[\rho(\bm{r})\right]}{\delta\psi(\bm{r})}=-\rho(\bm{r}).
\end{equation}
Ceci permet de voir la fonctionnelle $\Omega_{v}$ comme obtenue par
une transformée de Legendre généralisée, à partir de la fonctionnelle
intrinsèque. Les variables conjuguées $\psi(\bm{r})$ et $\rho(\bm{r})$
étant les analogues de $\mu$ et $\text{N}$ dans le cas homogène.

\section{Fonctions de corrélation directe et écriture de la fonctionnelle
d'excès\label{sec:Fonctions-de-corr=0000E9lations}}

La partie d'excès de la fonctionnelle est génératrice d'une hiérarchie
de fonctions de corrélation directe $c^{(i)}$. La fonction à une
particule est définie comme la première dérivée fonctionnelle du terme
d'excès par rapport à la densité,
\begin{equation}
c^{(1)}(\rho;\bm{r})=-\frac{\delta\beta{\cal F}_{\mathrm{exc}}\left[\rho(\bm{r})\right]}{\delta\rho(\bm{r})},\label{eq:c(1)}
\end{equation}
remarquons la dépendance à la densité de cette fonction.

En dérivant l'\ref{eq:F=00003DFid+Fexc} et en y injectant l'\ref{eq:c(1)},
on peut obtenir une expression de la densité d'équilibre,
\begin{equation}
\rho_{\mathrm{eq}}(\bm{r})=\frac{\text{e}^{\beta\mu}}{\Lambda^{3}}\exp\left[-\beta v(\bm{r})+c^{(1)}(\bm{r})\right]\label{eq:c(2)}
\end{equation}
En comparant ce résultat avec celui de la densité d'équilibre dans
le cas du gaz idéal isolé obtenu en \ref{sub:L'ensemble-Grand-Canonique},
on constate l'apparition d'un terme dû au potentiel extérieur. De
plus, l'effet des interactions entre particules est décrite par cette
fonction de corrélation à une particule.

La fonction de corrélation directe de paire est définie comme la dérivée
de la fonction à une particule,
\begin{equation}
c^{(2)}(\rho;\bm{r},\bm{r^{\prime}})=\frac{\delta c^{(1)}(\bm{r})}{\delta\rho(\bm{r}^{\prime})}=-\frac{\delta^{2}\beta{\cal F}_{\mathrm{exc}}\left[\rho(\bm{r})\right]}{\delta\rho(\bm{r}^{\prime})\delta\rho(\bm{r})}.
\end{equation}
C'est cette fonction de distribution de paire que l'on retrouve dans
une des équations clés de la théorie des liquides, l'équation de Ornstein-Zernike\cite{hansen_theory_2006}.

Les fonctions de corrélation d'ordre plus élevé sont, de la même manière,
générées par des dérivées fonctionnelles successives de la partie
d'excès.

Pour tenter de trouver une expression pour la partie d'excès, on peut
intégrer l'\ref{eq:c(1)} par rapport à la densité. Si $\rho_{0}(\bm{r})$
est la densité dans un état de référence du système (par exemple le
système homogène ou le fluide de sphères dures), et $c^{(1)}(\rho_{0};\bm{r})=c_{0}^{(1)}(\bm{r})$
la fonction de corrélation à une particule dans ce même état de référence,
on peut choisir un chemin d'intégration linéaire entre l'état de référence
et l'état final du système de densité $\rho(\bm{r})$ de sorte que
\begin{equation}
\rho(\bm{r})=\rho_{0}(\bm{r})+\lambda\Delta\rho(\bm{r}),
\end{equation}
avec $\Delta\rho(\bm{r})=\rho(\bm{r})-\rho_{0}(\bm{r})$ et $\lambda$
un paramètre d'intégration variant entre 0 et 1. On obtient par intégration
l'expression suivante pour la partie d'excès,
\begin{eqnarray}
{\cal F}_{\text{exc}}\left[\rho(\bm{r})\right]-{\cal F}_{\text{exc}}\left[\rho_{0}(\bm{r})\right] & = & \text{k}_{\text{B}}\text{T}\int_{0}^{1}\left[\iiint_{\mathbb{R}^{3}}\frac{\partial\rho(\bm{r};\lambda)}{\partial\lambda}c^{(1)}(\rho_{\lambda};\bm{r})\text{d}\bm{r}\right]\text{d}\lambda\nonumber \\
 & = & \text{k}_{\text{B}}\text{T}\int_{0}^{1}\left[\iiint_{\mathbb{R}^{3}}\Delta\rho(\bm{r})c^{(1)}(\rho_{\lambda};\bm{r})\text{d}\bm{r}\right]d\lambda.\label{eq:Fexc_path_lambda}
\end{eqnarray}
On peut évaluer la fonction de corrélation $c^{(1)}(\rho_{\lambda};\bm{r})$
par une intégration similaire à celle qui vient d'être utilisée, en
partant de l'\ref{eq:c(2)},
\begin{equation}
c^{(1)}(\rho_{\lambda};\bm{r})-c^{(1)}(0;\bm{r})=\int_{0}^{\lambda}\left[\iiint_{\mathbb{R}^{3}}\Delta\rho(\bm{r})c^{(2)}(\rho_{\lambda^{\prime}};\bm{r,r}^{\prime})\text{d}\bm{r}^{\prime}\right]\text{d}\lambda^{\prime}.\label{eq:c(1)_path_lambda}
\end{equation}
Finalement, en réinjectant l'\ref{eq:c(1)_path_lambda} dans l'\ref{eq:Fexc_path_lambda}
on obtient,%
\footnote{En utilisant le fait que $\int_{0}^{1}\left[\int_{0}^{\lambda}f(\lambda^{\prime})\mathrm{d}\lambda^{\prime}\right]\mathrm{d}\lambda=\int_{0}^{1}\left(1-\lambda\right)f(\lambda)\mathrm{d}\lambda$\cite{hansen_theory_2006}%
}
\begin{gather}
{\cal F}_{\text{exc}}\left[\rho(\bm{r})\right]-{\cal F}_{\text{exc}}\left[\rho_{0}(\bm{r})\right]=-\text{k}_{\text{B}}\text{T}\int\Delta\rho(\bm{r})c_{0}^{(1)}(\bm{r})\text{d}\bm{r}\nonumber \\
-\text{k}_{\text{B}}\text{T}\int_{0}^{1}(1-\lambda)\left[\iiint_{\mathbb{R}^{3}}\iiint_{\mathbb{R}^{3}}\Delta\rho(\bm{r}^{\prime})\Delta\rho(\bm{r})c^{(2)}(\rho_{\lambda};\bm{r},\bm{r}^{\prime})\text{d}\bm{r}\text{d}\bm{r}^{\prime}\right]\text{d}\lambda.\label{eq:Fexc_path_final}
\end{gather}
On signale que si le chemin d'intégration utilisé a été choisi pour
sa simplicité, le résultat obtenu est indépendant de ce chemin puisque
${\cal F}_{exc}$ est une fonctionnelle unique de la densité. On peut
alors exprimer le grand potentiel à l'aide des \ref{eq:Fexc_path_final},
\ref{eq:Omega=00003DF-int_rho_psi} et \ref{eq:Fid},
\begin{eqnarray}
\Omega_{v}\left[\rho(\bm{r})\right] & = & \text{k}_{\text{B}}\text{T}\iiint_{\mathbb{R}^{3}}\rho(\bm{r})\left[\ln\left(\rho(\bm{r})\Lambda^{3}\right)-1\right]\text{d}\bm{r}-\text{k}_{\text{B}}\text{T}\iiint_{\mathbb{R}^{3}}\Delta\rho(\bm{r})c_{0}^{(1)}(\bm{r})\text{d}\bm{r}\nonumber \\
 &  & +\text{k}_{\text{B}}\text{T}\iiint_{\mathbb{R}^{3}}\iiint_{\mathbb{R}^{3}}\Delta\rho(\bm{r}^{\prime})\Delta\rho(\bm{r})C^{(2)}(\bm{r},\bm{r}^{\prime})\text{d}\bm{r}\text{d}\bm{r}^{\prime}\nonumber \\
 &  & -\iiint_{\mathbb{R}^{3}}\rho(\bm{r})\psi(\bm{r})\text{d}\bm{r}+{\cal F}_{\mathrm{exc}}\left[\rho_{0}(\bm{r})\right],\label{eq:Omega_v exact expression}
\end{eqnarray}
avec,
\begin{equation}
C^{(2)}(\bm{r},\bm{r}^{\prime})=\int_{0}^{1}(1-\lambda)c^{(2)}(\rho_{\lambda};\bm{r},\bm{r}^{\prime})\text{d}\lambda.\label{eq:C(2)}
\end{equation}
Si on prend comme référence un fluide homogène qui possède le même
potentiel chimique que le fluide réel de densité $\rho_{b}$, son
grand potentiel peut se trouver en posant $\rho(\bm{r})=\rho_{b}$
dans l'\ref{eq:Omega_v exact expression},
\begin{equation}
\Omega_{b}=\text{k}_{\text{B}}\text{T}\iiint_{\mathbb{R}^{3}}\rho_{b}(\bm{r})\left[\ln\left(\rho_{b}\Lambda^{3}\right)-1\right]\text{d}\bm{r}+{\cal F}_{\mathrm{exc}}\left[\rho_{b}(\bm{r})\right]-\mu\iiint_{\mathbb{R}^{3}}\rho_{b}(\bm{r})\text{d}\bm{r}.
\end{equation}
En remarquant que dans le cas du fluide homogène la dérivée de l'\ref{eq:F=00003DFid+Fexc}
donne
\begin{eqnarray}
\frac{\delta{\cal F}\left[\rho(\bm{r})\right]}{\delta\rho(\bm{r})} & = & \frac{\delta{\cal F}_{\mathrm{id}}\left[\rho(\bm{r})\right]}{\delta\rho(\bm{r})}+\frac{\delta{\cal F}_{\mathrm{exc}}\left[\rho(\bm{r})\right]}{\delta\rho(\bm{r})}=\text{k}_{\text{B}}\text{T}\ln(\rho_{b}\Lambda^{3})-\text{k}_{\text{B}}\text{T}c_{b}(\bm{r})=\mu,
\end{eqnarray}
on peut reformuler de manière plus compacte l'\ref{eq:Omega_v exact expression}
en
\begin{eqnarray}
\Omega_{v}\left[\rho(\bm{r})\right] & = & \text{k}_{\text{B}}\text{T}\iiint_{\mathbb{R}^{3}}\left[\rho(\bm{r})\ln\left(\frac{\rho(\bm{r})}{\rho_{b}}\right)-\Delta\rho(\bm{r})\right]\text{d}\bm{r}+\text{k}_{\text{B}}\text{T}\iiint_{\mathbb{R}^{3}}\rho(\bm{r})v(\bm{r})\text{d}\bm{r}\nonumber \\
 & + & \text{k}_{\text{B}}\text{T}\iiint_{\mathbb{R}^{3}}\iiint_{\mathbb{R}^{3}}\Delta\rho(\bm{r}^{\prime})\Delta\rho(\bm{r})C^{(2)}(\bm{r\ ,r}^{\prime})\text{d}\bm{r}\text{d}\bm{r}^{\prime}+\Omega_{b}\label{eq:Omega_v usefull expression-1}\\
 & = & \Omega_{b}+{\cal F}_{\text{id}}[\rho(\bm{r})]+{\cal F}_{\text{ext}}[\rho(\bm{r})]+{\cal F}_{\text{exc}}[\rho(\bm{r})]
\end{eqnarray}
avec
\begin{equation}
{\cal F}_{\text{id}}[\rho(\bm{r})]=\text{k}_{\text{B}}\text{T}\iiint_{\mathbb{R}^{3}}\left[\rho(\bm{r})\ln\left(\frac{\rho(\bm{r})}{\rho_{b}}\right)-\Delta\rho(\bm{r})\right]\text{d}\bm{r}\label{eq:Fiddeltarho}
\end{equation}
\begin{equation}
{\cal F}_{\text{ext}}[\rho(\bm{r})]=\iiint_{\mathbb{R}^{3}}\rho(\bm{r})v(\bm{r})\text{d}\bm{r}\label{eq:Fext}
\end{equation}
\begin{equation}
{\cal F}_{\text{exc}}[\rho(\bm{r})]=\text{k}_{\text{B}}\text{T}\iiint_{\mathbb{R}^{3}}\iiint_{\mathbb{R}^{3}}\Delta\rho(\bm{r}^{\prime})\Delta\rho(\bm{r})C^{(2)}(\bm{r},\bm{r}^{\prime})\text{d}\bm{r}\text{d}\bm{r}^{\prime},\label{eq:Fexc}
\end{equation}
où $\Delta\rho(\bm{r})=\rho(\bm{r})-\rho_{b}$.

\subsection{Approximation du fluide homogène de référence (HRF)\label{sub:HRF approx}}

La partie d'excès décrite par l'\ref{eq:Fexc} est en fait difficilement
évaluable telle quelle car elle fait intervenir la fonction $C^{(2)}(\bm{r},\bm{r}^{\prime})$
qui, pour être calculée, nécessite de connaître la fonction de corrélation
à deux corps à toutes les densités entre la densité du fluide de référence
et la densité du système (pour toutes les valeurs de $\lambda$).
Pour simplifier considérablement le calcul de cette fonction on peut
considérer qu'elle est égale à celle du fluide homogène de référence
pour toutes les densités, soit
\begin{equation}
c^{(2)}(\rho_{\lambda};\bm{r},\bm{r}^{\prime})=c^{(2)}(\rho_{0};\bm{r},\bm{r}^{\prime})=c^{(2)}(\bm{r},\bm{r}^{\prime})
\end{equation}
En utilisant l'\ref{eq:C(2)} on obtient,
\begin{equation}
C^{(2)}(\bm{r},\bm{r}^{\prime})=-\frac{1}{2}c^{(2)}(\bm{r},\bm{r}^{\prime}).\label{eq:approx_c_HRF}
\end{equation}
Cette approximation est raisonnable si la densité reste proche de
celle du fluide homogène, ce qui peut ne pas être le cas.

On peut montrer que cette approximation est en fait équivalente à
l'approximation HNC (Hypernetted chain) employée dans l'étude des
fluides homogènes\cite{hansen_theory_2006}.

\newpage{}

\fbox{\begin{minipage}[t]{1\columnwidth}%
Il y a une autre façon de voir cette formulation comme équivalente
à HNC.

En repartant de l'\ref{eq:Fexc_path_final}, on peut écrire une expression
exacte de la fonction $c^{(2)}(\rho_{\lambda};\bm{r},\bm{r}^{\prime})$
comme une série des fonctions de corrélation directe du fluide de
référence d'ordres supérieurs.
\begin{eqnarray*}
c^{(2)}(\rho_{\lambda};\bm{r},\bm{r}^{\prime}) & = & c^{(2)}(\rho_{0};\bm{r},\bm{r}^{\prime})+\lambda\iiint_{\mathbb{R}^{3}}\Delta\rho(\bm{r}_{3})\frac{\delta c^{(2)}(\rho_{0};\bm{r},\bm{r}^{\prime})}{\delta\rho(\bm{r}_{3})}\text{d}\bm{r}_{3}\\
 &  & +\sum_{n=2}^{\infty}\frac{\lambda^{n}}{n!}\int_{\mathbb{R}}\cdot\cdot\int_{\mathbb{R}}\frac{\delta^{n}c^{(2)}(\rho_{0};\bm{r},\bm{r}^{\prime})}{\delta\rho(\bm{r}_{3})\cdot\cdot\delta\rho(\bm{r}_{n+2})}\Delta\rho(\bm{r}_{3})\cdot\cdot\Delta\rho(\bm{r}_{n+2})\text{d}\bm{r}_{3}\cdot\cdot\text{d}\bm{r}_{n+2}\\
c^{(2)}(\rho_{\lambda};\bm{r},\bm{r}^{\prime}) & = & c^{(2)}(\rho_{0};\bm{r},\bm{r}^{\prime})+\lambda\iiint_{\mathbb{R}^{3}}\Delta\rho(\bm{r}_{3})c^{(3)}(\rho_{0};\bm{r},\bm{r}^{\prime},\bm{r}_{3})\text{d}\bm{r}_{3}\\
 &  & \hspace{-3cm}+\sum_{n=2}^{\infty}\frac{\lambda^{n}}{n!}\int_{\mathbb{R}}\cdot\cdot\int_{\mathbb{R}}\Delta\rho(\bm{r}_{3})\cdot\cdot\Delta\rho(\bm{r}_{n+2})c^{(n)}(\rho_{0};\bm{r},\bm{r}^{\prime},\bm{r}_{3},\cdot\cdot,\bm{r}_{n+2})\text{d}\bm{r}_{3}\cdot\cdot\text{d}\bm{r}_{n+2}
\end{eqnarray*}

On voit donc que la fonction de corrélation de paire exacte du fluide
étudié dépend de toutes les fonctions de corrélation directe d'ordre
$n$ du fluide de référence. L'approximation HNC consiste à se limiter
aux corrélations d'ordre 2 dans l'équation de Ornstein-Zernicke et
donc à tronquer cette équation à l'ordre 2. On retrouve alors l'approximation
de l'\ref{eq:approx_c_HRF}.

En clair, l'approximation réalisée revient à négliger toutes les corrélations
d'ordre strictement supérieur à deux, c'est donc une approximation
importante.%
\end{minipage}}

Les équations clés de ce chapitre sur la théorie fonctionnelle de
la densité classique dans l'approximation du fluide de référence sont
récapitulées dans l'encart suivant

\fbox{\begin{minipage}[t]{1\columnwidth}%
\[
\Delta\rho(\bm{r})=\rho(\bm{r})-\rho_{b}
\]
\[
{\cal F}\left[\rho(\bm{r})\right]=\Omega_{v}\left[\rho(\bm{r})\right]-\Omega_{b}={\cal F}_{\text{id}}[\rho(\bm{r})]+{\cal F}_{\text{ext}}[\rho(\bm{r})]+{\cal F}_{\text{exc}}[\rho(\bm{r})]
\]
\[
{\cal F}_{\text{id}}[\rho(\bm{r})]=\text{k}_{\text{B}}\text{T}\iiint_{\mathbb{R}^{3}}\left[\rho(\bm{r})\ln\left(\frac{\rho(\bm{r})}{\rho_{b}}\right)-\Delta\rho(\bm{r})\right]\text{d}\bm{r}
\]
\[
{\cal F}_{\text{ext}}[\rho(\bm{r})]=\iiint_{\mathbb{R}^{3}}\rho(\bm{r})v(\bm{r})\text{d}\bm{r}
\]
\[
{\cal F}_{\text{exc}}[\rho(\bm{r})]=\text{k}_{\text{B}}\text{T}\iiint_{\mathbb{R}^{3}}\iiint_{\mathbb{R}^{3}}\Delta\rho(\bm{r}^{\prime})\Delta\rho(\bm{r})C^{(2)}(\bm{r},\bm{r}^{\prime})\text{d}\bm{r}\text{d}\bm{r}^{\prime}
\]
\end{minipage}}

\lhead[\chaptername~\thechapter]{\rightmark}

\rhead[\leftmark]{}

\lfoot[\thepage]{}

\cfoot{}

\rfoot[]{\thepage}

\chapter{\label{chap:MDFT_dup_multu}Une théorie de la fonctionnelle de la
densité moléculaire pour l'eau}

L'écriture de la fonctionnelle est adaptée à l'étude de la solvatation.
En effet, dans cette écriture est incluse la réponse du solvant à
une perturbation décrite par un potentiel extérieur $v_{\text{ext}}$.
Ce potentiel, dimensionné comme une densité d'énergie, peut avoir
plusieurs origines physiques : un champ électrique, une contrainte
mécanique, etc. Dans le cas de cette thèse, le potentiel extérieur
est une perturbation créée par un soluté.

\section{Modèle d'eau et de soluté}

\subsection{Modèle d'eau rigide}

Le modèle d'eau utilisé est le modèle SPC/E (Single Point Charge/Extended).
Ce modèle d'eau est rigide et se compose d'un site Lennard-Jones sur
l'oxygène et de trois charges partielles placées sur l'oxygène et
les deux hydrogènes. Les paramètres Lennard Jones et les charges partielles
sont indiquées dans le \ref{tab:Param=0000E8tre-du-mod=0000E8leSPC/E}

\begin{table}[h]

\noindent \begin{centering}
\begin{tabular}{|c|c|c|c|}
\hline 
site & $\epsilon$ (kJ.mol$^{-1}$) & $\sigma$ ($\textrm{\AA}$) & $q$ ($e$)\tabularnewline
\hline 
\hline 
O & 0.65 & 3.165 & -0.8476\tabularnewline
\hline 
H & 0.0 & 0.0 & 0.4238\tabularnewline
\hline 
\end{tabular}\protect\caption{Paramètres du modèle d'eau SPC/E\label{tab:Param=0000E8tre-du-mod=0000E8leSPC/E}}

\par\end{centering}

\end{table}

\begin{figure}[h]

\noindent \begin{centering}
\includegraphics[width=0.2\textwidth]{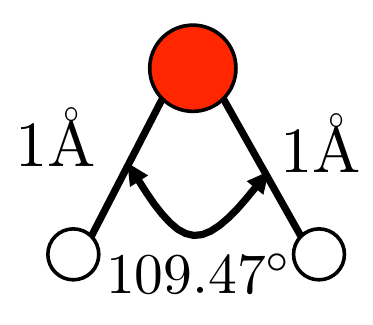}\protect\caption{Géométrie du modèle d'eau SPC/E, l'oxygène est en rouge, les hydrogènes
en blanc.\label{fig:G=0000E9om=0000E9trie-du-mod=0000E8leSPC/E}}

\par\end{centering}

\end{figure}

La géométrie du modèle est donnée en \ref{fig:G=0000E9om=0000E9trie-du-mod=0000E8leSPC/E}.

Avec cette géométrie et ces paramètres, le moment dipolaire d'une
molécule d'eau est de $\mu=$2.35$\,\text{D}$, orienté conventionnellement
des charges négatives (l'oxygène) vers les charges positives (le centre
de masse des hydrogènes). Ce modèle a été choisi car c'est un modèle
rigide, il ne possède donc pas de degré de liberté interne qui compliquerait
beaucoup l'écriture de la fonctionnelle. De plus, il est très utilisé
en simulation moléculaire car il reproduit correctement les fonctions
de distribution radiale, la densité et le coefficient de diffusion
de l'eau liquide obtenus expérimentalement\cite{berendsen_missing_1987}
tout en étant moins coûteux numériquement que des modèles plus évolués.
Du fait de son abondante utilisation on pourra comparer les résultats
de la cDFT à ceux obtenus par MD ou MC dans la littérature. Je n'ai
réalisé aucun calcul de dynamique moléculaire ou de Monte Carlo durant
ma thèse.

\subsection{Modèles de solutés rigides}

Les solutés sont eux aussi décrits par des modèles moléculaires rigides,
composés de différents sites Lennard-Jones et de charges partielles
ou entières. Le potentiel extérieur $v_{\mathrm{ext}}$ agissant sur
une molécule de solvant à la position $\bm{r}$ (on repère en fait
la position du seul site Lennard-Jones de notre modèle d'eau, l'oxygène)
avec une orientation $\bm{\Omega}$ est donc décrit par:

- un potentiel Lennard-Jones entre le solvant et tous les sites du
solutés ;

- un potentiel électrostatique provenant du champ électrique créé
par le soluté.

L'expression de ces potentiels extérieurs et leurs calculs sont décrits
dans le \ref{chap:impl=0000E9mentation}.

\section{Un traitement strictement dipolaire de la polarisation\label{sec:Fexcdip}}

\subsection{Écriture de la fonctionnelle}

\begin{figure}[h]

\noindent \begin{centering}
\includegraphics[width=0.4\linewidth]{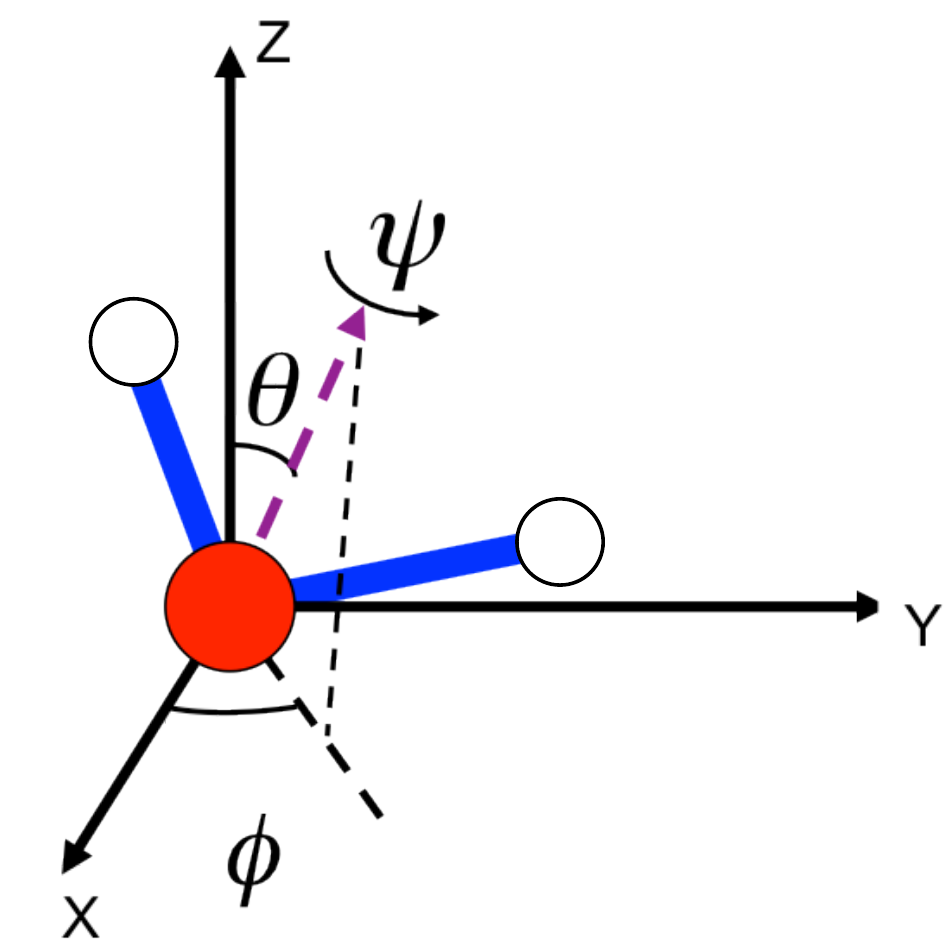}\protect\caption{Représentation des angles d'Euler décrivant l'orientation d'une molécule
d'eau dans le repère fixe du laboratoire.\label{fig:watererpere}}

\par\end{centering}

\end{figure}

Pour un solvant moléculaire, contrairement aux cas présentés dans
les chapitres précédents, le potentiel extérieur $v_{\mathrm{ext}}$
ne dépend plus uniquement des coordonnées spatiales $\bm{r}$, mais
également des orientations des molécules, définies dans le repère
fixe du laboratoire par les trois angles d'Euler $\ensuremath{\bm{\Omega}}=(\theta,\phi,\psi)$.
Ces angles sont définis en \ref{fig:watererpere}.

On minimise donc une fonctionnelle de la densité de solvant $\rho(\bm{r},\bm{\Omega})$
qui dépend également des orientations. Cependant, toutes les expressions
démontrées dans le \ref{chap:cDFT}, sont facilement généralisables
dans le cas de cette densité. 

Avant mon arrivée au laboratoire, le groupe avait déjà proposé une
fonctionnelle développée pour l'étude de la solvatation en milieu
aqueux dans l'approximation du fluide homogène de référence\cite{MDFT_ramirez_density_2002,MDFT_ramirez_density_2005,MDFT_ramirez_direct_2005,MDFT_gendre_classical_2009,MDFT_zhao_molecular_2011,MDFT_levesque_krfmt}.
Celle-ci consiste à partir d'une décomposition de la fonctionnelle
en trois termes
\begin{equation}
{\cal F}\left[\rho(\bm{r},\bm{\Omega})\right]={\cal F}_{\text{id}}\left[\rho(\bm{r},\bm{\Omega})\right]+{\cal F}_{\text{ext}}\left[\rho(\bm{r},\bm{\Omega})\right]+{\cal F}_{\text{exc}}\left[\rho(\bm{r},\bm{\Omega})\right].\label{eq:F=00003DFid+Fext+Fexc}
\end{equation}
Les deux premiers ont pour expression,
\begin{gather}
{\cal F}_{\text{id}}\left[\rho(\bm{r},\bm{\Omega})\right]=\nonumber \\
\text{k}_{\text{B}}\mathrm{T}\iiint_{\mathbb{R}^{3}}\int_{\theta=0}^{\pi}\int_{\phi=0}^{2\pi}\int_{\psi=0}^{2\pi}\left[\rho(\bm{r},\bm{\Omega})\ln\left(\frac{\rho(\bm{r},\bm{\Omega})}{\rho_{b}}\right)-\rho(\bm{r},\bm{\Omega})+\rho_{\mathrm{b}}\right]\text{d}\bm{r}\text{d}\bm{\Omega}\label{eq:FidRho(r,Omega)}
\end{gather}
et
\begin{equation}
{\cal F}_{\text{ext}}\left[\rho(\bm{r},\bm{\Omega})\right]=\iiint_{\mathbb{R}^{3}}\int_{\theta=0}^{\pi}\int_{\phi=0}^{2\pi}\int_{\psi=0}^{2\pi}\rho(\bm{r},\bm{\Omega})v_{\text{ext}}(\bm{r},\bm{\Omega})\text{d}\bm{r}\text{d}\bm{\Omega}\label{eq:Fext(rho(r,Omega))}
\end{equation}
Il s'agit simplement des réécritures des \ref{eq:Fiddeltarho} et
\ref{eq:Fexc} avec une densité qui dépend aussi des angles. Notons
que dans cette formule, $\rho_{\mathrm{b}}$ n'est pas la densité
en nombre du fluide homogène (que l'on notera à partir de maintenant
$n_{\mathrm{b}}$), mais la densité en nombre par unité d'angle solide,
$\rho_{\mathrm{b}}=n_{\mathrm{b}}/(8\pi^{2})$. 

Dans le cadre du développement par rapport au fluide homogène de référence,
la partie d'excès peut s'écrire à partir de la fonction de corrélation
directe du solvant homogène.
\begin{gather}
{\cal F}_{\text{exc}}\left[\rho(\bm{r},\bm{\Omega})\right]=\label{eq:Fexc_c(2,r1,r2,O1,O2)_HRF}\\
-\frac{\text{k}_{\text{B}}\text{T}}{2}\iiint_{\mathbb{R}^{3}}\iiint_{\mathbb{R}^{3}}\iiint_{8\pi^{2}}\iiint_{8\pi^{2}}\mathrm{\Delta}\rho(\bm{r}_{1},\bm{\Omega}_{1})c^{(2)}(\bm{r}_{1},\bm{r}_{2},\bm{\Omega}_{1},\bm{\Omega}_{2})\mathrm{\Delta}\rho(\bm{r}_{2},\bm{\Omega}_{2})\text{d}\bm{r}_{1}\text{d}\bm{\Omega}_{1}\text{d}\bm{r}_{2}\text{d}\bm{\Omega}_{2},\nonumber 
\end{gather}
où pour une écriture plus compacte on a défini : $\iiint_{8\pi^{2}}\equiv\int_{\theta=0}^{\pi}\int_{\phi=0}^{2\pi}\int_{\psi=0}^{2\pi}$.
L'expression donnée en \ref{eq:Fexc_c(2,r1,r2,O1,O2)_HRF} n'est pas
utilisée directement pour deux raisons. D'une part on ne connait pas
facilement la fonction de corrélation $c^{(2)}(\bm{r}_{1},\bm{r}_{2},\bm{\Omega}_{1},\bm{\Omega}_{2})$.
D'autre part cette fonctionnelle s'avère numériquement coûteuse à
calculer car elle requiert une intégration double sur le volume et
la sphère unité. Pour ces deux raisons, il n'est pas immédiatement
envisagé de calculer le terme d'excès en utilisant cette approche.

On peut réaliser un développement en invariants rotationnels de la
fonction de corrélation,
\begin{equation}
c^{(2)}\left(\bm{r}_{12},\bm{\Omega}_{1},\bm{\Omega}_{2}\right)=\sum_{m,n,l,\mu,\nu}c_{\mu\nu}^{mnl}\left(r\right)\Phi_{\mu\nu}^{mnl}\left(\bm{\Omega}_{1},\bm{\Omega}_{2},\bm{r}_{12}\right),
\end{equation}
où $\Phi_{\mu\nu}^{mnl}$ sont les invariants rotationnels déjà introduits
au \ref{chap:1}. Le développement en invariants rotationnels a été
formulé pour séparer les parties radiale et angulaire \cite{blum_invariant2_1972,blum_invariant_1972}. 

Dans l'hypothèse d'un solvant strictement dipolaire, c'est-à-dire
où l'on considère que la polarisation microscopique d'une molécule
d'eau peut être représentée par un dipôle orienté selon l'axe C$_{2}$
de la molécule, on peut, avec une bonne approximation\cite{blum_invariant_1972,patey_integral_1977,blum_invariant2_1972},
se limiter au développement à l'ordre 2 avec $\mu=\nu=0$. On ne considère
alors que les invariants suivants :
\begin{eqnarray*}
\mathrm{\Phi}^{000}(\bm{\Omega}_{1},\bm{\Omega}_{2},\bm{r}_{12}) & = & 1\\
\mathrm{\Phi}^{101}(\bm{\Omega}_{1},\bm{\Omega}_{2},\bm{r}_{12}) & = & -\bm{\Omega}_{1}\cdot\tilde{\bm{u}}_{12}\\
\mathrm{\Phi}^{011}(\bm{\Omega}_{1},\bm{\Omega}_{2},\bm{r}_{12}) & = & \bm{\Omega}_{2}\cdot\tilde{\bm{u}}_{12}\\
\mathrm{\Phi}^{110}(\bm{\Omega}_{1},\bm{\Omega}_{2},\bm{r}_{12}) & = & \bm{\Omega}_{1}\cdot\bm{\Omega}_{2}\\
\mathrm{\Phi}^{112}(\bm{\Omega}_{1},\bm{\Omega}_{2},\bm{r}_{12}) & = & 3(\bm{\Omega}_{1}\cdot\tilde{\bm{u}}_{12})(\bm{\Omega}_{2}\cdot\tilde{\bm{u}}_{12})-\bm{\Omega}_{1}\cdot\bm{\Omega}_{2},
\end{eqnarray*}
avec $\tilde{\bm{u}}_{12}=(\bm{r}_{2}-\bm{r}_{1})/\left\Vert \bm{r}_{2}-\bm{r}_{1}\right\Vert $
le vecteur unitaire reliant les positions des deux molécules. Cela
donne, si on réinjecte ce développement dans la partie d'excès : \foreignlanguage{english}{
\begin{gather}
{\cal F}_{\text{exc}}[n(\bm{r}),\bm{P}(\bm{r})]=\nonumber \\
-\frac{1}{2}\text{k}_{\text{B}}\text{T}\iiint_{\mathbb{R}^{3}}\iiint_{\mathbb{R}^{3}}\left[c_{000}(\left\Vert \bm{r}_{12}\right\Vert )\mathrm{\Delta}n(\bm{r}_{1})\mathrm{\Delta}n(\bm{r}_{2})\right]\text{d}\bm{r}_{1}\text{d}\bm{r}_{2}\nonumber \\
-\frac{1}{2\mu}\text{k}_{\text{B}}\text{T}\iiint_{\mathbb{R}^{3}}\iiint_{\mathbb{R}^{3}}\left[c_{101}(\left\Vert \bm{r}_{12}\right\Vert )\mathrm{\Delta}n(\bm{r}_{1})\bm{P}(\bm{r}_{2})\cdot\tilde{\bm{u}}_{12}\right]\text{d}\bm{r}_{1}\text{d}\bm{r}_{2}\nonumber \\
+\frac{1}{2\mu}\text{k}_{\text{B}}\text{T}\iiint_{\mathbb{R}^{3}}\iiint_{\mathbb{R}^{3}}\left[c_{011}(\left\Vert \bm{r}_{12}\right\Vert )\bm{P}(\bm{r}_{1})\mathrm{\Delta}n(\bm{r}_{2})\cdot\tilde{\bm{u}}_{12}\right]\text{d}\bm{r}_{1}\text{d}\bm{r}_{2}\nonumber \\
-\frac{1}{2\mu^{2}}\text{k}_{\text{B}}\text{T}\iiint_{\mathbb{R}^{3}}\iiint_{\mathbb{R}^{3}}\left[c_{110}(\left\Vert \bm{r}_{12}\right\Vert )\bm{P}(\bm{r}_{1})\cdot\bm{P}(\bm{r}_{2})\right]\text{d}\bm{r}_{1}\text{d}\bm{r}_{2}\nonumber \\
-\frac{1}{2\mu^{2}}\text{k}_{\text{B}}\text{T}\iiint_{\mathbb{R}^{3}}\iiint_{\mathbb{R}^{3}}c_{112}(\left\Vert \bm{r}_{12}\right\Vert )\left[3\left(\bm{P}(\bm{r}_{1})\cdot\tilde{\bm{u}}_{12}\right)\left(\bm{P}(\bm{r}_{2})\cdot\tilde{\bm{u}}_{12}\right)-\bm{P}(\bm{r}_{1})\cdot\bm{P}(\bm{r}_{2})\right]\text{d}\bm{r}_{1}\text{d}\bm{r}_{2}\nonumber \\
+{\cal F}_{\mathrm{cor}}[n(\bm{r}),\bm{P}(\bm{r})]\label{eq:Fexc_exacte}
\end{gather}
}où $n(\bm{r})$ est la densité de solvant à une particule qui ne
dépend plus que des coordonnées d'espace,
\begin{equation}
n(\bm{r})=\int_{\theta=0}^{\pi}\int_{\phi=0}^{2\pi}\int_{\psi=0}^{2\pi}\rho(\bm{r,\Omega})\text{d}\bm{\Omega},
\end{equation}
et $\bm{P}(\bm{r})$ le champ de polarisation dipolaire défini comme,\foreignlanguage{english}{
\begin{equation}
\bm{P(r)}=\mu\int_{\theta=0}^{\pi}\int_{\phi=0}^{2\pi}\int_{\psi=0}^{2\pi}\tilde{\bm{n}}\rho(\bm{r},\bm{\Omega})\text{d}\bm{\Omega},\label{eq:Pola_dip}
\end{equation}
}avec $\bm{\mu}$ le moment dipolaire d'une molécule d'eau de norme
$\mu$ et $\tilde{\bm{n}}=\bm{\mu}/\mu$ le vecteur orientation unitaire
de ce moment dipolaire.

Le premier terme du membre de droite de l'\ref{eq:Fexc_exacte} est
un terme qui corrèle les densités entre elles, il est donc lié au
facteur de structure. Les deux termes suivants sont des termes de
couplage entre polarisation et densité. Les deux termes suivants couplent
les densités de polarisation et peuvent donc être reliés aux permittivités
diélectriques non-locales du fluide pur.

Enfin, le dernier terme ${\cal F}_{\mathrm{cor}}[n(\bm{r}),\bm{P}(\bm{r})]$
est le terme correctif de cette approximation, qui est supposé ne
dépendre que de $n(\bm{r})$ et $\bm{P}(\bm{r})$. Il est défini comme
la différence entre la partie d'excès exacte et les termes du développement
à l'ordre 2. On l'appellera fonctionnelle de bridge par identification
au terme de bridge des équations intégrales.

\fbox{\begin{minipage}[t]{1\columnwidth}%
Dans la théorie des liquides pour un fluide homogène et isotrope,
la fonction de corrélation directe, $c^{(2)}$, et la fonction de
corrélation de paires $h$,
\begin{equation}
g(r_{12})=h(r_{12})+1,
\end{equation}

sont reliées par l'équation de Ornstein-Zernike.
\begin{equation}
h(r_{12})=c^{(2)}(r_{12})+n_{\mathrm{b}}\iiint_{\mathbb{R}^{3}}c^{(2)}(r_{13})h(r_{32})\text{d}\bm{r}_{3}.
\end{equation}
Si le potentiel d'interaction entre particules du fluide est un potentiel
de paire additif, $u(r_{12})$, une autre relation exacte peut être
écrite par expansion diagrammatique\cite{duh_integral_1995,hansen_theory_2006},
\begin{equation}
h(r)=\exp\left[-\beta u(r)+h(r)-c(r)+B(r)\right]-1,\label{eq:h=00003Dexp-bu...}
\end{equation}
où $B$ est appelée fonction bridge. Si cette fonction est connue
on a une expression exacte de la fonction de corrélation totale. Cependant,
cette fonction bridge n'a pas d'expression connue dans le cas général.
Une approximation de l'\ref{eq:h=00003Dexp-bu...} s'appelle une relation
de fermeture et en résolvant de manière conjointe cette équation de
fermeture et l'équation de Ornstein-Zernike on peut calculer les différentes
fonctions de corrélation. Une approximation classique, l'approximation
HNC, consiste à prendre cette fonction bridge nulle.

Si on suppose que ${\cal F}_{\mathrm{cor}}[n(\bm{r}),\bm{P}(\bm{r})]$
est nulle dans l'\ref{eq:Fexc_exacte}, on retrouve la fonctionnelle
d'excès dans l'approximation HRF qui est équivalente à l'approximation
HNC. On voit le rôle similaire que joue la fonctionnelle ${\cal F}_{\mathrm{cor}}[n(\bm{r}),\bm{P}(\bm{r})]$
dans la cDFT et la fonction bridge dans la théorie des liquides. C'est
le terme à ajouter à l'approximation HNC pour avoir une solution exacte. %
\end{minipage}}

En première approximation, on peut négliger les termes de couplage
entre la polarisation et la densité, ce qui revient à omettre les
termes qui dépendent de $c_{011}$ et $c_{101}$ dans l'\ref{eq:Fexc_exacte}.
On peut remarquer que les termes en $c_{110}$ et $c_{112}$ ont une
forme ressemblant au potentiel d'interaction dipôle-dipôle.

Les parties idéales et extérieures peuvent être calculées relativement
facilement. La partie d'excès, telle qu'elle est écrite en \ref{eq:Fexc_exacte},
est désormais beaucoup moins coûteuse à utiliser numériquement: contrairement
à l'\ref{eq:Fexc_c(2,r1,r2,O1,O2)_HRF}, elle ne requiert plus d'intégration
sur les angles. La double intégration sur l'espace est calculée efficacement
par transformées de Fourier rapides (FFT), voir la \ref{sec:Algorithme-du-code}.
Pour calculer la partie d'excès, il est nécessaire de connaître les
3 projections des fonctions de corrélation directe utilisées $c_{000,}\ c_{110},\ c_{112}$,
au lieu de la fonction de corrélation $c^{(2)}(\bm{r}_{1},\bm{r}_{2},\bm{\Omega}_{1},\bm{\Omega}_{2})$.
Cette fonction, qui dépend de douze variables est particulièrement
compliquée à déterminer par simulation numérique. Le problème est
donc considérablement simplifié puisqu'on le ramène à la détermination
de trois fonctions à une seule variable. La détermination de ces fonctions
peut se faire à partir de simulations de dynamique moléculaire de
deux manières, soit en résolvant l'équation de Ornstein-Zernike dans
l'espace de Fourier\cite{GendreThesis}, soit dans l'espace direct
en résolvant les équations de Baxter\cite{MDFT_ramirez_direct_2005,zhao_accurate_2013}
; ce qui a été fait pour l'eau au laboratoire avant mon arrivée. Ces
fonctions peuvent également être déterminées par des mesures expérimentales.

\begin{figure}[h]
\noindent \centering{}\includegraphics[width=0.6\linewidth]{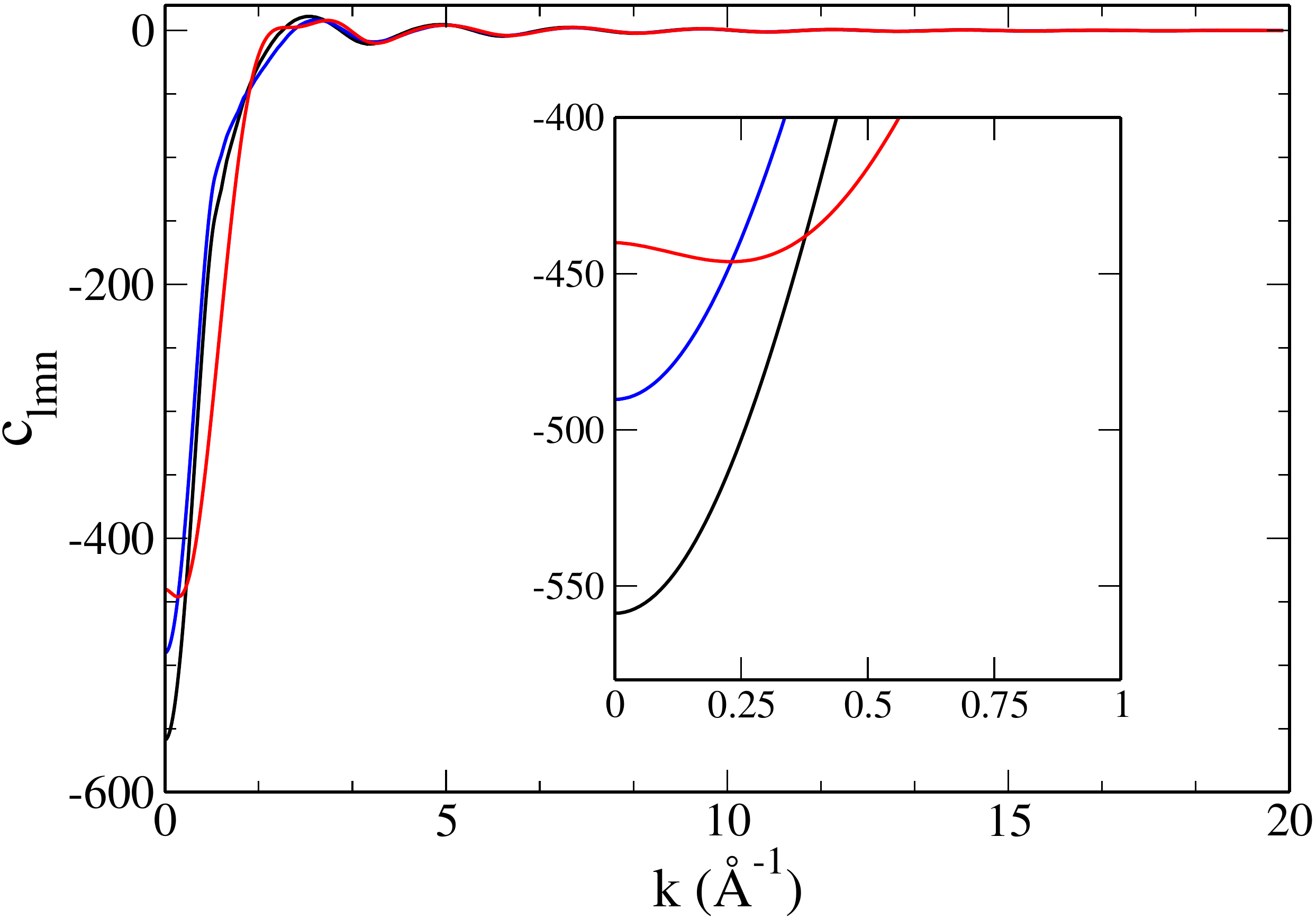}\protect\caption{Composantes des projections de la fonction de corrélation directe
de l'eau SPC/E, $c_{000}$ est en rouge, $c_{110}$ en noir et $c_{112}$
en bleu. Dans l'encart est représenté un zoom entre 0 et 1 $\textrm{\AA}^{-1}$.
\label{fig:clmn}}
\end{figure}

Dans le cas où on néglige les couplages entre polarisation et densité,
c'est-à-dire si on suppose que les termes $c_{011}$ et $c_{101}$
sont nuls, les transformées de Fourier de la fonction $c_{000}$ et
celle du facteur de structure sont directement liées\cite{hansen_theory_2006},
\begin{equation}
\hat{S}(\bm{k})=\frac{1}{1-n_{b}\hat{c}_{000}(\bm{k})}.
\end{equation}
Les transformées de Fourier de $c_{110}$ et $c_{112}$ sont quant
à elles reliées aux composantes transverse et longitudinale des permittivités
diélectriques\cite{raineri_static_1992,raineri_static_1993} :
\begin{equation}
\frac{n_{b}}{3}\left(\hat{c}_{110}(\bm{k})+2\hat{c}_{112}(\bm{k})\right)=1-\frac{4\beta\mu^{2}n_{b}}{3\epsilon_{0}\left(1-\hat{\epsilon}_{L}^{-1}(\bm{k})\right)}
\end{equation}
 et
\begin{equation}
\frac{n_{b}}{3}\left(\hat{c}_{110}(\bm{k})-\hat{c}_{112}(\bm{k})\right)=1-\frac{4\beta\mu^{2}n_{b}}{3\epsilon_{0}\left(\hat{\epsilon_{T}}(\bm{k})-1\right)}.
\end{equation}
En déterminant expérimentalement les deux composantes des permittivités
diélectriques et le facteur de structure, on peut ensuite calculer
ces fonctions de corrélation.

\fbox{\begin{minipage}[t]{1\columnwidth}%
Les fonctions de corrélation ne doivent être déterminées qu'une seule
fois pour chaque solvant. On peut ensuite étudier la solvatation de
n'importe quel soluté dans ce solvant puisque celui-ci n'intervient
que comme une perturbation, au travers du champ extérieur. %
\end{minipage}}

On va maintenant tester la qualité de cette fonctionnelle dipolaire
en étudiant les propriétés de solvatation de divers solutés en milieux
aqueux. Pour l'instant, on fait l'hypothèse simplificatrice que le
terme de bridge ${\cal F}_{\mathrm{cor}}$ est nul. On généralisera
par la suite au cas d'une fonctionnelle multipolaire tenant compte
de la distribution de charge complète des molécules d'eau.

\subsection{Structure de solvatation}

Nous allons tester la fonctionnelle précédemment introduite pour évaluer
sa précision pour la prédiction des propriétés structurales de solvatation
d'ions et de petites molécules en milieu aqueux. Les fonctions de
distribution radiale calculées à partir des densités d'équilibre obtenues
par minimisation de la fonctionnelle seront comparées aux résultats
exacts obtenus par dynamique moléculaire. Les études ont été réalisées
en utilisant les mêmes modèles de solutés et de solvants, \textit{i.e.},
les mêmes potentiels d'interaction dans les simulations MD et dans
les calculs MDFT.

\subsubsection{Application aux solutés apolaires}

\begin{figure}[h]
\noindent \centering{}\includegraphics[width=0.8\textwidth]{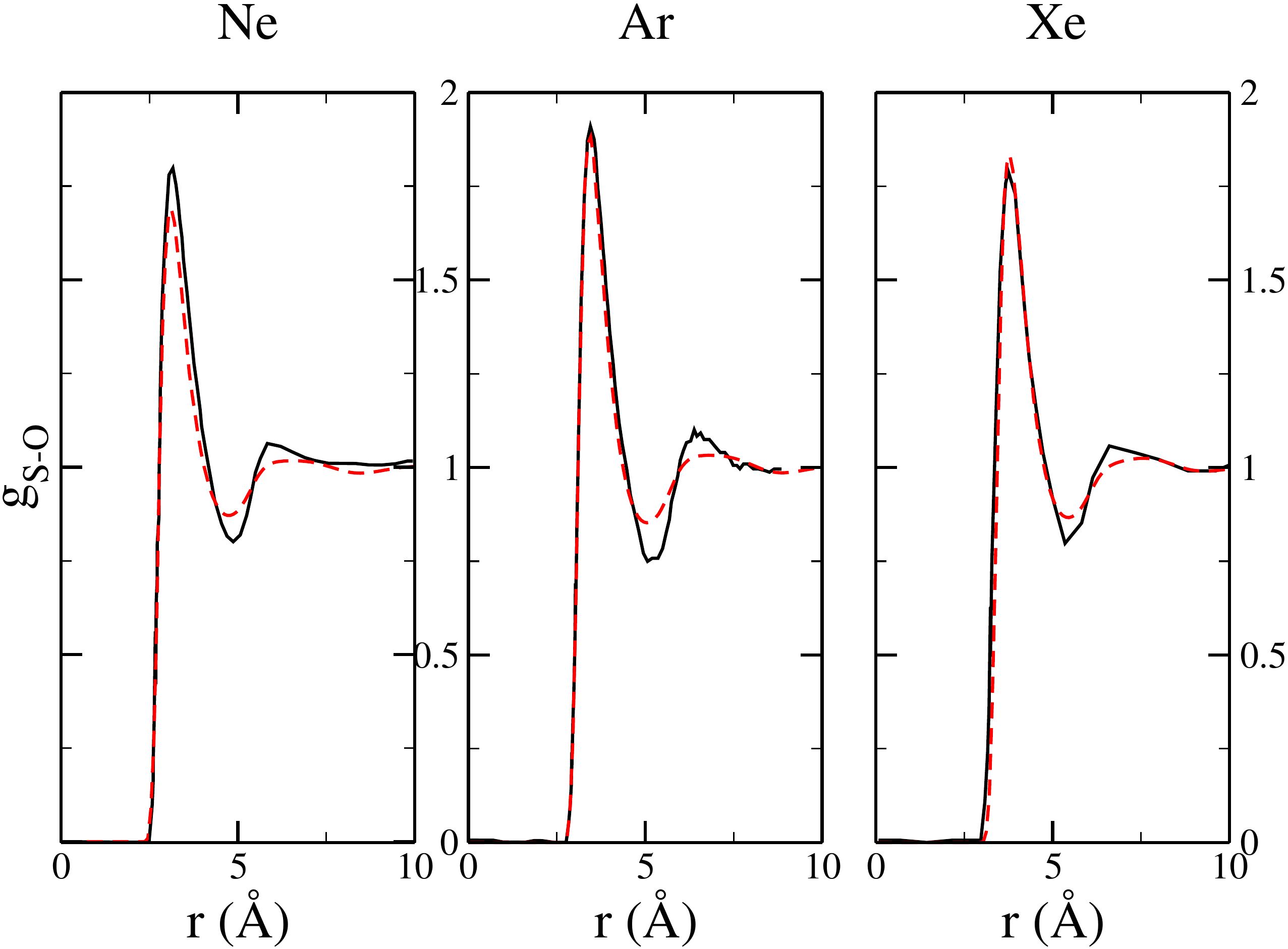}\protect\caption{Fonctions de distribution radiale entre l'oxygène de l'eau et les
atomes de néon, d'argon et de xénon. Les résultats obtenus par dynamique
moléculaire\cite{guillot_computer_1993,guillot_computersimulation_1991}
sont représentés par une ligne noire continue, tandis que les résultats
obtenus par MDFT sont présentés par une ligne rouge discontinue.\label{fig:rdf gaz rare dipol }}
\end{figure}

On étudie la solvatation en milieu aqueux supposé infiniment dilué
des atomes d'argon, néon et xénon. Ces atomes sont neutres et les
paramètres du champ de force Lennard-Jones\cite{guillot_computer_1993}
représentant l'interaction entre atomes de gaz rares et solvant sont
donnés dans le \ref{tab:Param gaz rare}. Les fonctions de distribution
radiale obtenues par minimisation de la fonctionnelle ainsi que celles
obtenues par dynamique moléculaire\cite{guillot_computer_1993,guillot_computersimulation_1991}
sont présentées en \ref{fig:rdf gaz rare dipol }. Dans les trois
cas, l'accord entre MDFT et MD est satisfaisant puisque les positions
des extrema de ces fonctions sont très bien reproduites, ainsi que
la hauteur du pic correspondant à la première couche de solvatation.
On sous-estime légèrement la hauteur du pic dû à la seconde couche
de solvatation, ainsi que la déplétion entre les deux maxima. On peut
remarquer que l'accord est meilleur à mesure que la taille des solutés
augmente.

\begin{table}[h]
\noindent \begin{centering}
\begin{tabular}{|c|c|c|c|}
\hline 
Atome & $\sigma\ (\lyxmathsym{\AA})$ & $\epsilon\ (\text{kJ.mo\ensuremath{l^{-1}}})$ & $q\ (e)$\tabularnewline
\hline 
\hline 
Ne & 3.035 & 0.15432 & 0.0\tabularnewline
\hline 
Ar & 3.415 & 1.03891 & 0.0\tabularnewline
\hline 
Xe & 3.975 & 1.78510 & 0.0\tabularnewline
\hline 
\end{tabular}\protect\caption{Champ de force pour les modèles de gaz rares étudiés. \label{tab:Param gaz rare}}

\par\end{centering}

\end{table}

\begin{figure}[h]
\noindent \begin{centering}
\includegraphics[width=0.8\textwidth]{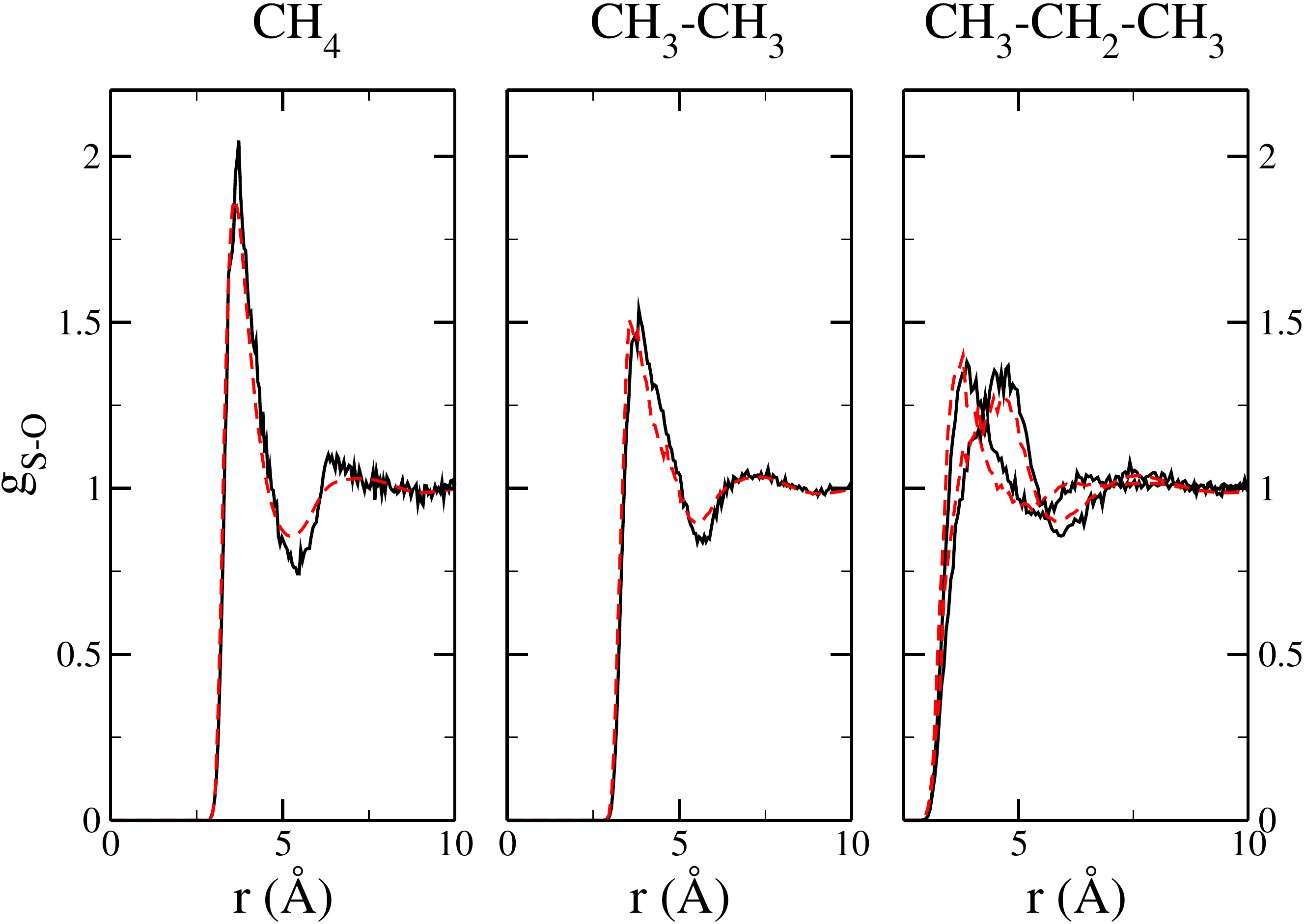}\protect\caption{Fonctions de distribution radiale entre l'oxygène de l'eau et les
groupements $\mathrm{CH_{n}}$ des trois premiers alcanes. Les résultats
obtenus par dynamique moléculaire\cite{jorgensen_optimized_1984}
sont représentés par une ligne noire continue, tandis que les résultats
obtenus par MDFT sont présentés par une ligne rouge discontinue.\label{fig:rdf alcane rare dipol }}

\par\end{centering}

\end{figure}

On s'intéresse ensuite à des solutés apolaires plus complexes car
non-sphériques en examinant la série des six premiers alcanes linéaires.
Les paramètres de potentiel d'interactions sont ceux du champ de force
OPLS (Optimized Intermolecular Potential Functions for Liquid Hydrocarbons)\cite{jorgensen_optimized_1984}
et sont donnés dans le \ref{tab:ParamOPLSalcane}. Dans ce modèle
les groupements $\mathrm{CH_{n}}$ sont représentés par une seule
sphère de Lennard-Jones. La comparaison entre les fonctions de distribution
radiale obtenues par MD et MDFT est donnée en \ref{fig:rdf alcane rare dipol }.
Seuls les résultats pour les trois premiers alcanes sont présentés.
Là encore, l'accord entre MD et MDFT est satisfaisant. On reproduit
correctement la position des première et deuxième sphères de solvatation
(maxima), ainsi que la hauteur de ces pics. La déplétion entre les
deux pics est encore une fois légèrement sous-estimée. 

\begin{table}[h]

\noindent \centering{}%
\begin{tabular}{|c|c|c|c|}
\hline 
Site & $\sigma\ (\textrm{\AA})$ & $\epsilon\ (\text{kJ.mol\ensuremath{{}^{-1}}})$ & $q\ (e)$\tabularnewline
\hline 
\hline 
$\mathrm{CH_{4}}$ (méthane) & 3.73 & 1.23 & 0.0\tabularnewline
\hline 
$\mathrm{CH_{3}}$(éthane) & 3.775 & 0.8661 & 0.0\tabularnewline
\hline 
$\mathrm{CH_{3}}$ & 3.905 & 0.732 & 0.0\tabularnewline
\hline 
$\mathrm{CH_{2}}$ & 3.905 & 0.494 & 0.0\tabularnewline
\hline 
\end{tabular}\protect\caption{Champ de force OPLS pour les alcanes étudiés. Les valeurs données
dans les deux dernières lignes sont communes au propane, au butane,
au pentane et à l'hexane.\label{tab:ParamOPLSalcane}}
\end{table}

\subsubsection{Application à des solutés hydrophiles\label{sub:solut=0000E9s hydrophyles sans F3B}}

La fonctionnelle dipolaire utilisée donne de bons résultats pour des
solutés neutres. On s'intéresse maintenant à la solvatation de molécules
chargées: l'eau et la N-méthylacétamide qui est un modèle de liaison
peptidique.

\begin{figure}[h]
\noindent \begin{centering}
\includegraphics[width=0.8\textwidth]{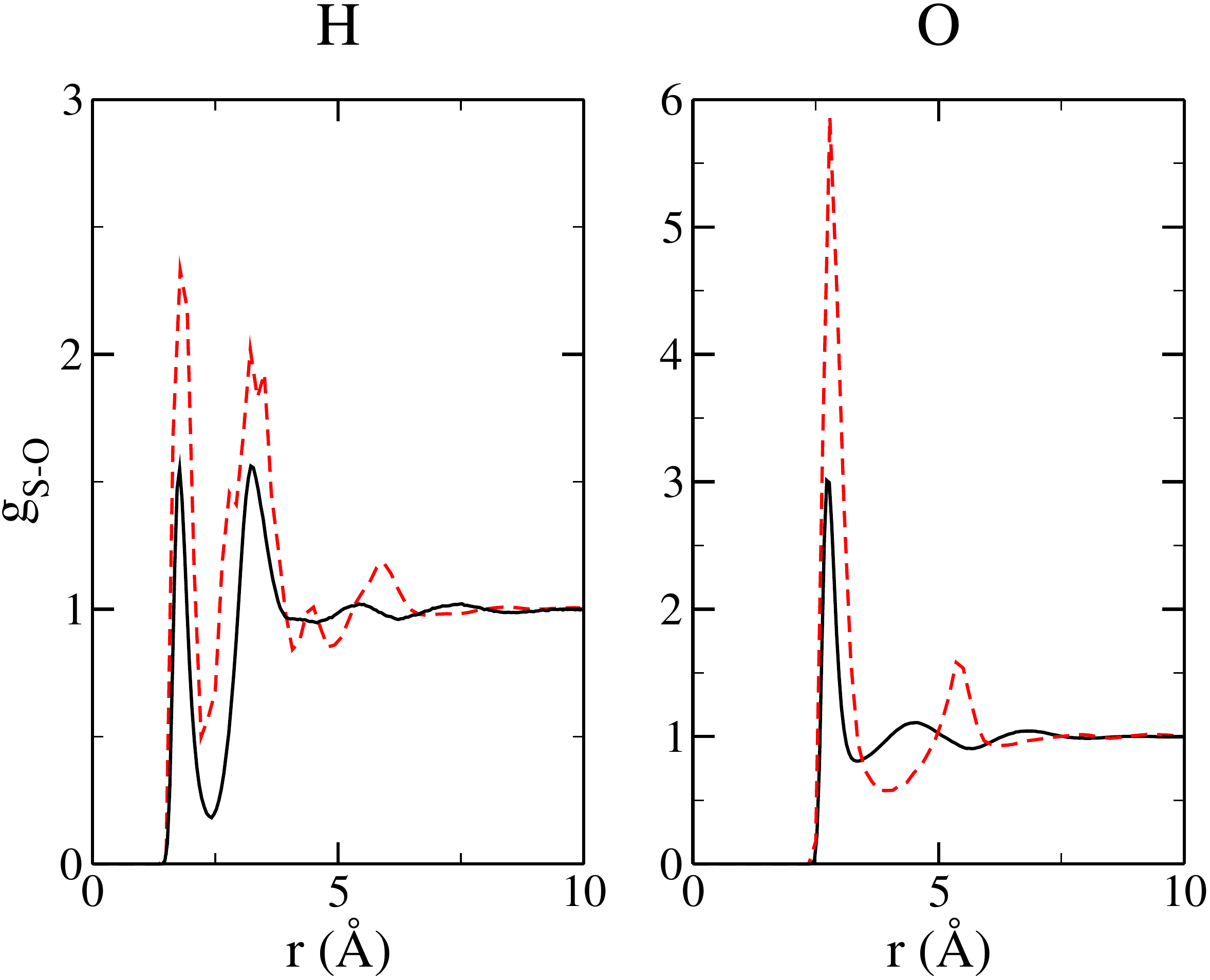}\protect\caption{Fonctions de distribution radiale entre l'oxygène de l'eau solvant
et l'oxygène et l'hydrogène de l'eau soluté. MD en trait noir plein,
MDFT en tirets rouges.\label{fig:rdf_water_dip}}

\par\end{centering}

\end{figure}

Le modèle utilisé pour le soluté eau est le même que celui utilisé
pour le solvant, SPC/E. Les fonctions de distribution radiale entre
l'oxygène du solvant et les oxygènes et hydrogènes de la molécule
d'eau soluté, sont présentées en \ref{fig:rdf_water_dip}. La minimisation
de la fonctionnelle ne permet pas de reproduire des résultats similaires
à ceux obtenus par MD.

En effet, dans la fonction de distribution radiale oxygène-oxygène
calculée par MD, le premier pic, correspondant à la première couche
de solvatation est plus haut et plus fin que pour un fluide de Lennard-Jones
(par exemple ceux de la \ref{fig:rdf gaz rare dipol }). 

Le second pic, qui correspond à la seconde couche de solvatation,
est plus proche du premier pic que dans le fluide de Lennard-Jones,
où il se trouve à peu près à deux fois la distance du premier pic.

Ce profil particulier peut s'expliquer par la géométrie tétraédrique
de l'eau liquide. Cet arrangement spatial particulier se voit aussi
sur la fonction de distribution radiale hydrogène-oxygène qui présente
deux pics marqués qui proviennent des deux hydrogènes des molécules
d'eau situées dans la première couche de solvatation. Si l'eau ne
possédait pas de structure spatiale localement tétraédrique on ne
distinguerait pas ces deux pics mais un seul, dû à une contribution
moyenne des hydrogènes.

Pour les résultats obtenus par minimisation fonctionnelle, la position
des pics de la première couche de solvatation est partiellement reproduite
mais leurs largeurs et leurs hauteurs ne sont pas satisfaisantes.
Les pics décrivant la seconde couche de solvatation sont déplacés
et surestimés. 

\begin{figure}[h]
\noindent \centering{}\includegraphics[width=1\textwidth]{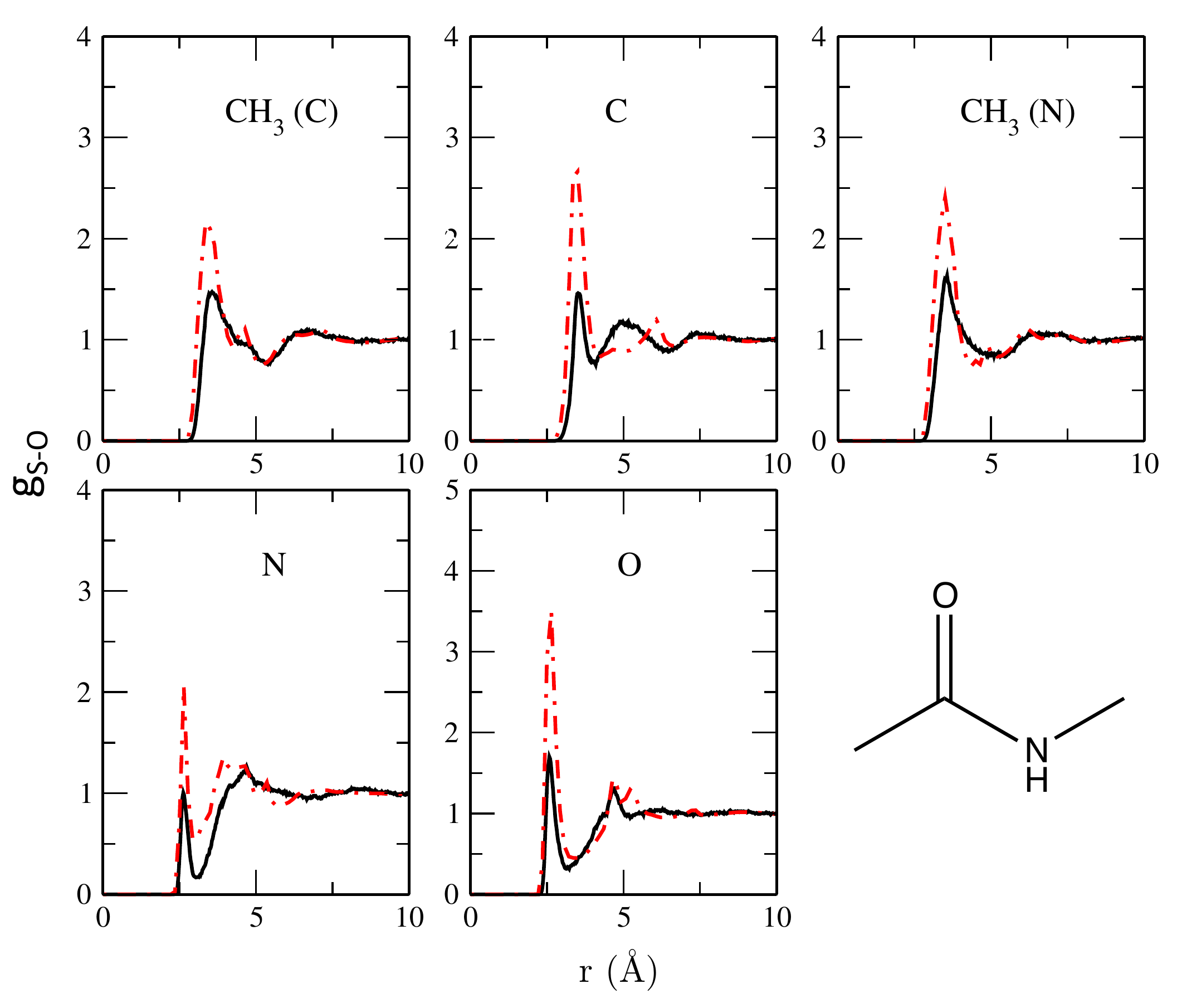}\protect\caption{Fonctions de distribution radiale entre l'oxygène de l'eau solvant
et les différents sites de la molécule NMA. La légende est la même
que précédemment. \label{fig:rdf_NMA_dip}}
\end{figure}

La fonction de distribution radiale oxygène-oxygène obtenue par MDFT
pour l'eau est en fait similaire à celle obtenue pour une sphère Lennard-Jones.
La théorie de la fonctionnelle de la densité présentée ici échoue
donc à reproduire l'ordre tétraédrique.
On peut espérer que ce problème soit inhérent à \og l'eau dans l'eau \fg{}.
On étudie donc la solvatation d'autres petites molécules polaires.
On donne ici l'exemple de la N-méthylacétamide (NMA) avec les paramètres
de potentiel d'interaction donnés dans le \ref{tab:ParaNMA}
\begin{table}[h]

\noindent \centering{}%
\begin{tabular}{|c|c|c|c|}
\hline 
Site & $\sigma\ (\textrm{\AA})$ & $\epsilon\ (\text{kJ.mol\ensuremath{{}^{-1}}})$ & $q\ (e)$\tabularnewline
\hline 
\hline 
$\text{CH\ensuremath{{}_{3}}\ (C)}$ & 3.91 & 0.160 & 0.00\tabularnewline
\hline 
C & 3.75 & 0.105 & 0.50\tabularnewline
\hline 
O & 2.96 & 0.210 & -0.50\tabularnewline
\hline 
N & 3.25 & 0.170 & -0.57\tabularnewline
\hline 
H & 0.00 & 0.000 & 0.37\tabularnewline
\hline 
$\text{CH\ensuremath{{}_{3}}\ (N)}$ & 3.80 & 0.170 & 0.20\tabularnewline
\hline 
\end{tabular}\protect\caption{Paramètres Lennard-Jones et charges partielles de la N-Méthylacétamide.\label{tab:ParaNMA}}
\end{table}
Les fonctions de distribution radiale obtenues par MD et minimisation
fonctionnelle sont données en \ref{fig:rdf_NMA_dip}. Les résultats
obtenus sont à rapprocher des précédents. Pour les sites peu ou non
chargés, c'est-à-dire les groupements méthyles, l'accord entre MD
et MDFT est correct quoique moins bon que dans le cas des solutés
apolaires, alors que pour les sites chargés, l'azote, l'oxygène et
le carbone de la fonction amide, les premiers pics sont surestimés
et les seconds sont situés à une distance trop grande par rapport
aux résultats de la MD, comme dans le cas de l'eau.

\begin{figure}[h]
\noindent \begin{centering}
\includegraphics[width=0.8\textwidth]{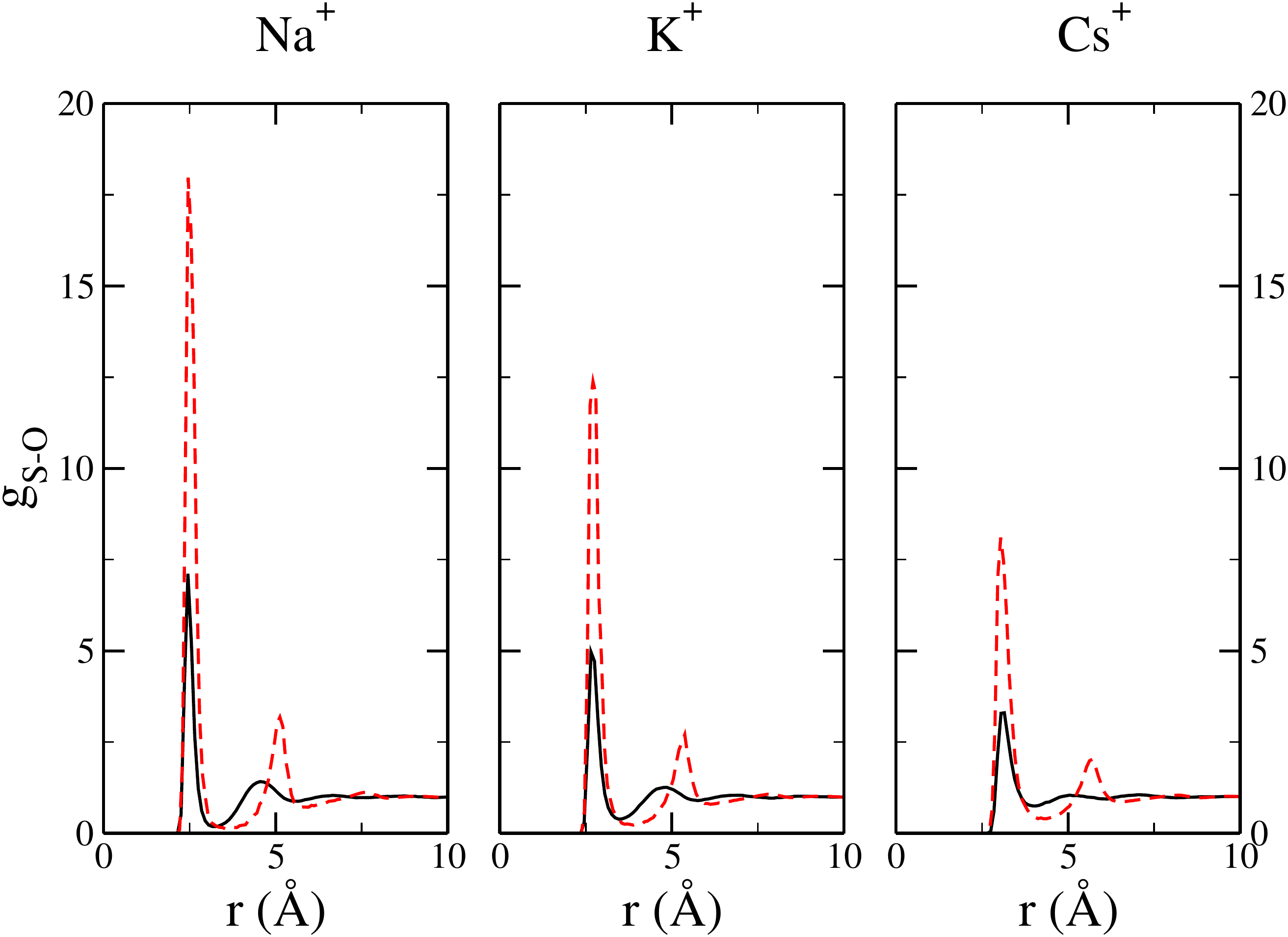}\protect\caption{Fonctions de distribution radiale entre l'oxygène de l'eau solvant
et trois cations alcalins : sodium, potassium et césium. Les résultats
des simulations MD sont en trait noir plein et ceux des calculs MDFT
en tirets rouges.\label{fig:rdf_alcalins_dip}}

\par\end{centering}

\end{figure}

\begin{figure}[h]
\noindent \centering{}\includegraphics[width=0.8\textwidth]{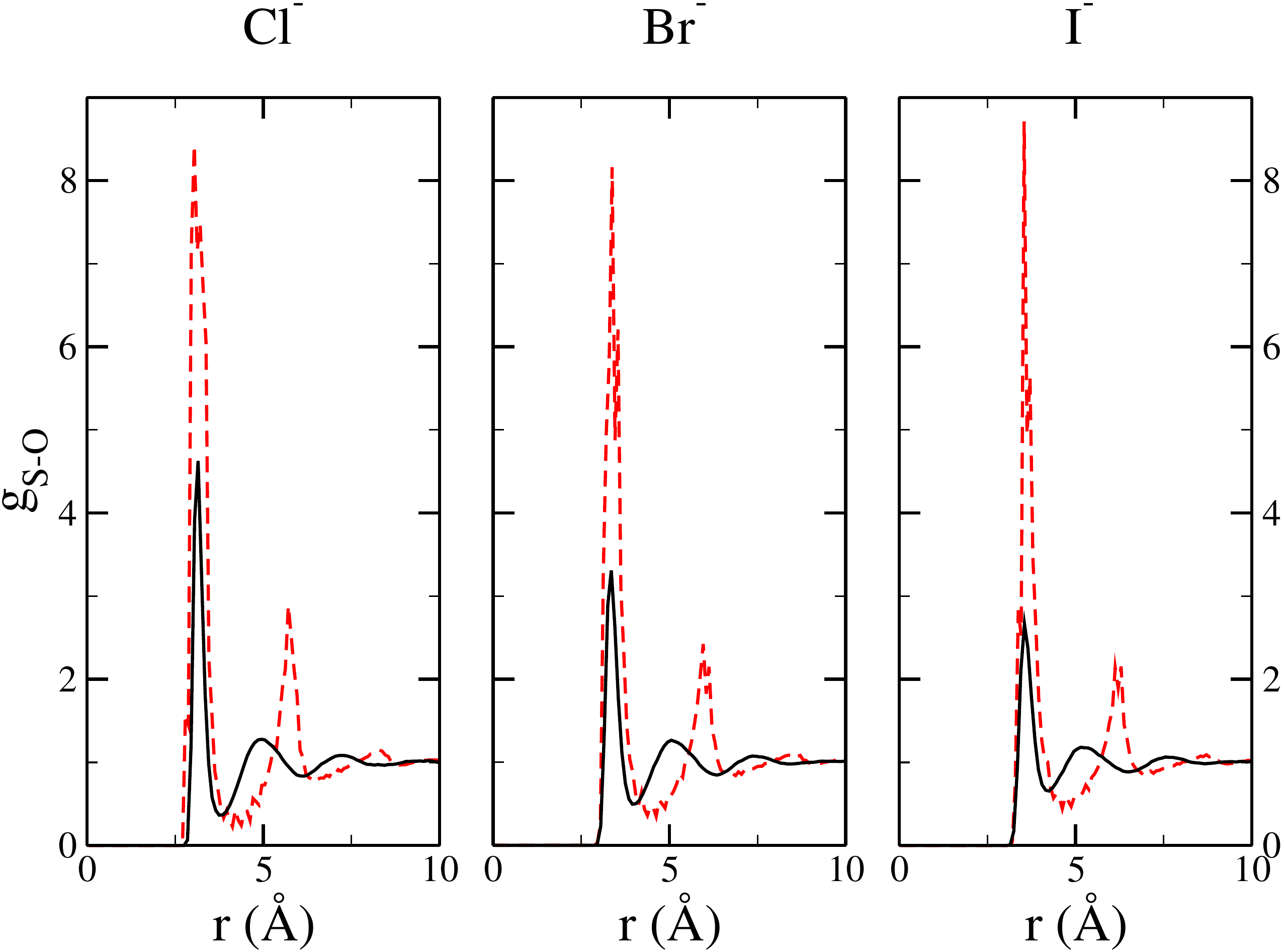}\protect\caption{Même figure que la \ref{fig:rdf_alcalins_dip} avec des solutés halogénures
: chlorure, bromure, iodure.\label{fig:rdf_halogenures_dip}}
\end{figure}

Ces sites sont engagés dans des liaisons hydrogène avec le solvant
ce qui crée également un arrangement spatial tétraédrique local autour
du soluté. Il est donc probable que le mauvais accord entre MD et
MDFT ait la même origine que pour le soluté eau.
Pour tester cette hypothèse, nous avons étudié des solutés fortement
hydrophiles : des cations alcalins et des anions halogénures. Nous
avons utilisé les potentiels d'interaction indiqués dans le \ref{tab:Para_ions}.
\begin{table}[h]
\noindent \centering{}%
\begin{tabular}{|c|c|c|c|}
\hline 
Site & $\sigma\ (\textrm{\AA})$ & $\epsilon\ (\text{kJ.mol\ensuremath{{}^{-1}}})$ & $q\ (e)$\tabularnewline
\hline 
\hline 
$\text{Na\ensuremath{{}^{+}}}$ & 2.581 & 0.129 & 1.0\tabularnewline
\hline 
$\mathrm{K}\text{\ensuremath{^{+}}}$ & 2.931 & 0.760 & 1.0\tabularnewline
\hline 
$\text{Cs\ensuremath{{}^{+}}}$ & 3.531 & 1.500 & 1.0\tabularnewline
\hline 
$\text{Cl\ensuremath{{}^{-}}}$ & 4.035 & 0.410 & -1.0\tabularnewline
\hline 
$\text{Br\ensuremath{{}^{-}}}$ & 4.581 & 0.499 & -1.0\tabularnewline
\hline 
$\mathrm{I}\text{\ensuremath{^{-}}}$ & 4.921 & 0.670 & -1.0\tabularnewline
\hline 
\end{tabular}\protect\caption{Paramètres Lennard-Jones et charges partielles des cations alcalins
et anions halogénures.\label{tab:Para_ions}}
\end{table}
Les fonctions de distribution radiale, obtenues par dynamique moléculaire
et par minimisation fonctionnelle sont montrées en \ref{fig:rdf_alcalins_dip}
et \ref{fig:rdf_halogenures_dip}. L'écart entre la MDFT et les résultats
exacts obtenus par MD est encore plus flagrant que dans le cas des
molécules polaires, puisque la MDFT surestime grandement le premier
pic de solvatation. Les pics dus aux deuxième et troisième couches
de solvatation sont eux aussi surestimés mais également situés à une
distance trop grande du soluté.

Pour résumer, une fonctionnelle dans l'approximation du fluide homogène
de référence avec un traitement de la polarisation strictement dipolaire,
et sans terme de bridge, donne d'excellents résultats quant à la structure
de solvatation des solutés apolaires. Cependant, elle ne permet pas
de reproduire correctement les structures de solvatation lorsque les
solutés sont constitués de sites portant une charge électrique importante.\clearpage{}

\subsection{Énergies de solvatation}

Nous allons désormais examiner les prédictions des énergies libres
de Helmholtz de solvatation en milieux aqueux par minimisation de
la fonctionnelle. Ces résultats seront cette fois encore comparés
à ceux obtenus par dynamique moléculaire. On rappelle qu'en dynamique
moléculaire il est nécessaire d'utiliser des techniques d'intégration
thermodynamique pour accéder aux énergies libres.

\subsubsection{Application aux solutés apolaires}

\begin{figure}[h]
\noindent \centering{}\includegraphics[width=0.8\textwidth]{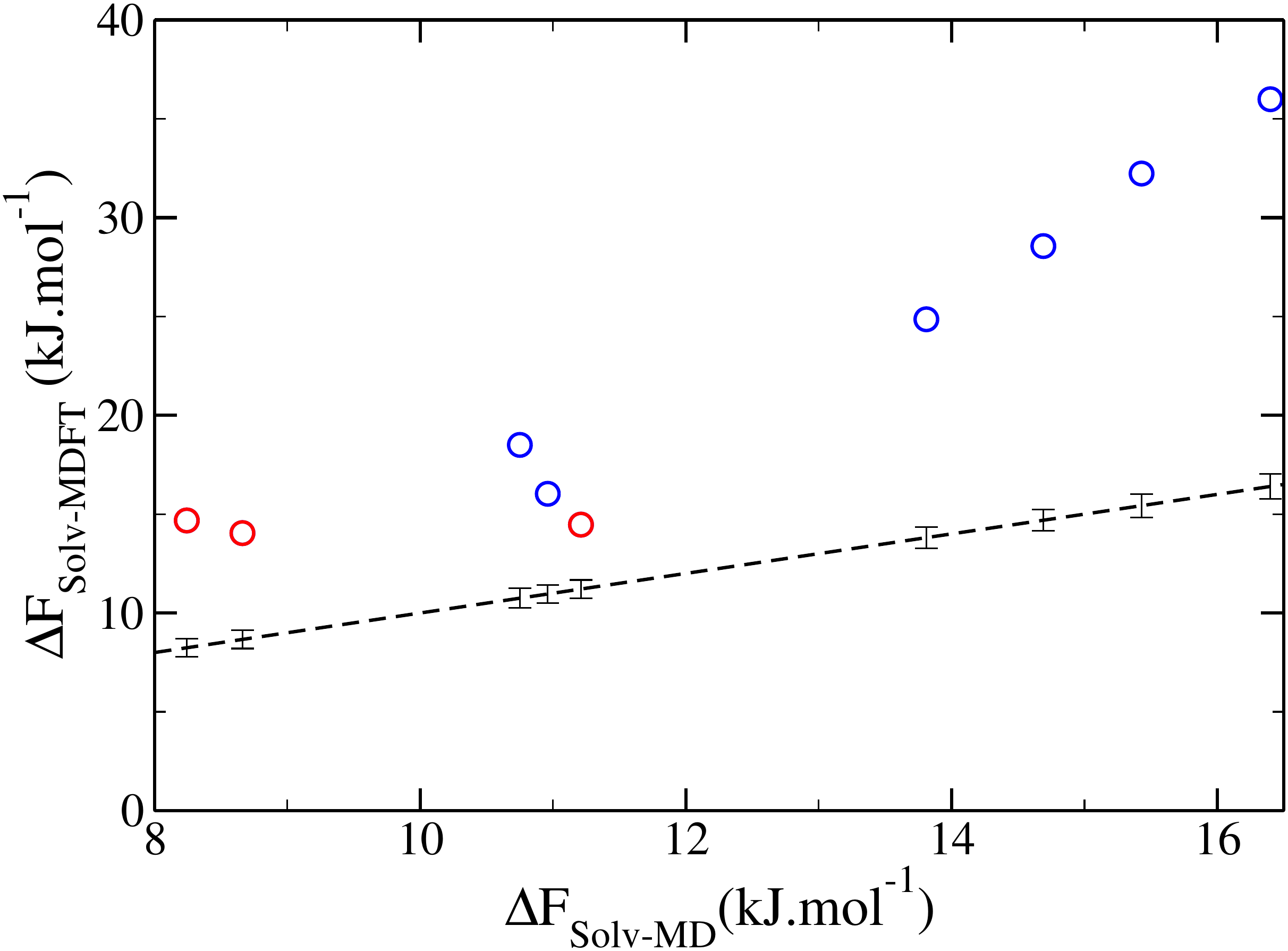}\protect\caption{Comparaison des énergies libres de solvatation obtenues par MD et
MDFT pour les gaz rares et les six premiers alcanes linéaires. La
droite en tirets noirs est le résultat obtenu par MD \cite{ashbaugh_hydration_1998,guillot_computer_1993}.
Les cercles sont les valeurs obtenues par minimisation de la fonctionnelle,
en rouge pour les gaz rares, en bleu pour les alcanes. Les barres
d'erreurs sont celles obtenues par MD.\label{fig:enercomp_neutre_dipol}}
\end{figure}

La comparaison des résultats obtenus par MD et MDFT est présentée
sur la \ref{fig:enercomp_neutre_dipol}. L'accord est loin d'être
satisfaisant puisque les énergies calculées par minimisation fonctionnelle
valent pratiquement le double de celles calculées par MD. On peut
s'interroger sur l'origine de ces mauvais résultats alors que la structure
de solvatation est très bien reproduite (voir \ref{fig:rdf gaz rare dipol }
et \ref{fig:rdf alcane rare dipol }). On peut penser que l'interaction
entre soluté et solvant est correctement estimée par la fonctionnelle,
ce qui explique notamment l'excellente reproduction de la première
couche de solvatation, mais que l'interaction solvant-solvant est
moins bien représentée. Ceci explique la légère différence au niveau
des couches de solvatation suivantes. 
Cette mauvaise estimation de l'énergie libre de solvatation provient
donc des approximations réalisées sur la partie d'excès,\textit{ }c'est-à-dire
l'approximation du fluide homogène de référence et le développement
à l'ordre strictement dipolaire de la fonction de corrélation directe.
En restant dans l'HRF, on peut essayer d'améliorer la fonctionnelle
en proposant une approximation du terme de bridge ${\cal F}_{\mathrm{cor}}\left[\rho(\bm{r},\bm{\Omega})\right]$
qui avait été supposé nul jusqu'à présent. Ce terme inconnu peut s'écrire
en développant de manière systématique la fonction de corrélation
directe du solvant pur en fonction des fonctions de corrélation à
$N$ corps, avec $N\geq3$. Cette approche n'a pas été réalisée car
elle nécessite la connaissance de ces fonctions de corrélation d'ordre
supérieur pour le liquide pur. En revanche, une approximation de ce
terme a été proposée au laboratoire\cite{MDFT_levesque_krfmt} : l'approximation
du bridge de sphères dures (HSB pour hard spheres bridge). Cette approximation
\cite{rosenfeld_free_1993,liu_site_2013,MDFT_zhao_new_2011,MDFT_zhao_correction_2011,oettel_integral_2005}
consiste à remplacer le terme de bridge, inconnu pour l'eau, par le
terme de bridge d'un fluide de sphères dures de même densité, qui
lui est connu exactement,
\begin{eqnarray}
{\cal F}_{\text{HSB}}\left[n(\bm{r})\right] & = & {\cal F}^{\text{HS}}\left[n(\bm{r})\right]-{\cal F}^{\text{HS}}\left[n_{b}\right]-\mu^{\text{HS}}\iiint_{\mathbb{R}^{3}}\Delta n(\bm{r})\text{d}\bm{r}\nonumber \\
 &  & +\frac{\text{k}\text{\ensuremath{_{B}}T}}{2}\iiint_{\mathbb{R}^{3}}\iiint_{\mathbb{R}^{3}}n\rho(\bm{r})c_{\mathrm{HS}}^{\text{(2)}}\left(\left\Vert \bm{r}-\bm{r}^{\prime}\right\Vert ;\rho_{b}\right)\Delta n(\bm{r}^{\prime})\text{d}\bm{r}\text{d}\bm{r}^{\prime}.\label{eq:F_HSB}
\end{eqnarray}
L'écriture de ce terme fait intervenir une fonctionnelle du fluide
de sphères dures ${\cal F}^{\text{HS}}$, dont des expressions de
très bonnes qualités sont connues \cite{rosenfeld_free-energy_1989,kierlik_free-energy_1990,kierlik_density-functional_1991}
(l'\ref{sec:FMT} traite de ces fonctionnelles de sphères dures).
Intervient également la fonction de corrélation directe du fluide
de sphères dures $c_{\mathrm{HS}}^{\text{(2)}}$, que l'on peut calculer
comme la dérivée fonctionnelle seconde de la fonctionnelle de sphères
dures par rapport à la densité. On voit que cette approximation revient
à remplacer tous les termes en $\Delta\rho^{n}$ avec $n\geq3$, supposés
nuls dans l'approximation du fluide homogène de référence, par ceux
du fluide de sphères dures.

\fbox{\begin{minipage}[t]{1\columnwidth}%
On peut voir cette approximation du bridge de sphères dures de deux
points de vue différents, selon qu'on corrige les approximations du
modèle approché développé ici (MDFT-HRF) ou au contraire qu'on modifie
un modèle de sphères dures pour le faire ressembler à un liquide réel:

- Dans le cas du fluide de sphères dures, la fonctionnelle est connue
avec une excellente approximation et il est possible de calculer la
fonction de corrélation directe d'ordre deux en utilisant l'\ref{eq:c(2)}.
On peut donc modifier la fonctionnelle de sphères dures en lui ajoutant
un \og léger goût \fg{} d'eau. Pour cela, on \og remplace \fg{}
la fonction de corrélation directe d'ordre deux par celle connue de
l'eau dans l'approximation du fluide de référence.

- Dans le cas d'un développement de Taylor de la fonctionnelle dans
le cadre de l'approximation du fluide de référence, on a \og supprimé \fg{}
tous les termes d'ordre supérieur à deux en densité, et ce terme de
bridge sphères dures sert à corriger cette approximation. Ces termes
d'ordres supérieurs sont à courte-portée donc essentiellement des
termes d'empilement, il ne semble pas absurde de les remplacer par
des termes sphères dures. %
\end{minipage}}

\begin{figure}[h]
\noindent \centering{}\includegraphics[width=0.8\textwidth]{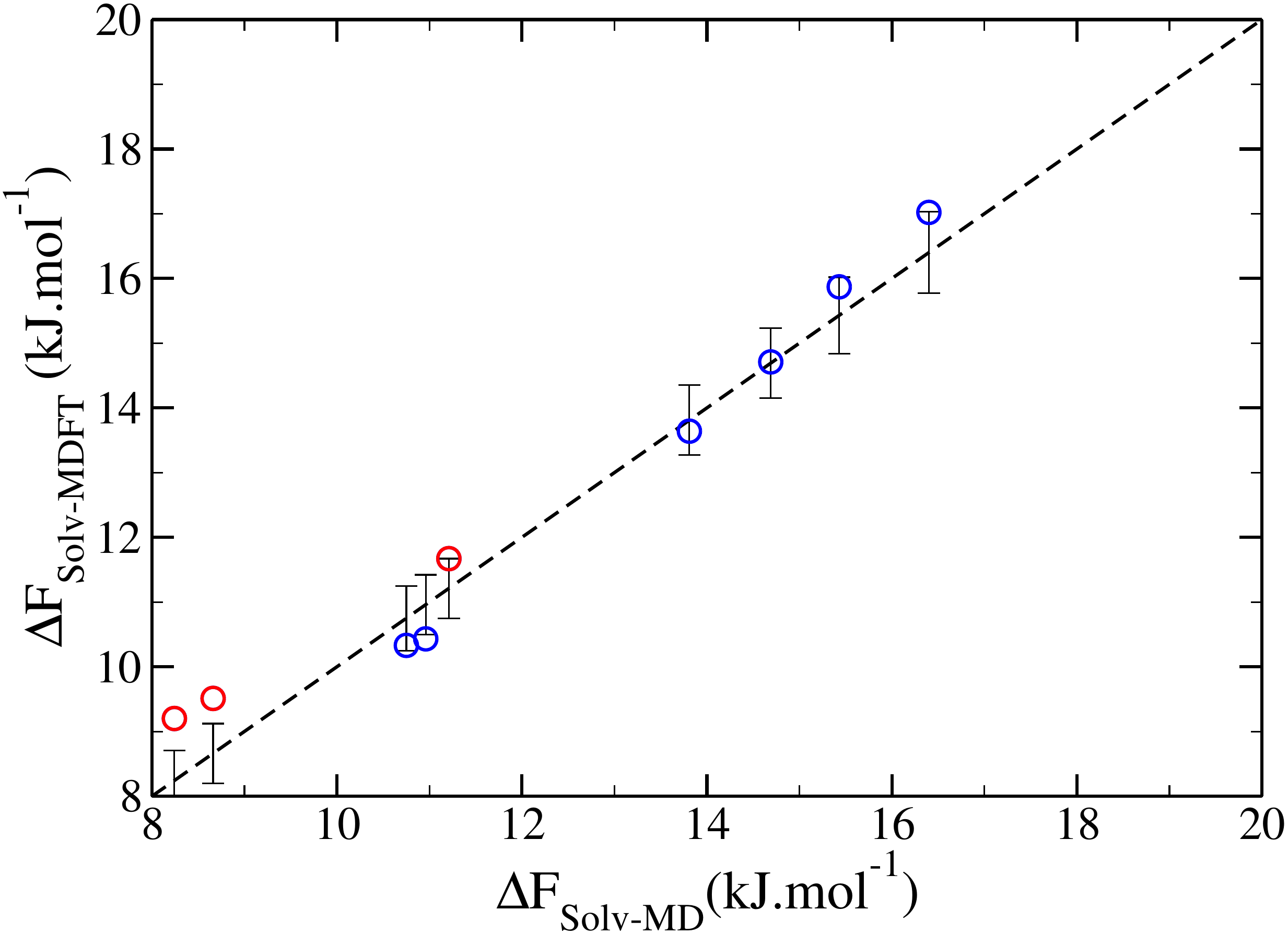}\protect\caption{Comparaison des énergies libres de solvatation obtenues par MD et
MDFT avec la correction de bridge sphères dures pour les gaz rares
et les six premiers alcanes linéaires. La droite en tirets noirs est
le résultat obtenu par MD \cite{ashbaugh_hydration_1998,guillot_computer_1993}.
Les cercles sont les valeurs obtenues par minimisation de la fonctionnelle,
en rouge pour les gaz rares, en bleu pour les alcanes. Les barres
d'erreurs sont celles obtenues par MD.\label{fig:enercomp_neutre_dipol+HSB}}
\end{figure}

\begin{figure}[h]
\noindent \centering{}\includegraphics[width=0.8\textwidth]{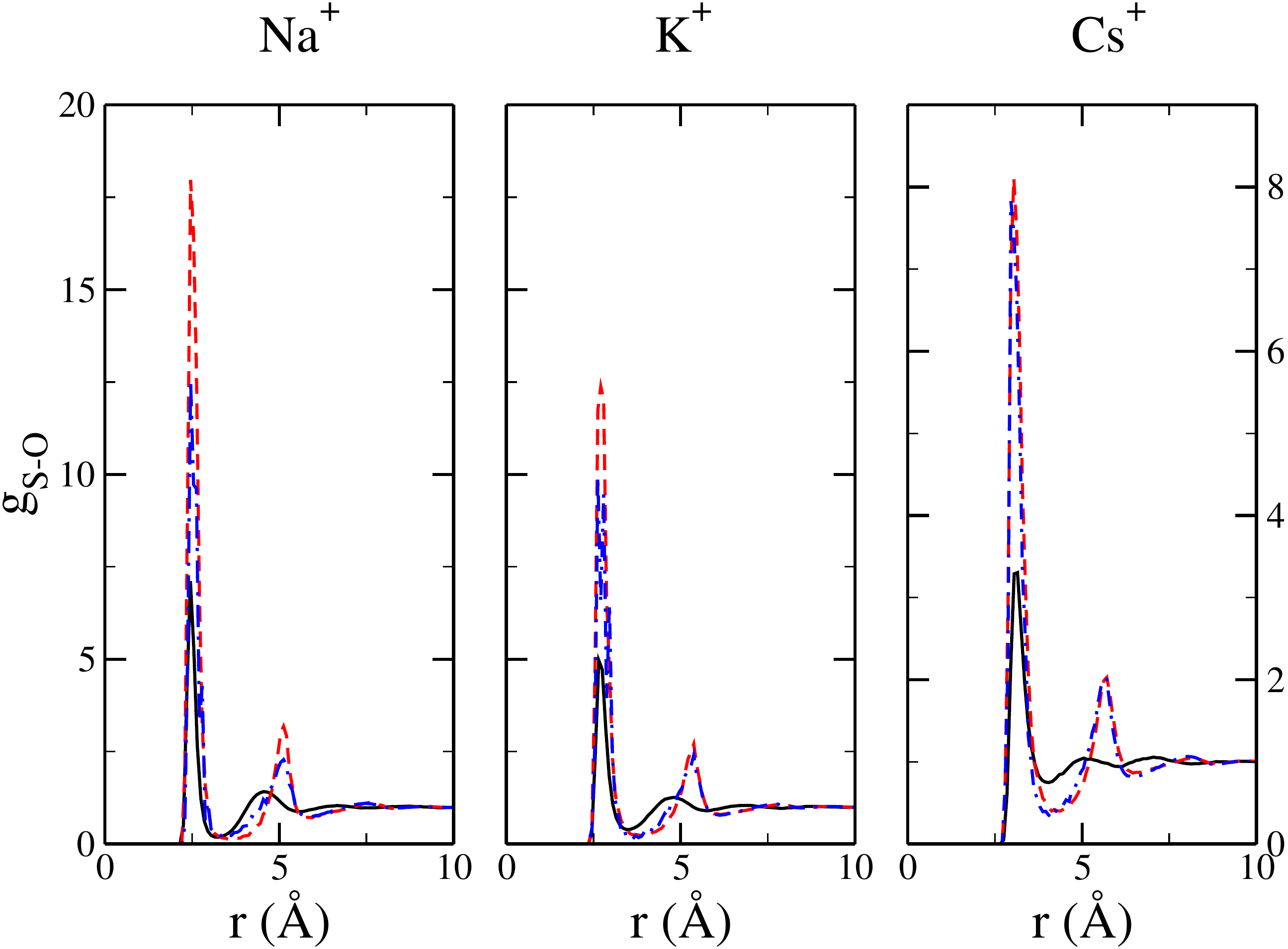}\protect\caption{Fonctions de distribution radiale entre l'oxygène de l'eau solvant
et trois cations alcalins, sodium, potassium et césium. Les résultats
des simulations MD sont en trait noir plein, ceux des calculs MDFT
en tirets rouges et ceux obtenus avec MDFT et la correction HSB en
tirets bleus.\label{fig:rdf_alcalins_dip-HSB}}
\end{figure}

Une figure analogue à la \ref{fig:enercomp_neutre_dipol} mais où
l'on a inclus ce bridge sphères dures est présentée en \ref{fig:enercomp_neutre_dipol+HSB}.
On constate que cette correction améliore grandement les valeurs des
énergies libres de solvatation obtenues sans modifier les bons résultats
des \ref{fig:rdf gaz rare dipol } et \ref{fig:rdf alcane rare dipol }.
Sur la \ref{fig:rdf_alcalins_dip-HSB} sont présentées les fonctions
de distribution radiale pour les cations alcalins obtenues par MD,
MDFT et MDFT avec la correction bridge sphères dures. Cette correction
améliore légèrement les fonctions de distribution radiale en réduisant
la hauteur du premier pic, mais l'accord reste néanmoins insuffisant.
Elle ne résout pas le manque d'ordre tétraédrique dans le modèle.

\clearpage{}

\section{Au delà de l'ordre dipolaire\label{sec:Fexcmulti}}

On s'est limité dans la partie précédente à un traitement strictement
dipolaire de la partie d'excès pour l'eau. Lors de ma thèse, j'ai
développé une écriture de cette partie d'excès ne restreignant plus
le traitement de la polarisation à l'ordre dipolaire. Cette formulation
est présentée ici.

\subsection{Écriture de la polarisation multipolaire}

La polarisation décrite dans l'\ref{eq:Pola_dip} est d'ordre purement
dipolaire. C'est-à-dire que l'on a restreint la symétrie de l'interaction
effective contenue dans la fonction $c$ à l'ordre 2. Une telle approximation
est valable pour un fluide purement dipolaire. Cependant, si on considère
le modèle d'eau SPC/E, on se rend compte que la distribution de charge
n'est pas purement dipolaire puisque le modèle est constitué de trois
charges partielles, voir \ref{fig:G=0000E9om=0000E9trie-du-mod=0000E8leSPC/E}.
L'approximation réalisée précédemment n'est pas justifiée à priori.

On peut définir une densité de charge $\sigma(\bm{r},\bm{\Omega})$
et de polarisation $\bm{\mu}(\bm{r},\bm{\Omega})$\cite{raineri_static_1992,raineri_static_1993}
pour une molécule d'eau SPC/E prise à l'origine du repère et ayant
une orientation $\bm{\Omega}$,\foreignlanguage{english}{
\begin{eqnarray}
\sigma(\bm{r},\bm{\Omega}) & = & \sum_{\text{m}}q_{\text{m}}\delta(\bm{r-s}_{m}(\bm{\Omega}))\nonumber \\
 & = & q_{O}\delta(\bm{r})+q_{H}\left[\delta(\bm{r-}\bm{s}_{H_{1}}(\bm{\Omega}))+\delta(\bm{r-}\bm{s}_{H_{2}}(\bm{\Omega}))\right],
\end{eqnarray}
} et\foreignlanguage{english}{
\begin{equation}
\bm{\mu}(\bm{r},\bm{\Omega})\bm{}=\sum_{\text{m}}q_{\text{m}}\bm{s}_{\text{m}}(\bm{\Omega})\intop_{0}^{1}\delta(\bm{r}-u\bm{s}_{\text{m}}(\bm{\Omega}))\text{d}u.
\end{equation}
}$\bm{s}_{\mathrm{m}}(\bm{\Omega})$ désigne la position du $\mathrm{m^{i\grave{e}me}}$
site du solvant quand la molécule possède l'orientation $\bm{\Omega}$,
et $q_{\text{m}}$ désigne la charge de ce site. On peut réécrire
la polarisation microscopique dans l'espace de Fourier,\foreignlanguage{english}{
\begin{eqnarray}
\bm{\mu}(\bm{k},\bm{\Omega}) & = & \sum_{\text{m}}q_{\text{m}}\bm{s}_{\text{m}}(\bm{\Omega})\intop_{0}^{1}e^{iu\bm{k.}\bm{s}_{m}(\bm{\Omega})}\text{d}u\nonumber \\
 & = & -i\sum_{\text{m}}q_{\text{m}}\frac{\bm{s}_{\text{m}}(\bm{\Omega})}{\bm{k.}\bm{s}_{\text{m}}(\bm{\Omega})}(e^{i\bm{k.}\bm{s}_{\text{m}}(\bm{\Omega})}-1)\nonumber \\
 & = & \sum_{\text{m}}q_{\text{m}}.\bm{s}_{\text{m}}(\bm{\Omega})+i\frac{q_{\text{m}}}{2}(\bm{k.}\bm{s}_{\text{m}}(\bm{\Omega}))\bm{s}_{\text{m}}(\bm{\Omega})+\mathcal{O}(\bm{k^{2}})\nonumber \\
 & = & \bm{\mu}(\bm{\Omega})+i\frac{q_{\text{m}}}{2}(\bm{k.}\bm{s}_{\text{m}}(\bm{\Omega}))\bm{s}_{\text{m}}(\bm{\Omega})+\mathcal{O}(\bm{k^{2}}).\label{eq:mu_k_omega}
\end{eqnarray}
}On reconnait le premier terme, qui est le terme purement dipolaire.
Les termes d'ordres supérieurs (c'est-à-dire le terme d'ordre un et
les termes d'ordres supérieurs qui ont été écrits, pour des raisons
de compacité, en $\mathcal{O}(\bm{k^{2}})$) proviennent de la distribution
de charge non dipolaire de la molécule. On peut vérifier que c'est
en fait la densité de polarisation associée à la distribution de charge
d'une molécule de solvant centrée à l'origine via l'équation de Poisson
locale:
\begin{equation}
\sigma(\bm{r},\bm{\Omega})=-\nabla\cdot\bm{\mu}(\bm{r},\bm{\Omega}),
\end{equation}
 soit dans l'espace de Fourier,
\begin{equation}
\sigma(\bm{k},\bm{\Omega})=-i\bm{\mu}(\bm{k},\bm{\Omega}).
\end{equation}
Dans le cas d'un solvant constitué d'un seul site Lennard-Jones (comme
l'eau SPC/E), le Hamiltonien qui décrit le système de N molécules
d'eau en présence d'un soluté s'écrit
\begin{equation}
{\cal H}={\cal K}+{\cal U}+\iiint_{\mathbb{R}^{3}}\tilde{n}(\bm{r})v_{\text{LJ}}(\bm{r})\text{d}\bm{r}-\iiint_{\mathbb{R}^{3}}\tilde{\bm{P}}(\bm{r})\bm{.E}(\bm{r})\text{d}\bm{r}\label{eq:Hwater_multi}
\end{equation}
$v_{\text{LJ}}$ désigne le potentiel extérieur Lennard-Jones dû à
l'interaction soluté-solvant et $\bm{E}$ le champ électrostatique
créé par le soluté en un point $\bm{r}$. $\tilde{\rho}(\bm{r},\bm{\Omega})$
et $\tilde{n}(\bm{r})$ sont les densités de solvant microscopiques
et $\tilde{\bm{P}}(\bm{r})$ la densité de polarisation microscopique,
c'est-à-dire calculées pour une configuration des molécules donnée.
Ces grandeurs s'écrivent :
\begin{gather}
\tilde{\rho}(\bm{r},\bm{\Omega})=\sum_{\text{i}=1}^{\text{N}}\delta(\bm{r}-\bm{r}_{\text{i}})\delta(\bm{\Omega}-\bm{\Omega}_{\mathrm{i}})\\
\tilde{n}(\bm{r})=\sum_{\text{i}=1}^{\text{N}}\delta(\bm{r}-\bm{r}_{\text{i}})=\iiint_{8\pi^{2}}\tilde{\rho}(\bm{r},\bm{\Omega})\mathrm{d}\bm{\Omega}\\
\tilde{\bm{P}}(\bm{r})=\sum_{\text{i}=1}^{N}\bm{\mu}(\bm{r}-\bm{r}_{\text{i}},\bm{\Omega}_{\text{i}})=\iiint_{\mathbb{R}^{3}}\int_{8\pi^{2}}\bm{\mu(r-r^{\prime},\Omega)}\tilde{\rho}(\bm{r}^{\prime},\boldsymbol{\Omega})\text{d}\bm{r}^{\prime}\text{d}\bm{\Omega}.
\end{gather}
On peut alors définir des densités de particule et de polarisation
$n(\bm{r})=\left\langle \tilde{n}(\bm{r})\right\rangle $ et $\bm{P}(\bm{r})=\left\langle \tilde{\bm{P}}(\bm{r})\right\rangle $.
On peut montrer, en suivant une démonstration analogue à celle d'Evans
\cite{evans79}, que le grand-potentiel, que l'on notera désormais
$\Theta$ pour éviter la confusion avec les orientations, peut s'exprimer
comme une fonctionnelle de ces densités de particule et de polarisation.
Ce grand potentiel atteint son minimum pour les densités de particule
et polarisation à l'équilibre thermodynamique. La démonstration est
donnée en \ref{sec:gdPotmini}.

On va utiliser l'approximation du fluide homogène de référence décrite
en \ref{sub:HRF approx}. On définit alors de manière identique la
fonctionnelle d'énergie libre ${\cal F}\left[n(\bm{r}),\bm{P}(\bm{r})\right]=\Theta\left[n(\bm{r}),\bm{P}(\bm{r})\right]-\Theta_{0}$,
qui est également une fonctionnelle des densités de particule et de
polarisation. 

\begin{figure}[h]
\noindent \centering{}\includegraphics[width=0.8\textwidth]{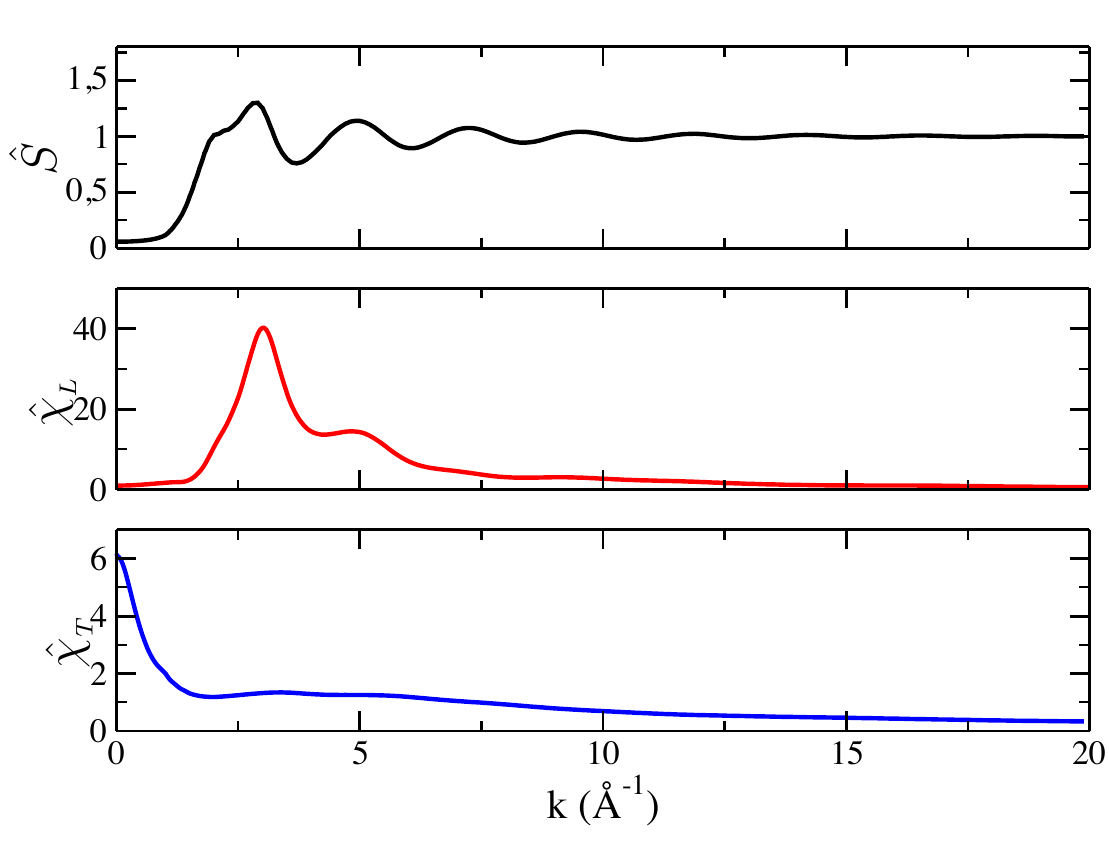}\protect\caption{Facteur de structure et composantes longitudinale et transverse des
susceptibilités de l'eau SPC/E, calculées avec les formules de Bopp
et collaborateurs\cite{bopp_frequency_1998}.\label{fig:S_and_chi}}
\end{figure}

On introduit la fonctionnelle intrinsèque, qui est la partie ne dépendant
pas des champs extérieurs:
\begin{equation}
{\cal F}\left[n(\bm{r}),\bm{P}(\bm{r})\right]={\cal F}_{\text{int}}\left[n(\bm{r}),\bm{P}(\bm{r})\right]+\iiint_{\mathbb{R}^{3}}n(\bm{r})v_{\mathrm{LJ}}(\bm{r})\text{d}\bm{r}-\iiint_{\mathbb{R}^{3}}\bm{P}(\bm{r})\cdot\bm{E}(\bm{r})\text{d}\bm{r}
\end{equation}
Cette partie intrinsèque est celle due au solvant seul. Elle contient
comme précédemment une partie entropique et une partie d'interaction
entre molécules de solvant. On aimerait pouvoir scinder ce terme en
un terme idéal et un terme d'excès comme dans l'\ref{eq:F=00003DFid+Fexc}.
Cependant, on ne connait pas d'expression de l'énergie idéale en fonction
des densités de particule et de polarisation. On va donc se ramener
à une expression connue et écrire la contribution idéale de la partie
intrinsèque comme dépendant de la densité de particule \og totale \fg{}
$\rho(\bm{r},\bm{\Omega})$, c'est-à-dire utiliser l'expression de
l'\ref{eq:Fiddeltarho}. On peut montrer, dans le cadre de la théorie
de la réponse linéaire, que la fonctionnelle intrinsèque peut s'écrire,
sous l'hypothèse d'un développement quadratique, en fonction des corrélations
entre densité de particule et densité de polarisation. La démonstration
de cette écriture, dans le cas plus simple d'une fonctionnelle dépendant
d'une seule variable, est proposée dans le prochain encart. Une démonstration
parfaitement similaire est réalisable dans le cas présent. On peut
donc écrire la fonctionnelle d'excès comme,\foreignlanguage{english}{\textit{
\begin{eqnarray}
\beta{\cal F}_{\mathrm{exc}}[\Delta n(\bm{r}),\bm{P}(\bm{r})] & = & \beta{\cal F}_{\mathrm{int}}[\Delta n(\bm{r}),\bm{P}(\bm{r})]-\beta{\cal F}_{\mathrm{id}}[\Delta n(\bm{r}),\bm{P}(\bm{r})]\\
\beta{\cal F}_{\mathrm{exc}}[\Delta n(\bm{r}),\bm{P}(\bm{r})] & = & \frac{1}{2}\iiint_{\mathbb{R}^{3}}\iiint_{\mathbb{R}^{3}}S^{-1}(r_{12})\Delta n(\bm{r}_{1})\Delta n(\bm{r}_{2})\mathrm{d}\bm{r}_{1}\mathrm{d}\bm{r}_{2}\nonumber \\
 &  & +\frac{1}{8\pi\epsilon_{0}}\iiint_{\mathbb{R}^{3}}\iiint_{\mathbb{R}^{3}}\chi_{L}^{-1}(r_{12})\bm{P}_{\mathrm{L}}(\bm{r}_{1})\bm{P}_{\mathrm{L}}(\bm{r}_{2})\mathrm{d}\bm{r}_{1}\mathrm{d}\bm{r}_{2}\nonumber \\
 &  & +\frac{1}{8\pi\epsilon_{0}}\iiint_{\mathbb{R}^{3}}\iiint_{\mathbb{R}^{3}}\chi_{T}^{-1}(r_{12})\bm{P}_{\mathrm{T}}(\bm{r}_{1})\bm{P}_{\mathrm{T}}(\bm{r}_{2})\mathrm{d}\bm{r}_{1}\mathrm{d}\bm{r}_{2}\nonumber \\
 &  & -\mathrm{k_{B}T}\iiint_{\mathbb{R}^{3}}\frac{\Delta n(\bm{r})^{2}}{2n_{0}}\mathrm{d}\bm{r}-\mathrm{k_{B}T}\iiint_{\mathbb{R}^{3}}\frac{3}{2\mu_{0}n_{0}^{2}}\bm{P}\left(\bm{r}\right)^{2}\mathrm{d}\bm{r}\nonumber \\
 &  & +\beta{\cal F}_{\mathrm{cor}}[\Delta n(\bm{r}),\bm{P}(\bm{r})].\label{eq:Fexcmulti}
\end{eqnarray}
}}Dans cette expression, on a négligé les termes de couplage entre
densité et polarisation. Les trois premiers termes viennent du développement
quadratique de la partie intrinsèque. On a séparé ici les composantes
longitudinale et transverse de la polarisation, qui sont obtenues
dans l'espace de Fourier,
\begin{equation}
\hat{\bm{P}_{\mathrm{L}}}(\bm{k})=\frac{\left(\hat{\bm{P}}(\bm{k})\cdot\bm{k}\right)\bm{k}}{\left\Vert \bm{k}\right\Vert {}^{2}},\hspace{3cm}\hat{\bm{P}_{\mathrm{T}}}(\bm{k})=\hat{\bm{P}}(\bm{k})-\hat{\bm{P}_{\mathrm{L}}}(\bm{k})\label{eq:PL/PT(k)}
\end{equation}
$S^{-1}$, $\chi_{L}^{-1}$ et $\chi_{T}^{-1}$ sont définies comme
les transformées de Fourier inverses des inverses du facteur de structure
$\hat{S}$ et des susceptibilités diélectriques longitudinale et transverse
$\hat{\chi}_{L}$ et $\hat{\chi}_{T}$. Soulignons que ces fonctions
sont liées aux fluctuations de densité et de polarisation du solvant
pur. Ces fonctions ont été calculées par dynamique moléculaire dans
l'espace de Fourier, grâce à la méthode introduite par Bopp et collaborateurs
\cite{bopp_static_1996,bopp_frequency_1998}. Ces fonctions sont présentées
en \ref{fig:S_and_chi}.

On mentionne que les susceptibilités diélectriques longitudinale et
transverse peuvent être reliées aux composantes longitudinale et transverse
des constantes diélectriques par les relations,
\begin{equation}
\hat{\chi}_{L}(k)=\frac{1}{1-\hat{\epsilon}_{L}(k)},\hspace{3cm}\hat{\chi_{T}}(k)=\frac{\hat{\epsilon}_{T}(k)-1}{4\pi},
\end{equation}
de telle sorte que leurs valeurs aux petites valeurs de $k$, qui
sont mal estimées par les simulations à cause d'effet de tailles finies,
peuvent être extrapolées puisque l'on connait la valeur macroscopique
des constantes diélectriques $\epsilon_{T}(0)=\epsilon_{L}(0)=71$
pour l'eau SPC/E\cite{rami_reddy_dielectric_1989}.

Les deux termes suivants de l'\ref{eq:Fexcmulti} sont les termes
linéarisés de la partie idéale, que l'on retire pour ne pas les compter
deux fois. Le dernier terme est un terme correctif inconnu qui contient
à priori tous les ordres supérieurs à deux en $\Delta n$ et $\bm{P}$.
Ce terme est l'équivalent du terme de bridge dans le développement
dipolaire de la fonctionnelle présenté plus haut.

Soulignons que dans tout ce développement, les termes de couplage
entre densité de particule et polarisation ont été négligés pour être
cohérent avec l'approche dipolaire réalisée précédemment. Cependant
une expression de la fonctionnelle d'excès incluant ces termes est
proposée en \ref{sec:Couplage-densit=0000E9-polarisation}.

On peut donner une expression alternative de la fonctionnelle en fonction
de la densité de charge de molécules de solvant $\rho_{c}$. La densité
de charge du solvant s'exprime également en fonction de la densité
de charge microscopique:
\begin{equation}
\rho_{c}(\bm{r})=\left\langle \tilde{\rho}_{c}(\bm{r})\right\rangle ,
\end{equation}
avec
\begin{equation}
\tilde{\rho}_{c}(\bm{r})=\iiint_{8\pi^{2}}\iiint_{\mathbb{R}^{3}}\sigma(\bm{r}-\bm{r}^{\prime},\bm{\Omega})\tilde{\rho}(\bm{r}^{\prime},\bm{\Omega})\mathrm{d}\bm{r}\mathrm{d}\bm{\Omega}.
\end{equation}
 La densité de charge est également reliée à la polarisation via une
équation de Poisson,
\begin{equation}
\rho_{c}(\bm{r})=-\nabla\cdot\bm{P}(\bm{r}).
\end{equation}

La fonctionnelle se réécrit en fonction de $\rho_{c}$,\foreignlanguage{english}{\textit{
\begin{eqnarray}
\beta{\cal F}_{\mathrm{exc}}[\Delta n(\bm{r}),\rho_{c}(\bm{r})] & = & \frac{1}{2}\iiint_{\mathbb{R}^{3}}\iiint_{\mathbb{R}^{3}}S^{-1}(r_{12})\Delta n(\bm{r}_{1})\Delta n(\bm{r}_{2})\mathrm{d}\bm{r}_{1}\mathrm{d}\bm{r}_{2}\nonumber \\
 &  & +\frac{1}{8\pi\epsilon_{0}}\iiint_{\mathbb{R}^{3}}\iiint_{\mathbb{R}^{3}}S_{cc}^{-1}(r_{12})\rho_{c}(\bm{r}_{1})\rho_{c}(\bm{r}_{2})\mathrm{d}\bm{r}_{1}\mathrm{d}\bm{r}_{2}\nonumber \\
 &  & +\frac{1}{8\pi\epsilon_{0}}\iiint_{\mathbb{R}^{3}}\iiint_{\mathbb{R}^{3}}\chi_{T}^{-1}(r_{12})\bm{P}_{\mathrm{T}}(\bm{r_{1}})\bm{P}_{\mathrm{T}}(\bm{r_{2}})\mathrm{d}\bm{r}_{1}\mathrm{d}\bm{r}_{2}\nonumber \\
 &  & -\mathrm{k_{B}T}\iiint_{\mathbb{R}^{3}}\frac{\Delta n(\bm{r})^{2}}{2n_{0}}\mathrm{d}\bm{r}-\mathrm{k_{B}T}\iiint_{\mathbb{R}^{3}}\frac{3}{2\mu_{0}n_{0}^{2}}\bm{P}\left(\bm{r}\right)^{2}\mathrm{d}\bm{r}\nonumber \\
 &  & +\beta{\cal F}_{\mathrm{cor}}[\Delta n(\bm{r}),\bm{P}(\bm{r})].
\end{eqnarray}
}}On a introduit le facteur de structure charge-charge défini comme
\begin{equation}
k^{2}S_{cc}(k)=\chi_{L}(k).
\end{equation}

\begin{framed}%
On montre ici, dans le cas d'une fonctionnelle ne dépendant que de
la densité à une particule $n(\bm{r})$, comment on peut proposer
une expression de la fonctionnelle d'excès.

On suppose qu'on décrit le système dans un état proche de celui du
fluide de référence à la densité homogène $n_{\mathrm{b}}$. On étudie
une petite modulation de la densité de particule $\Delta n(\bm{r})=n(\bm{r})-n_{\mathrm{b}}$.
On suppose que cette modulation provient de la présence d'un petit
potentiel extérieur $\delta\phi$.

Ceci revient à supposer que le hamiltonien décrivant le système est
égal au hamiltonien du fluide homogène plus cette petite perturbation,
\begin{equation}
{\cal H}={\cal H}_{b}+\delta\phi(\bm{r}).
\end{equation}
Cette perturbation étant faible, on peut supposer que la réponse de
la densité peut être décrite, dans le cadre de la théorie de la réponse
linéaire, en fonction de la fonction de réponse $\chi(\bm{r},\bm{r}^{\prime})$,
\begin{equation}
\Delta n(\bm{r})=\iiint_{\mathbb{R}^{3}}\chi(\bm{r},\bm{r}^{\prime})\delta\phi(\bm{r}^{\prime})\text{d}\bm{r}^{\prime}\label{eq:lien chi_deltan}
\end{equation}
La fonction de réponse vaut par définition, 
\begin{equation}
\chi(\bm{r},\bm{r}^{\prime})=\left.\frac{\delta n(\bm{r})}{\delta\phi(\bm{r}^{\prime})}\right|_{\delta\phi=0}=-\left.\frac{\delta n(\bm{r})}{\delta\psi(\bm{r}^{\prime})}\right|_{\delta\phi=0}=-\beta\left\langle \Delta\tilde{n}(\bm{r})\Delta\tilde{n}(\bm{r}^{\prime})\right\rangle .
\end{equation}

L'égalité de gauche se trouve par dérivée fonctionnelle de l'\ref{eq:lien chi_deltan}.
Le lien avec les fluctuations provient du théorème de fluctuation-dissipation\cite{kubo2012statistical}.

La fonction de réponse est donc égale aux corrélations densité-densité
du système non perturbé. Si l'on prend la transformée de Fourier de
l'\ref{eq:lien chi_deltan}, on trouve la relation suivante qui relie
fonction de réponse, densité et perturbation
\begin{equation}
\Delta\hat{n}(\bm{k})=\hat{\chi}(\bm{k})\delta\hat{\phi}(\bm{k}).\label{eq:lien chi deltanFourier}
\end{equation}
Si on suppose que ${\cal F}_{\mathrm{int}}$ est quadratique en la
modulation de densité, alors, 
\begin{equation}
{\cal F}_{\mathrm{int}}[\Delta n]=\frac{1}{2}\iiint_{\mathbb{R}^{3}}\iiint_{\mathbb{R}^{3}}\Delta n(\bm{r})X_{0}(\left\Vert \bm{r}-\bm{r}^{\prime}\right\Vert )\Delta n(\bm{r}^{\prime})\text{d}\bm{r}\text{d}\bm{r}^{\prime}+{\cal O}(\Delta n^{3}).\label{eq:Fint_dvlptquad}
\end{equation}
Puisque pour un potentiel extérieur constant la fonctionnelle doit
être minimale pour une densité uniforme, il n'y a pas de terme d'ordre
1 en $\Delta n$.

La fonction $X_{0}(\bm{r},\bm{r}^{\prime})$ est une grandeur caractéristique
du fluide de référence et ne dépend donc que de $\left\Vert \bm{r}-\bm{r}^{\prime}\right\Vert $.

Si on réécrit l'\ref{eq:Fint_dvlptquad} dans l'espace réciproque,
en transformées de Fourier discrète, celle-ci devient,
\begin{equation}
{\cal F}_{\text{int}}[\Delta n]=\frac{1}{2V}\sum_{\bm{k}}\Delta\hat{n}(\bm{k})\hat{X_{0}}(\bm{k})\Delta\hat{n}(-\bm{k})d\bm{k}+{\cal O}(\Delta\rho^{3}).\label{eq:Fint_dvlptquadFourrier}
\end{equation}
Si on applique le principe variationel à cette équation on trouve,
en utilisant \ref{eq:dF/drho=00003Dpsi},
\begin{equation}
\Delta\hat{n}(\bm{k})\hat{X_{0}}(\bm{k})=-\Delta\phi(\bm{k})\label{eq:equaX0}
\end{equation}
Ce qui, en comparant l'\ref{eq:lien chi deltanFourier} et l'\ref{eq:equaX0},
donne
\begin{equation}
\hat{X_{0}}(\bm{k})=-\hat{\chi}^{-1}(\bm{k})
\end{equation}
Or, on peut relier la fonction de corrélation $\chi$ au facteur de
structure $S$\cite{hansen_theory_2006} :
\begin{equation}
\hat{\chi}(\bm{k})=-\beta n_{b}\hat{S}(\bm{k})=\frac{-\beta n_{b}}{1-n_{b}\hat{c}(\bm{k})}
\end{equation}
On a donc dans le cas présent :
\begin{eqnarray}
{\cal F}_{\text{exc}}[\Delta n] & = & {\cal F}_{\text{int}}[n]-{\cal F}_{\text{id}}[n]\nonumber \\
 & = & \frac{1}{2}\iiint_{\mathbb{R}^{3}}\iiint_{\mathbb{R}^{3}}\Delta n(\bm{r})\chi^{-1}(\left\Vert \bm{r}-\bm{r}^{\prime}\right\Vert )\Delta n(\bm{r}^{\prime})\text{d}\bm{r}\text{d}\bm{r}^{\prime}-{\cal F}_{\text{id}}[n]\nonumber \\
 & = & -\frac{\text{k}_{\mathrm{B}}\text{T}}{2}\iiint_{\mathbb{R}^{3}}\iiint_{\mathbb{R}^{3}}\Delta n(\bm{r})S(\left\Vert \bm{r}-\bm{r}^{\prime}\right\Vert )\Delta n(\bm{r}^{\prime})\text{d}\bm{r}\text{d}\bm{r}^{\prime}-{\cal F}_{\text{id}}[n]
\end{eqnarray}
Si on linéarise le terme idéal, on obtient :
\begin{equation}
{\cal F}_{\text{exc}}[\Delta n]=-\frac{\text{k}\text{\ensuremath{_{B}}T}}{2}\iiint_{\mathbb{R}^{3}}\iiint_{\mathbb{R}^{3}}\Delta n(\bm{r})c(\left\Vert \bm{r}-\bm{r}^{\prime}\right\Vert )\Delta n(\bm{r}^{\prime})\text{d}\bm{r}\text{d}\bm{r}^{\prime}.
\end{equation}

On retrouve bien l'expression de la partie d'excès dans le cadre de
l'approximation du fluide homogène de référence.\end{framed}

\subsection{Structures de solvatation}

Comme attendu, les structures de solvatation des solutés neutres,
alcanes et gaz rares, sont inchangées à l'ordre multipolaire puisque
ceux-ci ne créent pas de champ électrique.

\subsubsection{Les solutés polaires}

\begin{figure}[h]
\noindent \centering{}\includegraphics[width=0.8\textwidth]{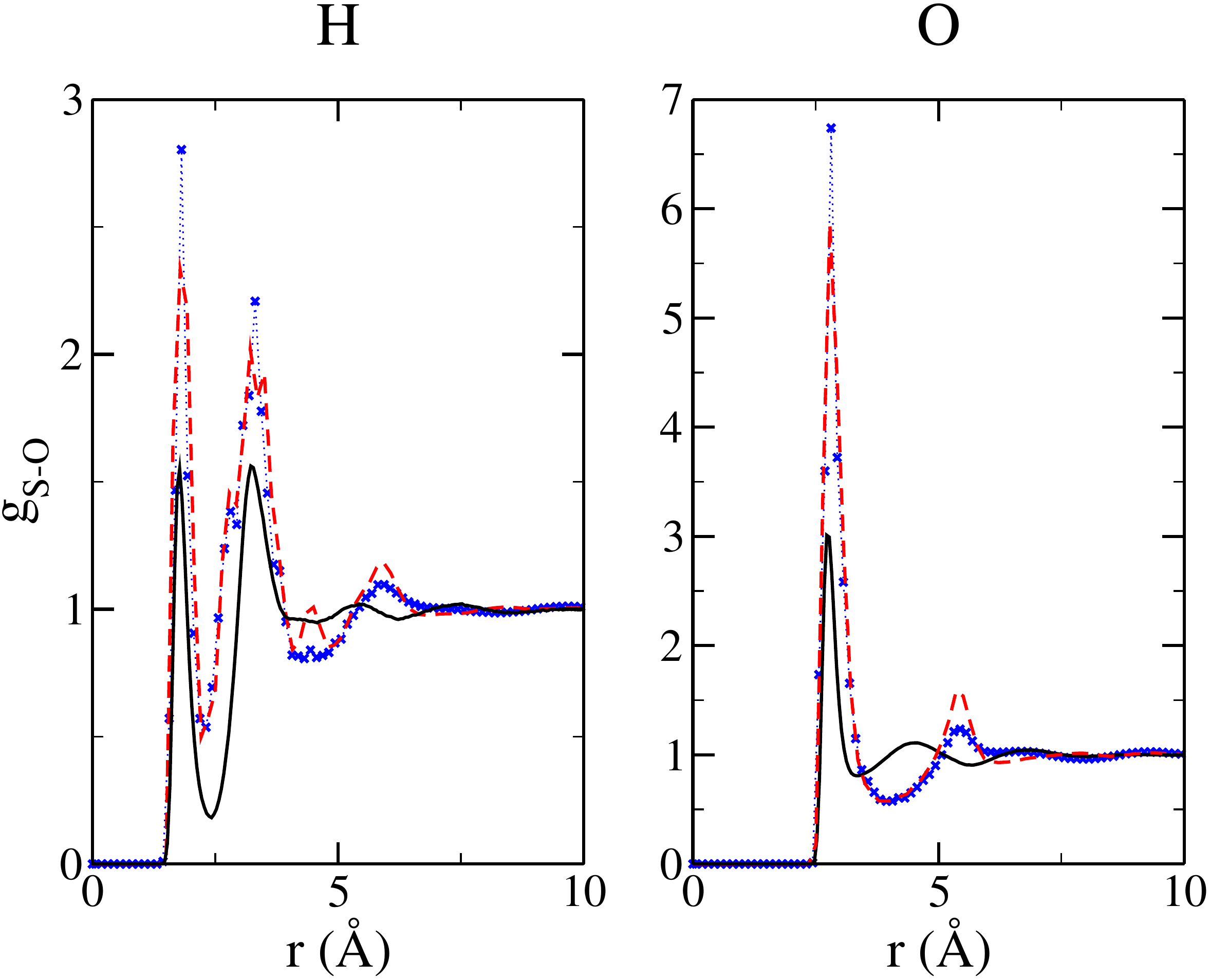}\protect\caption{Fonctions de distribution radiale entre l'oxygène de l'eau solvant
et l'oxygène et l'hydrogène de l'eau soluté. Les résultats calculés
par MD sont en trait noir plein, tandis que ceux obtenus par minimisation
de la fonctionnelle multipolaire sont représentés par des croix bleues
reliées par des pointillés. Pour mémoire, on a laissé les résultats
obtenus avec la fonctionnelle dipolaire en traits rouges discontinus.
\label{fig:rdf_water_multi}}
\end{figure}

\begin{figure}[h]
\noindent \centering{}\includegraphics[width=0.78\textwidth]{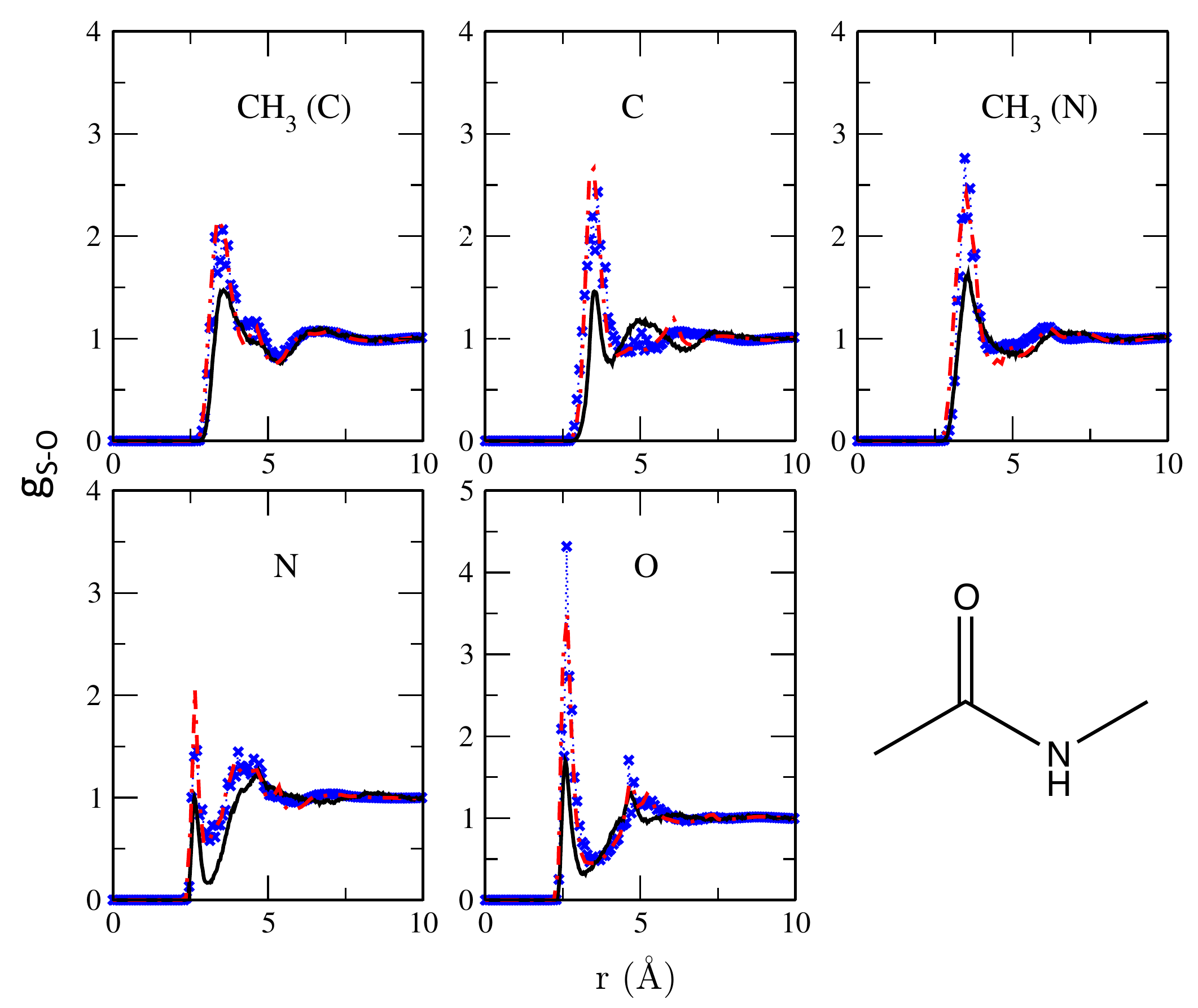}\protect\caption{Fonctions de distribution radiale entre l'oxygène de l'eau solvant
et les différents sites de la molécule NMA. La légende est la même
que sur la \ref{fig:rdf_water_multi}. \label{fig:rdf_NMA_multi}}
\end{figure}

On présente, sur les \ref{fig:rdf_water_multi} et \ref{fig:rdf_NMA_multi},
les fonctions de distribution radiale obtenues par simulation de dynamique
moléculaire et par minimisation des fonctionnelles dipolaires et multipolaires
des \ref{sec:Fexcdip} et \ref{sec:Fexcmulti}, pour l'eau et la NMA,
avec les champs de force des \ref{tab:Param=0000E8tre-du-mod=0000E8leSPC/E}
et \ref{tab:ParaNMA}.
On remarque que l'approche multipolaire ne modifie que peu l'allure
des fonctions de distribution radiale obtenues par MDFT dipolaire
pour les molécules polaires. Il y a une très légère amélioration de
la hauteur des pics dus à la seconde couche de solvatation dont l'intensité
diminue pour l'eau et la NMA.

\begin{figure}[h]
\noindent \centering{}\includegraphics[width=0.78\textwidth]{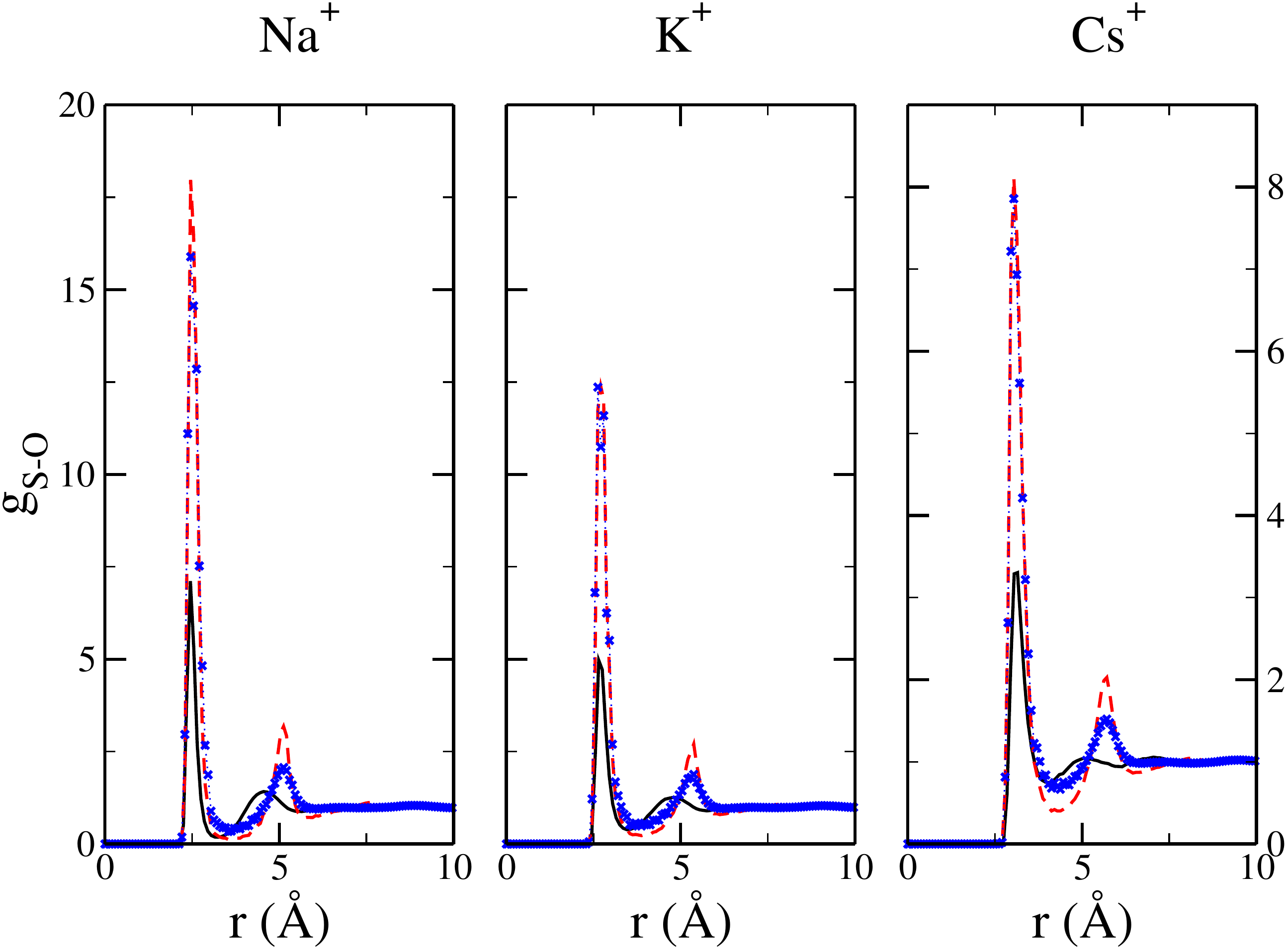}\protect\caption{Fonctions de distribution radiale entre l'oxygène de l'eau solvant
et trois cations alcalins. La légende est la même que sur la \ref{fig:rdf_water_multi}.
\label{fig:rdf_alkalin multi}}
\end{figure}

Cette tendance est confirmée pour les ions. On étudie par exemple
les cations alcalins dans la \ref{fig:rdf_alkalin multi}, où la hauteur
du premier pic et la déplétion entre le premier et le second pic sont
nettement diminués.

On arrive donc à une conclusion un peu décevante: un traitement rigoureux
de l'électrostatique au delà d'une approximation dipolaire n'améliore
pas sensiblement les résultats. Une théorie bilinéaire en $n(\bm{r})$
et $\bm{P}(\bm{r})$ conduit à une mésestimation de l'ordre tétraédrique.
Pallier ce défaut est l'objet du chapitre suivant.

\lhead[\chaptername~\thechapter]{\rightmark}

\rhead[\leftmark]{}

\lfoot[\thepage]{}

\cfoot{}

\rfoot[]{\thepage}

\chapter{Correction à trois corps\label{chap:F3B}}

\section{Retour sur les solutés chargés}

Nous avons vu dans le \ref{chap:MDFT_dup_multu} que l'on peut écrire
une théorie de la fonctionnelle de la densité pour l'eau dans l'approximation
du fluide homogène de référence. Cette fonctionnelle donne des résultats
satisfaisants pour les solutés apolaires mais pas pour les solutés
polaires qui ne sont pas en accord avec la MD.

\begin{figure}
\noindent \centering{}\includegraphics[width=0.6\textwidth]{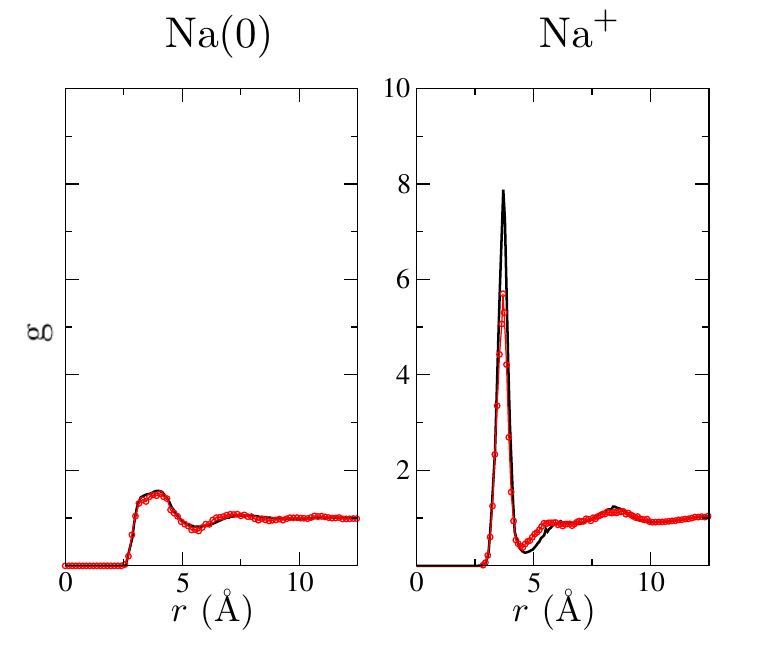}\protect\caption{Fonctions de distribution radiale entre le centre de masse de la molécule
d'acétonitrile et un atome (à gauche) et un ion sodium (à droite)
\cite{MDFT_gendre_classical_2009}. MD en rouge, MDFT en noir. \label{fig:g(r)diversCinCH3CN}}
\end{figure}

\begin{figure}
\noindent \centering{}\includegraphics[width=0.8\textwidth]{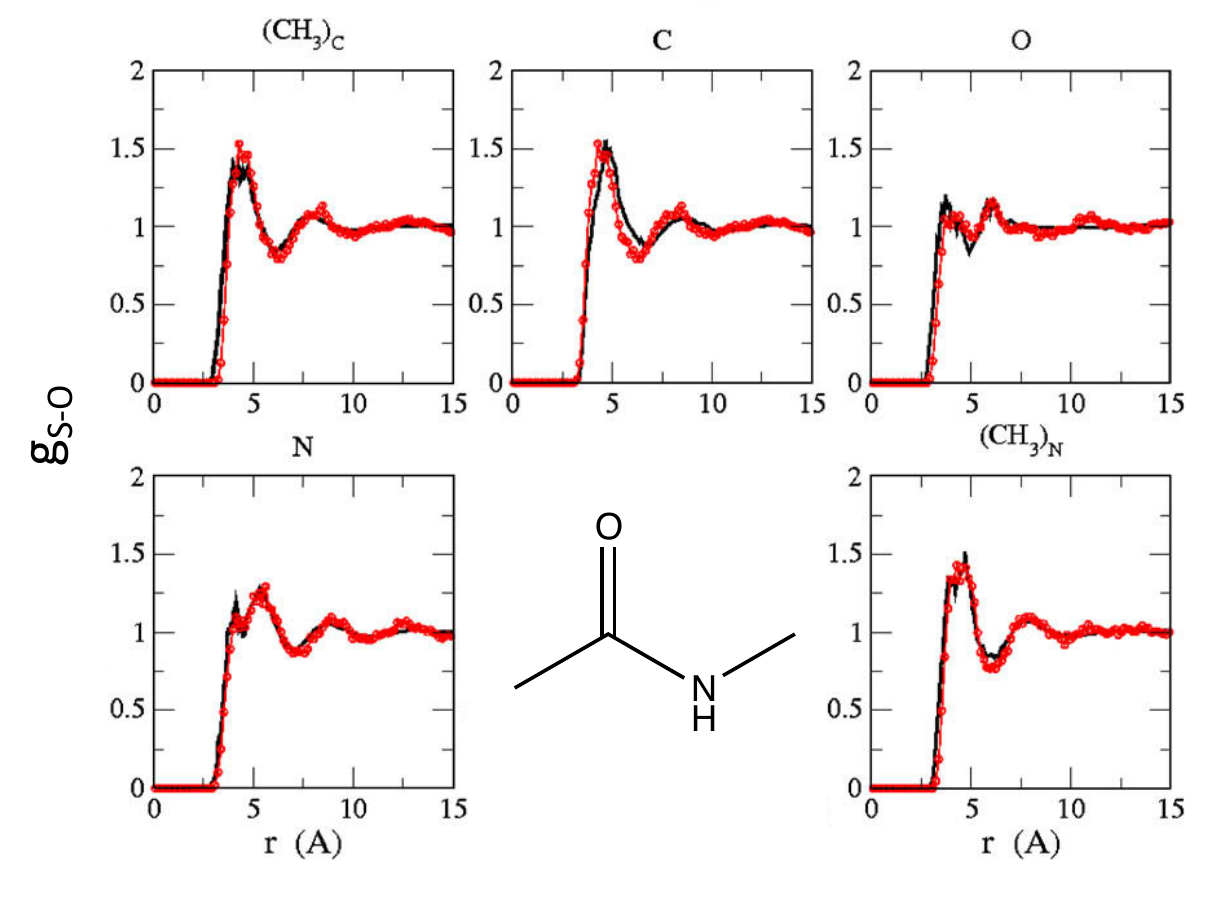}\protect\caption{Fonctions de distribution radiale entre le centre de masse de la molécule
d'acétonitrile et les différents sites de la molécule de NMA\cite{MDFT_gendre_classical_2009}.
MD en rouge, MDFT en noir.\label{fig:g(r)NMACinCH3CN}}
\end{figure}

Cette théorie avait pourtant déjà été utilisée avec succès pour l'étude
de molécules polaires et d'ions dans le solvant acétonitrile\cite{MDFT_gendre_classical_2009,MDFT_zhao_molecular_2011}
: par exemple, les fonctions de distribution radiale entre l'acétonitrile
et un atome neutre ou un cation sont présentées en \ref{fig:g(r)diversCinCH3CN},
ou avec la molécule de N-méthylacétamide (NMA) en \ref{fig:g(r)NMACinCH3CN}.
Les pics des fonctions de distribution radiale dus aux première et
deuxième couches de solvation sont très bien reproduits par la MDFT.
De plus, les énergies de solvatation pour la série des anions halogénures
sont en accord avec les mesures expérimentales\cite{MDFT_gendre_classical_2009}. 

Pourquoi les résultats pour la solvatation des solutés chargés dans
l'eau diffèrent autant de la MD, alors que l'accord est quasi-parfait
pour l'acétonitrile? Ces deux solvants sont polaires (1.85 D pour
l'eau, 3.9 D pour l'acétonitrile). L'eau est un solvant protique ce
qui en fait un bon donneur de liaisons hydrogènes. De plus, l'eau
possède un caractère accepteur de liaisons hydrogènes. L'acétonitrile
en revanche, ne développe pas de liaisons hydrogènes. Sur la \ref{fig:rdf_alcalins_dip-HSB},
le premier pic de la fonction de distribution radiale est à la bonne
position mais il est surestimé. Le second pic est lui-aussi surestimé
et situé à une distance trop éloignée par rapport au soluté. Ce second
pic est en réalité très similaire à celui obtenu avec le solvant acétonitrile
sur la \ref{fig:g(r)diversCinCH3CN}. Au contraire lorsque les sites
de soluté sont neutres comme sur la \ref{fig:rdf_alcalins_dip-HSB},
ils ne peuvent interagir avec le solvant par liaisons hydrogène, dans
ce cas l'accord entre MD et MDFT est bon. C'est donc l'interaction
par des liaisons hydrogènes de l'eau avec les solutés chargés qui
explique le désaccord entre MD et MDFT.

\begin{figure}
\noindent \centering{}\includegraphics[width=0.4\textwidth]{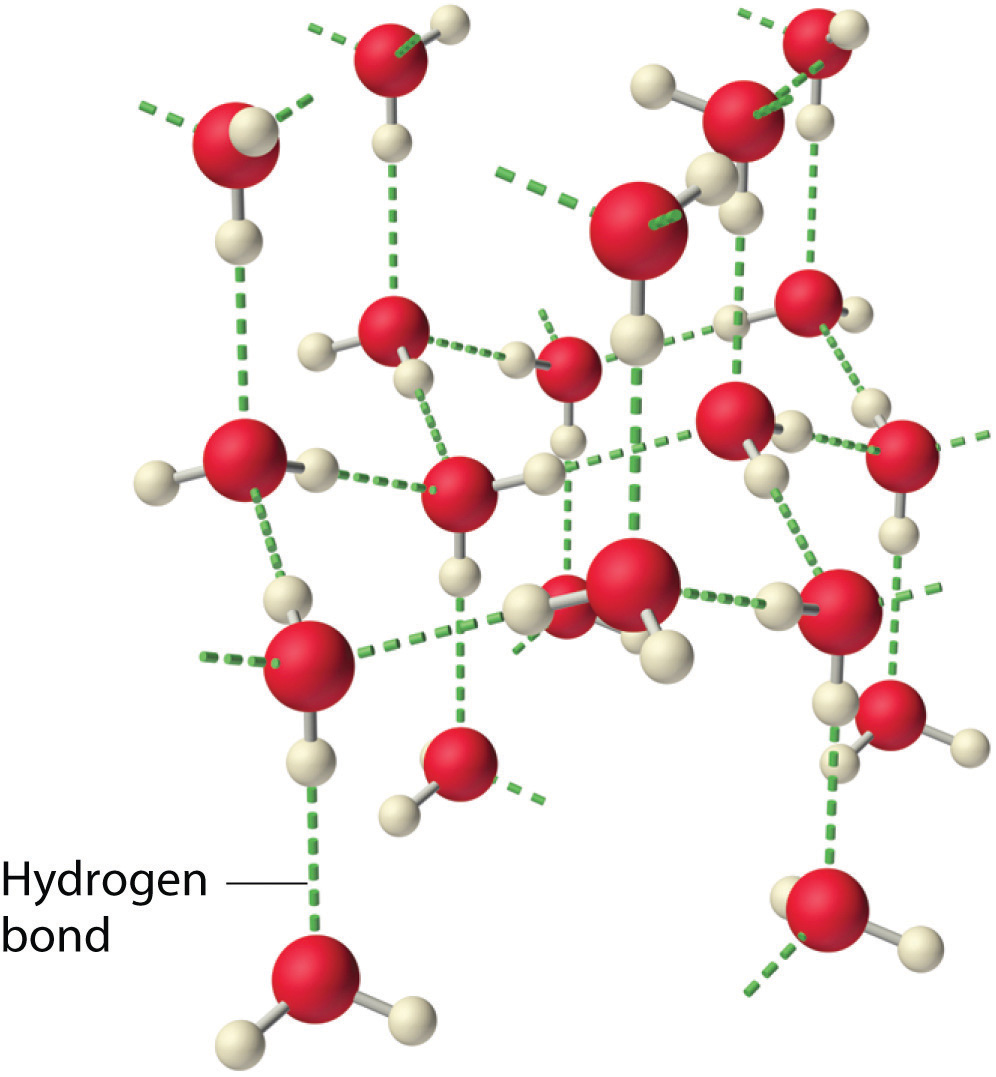}\protect\caption{Illustration de l'arrangement tétraédrique de l'eau liquide, d'après
\cite{averill_chemistry:_2007}\label{tetraedralstructurewater}. }
\end{figure}

Les liaisons hydrogènes, dans l'eau, provoquent un arrangement tétraédrique
local autour des solutés chargés. Cet arrangement particulier, illustré
sur la \ref{tetraedralstructurewater}, a deux conséquences sur les
fonctions de distribution radiale. Le nombre de premiers voisins est
réduit à quatre : le pic dû à la première sphère de solvatation est
donc plus faible que dans un fluide non-associé. De plus, les molécules
de la seconde sphère de solvatation peuvent se rapprocher plus du
soluté car l'encombrement dû aux molécules de la première couche de
solvatation est plus faible : le pic dû à la seconde couche de solvatation
est plus proche soluté.

\section{Terme de correction à 3 corps}

Les résultats présentés dans la partie précédente pour la solvatation
dans l'eau montrent que les fonctionnelles d'excès utilisées ne parviennent
pas à reproduire la structure tétraédrique du solvant autour des solutés
chargés. Ceci peut en partie s'expliquer par un développement systématique
de la fonction de corrélation directe du solvant pur qui a été limité
à l'ordre deux. On pourrait imaginer améliorer la fonctionnelle en
incluant les corrélations d'ordre supérieur, mais cette idée n'a pas
été retenue pour des raisons numériques. De plus, il est compliqué
d'obtenir avec précision les termes supérieurs du développement de
la fonction de corrélation directe. Une des premières tentatives pour
corriger ce problème a été l'utilisation du \og bridge \fg{} de
sphères dures décrit à l'\ref{eq:F_HSB}. L'inclusion de ce terme
n'a que peu amélioré les résultats, comme on peut le voir sur la \ref{fig:rdf_alcalins_dip-HSB}.
Le seul ion pour lequel l'ajout de ce terme HSB a un effet notable
est le cation sodium. Ceci peut s'expliquer par le fait que dû à son
importante densité de charge (qu'on peut définir ici comme le rapport
entre la charge et le volume de la sphère Lennard-Jones), il attire
particulièrement les molécules de solvant dans sa première sphère
de solvatation. Il en découle un premier pic très intense. L'ajout
d'un bridge de sphères dures empêche un empilement très élevé des
molécules de solvant, ce qui tend à faire diminuer ce premier pic.
Cet effet est déjà moins visible pour le cation potassium et est inexistant
pour le cation césium. Cette correction rajoute des termes de corrélation
d'ordre supérieur d'un fluide de sphères dures. Ce sont des termes
d'empilement isotropes qui ne permettent pas de reproduire un ordre
tétraédrique local. Cette correction ne peut donc pas modifier la
position du second pic de la fonction de distribution radiale.

En conséquence, nous avons décidé d'introduire un terme correctif
qui renforce l'ordre tétraédrique local entre molécules d'eau et solutés
chargés. Ce terme correctif s'inspire d'un modèle d'eau gros-grains
(MW) développé par Molinero et Moore\cite{molinero_water_threebody_2009}.
C'est une adaptation d'un potentiel pour des fluides tétracoordinés
comme le silicium, qui a été originellement proposé par Stillinger
et Weber\cite{PhysRevB.31.5262}. 

Dans ce modèle, l'interaction entre molécules de solvant se fait à
l'aide de potentiels d'interaction de deux types : un potentiel Lennard-Jones
à deux corps et un potentiel à deux corps qui pénalise les arrangements
non tétraédrique de la forme :
\begin{equation}
\phi_{3}(\bm{r}_{ij},\bm{r}_{ik},\theta_{ijk})=\lambda_{ijk}\epsilon\left(\cos\theta_{ijk}-\cos\theta_{0}\right)^{2}f(\left\Vert \bm{r}_{ij}\right\Vert )f(\left\Vert \bm{r}_{ik}\right\Vert ).\label{eq:Vmolinero}
\end{equation}
C'est un potentiel d'interaction entre des sites positionnés en $\bm{r}_{i},\bm{r}_{j}$
et $\bm{r}_{k}$, avec $\theta_{ijk}$ l'angle entre ces trois sites
et $\left\Vert \bm{r}_{ij}\right\Vert $ la distance entre les sites
$i$ et $j$. Les fonctions $f$ sont des fonctions qui limitent la
portée de ce potentiel. $\theta_{0}=109.4\text{\textdegree}$ désigne
l'angle tétraédrique. $\epsilon$ est la profondeur du potentiel Lennard-Jones
et les paramètres $\lambda_{ijk}$ servent à moduler l'importance
relative de ce potentiel d'interaction à trois corps par rapport au
potentiel à deux corps. Ce potentiel crée une pénalité énergétique
d'autant plus forte que trois molécules de solvant suffisamment proches
forment un angle différent de l'angle tétraédrique $\theta_{0}$. 

Grâce à ce modèle d'eau simplifié, Molinero et Moore parviennent à
reproduire de nombreuses propriétés de l'eau et notamment les propriétés
structurales et de densité. 

Le terme de correction ajouté à la fonctionnelle pour renforcer l'ordre
tétraédrique local inspiré de ce potentiel de Molinero et Moore est
donc :
\begin{equation}
\mathrm{{\cal F}_{cor}^{3B}[}n(\bm{r})]=\mathrm{{\cal F}_{cor}^{3B-1S}}[n(\bm{r})]+\mathrm{{\cal F}_{cor}^{3B-2S}}[n(\bm{r})]+\mathrm{{\cal F}_{cor}^{3B-ww}[}n(\bm{r})].\label{eq:F3B}
\end{equation}
Avec : 
\begin{gather}
\hspace{-1cm}\beta\mathrm{{\cal F}_{cor}^{3B-1S}}[n(\bm{r})]=\frac{1}{2}\sum_{m}\lambda_{m}^{1\mathrm{S}}\iiint_{\mathbb{R}^{3}}\iiint_{\mathbb{R}^{3}}f_{m}(r_{m2})f_{m}(r_{m3})\left(\frac{\bm{r}_{m2}\cdot\bm{r}_{m3}}{r_{m2}r_{m3}}-\cos\theta_{0}\right)^{2}n(\bm{r}_{2})n(\bm{r}_{3})\mathrm{d}\bm{r}_{2}\mathrm{d}\bm{r}_{3},\label{eq:F3B1S}\\
\hspace{-1cm}\beta\mathrm{{\cal F}_{cor}^{3B-2S}}[n(\bm{r})]=\frac{1}{2}\sum_{m}\lambda_{m}^{2\mathrm{S}}\iiint_{\mathbb{R}^{3}}\iiint_{\mathbb{R}^{3}}f_{m}(r_{m2})f_{w}(r_{23})\left(\frac{\bm{r}_{m2}\cdot\bm{r}_{23}}{r_{m2}r_{23}}-\cos\theta_{0}\right)^{2}n(\bm{r}_{2})n(\bm{r}_{3})\mathrm{d}\bm{r}_{2}\mathrm{d}\bm{r}_{3},\label{eq:F3B2ndS}\\
\hspace{-1cm}\beta\mathrm{{\cal F}_{cor}^{3B-ww}}[n(\bm{r})]=\frac{1}{2}\lambda_{w}\iiint_{\mathbb{R}^{3}}\iiint_{\mathbb{R}^{3}}n(\bm{r}_{1})d\bm{r}_{1}\iiint_{\mathbb{R}^{3}}f_{w}(r_{12})f_{w}(r_{13})\left(\frac{\bm{r}_{12}\cdot\bm{r}_{13}}{r_{12}r_{13}}-\cos\theta_{0}\right)^{2}n(\bm{r}_{2})n(\bm{r}_{3})\mathrm{d}\bm{r}_{2}\mathrm{d}\bm{r}_{3},\label{eq:F3Bww}
\end{gather}
 où $n$ désigne la densité de solvant moyennée sur les angles. Les
expressions des dérivés fonctionnelles sont données en \ref{sec:D=0000E9riv=0000E9es-des-fonctionnelles}.

\begin{figure}
\noindent \centering{}\includegraphics[width=0.6\textwidth]{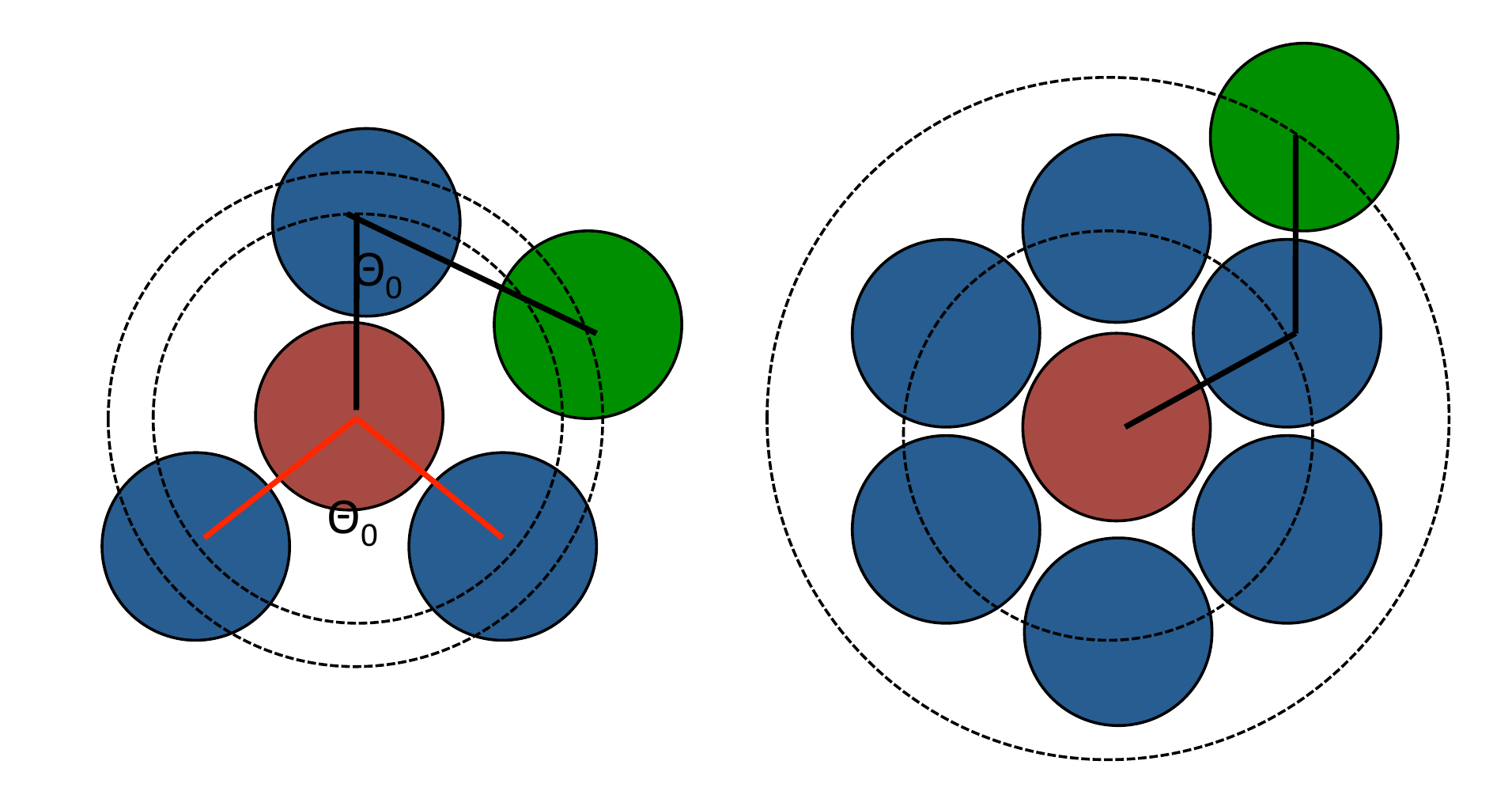}\protect\caption{Schéma de la structure de solvatation autour d'un soluté (en rouge)
dans un fluide associé (à gauche) et non-associé (à droite). Les molécules
de la première couche de solvation sont représentées en bleu, celles
de la deuxième en vert. L'angle tétraédrique $\theta_{0}$ favorisé
par l'utilisation de l'\ref{eq:F3B1S} est en rouge, celui favorisé
par l' \ref{eq:F3B2ndS} est en noir.  \label{fig:shemaF3B}}
\end{figure}

La correction se décompose en la somme de trois termes car il existe
un arrangement tétraédrique entre les molécules de solvant mais aussi
entre le soluté et le solvant. Le terme de l'\ref{eq:F3Bww} correspond
à un terme d'interaction entre molécules de solvant, c'est donc le
terme équivalent au potentiel du modèle d'eau MW de l'\ref{eq:Vmolinero}.
Le terme de l'\ref{eq:F3B1S} renforce l'arrangement tétraédrique
entre les sites des solutés et deux molécules d'eau se trouvant dans
la première couche de solvatation. Celui de l'\ref{eq:F3B2ndS} renforce
l'ordre tétraédrique entre les sites de solutés et deux molécules
d'eau, l'une se trouvant dans la première couche de solvatation et
l'autre dans la seconde. Les arrangements spatiaux favorisés par ces
deux derniers termes sont illustrés sur la \ref{fig:shemaF3B}.

\begin{figure}
\noindent \centering{}\includegraphics[width=0.6\textwidth]{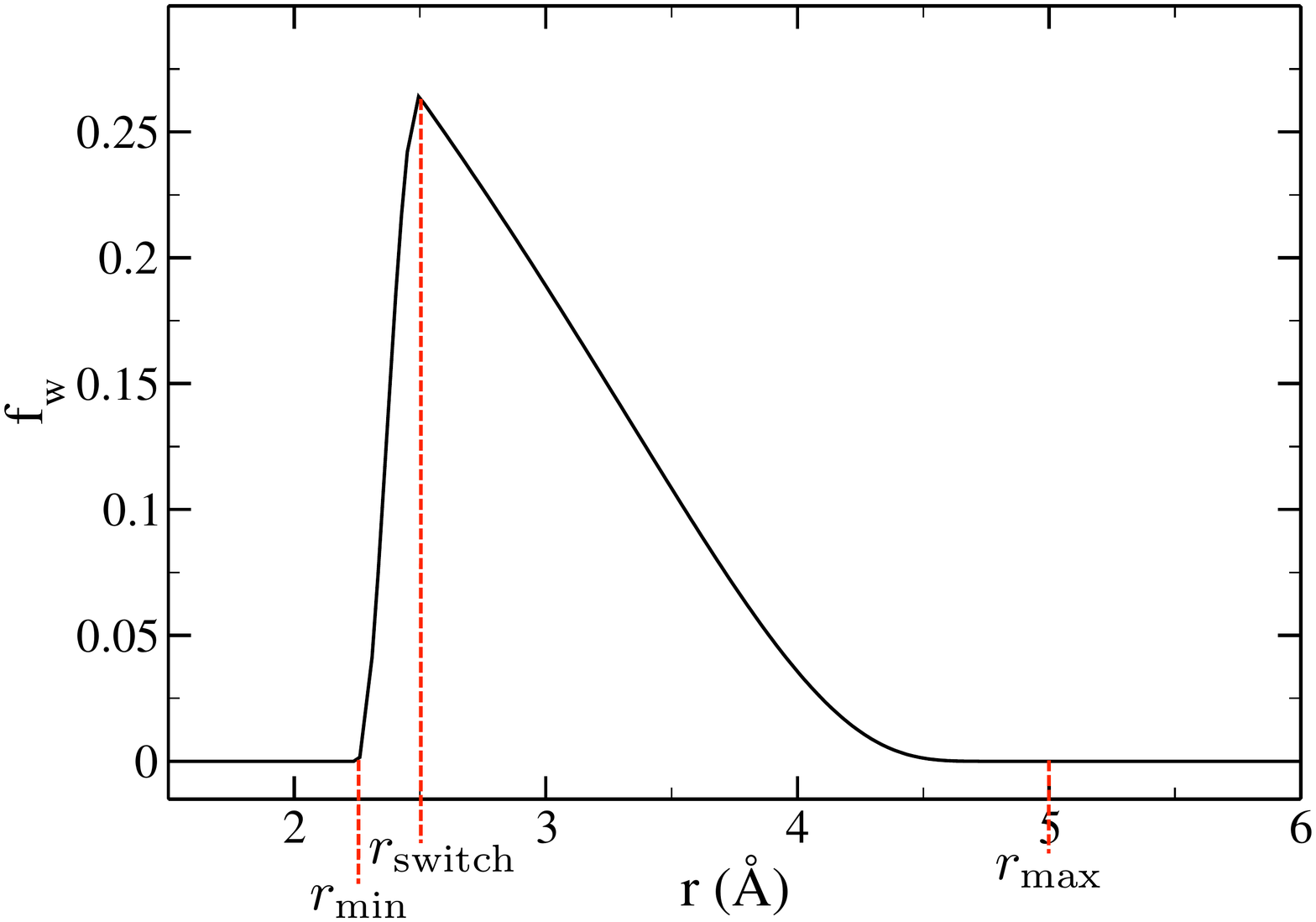}\protect\caption{Allure de la fonction $f_{w}$ utilisée dans la correction à trois
corps en fonction de la distance $r$ entre deux sites. Les paramètres
utilisés sont $r_{\mathrm{min}}=2.25\,\textrm{\AA}$, $r_{\mathrm{max}}=5.0\ \textrm{\AA}$
et $r_{\mathrm{switch}}=2.5\ \textrm{\AA}$ . Cette fonction est maximale
pour $r=r_{\mathrm{switch}}$ qui est choisi à une valeur typique
de la distance entre oxygènes de deux molécules d'eau partageant une
liaison hydrogène dans l'eau bulk.\label{fig:fww}}
\end{figure}

Une sommation est réalisée sur les $m$ sites du soluté. $\lambda_{m}^{1\mathrm{S}}$
et $\lambda_{m}^{2\mathrm{S}}$ permettent, pour chaque site du soluté,
de moduler l'importance des termes des première et deuxième couches
de solvatation. Typiquement, pour un site très chargé les valeurs
de $\lambda_{m}^{1\mathrm{S}}$ et $\lambda_{m}^{2\mathrm{S}}$ sont
élevées, pour un site peu chargé elles sont faibles. Les fonctions
$f_{m}$ et $f_{w}$ sont, ici aussi, des fonctions à courte portée,
elles sont définies comme,
\begin{gather}
f(r;r_{\mathrm{min}},r_{\mathrm{max}},r_{\mathrm{swap}})=S(r;r_{\mathrm{min}},r_{\mathrm{switch}})\exp\left[\frac{2}{3}\frac{r_{\mathrm{max}}}{r-r_{\mathrm{max}}}\right]\text{si \ensuremath{r<r_{\mathrm{max}}}et \ensuremath{0}sinon}.
\end{gather}
$S(r;r_{\mathrm{min}},r_{\mathrm{switch}})$ est une fonction cubique,
construite de telle sorte qu'elle soit nulle pour $r<r_{\mathrm{min}}$
et égale à $1$ pour $r>\mathrm{r_{switch}}$. Cette fonction a pour
but de couper l'interaction à trois corps à l'intérieur des particules,
de la rendre maximale en $r_{\mathrm{switch}}$ et de la faire décroitre
jusqu'à ce qu'elle soit nulle en $r>r_{\mathrm{max}}$. L'allure de
la fonction $f$ est donnée en \ref{fig:fww}, avec les paramètres
utilisés dans le code.

Les termes relatifs aux interactions entre molécules de solvant et
ions ont déjà été proposés dans la ref.~\cite{MDFT_zhao_molecular_2011}
mais ces termes étaient évalués dans l'espace direct ce qui rendait
leur calcul coûteux. J'ai développé le calcul des différents termes
de l'\ref{eq:F3B} dans l'espace de Fourier, pour un coût en $\mathrm{N\log N}$
plutôt que $\mathrm{N}^{3}$.

Si on introduit les produits de convolution suivants, calculables
par le produit de leurs transformées de Fourier, 
\begin{equation}
F_{\alpha\beta}(\bm{r}_{1})=\iiint_{\mathbb{R}^{3}}f_{w}(r_{12})\frac{\alpha_{12}\beta_{12}}{r_{12}^{2}}n(\bm{r}_{2})\mathrm{d}\bm{r}_{2}\text{, avec \ensuremath{\alpha\text{ et }\beta=x}ou \ensuremath{y}ou \ensuremath{z}},
\end{equation}
\begin{equation}
\bm{F}(\bm{r}_{1})=\iiint_{\mathbb{R}^{3}}f_{w}(r_{12})\frac{\bm{r}_{12}}{r_{12}}n(\bm{r}_{2})\mathrm{d}\bm{r}_{2},
\end{equation}
\begin{equation}
F_{0}(\bm{r}_{1})=\iiint_{\mathbb{R}^{3}}f_{w}(r_{12})n(\bm{r}_{2})\mathrm{d}\bm{r}_{2},
\end{equation}
l'\ref{eq:F3Bww} se réécrit en fonction de ces produits de convolution,
\begin{equation}
\beta\mathrm{{\cal F}_{cor}^{3B-ww}}[n(\bm{r})]=\frac{1}{2}\lambda_{w}\iiint_{\mathbb{R}^{3}}n(\bm{r}_{1})\left[\sum_{\left(\alpha,\beta\right)\in\left\{ x,y,z\right\} ^{2}}F_{\alpha\beta}(\bm{r}_{1})^{2}+\cos^{2}\theta_{0}F_{0}(\bm{r}_{1})^{2}-2\cos\theta_{0}\bm{F}(\bm{r}_{1})\cdot\bm{F}(\bm{r}_{1})\right]\mathrm{d}\bm{r}_{1}.
\end{equation}
On définit également plusieurs intégrales permettant de calculer les
termes d'interaction à trois corps entre soluté et solvant. On souligne
la présence de l'exposant $m$ qui rappelle que ces produits de convolution
sont définis pour chaque site de soluté,
\begin{equation}
\mathrm{H}_{\alpha\beta}^{m}=\iiint_{\mathbb{R}^{3}}f_{m}(r_{m2})\frac{\alpha_{m2}\beta_{m2}}{r_{m2}^{2}}n(\bm{r}_{2})\mathrm{d}\bm{r}_{2},\ \text{, avec \ensuremath{\alpha\text{ et }\beta=x}ou \ensuremath{y}ou \ensuremath{z}},
\end{equation}
\begin{equation}
\bm{\mathrm{H}}{}^{m}=\iiint_{\mathbb{R}^{3}}f_{m}(r_{m2})\frac{\bm{r}_{m2}}{r_{m2}}n(\bm{r}_{2})\mathrm{d}\bm{r}_{2},
\end{equation}
\begin{equation}
\mathrm{H}_{0}^{m}=\iiint_{\mathbb{R}^{3}}f_{w}(r_{m2})n(\bm{r}_{2})\mathrm{d}\bm{r}_{2}.
\end{equation}

On calcule ces intégrales dans l'espace direct, sur un petit volume
autour des sites $m$ car les fonctions $f$ sont à courte portée.

Le terme de l'\ref{eq:F3B1S}, jouant sur la première couche de solvatation,
s'écrit alors, 
\begin{equation}
\beta\mathrm{{\cal F}_{cor}^{3B-1S}}[n(\bm{r})]=\sum_{m}\lambda_{m}^{1\mathrm{S}}\left[\sum_{\left(\alpha,\beta\right)\in\left\{ x,y,z\right\} ^{2}}\left(\mathrm{H}_{\alpha\beta}^{m}\right)^{2}+\left(\cos\theta_{0}\mathrm{H}_{0}^{m}\right)^{2}-2\cos\theta_{0}\bm{\mathrm{H}}{}^{m}\cdot\bm{\mathrm{H}}{}^{m}\right].
\end{equation}
Le terme de l'\ref{eq:F3B2ndS}, jouant sur la seconde couche de solvatation,
peut se réécrire en utilisant les produit de convolution $F_{0}$,
$\bm{F}$ et $F_{\alpha\beta}$ que l'on vient d'introduire,
\begin{gather}
\beta\mathrm{{\cal F}_{cor}^{3B-2S}}[n(\bm{r})]=\sum_{\left(\alpha,\beta\right)\in\left\{ x,y,z\right\} ^{2}}\iiint_{\mathbb{R}^{3}}F_{\alpha\beta}(\bm{r}_{1})G_{\alpha\beta}(\bm{r}_{1})\mathrm{d}\bm{r}_{1}+\cos^{2}\theta_{0}\iiint_{\mathbb{R}^{3}}F_{0}(\bm{r}_{1})G_{0}(\bm{r}_{1})\mathrm{d}\bm{r}_{1}\nonumber \\
\hspace{4cm}-2\cos\theta_{0}\iiint_{\mathbb{R}^{3}}\bm{F}(\bm{r}_{1})\cdot\bm{G}(\bm{r}_{1})\mathrm{d}\bm{r}_{1},
\end{gather}
où on a introduit les notations suivantes,
\begin{equation}
G_{\alpha\beta}(\bm{r}_{1})=\sum_{m}\lambda_{m}^{2\mathrm{S}}f_{m}(r_{m1})\frac{\alpha_{m1}\beta_{m1}}{r_{m1}^{2}}n(\bm{r}_{1}),
\end{equation}
\begin{equation}
\bm{G}(\bm{r}_{1})=\sum_{m}\lambda_{m}^{2\mathrm{S}}f_{m}(r_{m1})\frac{\bm{r}_{m1}}{r_{m1}}n(\bm{r}_{1}),
\end{equation}
\begin{equation}
G_{0}(\bm{r}_{1})=\sum_{m}\lambda_{m}^{2\mathrm{S}}f_{m}(r_{m1})n(\bm{r}_{1}).
\end{equation}

Ces huit fonctions $G$ sont calculées dans l'espace direct.

\begin{figure}
\noindent \centering{}\includegraphics[width=0.6\textwidth]{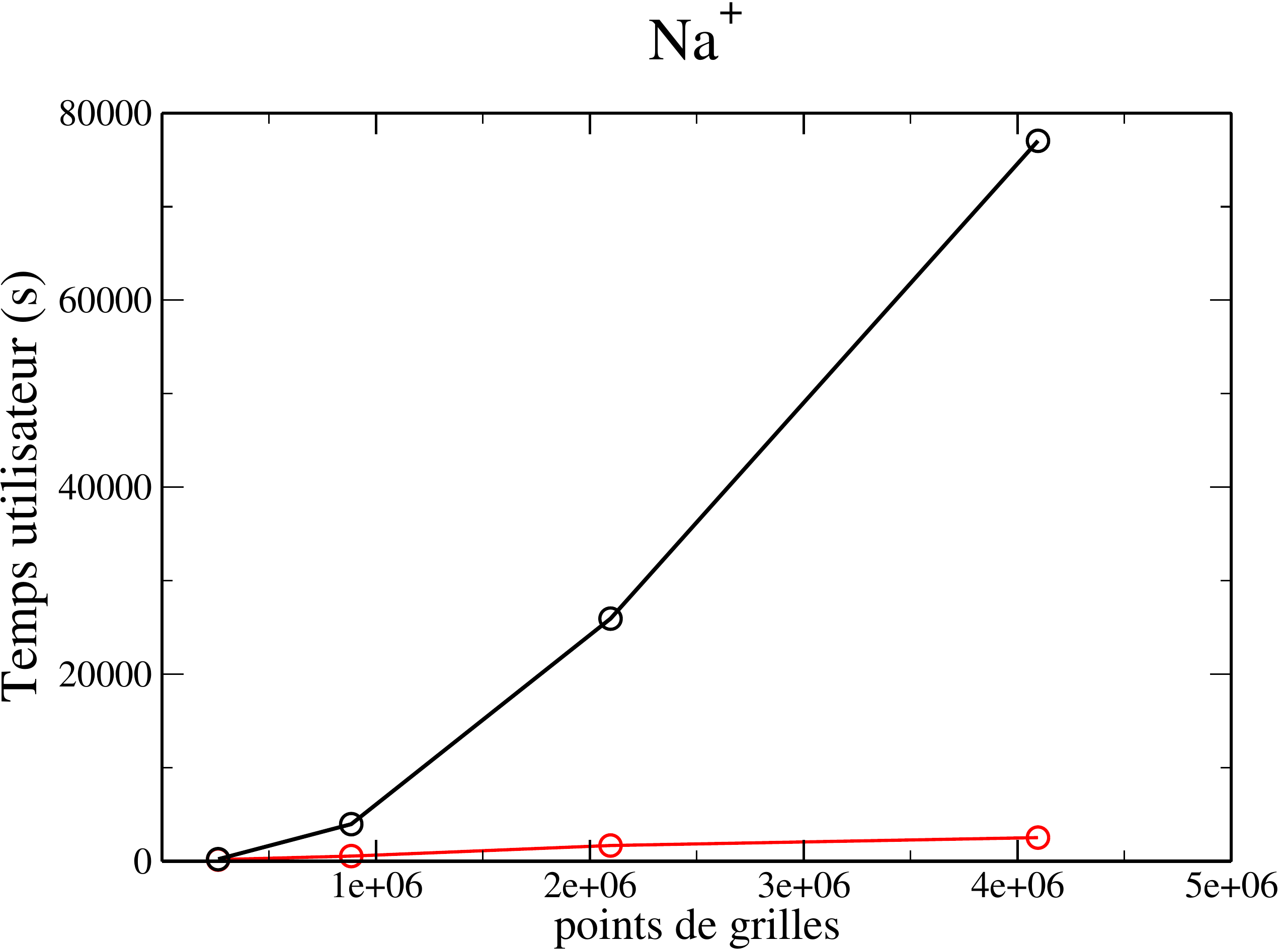}\protect\caption{Temps de calcul utilisateur nécessaire à la minimisation d'une fonctionnelle
utilisant le terme à trois corps, soit calculé dans l'espace direct
(en noir) soit à l'aide de transformées des Fourier discrètes (en
rouge), dans le cas de $\mathrm{Na^{+}}$. On compare les résultats
pour une boite cubique d'arête $25\ \textrm{\AA}$, avec un maillage
de $64^{3},\ 96^{3},\ 128^{3},\ 160^{3}$ points. \label{fig:Comp_OldNew_F3B_temps}}
\end{figure}

On insiste sur le fait qu'au lieu d'avoir à calculer des doubles intégrales
sur le volume dans l'espace direct, on se ramène au calcul des dix
produits de convolution $F_{0},\ F_{\alpha\beta}$ et $\bm{F}$, soit
vingt transformées de Fourier (directes et inverses). Il suffit ensuite
de calculer des intégrales simples sur le volume. Sur la \ref{fig:Comp_OldNew_F3B_temps},
on compare l'efficacité des deux routines pour le cation alcalin sodium.
L'une calcule ce terme à trois corps dans l'espace direct, l'autre
en utilisant le nouvel algorithme.  On constate que l'utilisation
de transformées de Fourier réduit considérablement le temps de calcul
du terme à trois corps et, a fortiori, le temps de calcul global de
la minimisation fonctionnelle. La formulation utilisant les transformées
de Fourier introduite est donc parfaitement adaptée à l'étude d'un
soluté possédant un ou plusieurs sites chargés.

\section{Application à la solvatation}

Dans l'\ref{eq:F3B1S} et l'\ref{eq:F3B2ndS}, les paramètres $\lambda_{m}^{1\mathrm{S}}$
et $\lambda_{m}^{2\mathrm{S}}$ permettent de moduler l'importance
du terme de correction à trois corps. Lorsque les sites de solvant
ne sont pas des sites susceptibles d'être engagés dans une liaison
hydrogène, ces deux paramètres sont choisis nuls : $\lambda_{m}^{1\mathrm{S}}=\lambda_{m}^{2\mathrm{S}}=0$.
Il n'y a donc aucune modification des résultats structuraux ou énergétiques
pour ces solutés, notamment pour les alcanes et les gaz rares présentés
dans la \ref{sec:The-next-section}. Pour les sites chargés, il est
évident qu'il faudrait choisir ces termes en fonction de la force
des liaisons hydrogènes mettant en jeu le site considéré. 

Cependant, ce travail de thèse a pour but de prouver que le principe
de l'utilisation de la théorie de la fonctionnelle de la densité au
cas de l'eau est viable. L'optimisation de tous les paramètres mis
en jeu, notamment dans ce cas particulier, est repoussée à plus tard,
quand on sera satisfait de la fonctionnelle utilisée.

Par simplicité, et pour ne pas faire un ajustement des paramètres
sur les résultats que l'on cherche à obtenir, l'hypothèse simplificatrice
suivante a été faite: pour les sites ayant une charge entière, typiquement
les ions monovalents, on choisit $\mathrm{\mathrm{k_{B}}T}\lambda_{m}^{1\mathrm{S}}=\mathrm{\mathrm{k_{B}}T}\lambda_{m}^{2\mathrm{S}}=100\ \mathrm{kJ.mol^{-1}}$.
Pour les sites impliqués dans des liaisons hydrogènes portant des
charges partielles (par exemple l'oxygène d'une molécule d'eau), on
choisit $\mathrm{\mathrm{k_{B}}T}\lambda_{m}^{1\mathrm{S}}=\mathrm{\mathrm{k_{B}}T}\lambda_{m}^{2\mathrm{S}}=75\ \mathrm{kJ.mol^{-1}}$.
Le choix des valeurs pour les ions a été fait pour obtenir le meilleur
profil de distribution radiale possible pour l'ion chlorure tandis
que les paramètres pour les molécules neutres permettent d'obtenir
le meilleur profil de distribution radiale possible pour l'eau.

\subsection{Structure}

\subsubsection{Les ions}

\begin{figure}
\noindent \centering{}\includegraphics[width=0.8\textwidth]{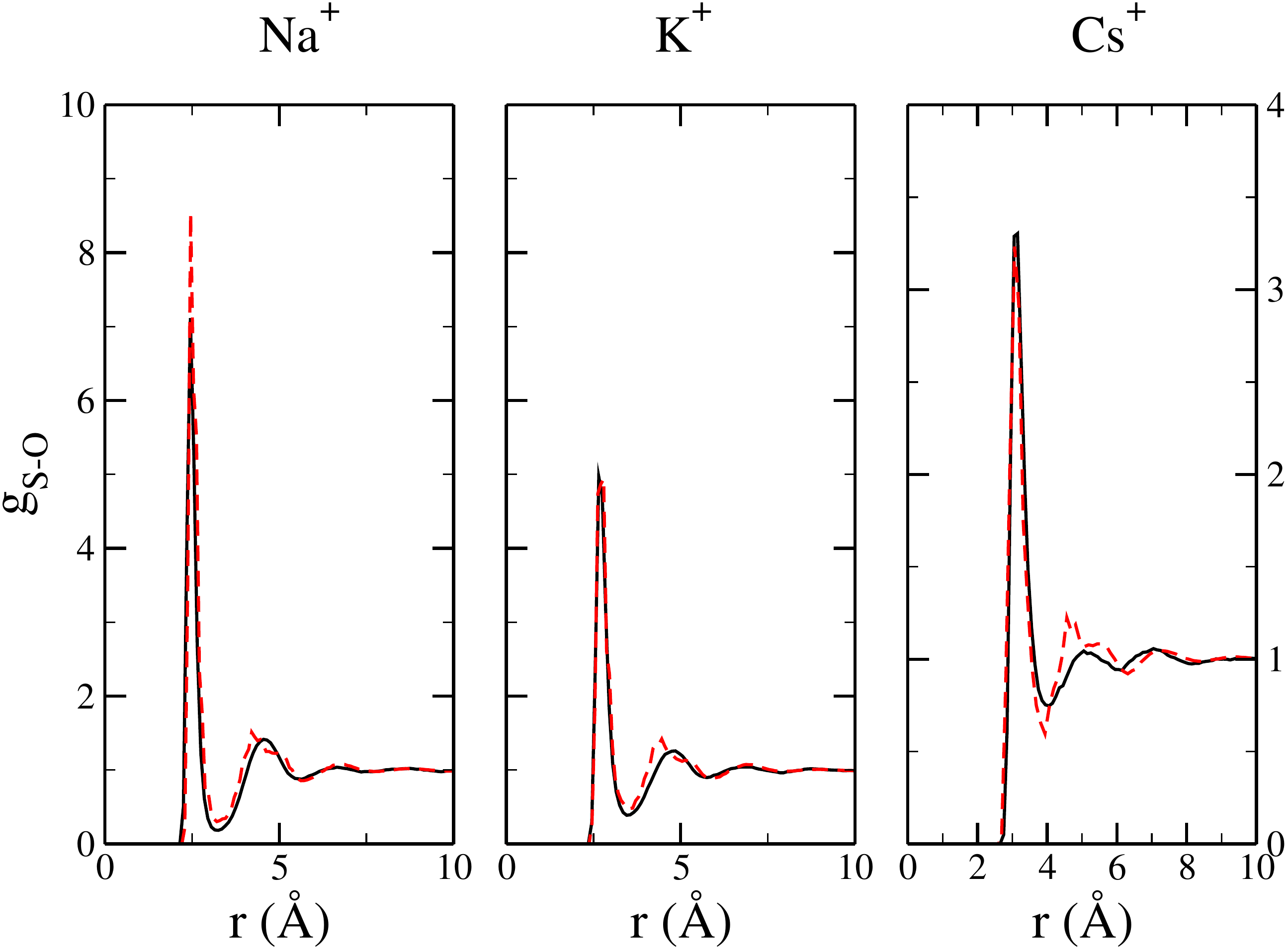}\protect\caption{Fonctions de distribution radiale entre l'oxygène de l'eau et trois
cations alcalins, sodium, potassium et césium. Les résultats des simulations
MD sont en trait noir plein et ceux des calculs MDFT, incluant la
correction trois corps, en tirets rouges.\label{fig:rdf_alcalins_multi_F3B}}
\end{figure}

\begin{figure}
\noindent \centering{}\includegraphics[width=0.8\textwidth]{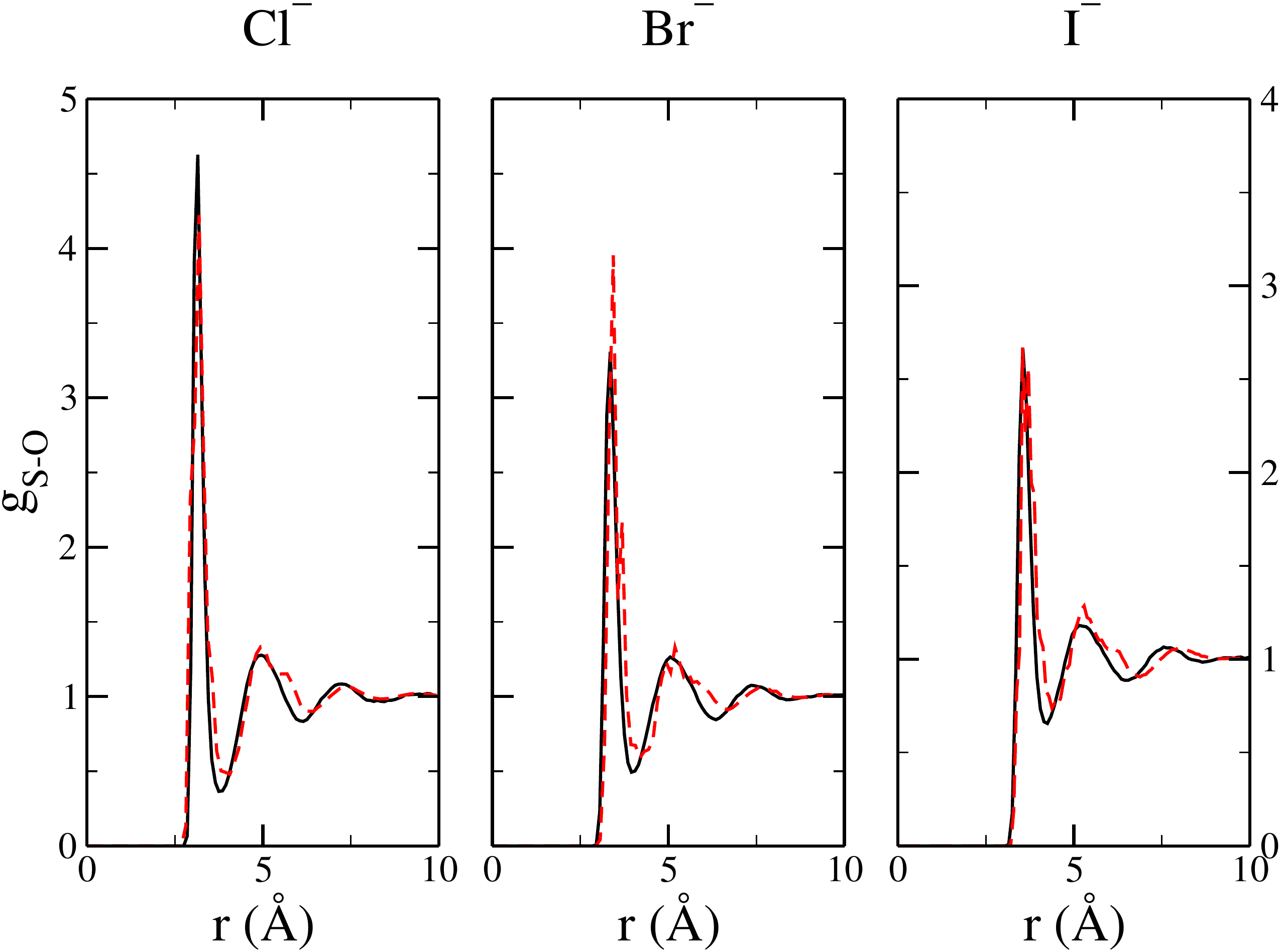}\protect\caption{Fonctions de distribution radiale entre le solvant et trois anions
halogénures, chlorure, bromure et iodure. La légende est la même que
sur la \ref{fig:rdf_alcalins_multi_F3B}.\label{fig:rdf_halogenures_multi_F3B}}
\end{figure}

On donne les fonctions de distribution radiale obtenues par dynamique
moléculaire et par minimisation fonctionnelle pour des cations alcalins
sur la \ref{fig:rdf_alcalins_multi_F3B}, et pour des halogénures
sur la \ref{fig:rdf_halogenures_multi_F3B}. On utilise la fonctionnelle
multipolaire et la correction à trois corps. Ces résultats sont donc
à comparer avec la \ref{fig:rdf_alkalin multi} pour les alcalins.
Pour les halogénures, on peut comparer avec la \ref{fig:rdf_halogenures_dip}
puisque l'utilisation de la fonctionnelle multipolaire modifie peu
les fonctions de distribution radiale. On constate que l'introduction
de ce terme à trois corps améliore notablement l'allure des fonctions
de distribution radiale. La hauteur du premier pic est désormais correcte,
la troisième couche de solvatation est très bien reproduite en position
et en intensité. Le second pic, dû à la seconde couche de solvatation
est également amélioré même si on surestime toujours légèrement l'intensité
pour des distances juste inférieures au maximum. On peut donc conclure
que c'est bien l'arrangement local tétraédrique des molécules d'eau
autour des solutés chargés qui est responsable de l'allure caractéristique
de leurs fonctions de distribution radiale.

L'écart par rapport à la MD pour le second pic empire à mesure que
l'on descend dans le tableau périodique mais les paramètres $\lambda_{m}^{1\mathrm{S}}$
et $\lambda_{m}^{2\mathrm{S}}$ ont été choisis pour le sodium et
le chlorure. Ces ions se lient fortement avec l'eau par liaisons hydrogènes.
Quand on descend dans le tableau périodique, les rayons ioniques augmentent
et la densité de charge diminue. Le potassium et le bromure et à fortiori
l'iodure et le césium, vont donc faire des liaisons hydrogènes moins
fortes avec l'eau. On a vérifié que des paramètres $\lambda_{m}^{1\mathrm{S}}$
et $\lambda_{m}^{2\mathrm{S}}$ plus faibles sont mieux adaptés pour
ces ions plus gros. À ce stade, il est cependant satisfaisant de pouvoir
utiliser les mêmes paramètres pour tous les ions monovalents.

En trouvant une façon de choisir ces paramètres à priori en fonction
de la force des liaisons hydrogènes mises en jeu, on peut s'attendre
à reproduire correctement l'évolution de l'allure des fonctions de
distribution radiale dans un groupe du tableau périodique.

\subsubsection{Les solutés neutres}

En ce qui concerne les solutés neutres, on a vu en \ref{sub:solut=0000E9s hydrophyles sans F3B}
que les profils de solvatation des sites peu chargés sont bien reproduits
par minimisation de la fonctionnelle avec un traitement multipolaire
de la polarisation. Ce n'est pas le cas pour les sites chargés.

\begin{figure}
\noindent \centering{}\includegraphics[width=0.8\textwidth]{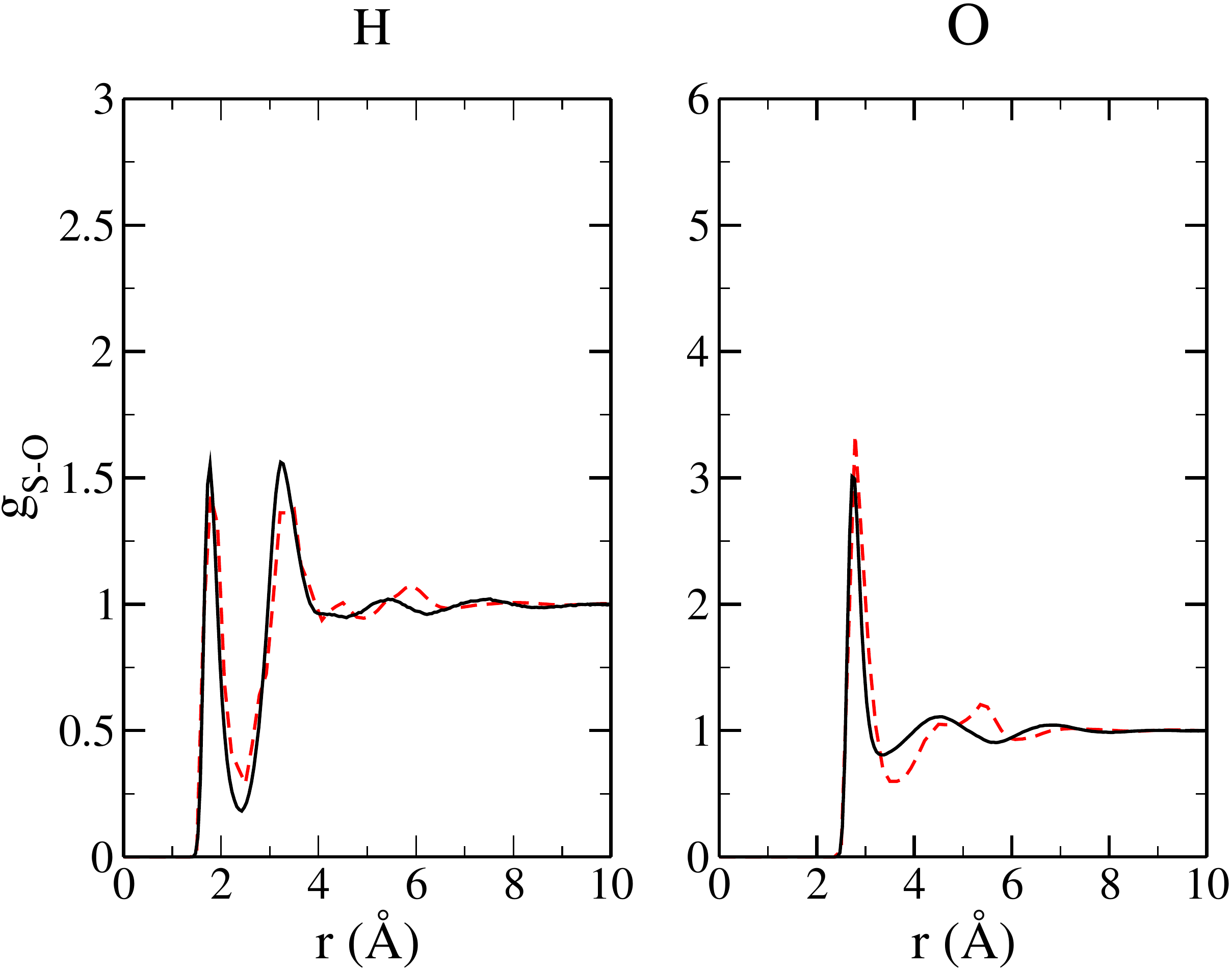}\protect\caption{Fonctions de distribution radiale entre l'oxygène de l'eau solvant
et l'oxygène et l'hydrogène de l'eau soluté. MD en traits noirs pleins
et MDFT incluant le terme correctif à trois corps en traits rouges
discontinus. \label{fig:rdf_water_dip-avec-F3B}}
\end{figure}

On observe sur la \ref{fig:rdf_water_dip-avec-F3B} que l'ajout du
terme correctif à trois corps améliore considérablement l'accord entre
les fonctions de distribution radiale obtenues par dynamique moléculaire
et celles obtenues avec le code mdft pour le soluté eau SPC/E. La
première couche de solvatation de l'eau est responsable des deux premiers
pics de la fonction de distribution radiale entre l'hydrogène du soluté
et l'oxygène du solvant, et du premier pic de celle entre l'oxygène
du soluté et celui du solvant. Ces pics sont désormais très bien reproduits
par minimisation fonctionnelle. Leurs positions et leurs hauteurs
sont désormais semblables pour les deux méthodes, MD et MDFT. On surestime
toujours, mais moins, la déplétion entre la première et la deuxième
couche de solvatation pour $g_{OO}$.

Dans le cas de $g_{OO}$ tout comme comme dans celui des ions, on
constate une légère amélioration de la prédiction de la seconde couche
de solvatation, avec une réduction du pic situé vers 5.5 $\textrm{\AA}$.
Ce pic était situé trop loin de l'atome d'oxygène du soluté. Un nouveau
pic apparait vers 4.5 $\textrm{\AA}$, en accord avec la MD. L'effet
du terme à trois corps est clair : il renforce l'ordre tétraédrique,
réduit la distance entre soluté et molécules d'eau de la deuxième
couche de solvatation, augmentant ainsi la densité pour des zones
de l'espace situées vers 4.5 $\textrm{\AA}$. C'est ce phénomène qui
est responsable de ce nouveau pic. La répulsion entre molécules d'eau
a pour conséquence de réduire les densités au voisinage de cette zone
de plus haute densité à 4.5 $\textrm{\AA}$. En conséquence, la densité
diminue là où se situaient les zones de hautes densités dues à la
seconde couche de solvatation précédemment, et le pic situé vers $5.5\ \textrm{\AA}$
diminue.

Le troisième pic, dû la troisième couche de solvatation est lui aussi
grandement amélioré par l'introduction du terme à trois corps. C'est
une conséquence directe de la modification des zones de haute densité
de la seconde couche de solvatation. Ces progrès pour les deuxième
et troisième couches de solvatation sont moins visibles sur la fonction
de distribution radiale entre l'oxygène du solvant et l'hydrogène
du soluté. 

\begin{figure}
\noindent \centering{}\includegraphics[width=0.6\textwidth]{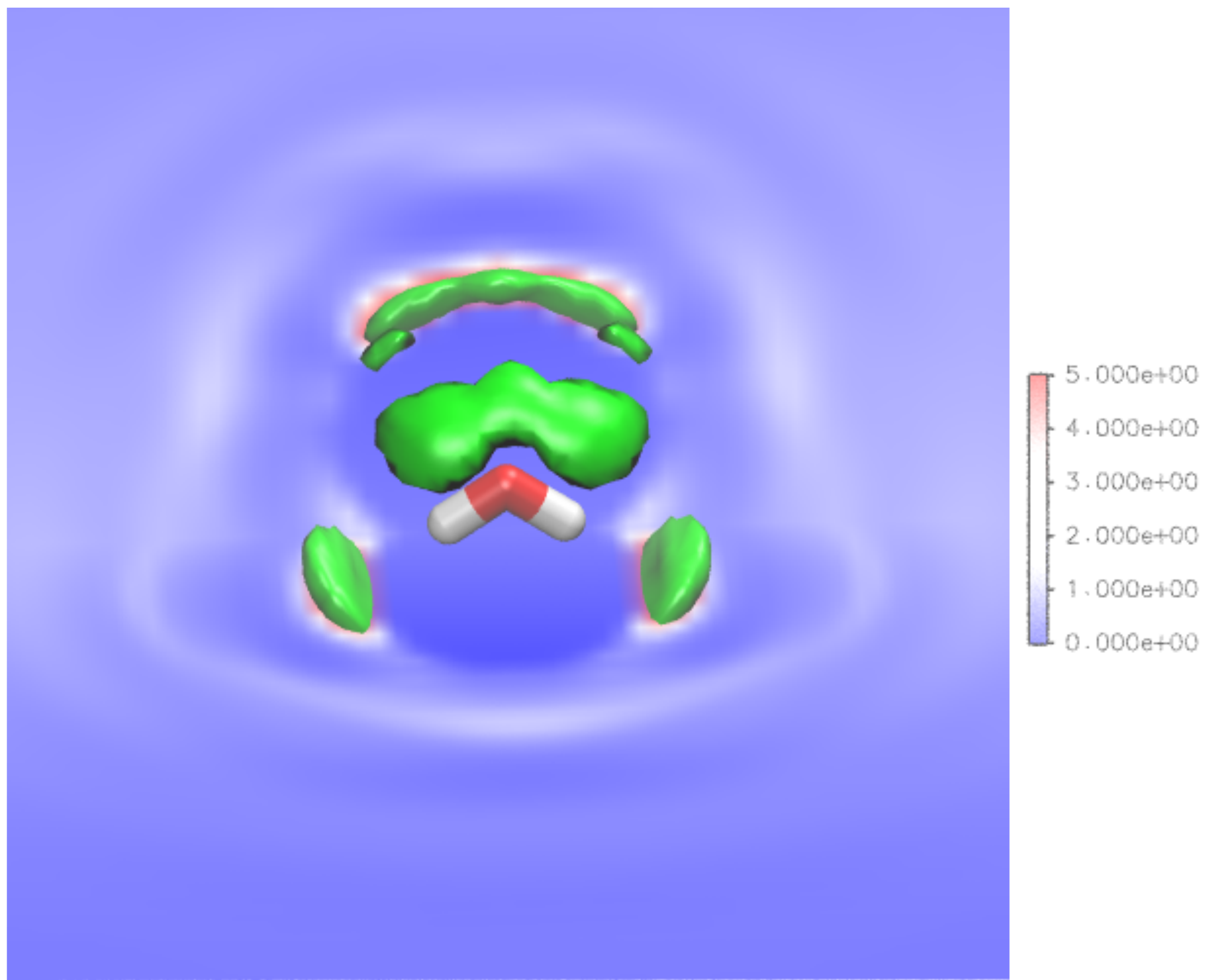}\protect\caption{Carte de densité autour de la molécule d'eau obtenue en incluant le
terme correctif à trois corps. Des isosurfaces de haute densité sont
représentées en vert ($n(\bm{r})/n_{b}$=3) pour aider à la visualisation.\label{fig:waterF3Bmap}}
\end{figure}

Sur la \ref{fig:waterF3Bmap}, on représente des cartes de densité
autour du soluté eau, ainsi que des isosurfaces de haute densité,
($n(\bm{r})/n_{b})$=3. On constate que les régions de l'espace où
on s'attend à observer des liaisons hydrogènes coïncident avec les
zones de haute densité. On remarque aussi que les régions qui jouxtent
ces zones de haute densité subissent une déplétion. De telles cartes
de densité sont compliquées à obtenir par MD ou MC car elles requièrent
de faire des histogrammes de densité. Elles sont au contraire obtenues
directement par le code mdft, la densité étant la variable naturelle
de la théorie.

\begin{figure}
\noindent \centering{}\includegraphics[width=0.8\textwidth]{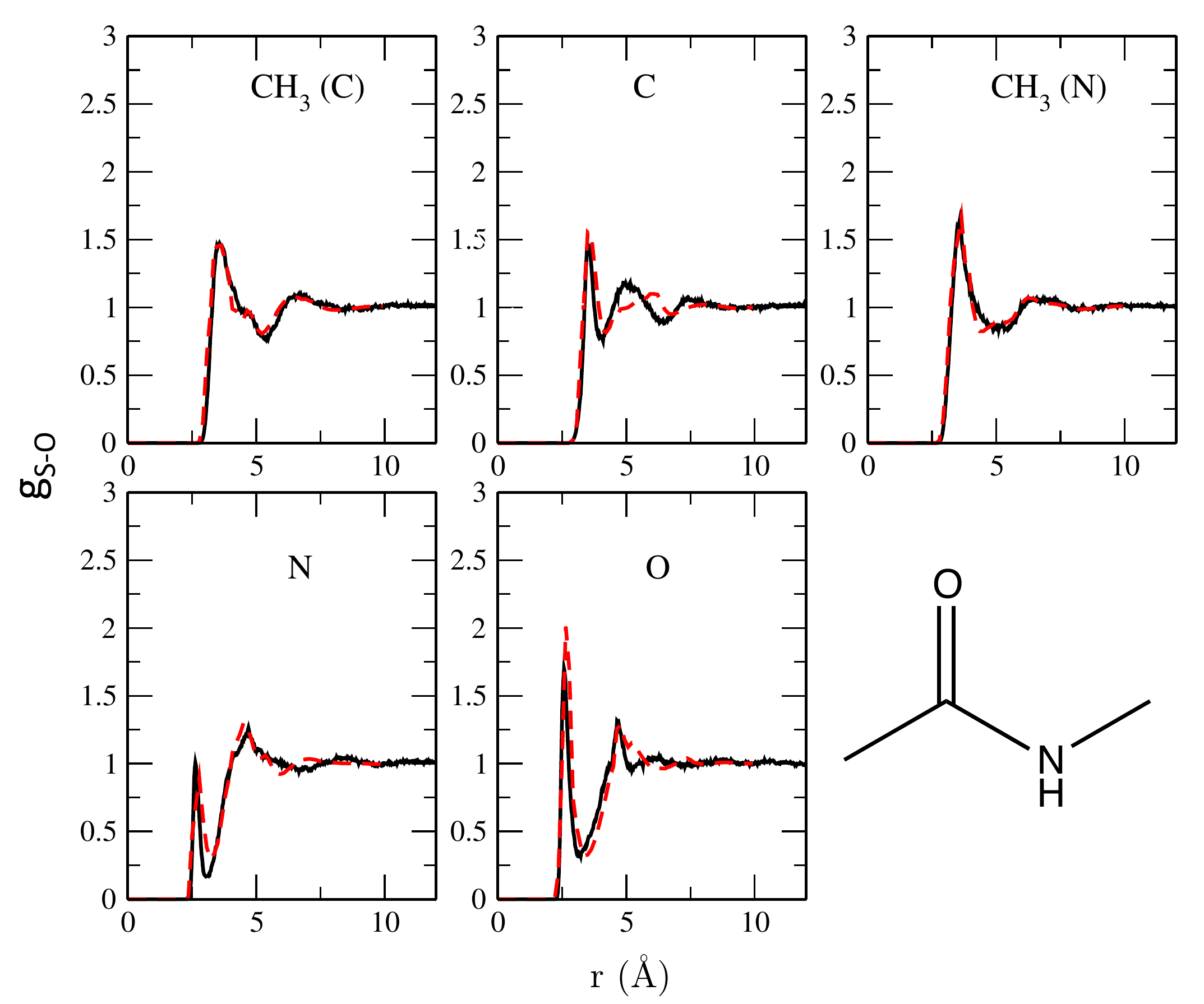}\protect\caption{Fonctions de distribution radiale entre l'oxygène de l'eau solvant
et les différents sites de la molécule NMA. La légende est la même
que sur la figure \ref{fig:rdf_water_dip-avec-F3B}. \label{fig:rdf_NMA_dip-avec F3B}}
\end{figure}

Cette amélioration réalisée grâce à l'inclusion du terme à trois corps
est confirmée par les fonctions de distribution radiale obtenues pour
la N-méthylacétamide présentées en \ref{fig:rdf_NMA_dip-avec F3B}.
Les pics des première et deuxième couches de solvatation des sites
impliqués dans des liaisons hydrogènes, l'oxygène et l'azote, sont
améliorés en terme de hauteur et de position. Les fonctions de distribution
radiale des sites ne formant pas de liaisons hydrogènes sont elles-aussi
considérablement améliorées. Ceci s'explique par le fait que les modèles
de molécule utilisés sont rigides. En conséquence, la modification
de l'environnement proche d'un site modifie directement l'allure de
la fonction de distribution radiale calculée depuis un autre site. 

\begin{figure}[H]
\noindent \centering{}\includegraphics[width=0.6\textwidth]{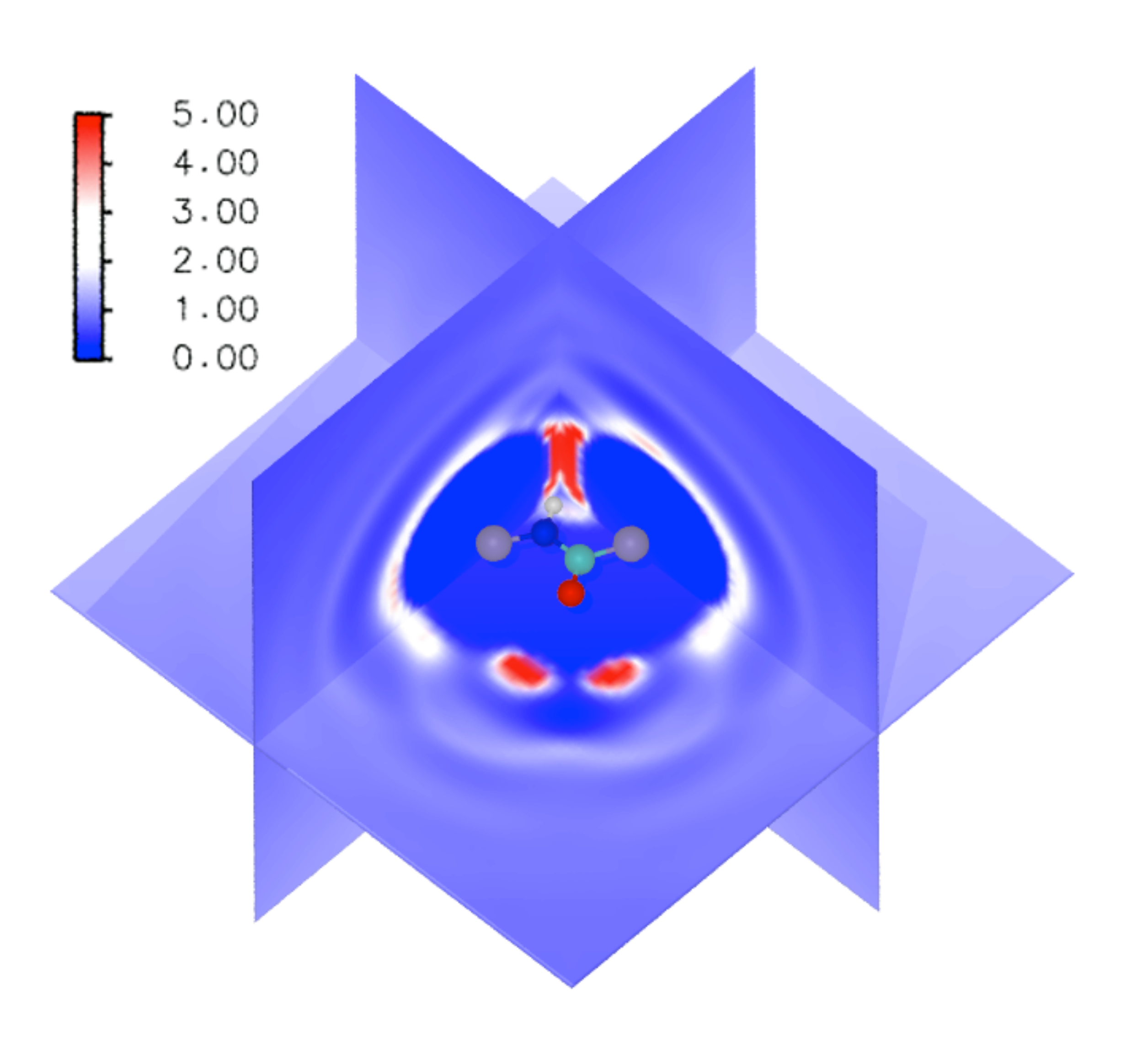}\protect\caption{Densité 3D du solvant eau autour de la molécule fixe de NMA obtenue
par MDFT. \label{fig:rdf_NMA_dip-avec F3B-density-map}}
\end{figure}

L'utilisation du terme à trois corps permet de renforcer la coordination
tétraédrique sur les sites engagés dans des liaisons hydrogènes sur
la \ref{fig:rdf_NMA_dip-avec F3B-density-map}. Des zones de plus
haute densité, en rouge, apparaissent clairement autour du groupe
carbonyle et de la liaison azote-hydrogène de la molécule NMA.

\fbox{\begin{minipage}[t]{1\columnwidth}%
L'inclusion d'un terme \textit{ad-hoc} qui renforce l'ordre tétraédrique
entre un site donneur ou accepteur de liaisons hydrogènes et les molécules
d'eau du solvant améliore considérablement les structures de solvatation
obtenues par minimisation fonctionnelle des solutés polaires. En réalité
ce terme modifie le potentiel d'interaction entre soluté et solvant
et peut être considéré comme un terme extérieur plutôt que d'excès.
\textit{Stricto sensu} les potentiels d'interaction utilisés pour
les calculs MD et MDFT ne sont alors plus les mêmes.%
\end{minipage}}

\lhead[\chaptername~\thechapter]{\rightmark}

\rhead[\leftmark]{}

\lfoot[\thepage]{}

\cfoot{}

\rfoot[]{\thepage}

\chapter{Hydrophobicité et couplage multi-échelle \label{chap:hydro}}

La théorie de la fonctionnelle de la densité classique MDFT présentée
dans les chapitres précédents est construite à un niveau de description
moléculaire. Elle donne des résultats sensiblement comparables à ceux
des simulations moléculaires pour des petits solutés hydrophobes.
Les propriétés de solvatation dans l'eau des petits solutés hydrophobes
(quelques angströms) sont très différentes de celles des solutés de
taille plus importante (nanométrique)\cite{lum_hydrophobicity_1999}.
Cette différence de comportement s'explique par une modification du
phénomène physique gouvernant la mise en solution lorsque la taille
du soluté augmente. On va d'abord décrire les propriétés de solvatation
de solutés hydrophobes à ces deux échelles, avant de proposer un moyen
d'introduire dans la fonctionnelle la solvatation des solutés hydrophobes
aux échelles microscopiques et mésoscopiques. Cette modification de
la fonctionnelle constitue une illustration de la possibilité d'utiliser
la MDFT pour étudier des problèmes multi-échelles. L'objectif est
de pouvoir coupler la MDFT, qui permet une description rapide du solvant
au niveau moléculaire, avec des théories mésoscopiques utilisant aussi
la densité comme l'hydrodynamique.

\section{La solvatation des solutés hydrophobes aux échelles microscopique
et mésoscopique}

On nomme hydrophobes les espèces peu solubles dans l'eau. Les solutés
hydrophobes apolaires et aprotiques ne peuvent pas créer de liaisons
hydrogènes avec les molécules d'eau. De ce fait, la réorganisation
des molécules de solvant autour des solutés vise à conserver au maximum
le nombre de liaisons hydrogènes entre molécules d'eau.

\begin{figure}
\noindent \begin{centering}
\includegraphics[width=0.3\textwidth]{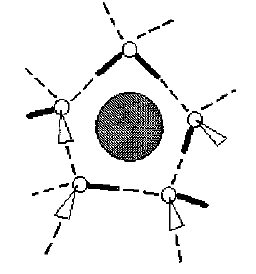}\protect\caption{Schéma de la structure des molécules d'eau autour d'un petit soluté
hydrophobe. Les tirets indiquent des liaisons hydrogènes (figure extraite
de \cite{lum_hydrophobicity_1999})\label{fig:small_solute_hydro}. }

\par\end{centering}

\end{figure}

Comme schématisé sur la \ref{fig:small_solute_hydro}, l'ordre spatial
des molécules d'eau autour des petits solutés permet de ne pas perdre
de liaisons hydrogènes entre molécules de solvant. Le coût de cette
réorganisation est principalement entropique et explique la faible
solubilité des petits solutés apolaires dans l'eau. On peut montrer
que l'énergie libre de solvatation d'un tel soluté hydrophobe est
proportionnel au volume exclu, c'est-à-dire au volume effectif du
soluté\cite{chandler_lectures_2011,lum_hydrophobicity_1999}.

La mise en solution des solutés de plus grande taille peut causer
un démouillage, c'est-à-dire qu'il peut se former à la surface du
soluté une interface liquide-gaz\cite{stillinger_structure_1973}.
Les molécules d'eau en surface du soluté perdent une partie de leurs
liaisons hydrogènes, le coût énergétique de la solvatation est essentiellement
enthalpique. On peut montrer que l'énergie de solvatation de ces grands
solutés est proportionnelle à leur surface. 

\begin{figure}
\noindent \centering{}\includegraphics[width=0.6\textwidth]{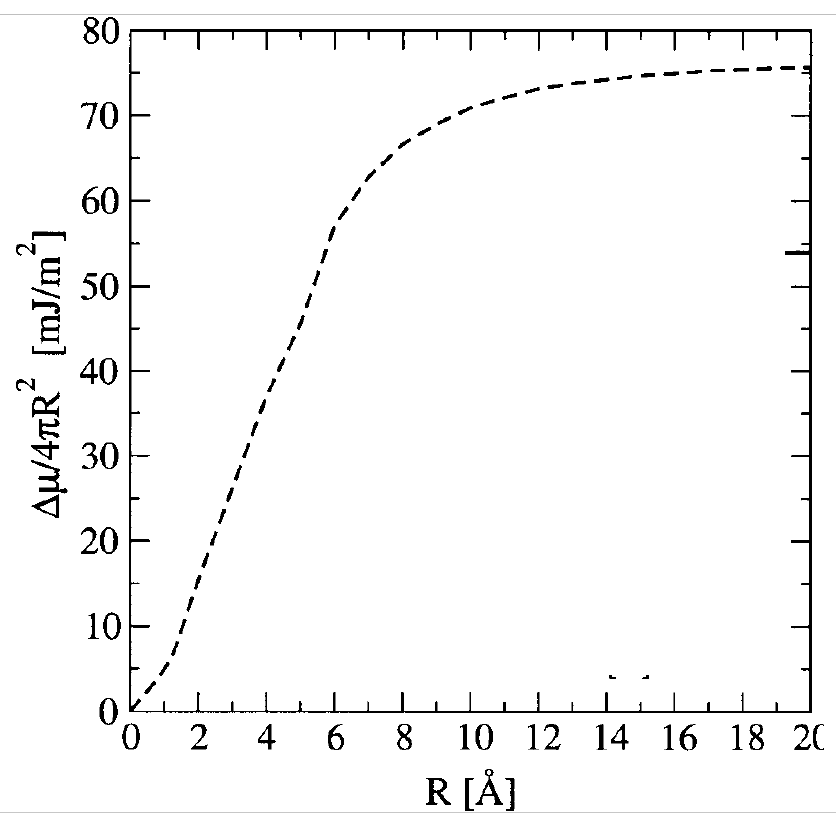}\protect\caption{Energie libre de solvatation par unité de surface, notée $\Delta\mu$
pour un soluté sphère dure de rayon R, à 298$\ $K dans l'eau SPC/E
d'après \cite{huang_hydrophobic_2002}\label{fig:ener_chandler_spce}. }
\end{figure}

Dans la référence \cite{huang_hydrophobic_2002}, Chandler et ses
collaborateurs illustrent ce changement de comportement en calculant
l'énergie de solvatation par unité surface d'un soluté sphère dure
dans de l'eau SPC/E, ces résultats sont donnés en \ref{fig:ener_chandler_spce}.
Pour des solutés de faible rayon, l'énergie libre de solvatation est
proportionnelle au volume du soluté. Pour des solutés plus gros l'énergie
libre de solvatation est proportionnelle à la surface du soluté.

La solvatation des solutés hydrophobes est gouvernée par des phénomènes
physiques différents à des échelles différentes. Nous allons voir
que si le comportement aux grandes échelles n'est pas inclus à priori
dans le formalisme MDFT développé précédemment, il est néanmoins possible
de l'inclure par une approche multi-échelle.

\section{La théorie MDFT/HRF avec HSB}

On s'intéresse à deux systèmes hydrophobes déjà étudiés par simulation
numérique (MD ou MC). Chandler, dans les réf. \cite{huang_hydrophobic_2002,gaussian_field_varilly_improved_2011},
a étudié la solvatation d'une sphère dure dans l'eau SPC/E, tandis
que Hansen et Dzubiella dans la ref. \cite{DHS_dzubiella_competition_2004}
ont étudié la solvatation d'une sphère molle purement répulsive. Cette
sphère interagit avec le solvant avec le potentiel répulsif de l'\ref{eq:VpHansen_gl},
en $r^{-12}$.

\begin{figure}
\noindent \centering{}%
\begin{tabular}{cc}
\includegraphics[width=0.5\textwidth]{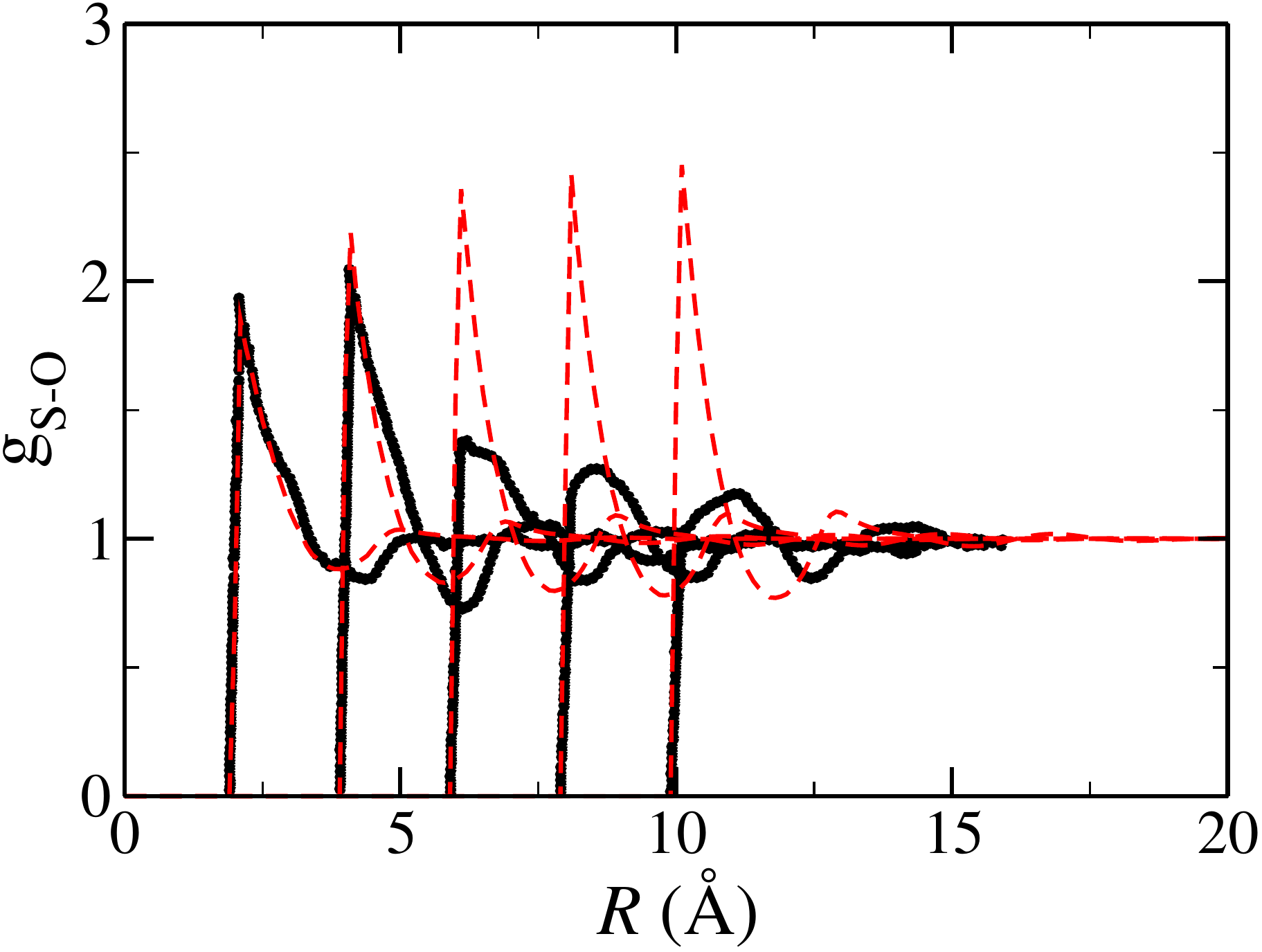} & \includegraphics[width=0.5\textwidth]{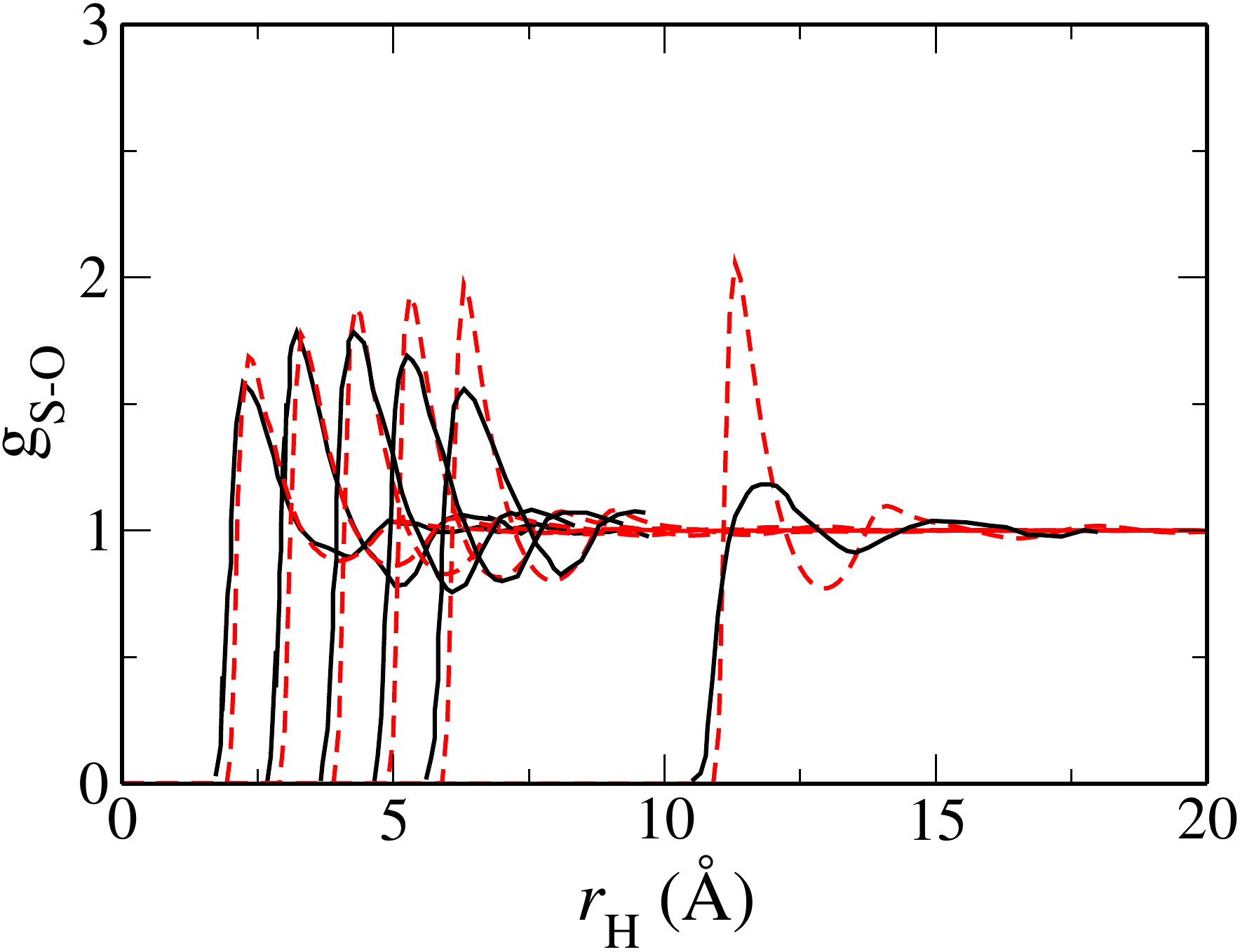}\tabularnewline
\end{tabular}\protect\caption{Comparaison des fonctions de distribution radiale:\protect \\
 -entre un soluté sphère dure de rayon $R$ \label{fig:rdf_Chandler+Hansen_MDFT_HSB}
et l'eau à gauche ;\protect \\
 -entre un soluté sphère molle de rayon $r_{\text{H}}$ et l'eau à
droite.\protect \\
 Les résultats obtenus par MDFT avec le bridge de sphères dures sont
en tirets rouges, ceux obtenus par simulation Monte-Carlo\cite{huang_hydrophobic_2002}
ou par dynamique moléculaire\cite{DHS_dzubiella_competition_2004}
sont en noir.}
\end{figure}

Ces deux systèmes ont été étudiés avec la même fonctionnelle que celle
utilisée pour les alcanes dans la \ref{sec:The-next-section}, avec
le terme de bridge de sphères dures. Dans la \ref{fig:rdf_Chandler+Hansen_MDFT_HSB},
on compare les fonctions de distribution radiale obtenues par MDFT
à celles obtenues par MD pour ces deux systèmes, en fonction de la
taille des sphères.

\begin{figure}
\noindent \centering{}\includegraphics[width=0.6\textwidth]{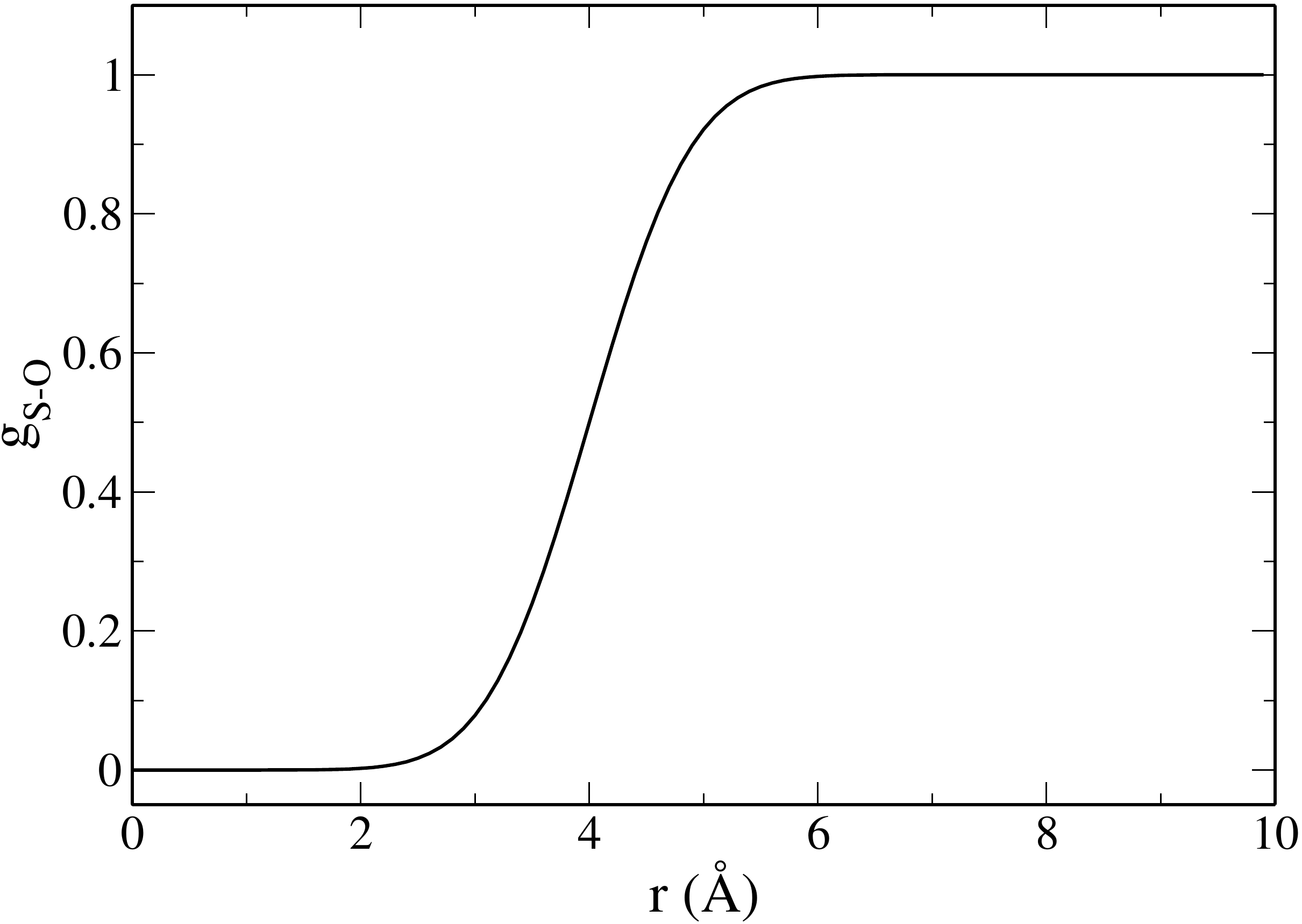}\protect\caption{Allure de la fonction de distribution radiale entre soluté et solvant
dans le cas d'un démouillage\label{fig:erf}. }
\end{figure}

En l'absence de démouillage les fonctions de distribution radiale
ont une allure similaire à celles rencontrées pour les gaz rares.
Lorsqu'il y a démouillage, les fonctions de distribution radiale ont
une allure sigmoïdale, voir la \ref{fig:erf}.

Le changement de régime apparait clairement sur les fonctions de distribution
radiale obtenues par simulation numérique pour les deux systèmes étudiés.
On observe d'abord une augmentation de la hauteur du pic de la première
couche de solvatation pour des rayons inférieurs à 5$\ \textrm{\AA}$,
puis la hauteur de ce pic diminue pour des rayons supérieurs. Dans
les deux cas, les fonctions de distribution radiale obtenues par minimisation
fonctionnelle ne présentent pas ce changement de comportement : les
pics de la première couche de solvatation augmentent avec le rayon
de la sphère pour toutes les valeurs considérées.

\begin{figure}
\noindent \begin{centering}
\begin{tabular}{cc}
\includegraphics[width=0.5\textwidth]{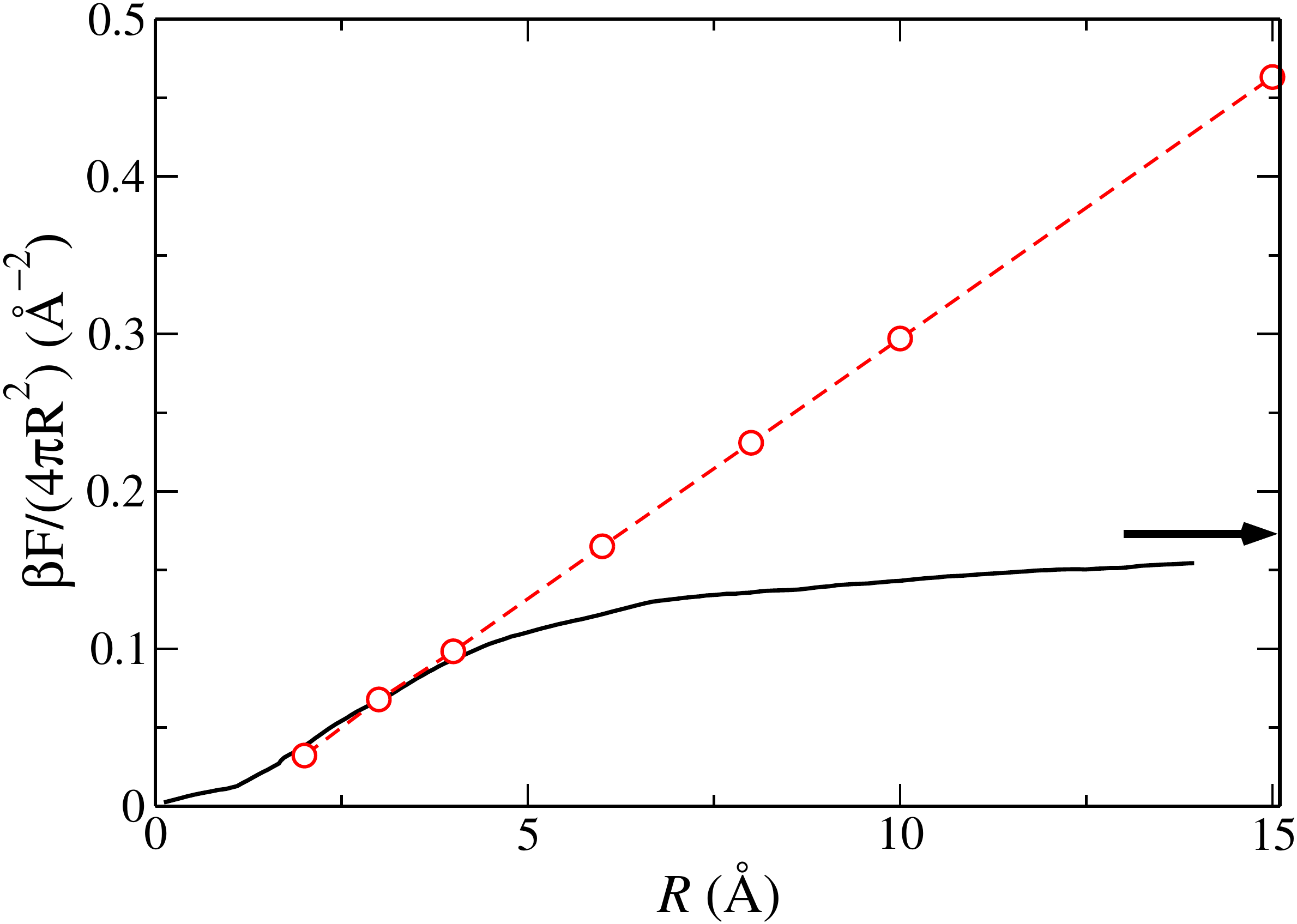} & \includegraphics[width=0.5\textwidth]{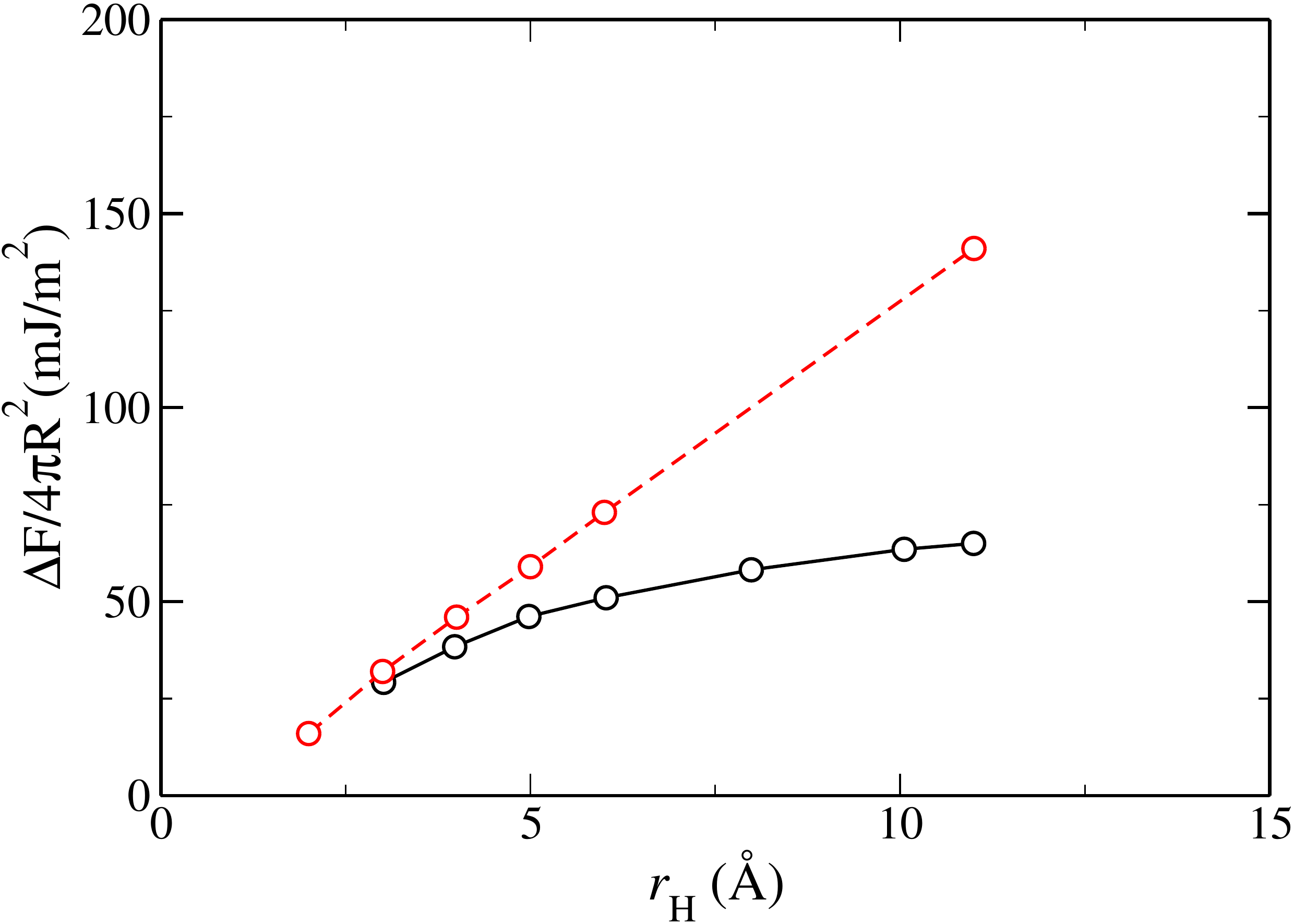}\tabularnewline
\end{tabular}
\par\end{centering}

\noindent \centering{}\protect\caption{Energie libre de solvatation par unité de surface pour les systèmes
sphères dures (à gauche) et sphères molles (à droite). Les résultats
obtenus avec le code MDFT sont les cercles rouges, ceux obtenus par
MC et MD sont en noir\label{fig:ener_chandler+HANSEN_spce-MDFT}. }
\end{figure}

Les énergies libres de solvatation sont présentées en \ref{fig:ener_chandler+HANSEN_spce-MDFT}.
On observe à nouveau un changement de régime sur les courbes obtenues
par simulations numériques. Avant $5\ \textrm{\AA}$ l'énergie de
solvatation est proportionnelle au volume. Pour des rayons plus grands
la pente diminue pour finalement tendre vers un palier égal à la tension
de surface liquide-gaz. L'énergie libre est alors proportionnelle
à la surface. La limite indiquée par la flèche noire correspond à
la valeur théorique de la tension de surface liquide-gaz de l'eau.
Les énergies libres obtenues par MDFT sont en revanche proportionnelles
au volume sur toute la plage des rayons considérés.

Il apparait clair que la théorie de la fonctionnelle de la densité
utilisée est adaptée à l'étude de solutés hydrophobes de petite taille
mais ne permet pas de reproduire le comportement à échelle mésoscopique
de ces solutés.

\section{Description de l'hydrophobicité à différentes échelles dans MDFT}

\subsection{Théorie}

On introduit ici une nouvelle physique dans la fonctionnelle qui permet
de reproduire les propriétés structurale et énergétique de solvatation
des solutés hydrophobes\cite{jeanmairet_molecular_2013} de dimensions
microscopique et mésoscopique. 

Hughes a proposé \cite{hughes_classical_2013} une théorie de la fonctionnelle
de la densité classique basée sur la fonctionnelle sphères dures dans
l'approche FMT (fundamental measure theory) \cite{rosenfeld_free-energy_1989}
additionnée d'interactions attractives basées sur la théorie statistique
des fluides associés (SAFT). SAFT est une équation d'état construite
pour décrire les fluides associés. Par rapport à cela, notre théorie
à l'avantage d'inclure en plus le traitement de la polarisation et
donc l'interaction avec des solutés chargés.

On part de la fonctionnelle multipolaire décrite dans la \ref{sec:Fexcmulti}.
Comme les solutés étudiés ici sont hydrophobes et que l'on néglige
un éventuel couplage entre polarisation et densité, les champs de
polarisation longitudinale et transverse sont nuls. La fonctionnelle
se réduit alors à,

\begin{flalign}
{\cal F}_{\text{exc}}[n(\bm{r})]= & -\frac{\text{k}_{\text{B}}\text{T}}{2}\iiint_{\mathbb{R}^{3}}\Delta n(\bm{r})^{2}\text{d}\bm{r}+{\cal F}_{\text{cor}}[n(\bm{r})]\nonumber \\
 & +\frac{\text{k}_{\text{B}}\text{T}}{2}\iiint_{\mathbb{R}^{3}}\iiint_{\mathbb{R}^{3}}S^{-1}(n_{\mathrm{b}};\left\Vert \bm{r}-\bm{r}^{\prime}\right\Vert )\Delta n(\bm{r})\Delta n(\bm{r}^{\prime})\text{d}\bm{r}\text{d}\bm{r}^{\prime}
\end{flalign}
ou bien,
\begin{equation}
{\cal F}_{\text{exc}}[n(\bm{r})]=-\frac{\text{k}_{\text{B}}\text{T}}{2}\iiint_{\mathbb{R}^{3}}\iiint_{\mathbb{R}^{3}}c_{000}(n_{\mathrm{b}};\left\Vert \bm{r}-\bm{r}^{\prime}\right\Vert )\Delta n(\bm{r})\Delta n(\bm{r}^{\prime})\text{d}\bm{r}\text{d}\bm{r}^{\prime}+{\cal F}_{\text{cor}}[n(\bm{r})],\label{eq:FexcPolstartingHydro}
\end{equation}

où on rappelle que $\Delta n(\bm{r})=n(\bm{r})-n_{\mathrm{b}}$. On
retrouve les termes quadratiques qui proviennent du développement
autour du fluide de référence, que l'on peut écrire, de manière équivalente,
en fonction de l'inverse du facteur de structure, ou de la fonction
de corrélation directe de l'eau %
\footnote{Facteur de structure et fonction de corrélation sont liés dans l'espace
de Fourier en l'absence de polarisation, $n_{\mathrm{b}}\hat{c}_{000}(n_{b};k)=1-\hat{S}^{-1}(n_{\mathrm{b}};k)$%
}. Tous les termes d'ordre supérieur à deux sont rassemblés dans la
fonctionnelle inconnue ${\cal F}_{\text{cor}}$. On approxime ce terme,
comme dans d'autres articles\cite{rosenfeld_free_1993,liu_site_2013,MDFT_zhao_new_2011,MDFT_zhao_correction_2011,oettel_integral_2005}
et comme dans le \ref{chap:MDFT_dup_multu}, par un bridge pour un
fluide de référence sphères dures de rayon $\text{R}_{0}$ et de même
densité $n_{\mathrm{b}}$ que le fluide étudié. 
\begin{eqnarray}
{\cal F}_{\text{cor}}[n(\bm{r})] & = & {\cal F}_{\text{exc}}^{\text{HS}}[n(\bm{r})]-{\cal F}_{\text{exc}}^{\text{HS}}[n_{b}]-\mu_{\text{exc}}^{\text{HS}}\iiint_{\mathbb{R}^{3}}\Delta n(\bm{r})\text{d}\bm{r}\nonumber \\
 &  & +\frac{\text{k}_{\text{B}}\text{T}}{2}\iiint_{\mathbb{R}^{3}}\iiint_{\mathbb{R}^{3}}c_{000}^{\text{HS}}(n_{b};\left\Vert \bm{r}-\bm{r}^{\prime}\right\Vert )\Delta n(\bm{r})\Delta n(\bm{r}^{\prime})\text{d}\bm{r}\text{d}\bm{r}^{\prime}.\label{eq:HSB}
\end{eqnarray}

On rappelle que ce terme revient à remplacer les termes de corrélation
d'ordre supérieur ou égaux à trois par ceux d'un fluide de sphères
dures.

Puisque l'on veut décrire des phénomènes physiques à deux échelles
différentes, microscopique et mésoscopique, on veut pouvoir séparer
la fonctionnelle en une partie courte-distance et une partie longue-distance.
Pour cela, on définit une densité gros grains $\bar{n}(\bm{r})$,
\begin{equation}
\bar{n}(\bm{r})=\iiint_{\mathbb{R}^{3}}G(\left\Vert \bm{r}-\bm{r}^{\prime}\right\Vert )n(\bm{r}^{\prime})\text{d}\bm{r}^{\prime},\label{eq:nbar_def}
\end{equation}
où $G$ est une fonction de convolution. 

Le produit de convolution de l'\ref{eq:nbar_def} se réécrit plus
simplement dans l'espace de Fourier,
\begin{equation}
\bar{\hat{n}}(\bm{k})=\hat{G}(k)\hat{n}(\bm{k}).
\end{equation}

Pour simplifier l'explication qui suit supposons que cette fonction
$G$ est la fonction de Heaviside $\Theta(k_{c}-k)$; c'est-à-dire
une fonction égale à 1 si $k<k_{c}$ et 0 sinon ; $k_{c}^{-1}$ est
de l'ordre de quelques dimensions caractéristiques de la molécule
d'eau, c'est à dire quelques angströms.

Dans ce cas,
\begin{equation}
\bar{\hat{n}}(\bm{k})\left(\hat{n}(\bm{k})-\bar{\hat{n}}(\bm{k})\right)=0,\ \forall\bm{k}.\label{eq:nbar(n-nbar)=00003D0}
\end{equation}
On peut alors rassembler les termes quadratiques en $\Delta n$ des
\ref{eq:FexcPolstartingHydro} et \ref{eq:HSB} en un seul terme définissant
une énergie libre attractive.
\begin{equation}
{\cal F}_{\text{exc}}^{\text{VdW}}[n(\bm{r})]=-\frac{\text{k}_{\text{B}}\text{T}}{2}\iiint_{\mathbb{R}^{3}}\iiint_{\mathbb{R}^{3}}c_{000}^{\text{VdW}}(n_{b};\left\Vert \bm{r}-\bm{r}^{\prime}\right\Vert )\Delta n(\bm{r})\Delta n(\bm{r}^{\prime})\text{d}\bm{r}\text{d}\bm{r}^{\prime},
\end{equation}

avec $c_{000}^{\text{VdW}}(n_{b};\left\Vert \bm{r}-\bm{r}^{\prime}\right\Vert )=c_{000}(n_{b};\left\Vert \bm{r}-\bm{r}^{\prime}\right\Vert )-c_{000}^{\text{HS}}(n_{b};\left\Vert \bm{r}-\bm{r}^{\prime}\right\Vert )$.
On a alors un terme qui rappelle l'équation de Van-der-Waals puisque
le fluide est traité comme un fluide de sphères dures auquel on ajoute
ce terme globalement attractif, c'est pourquoi on appellera par la
suite la fonctionnelle utilisée ici correction de Van-der-Waals. 

En utilisant l'\ref{eq:nbar(n-nbar)=00003D0}, on peut décomposer
cette fonctionnelle en une partie courte portée et une partie longue
portée:
\begin{eqnarray}
{\cal F}_{\text{exc}}^{\text{VdW}}[n(\bm{r})] & =- & \frac{\text{k}_{\text{B}}\text{T}}{2}\iiint_{\mathbb{R}^{3}}\iiint_{\mathbb{R}^{3}}c_{000}^{\text{VdW}}(n_{\mathrm{b}};\left\Vert \bm{r}-\bm{r}^{\prime}\right\Vert )\Bigl(n(\bm{r})-\bar{n}(\bm{r})\Bigr)\Bigl(n(\bm{r}^{\prime})-\bar{n}(\bm{r}^{\prime})\Bigr)\text{d}\bm{r}\text{d}\bm{r}^{\prime}\nonumber \\
 &  & -\frac{\text{k}_{\text{B}}\text{T}}{2}\iiint_{\mathbb{R}^{3}}\iiint_{\mathbb{R}^{3}}c_{000}^{\text{VdW}}(n_{\mathrm{b}};\left\Vert \bm{r}-\bm{r}^{\prime}\right\Vert )\Delta\bar{n}(\bm{r})\Delta\bar{n}(\bm{r}^{\prime})\text{d}\bm{r}\text{d}\bm{r}^{\prime}.\label{eq:FexcVdW}
\end{eqnarray}
Comme la fonction $c_{000}^{\text{VdW}}$ est à courte portée et que
la densité gros grains $\bar{n}$ varie lentement, la seconde intégrale,
à longue portée, de l'\ref{eq:FexcVdW} peut s'exprimer dans l'espace
de Fourier comme,
\begin{gather}
\iiint_{\mathbb{R}^{3}}c_{000}^{\text{VdW}}(n_{b};k)\Delta\bar{\hat{n}}(\bm{k})\Delta\bar{\hat{n}}(-\bm{k})\text{d}\bm{k}=\nonumber \\
\iiint_{\mathbb{R}^{3}}\left(c_{000}^{\text{VdW}}(k=0)+\left.k^{2}\frac{\text{d}^{2}c_{000}^{\text{VdW}}}{\text{d}k^{2}}\right|_{k=0}+\dots\right)\Delta\bar{\hat{n}}(\bm{k})\Delta\bar{\hat{n}}(-\bm{k})\text{d}\bm{k}=\nonumber \\
\text{a}\iiint_{\mathbb{R}^{3}}\Delta\bar{\hat{n}}(\bm{k})\Delta\bar{\hat{n}}(-\bm{k})\text{d}\bm{k}-\text{m}\iiint_{\mathbb{R}^{3}}k^{2}\Delta\bar{\hat{n}}(\bm{k})\Delta\bar{\hat{n}}(-\bm{k})\text{d}\bm{k}+\dots=\nonumber \\
\text{a}\iiint_{\mathbb{R}^{3}}\Delta\bar{n}(\bm{r})^{2}\text{d}\bm{r}-\text{m}\iiint_{\mathbb{R}^{3}}\Bigl(\nabla\bar{n}(\bm{r})\Bigr)^{2}\text{d}\bm{r}+\dots,
\end{gather}
où a et m sont des réels positifs. Le terme en gradient est un terme
similaire à celui de l'équation de Cahn-Hilliard. Il pénalise la création
d'inhomogénéité, en particulier d'interface.

\begin{figure}
\noindent \begin{centering}
\includegraphics[width=0.6\textwidth]{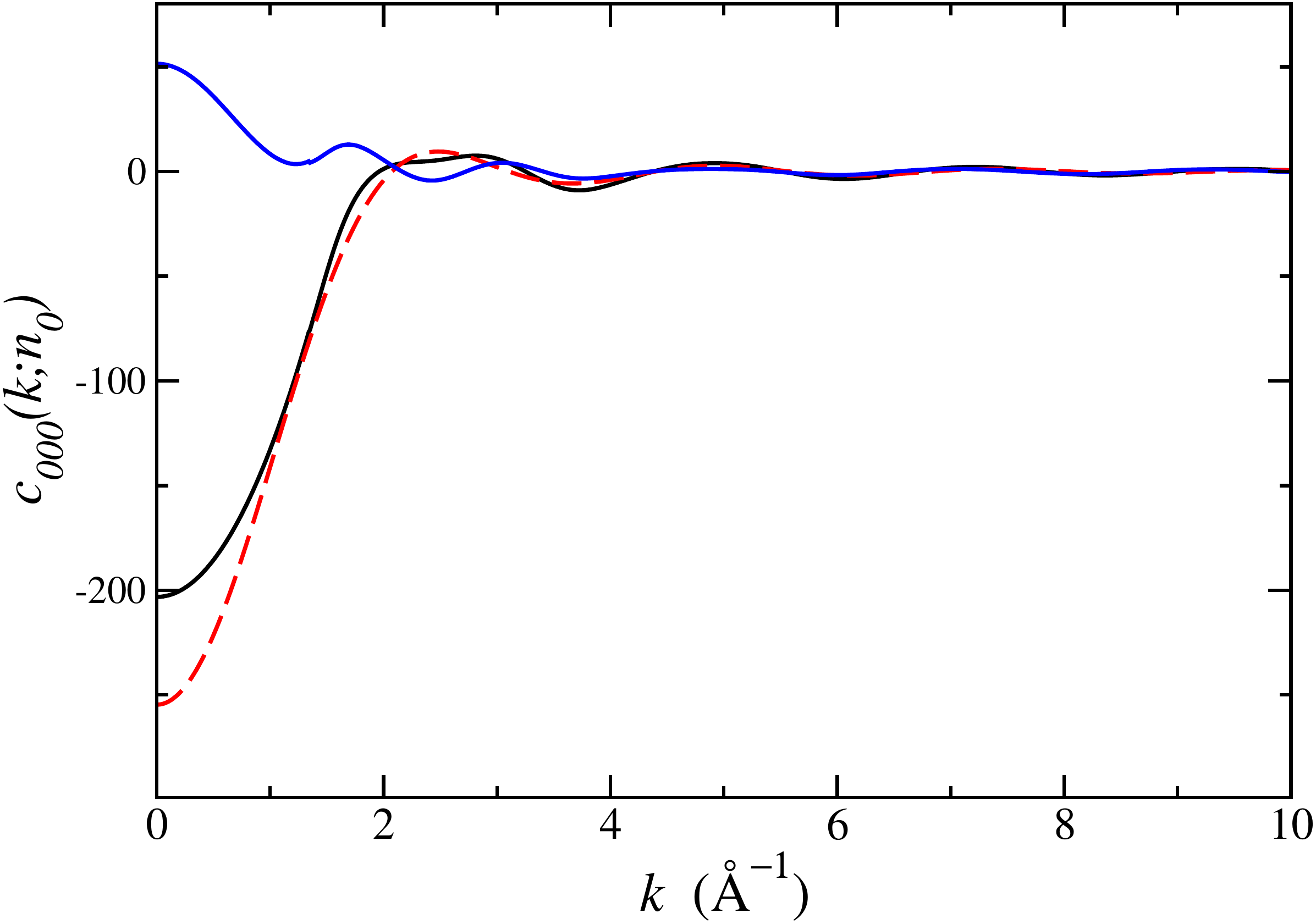}
\par\end{centering}

\noindent \centering{}\protect\caption{Comparaison des fonctions de corrélation directe pour l'eau (en noir)
et pour un fluide de sphères dures de rayon $R_{0}=1.27\ \textrm{\AA}$
(en rouge) à la même densité $n_{b}=0.0333\ \textrm{\AA}^{-3}$. La
courbe bleue est la différence entre ces deux fonctions, c'est-à-dire
la fonction de corrélation \og attractive \fg{} $c_{000}^{\text{VdW}}$.
On peut vérifier que $\mathrm{a}=c_{000}^{\text{VdW}}(k=0)$ et $\mathrm{m}=-\left.k^{2}\frac{\text{d}^{2}c_{000}^{\text{VdW}}}{\text{d}k^{2}}\right|_{k=0}$\label{fig:cVDW}
sont positifs. }
\end{figure}

On peut vérifier le signe de ces paramètres sur la figure \ref{fig:cVDW}
où sont représentées les fonctions de corrélation pour l'eau et le
fluide de sphères dures ainsi que la fonction $c_{000}^{\text{VdW}}$.
Bien que ces deux termes soient fixés par la théorie FMT et par la
fonction de corrélation directe de l'eau utilisée comme input, on
considérera ci-après que ce sont deux paramètres phénoménologiques
que l'on s'autorise à faire varier.

En rassemblant les développements réalisés précédemment on aboutit
à la fonctionnelle suivante
\begin{eqnarray}
{\cal F}_{\text{exc}}^{\text{VdW}}[n(\bm{r})] & = & -\frac{\text{k}_{\text{B}}\text{T}}{2}\left[\text{a}\iiint_{\mathbb{R}^{3}}\Delta\bar{n}(\bm{r})^{2}\text{d}\bm{r}-\text{m}\iiint_{\mathbb{R}^{3}}(\nabla\bar{n}(\bm{r}))^{2}\text{d}\bm{r}\right]\nonumber \\
 &  & +\frac{\text{k}_{\text{B}}\text{T}}{2}\iiint_{\mathbb{R}^{3}}\iiint_{\mathbb{R}^{3}}c_{000}^{\text{VdW}}(n_{b};|\bm{r}-\bm{r}^{\prime}|)(n(\bm{r})-\bar{n}(\bm{r}))(n(\bm{r}^{\prime})-\bar{n}(\bm{r}^{\prime}))\text{d}\bm{r}\text{d}\bm{r}^{\prime}\nonumber \\
 &  & +{\cal F}_{\text{exc}}^{\text{HS}}[n(\bm{r})]-{\cal F}_{\text{exc}}^{\text{HS}}[n_{b}]-\mu_{\text{exc}}^{\text{HS}}\iiint_{\mathbb{R}^{3}}\Delta n(\bm{r})\text{d}\bm{r}.
\end{eqnarray}

Pour la fonctionnelle sphères dures on utilise la fonctionnelle FMT
scalaire de Kierlik et Rosinberg (KR-FMT)\cite{kierlik_free-energy_1990,kierlik_density-functional_1991}
décrite dans l'\ref{sec:FMT}.

Lorsque $n=\bar{n}$, le terme courte portée (celui en $n-\bar{n}$)
s'annule et on peut approximer la fonctionnelle sphères dures par
sa forme locale, soit celle de Percus-Yevick (PY), soit celle de Carnahan-Starling
(CS) selon la forme de fonctionnelle KR-FMT choisie\cite{MDFT_levesque_krfmt}.
\begin{equation}
{\cal F}_{\text{exc}}^{\text{HS}}[\bar{n}(\bm{r})]=\iiint_{\mathbb{R}^{3}}f_{\text{exc}}^{\text{HS}}(\bar{n}(\bm{r}))\bar{n}(\bm{r})\text{d}\bm{r},
\end{equation}

avec si on choisit par exemple celle de Carnahan-Starling,
\begin{equation}
f_{\text{exc}}^{\text{HS}}(n(\bm{r}))=\frac{\eta(4-3\eta)}{(1-\eta)^{2}},
\end{equation}

où $\eta=4\pi R_{0}^{3}n/3$ est la fraction d'empilement. Il est
montré dans la ref. \cite{MDFT_levesque_krfmt} que le choix de PY
ou CS donne des résultats très proches pour l'état fluide. On peut
alors écrire la partie intrinsèque (terme idéal et terme d'excès)
de ${\cal F}[\bar{n}(\bm{r})]$ comme,
\begin{equation}
{\cal F}_{\text{int}}[\bar{n}(\bm{r})]=\text{k}_{\text{B}}\text{T}\iiint_{\mathbb{R}^{3}}\left[F_{\text{VdW}}(\bar{n}(\bm{r}))+\frac{1}{2}\text{m}(\nabla\bar{n}(\bm{r}))^{2}\right]\text{d}\bm{r},
\end{equation}
avec,
\begin{eqnarray}
F_{\text{VdW}}(\bar{n}(\bm{r})) & = & \bar{n}(\bm{r})\ln\left(\frac{\bar{n}(\bm{r})}{n_{b}}\right)-\Delta\bar{n}(\bm{r})+\bar{n}(\bm{r})f_{\text{exc}}^{\text{HS}}(\bar{n}(\bm{r}))-n_{b}f_{\text{exc}}^{\text{HS}}(n_{b})\nonumber \\
 &  & -\left(f_{\text{exc}}^{\text{HS}}(n_{\mathrm{b}})+n_{\mathrm{b}}\left.\frac{\text{d}f_{\text{exc}}^{\text{HS}}}{\text{d}n}\right|_{n=n_{\mathrm{b}}}\right)\Delta\bar{n}(\bm{r})-\frac{1}{2}\text{a}\Delta\bar{n}(\bm{r})^{2}.\label{eq:FvDW}
\end{eqnarray}

\begin{figure}
\noindent \begin{centering}
\includegraphics[width=0.6\textwidth]{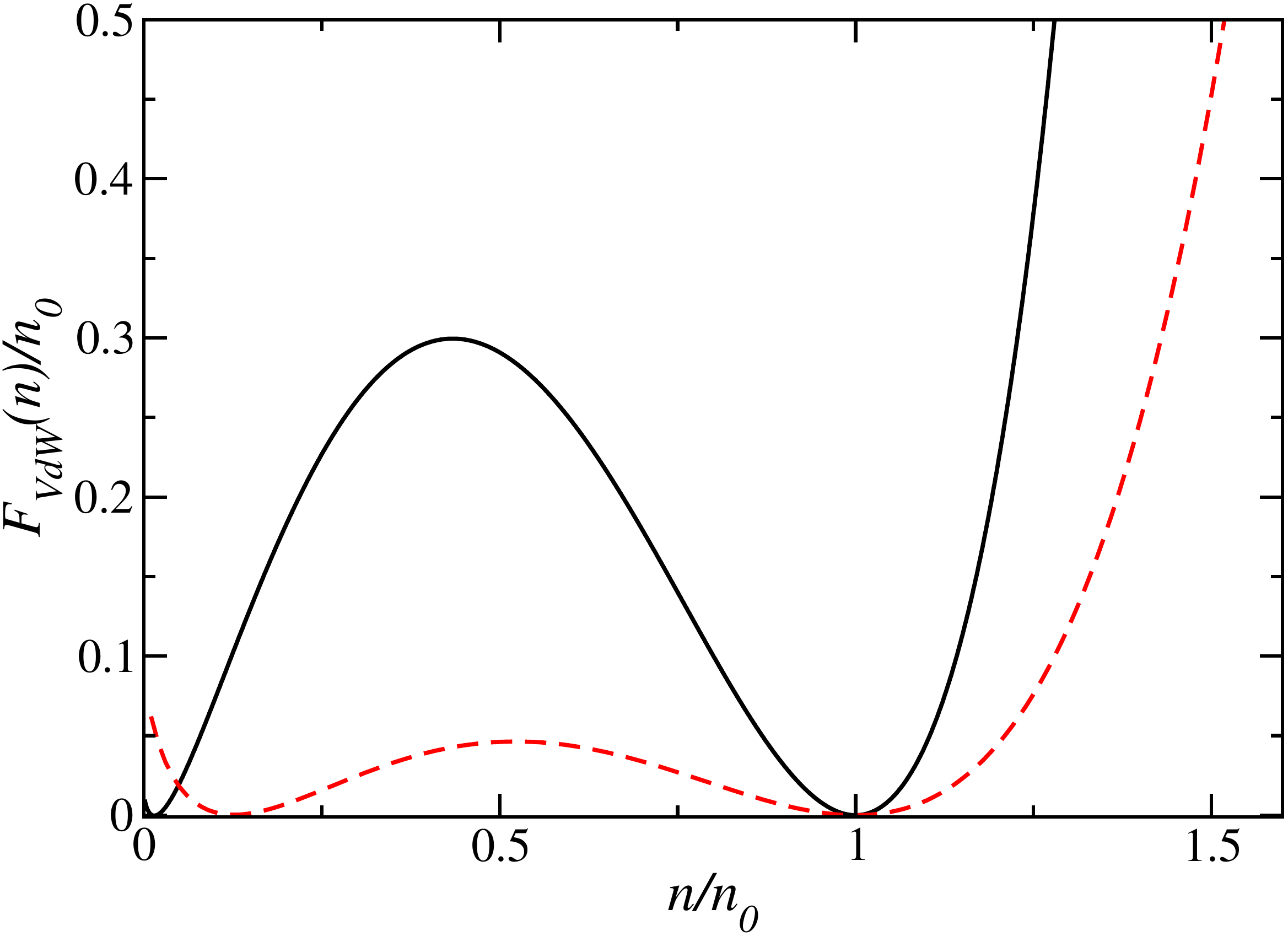}
\par\end{centering}

\noindent \centering{}\protect\caption{Fonctions de Van-der-Waals pour les densités gros grains, décrites
par l'\ref{eq:FvDW}, proches de la coexistence liquide gaz. La courbe
noire correspond à un rayon de sphères dures du fluide de référence
de $1.42\ \textrm{\AA}$ et la courbe rouge à un rayon de $1.27\ \textrm{\AA}$\label{fig:FVdW}.}
\end{figure}

L'eau liquide dans les conditions ambiantes est proche de la coexistence
liquide-gaz. Il est donc important de choisir un paramètre a tel que
les énergies libres associées au gaz et au liquide soient égales,
ce qui revient à imposer que $F_{\text{VdW}}(n(\bm{r}))$ ait deux
minima équivalents, l'un à la densité du liquide $n_{\mathrm{b}}$,
l'autre à une densité très faible, celle du gaz. On représente une
telle courbe sur la \ref{fig:FVdW} pour un liquide de densité $n_{\mathrm{b}}=0.0333\ \textrm{\AA}^{-3}$
et pour des valeurs de rayons de sphères dures $R_{0}=1.27\ \textrm{\AA}$
et $R_{0}=1.42\ \textrm{\AA}$. Ces deux valeurs ont déjà été utilisées
dans la littérature pour étudier un fluide de sphères dures mimant
le cœur dur de l'eau\cite{zhao_new_2011,DHS_oleksy_wetting_2010}.
Pour ces deux rayons, les minima ont des valeurs similaires pour des
paramètres valant respectivement $n_{\mathrm{b}}$a$=6.7$ et $n_{\mathrm{b}}$a$=12.5$.
Avec cette forme simple de l'équation de Van-der-Waals, on ne peut
pas choisir indépendamment la hauteur de la barrière et la densité
de la phase gaz, c'est cette version de la correction hydrophobe qui
a été utilisée dans la ref. \cite{jeanmairet_molecular_2013}. Le
paramètre m semble lui n'avoir que peu d'influence. Dans le cadre
du modèle de Chandler\cite{lum_hydrophobicity_1999,huang_hydrophobic_2002,gaussian_field_varilly_improved_2011}
de la théorie de Cahn-Hilliard pour une interface liquide-gaz, le
profil de densité d'un fluide près d'un soluté hydrophobe évolue comme
une tangente hyperbolique de la distance, et l'énergie libre de Van-der-Waals
est, 
\begin{equation}
F_{\text{VdW}}(n(\bm{r}))=\frac{6\gamma}{d}(n(\bm{r})-n_{l})^{2}(n(\bm{r})-n_{g})^{2},
\end{equation}
où $\gamma$ est la tension de surface liquide-gaz, $n_{l}$ et $n_{g}$
sont les densités d'équilibre du liquide et du gaz respectivement.
Si on compare le profil de l'énergie de Van-de-Waals de notre théorie
avec celle issue du modèle de Chandler on se rend compte que pour
pouvoir fixer la position du second minimum ainsi que la hauteur de
la barrière, il est nécessaire d'ajouter un terme cubique et un terme
d'ordre quatre en $\Delta\bar{n}$. Ce développement a été réalisé
en collaboration avec Stojanovi\'c dans un article à paraitre\cite{Stojanovic_2014}.

Il suffit ensuite de procéder à une minimisation fonctionnelle comme
on l'a fait précédemment. 

\fbox{\begin{minipage}[t]{1\columnwidth}%
Soulignons ici que cette théorie a l'avantage d'être auto-cohérente:
même si l'énergie libre est écrite comme une fonctionnelle de $n$
et de $\bar{n}$, la minimisation est toujours conduite uniquement
sur $n$. $\bar{n}$ est elle-même une fonctionnelle de $n$ dans
l'\ref{eq:nbar_def}.%
\end{minipage}}

Si la dérivation a été faite avec une fonction de convolution de Heaviside,
nous avons utilisé ici une fonction gaussienne avec une largeur à
mi-hauteur de $4\ $$\textrm{\AA}$.

\subsection{Résultats}

Les résultats discutés ici ont été obtenus avec une énergie libre
de Van-der-Waals limitée à l'ordre deux.

\begin{figure}
\noindent \centering{}%
\begin{tabular}{cc}
\includegraphics[width=0.5\textwidth]{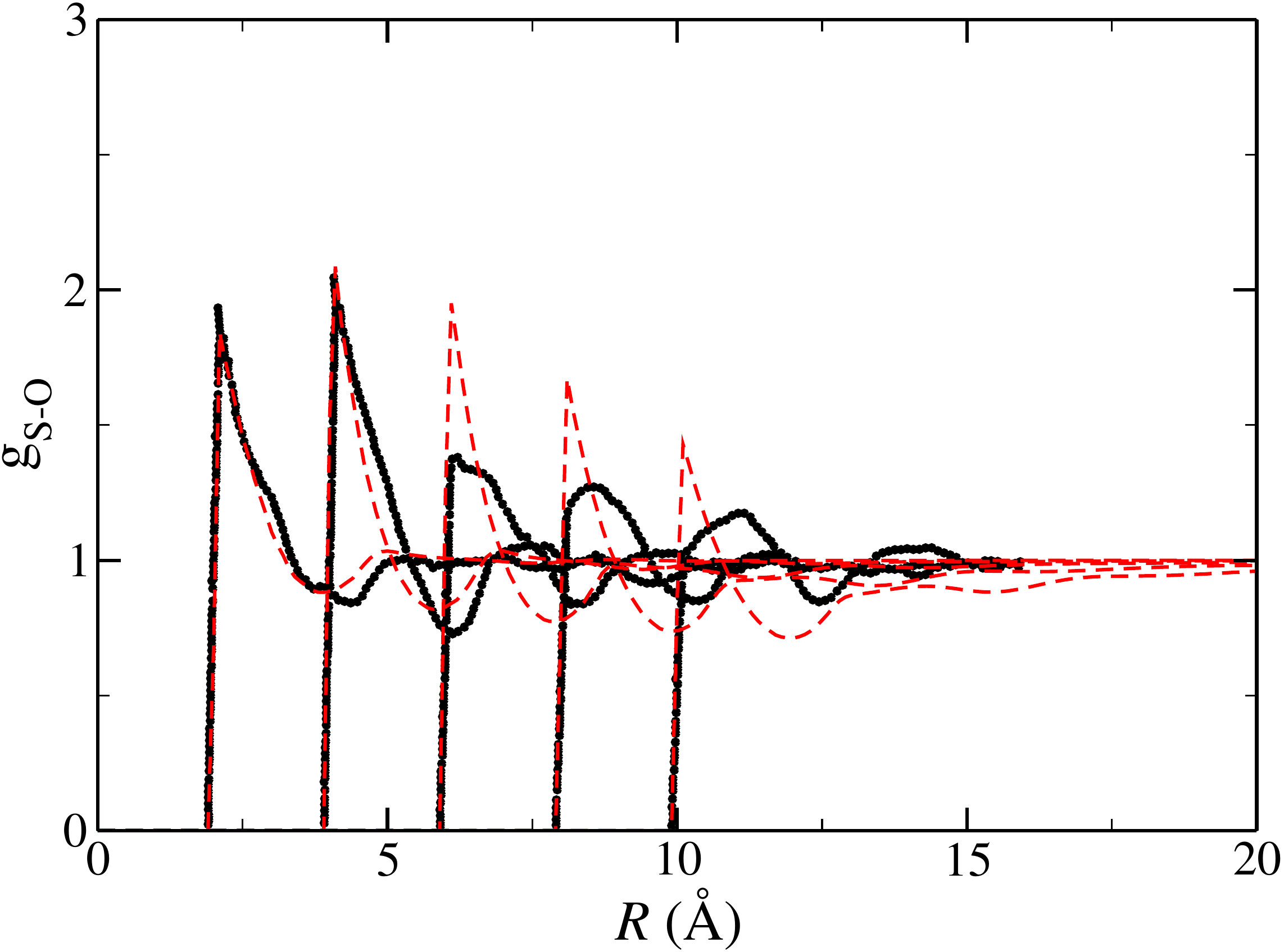} & \includegraphics[width=0.5\textwidth]{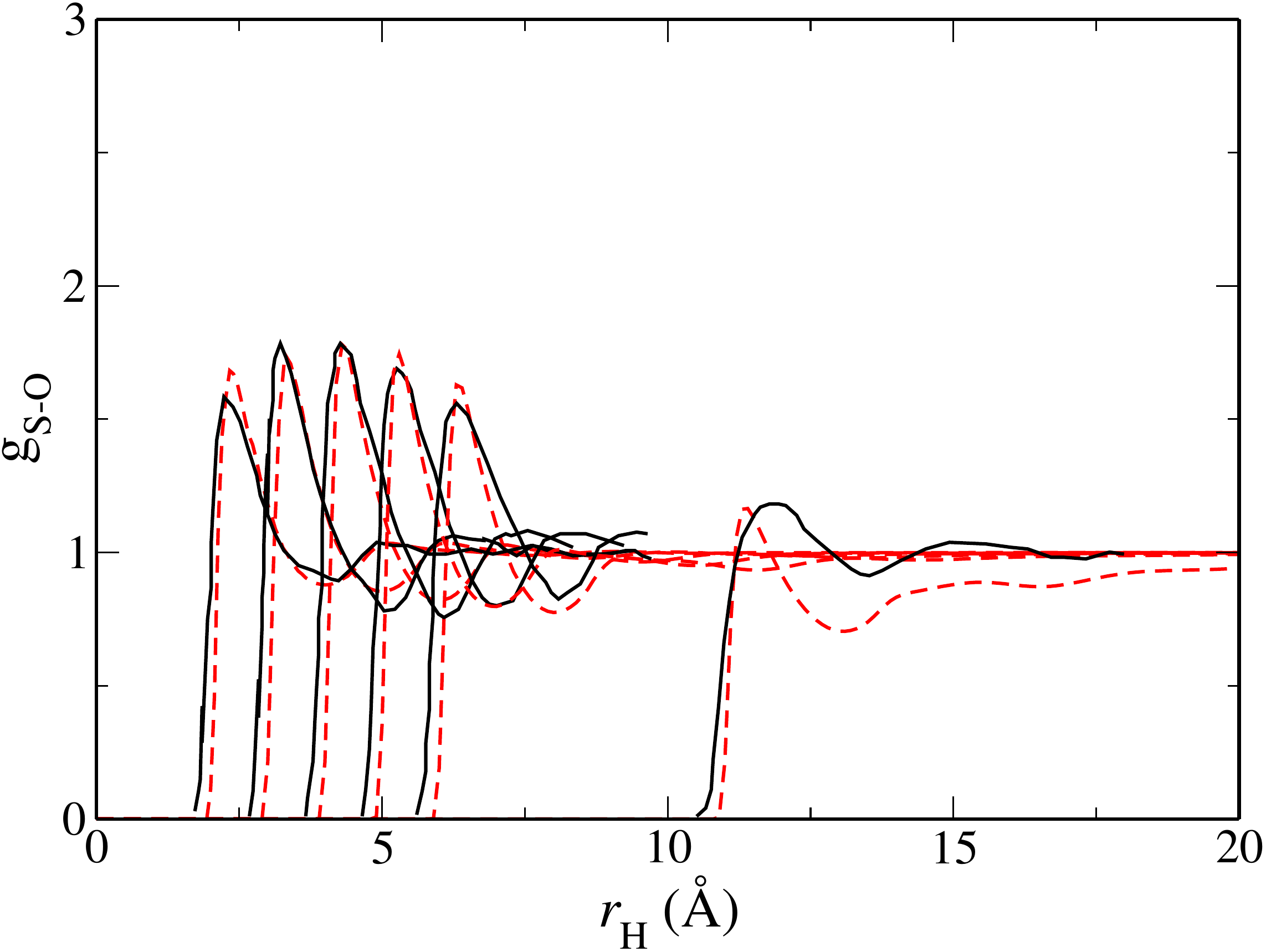}\tabularnewline
\end{tabular}\protect\caption{Comparaison des fonctions de distribution radiale:\protect \\
 -entre un soluté sphère dure de rayon $R$ \label{fig:rdf_Chandler+Hansen_MDFT_VdW}
et l'eau à gauche ;\protect \\
 -entre sphère molle de rayon $r_{\text{H}}$ et l'eau à droite.\protect \\
Les résultats obtenus par MDFT avec la correction hydrophobe sont
en tirets rouges, ceux obtenus par simulation Monte-Carlo\cite{huang_hydrophobic_2002}
ou par dynamique moléculaire\cite{DHS_dzubiella_competition_2004}
sont en noir.}
\end{figure}

L'effet de l'ajout de cette correction de l'hydrophobicité sur les
fonctions de distribution radiale pour les systèmes de sphères dures
et de sphères molles est donné en \ref{fig:rdf_Chandler+Hansen_MDFT_VdW}. 

\begin{figure}
\noindent \begin{centering}
\begin{tabular}{cc}
\includegraphics[width=0.5\textwidth]{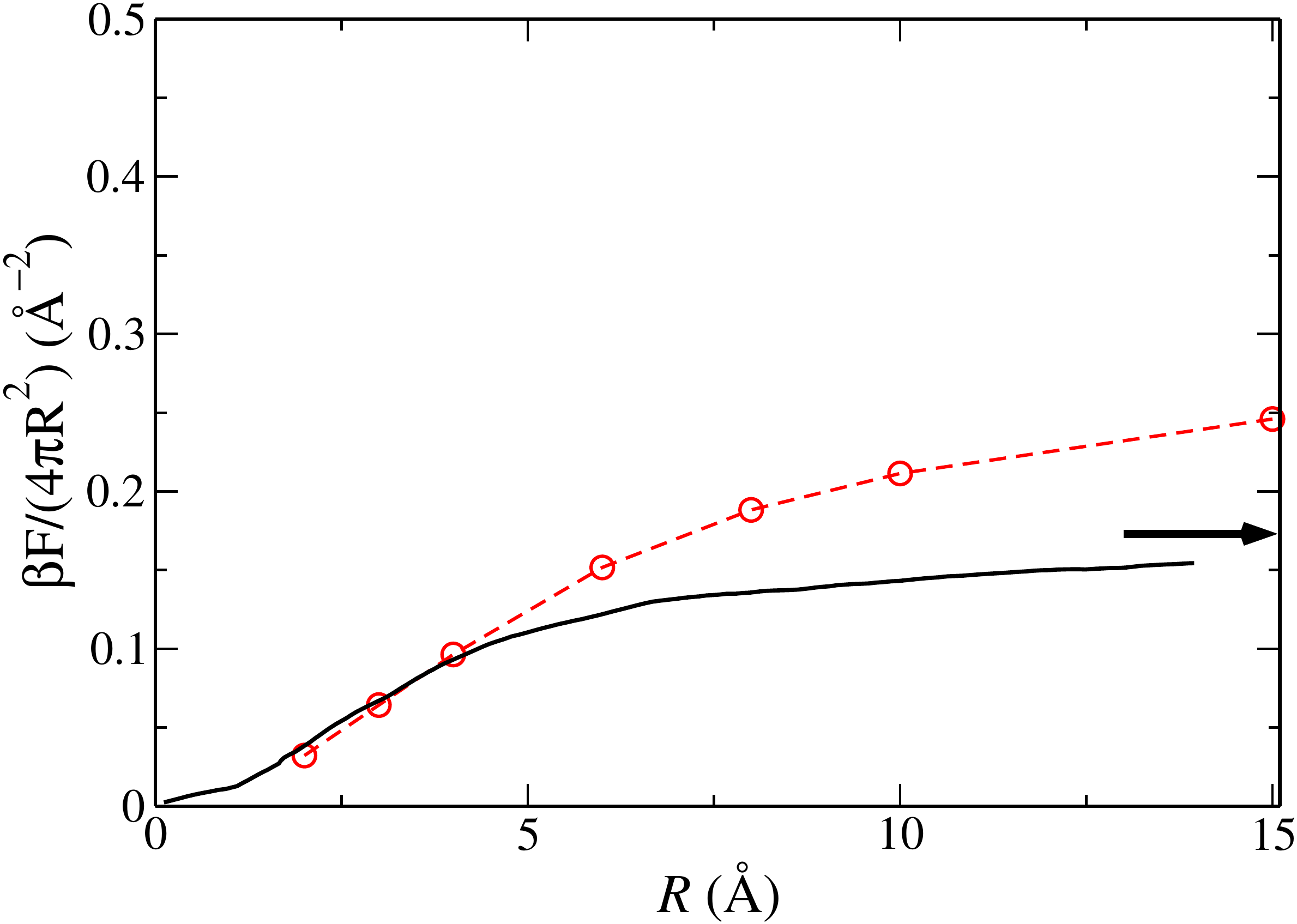} & \includegraphics[width=0.5\textwidth]{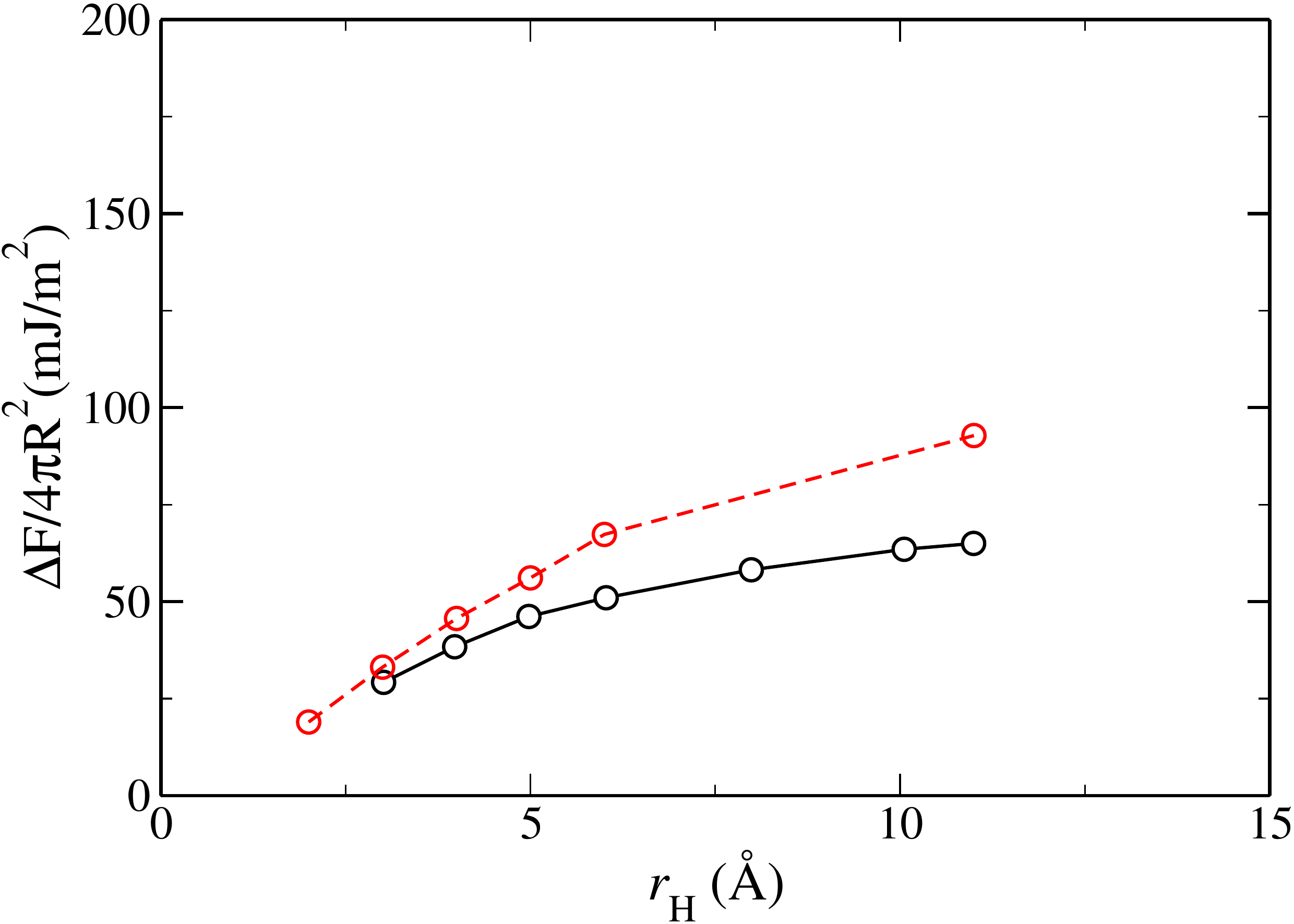}\tabularnewline
\end{tabular}
\par\end{centering}

\noindent \centering{}\protect\caption{Energie libre de solvatation par unité de surface pour les systèmes
sphères dures (à gauche) et sphères molles (à droite). Les résultats
obtenus avec le code mdft avec la correction hydrophobe sont en cercles
rouges, ceux obtenus par MC et MD sont en noir\label{fig:ener_chandler+HANSEN_spce-MDFT-VdW}. }
\end{figure}

Cette correction hydrophobe améliore les résultats par rapport à la
simple inclusion du bridge sphères dures. Avec cette description de
l'hydrophobicité à longue portée, on observe bien la diminution du
pic de la première couche de solvatation à partir d'un rayon de $5\ \textrm{\AA}$,
pour les deux systèmes considérés. Cependant, la correction n'est
pas parfaite puisque pour les sphères dures de rayon $R>4\ \textrm{\AA}$,
les pics apparaissent trop grands et trop piqués. De plus, pour les
solutés de rayon supérieur à $6\ \textrm{\AA}$, la déplétion après
le premier maximum est surestimée. Cette observation est confirmée
sur les sphères molles où les pics dus à la première couche de solvatation
sont mieux reproduits mais où on surestime toujours la déplétion après
le premier pic pour des solutés de rayon supérieur à 6 $\textrm{\AA}$.
On donne aussi sur la \ref{fig:ener_chandler+HANSEN_spce-MDFT-VdW}
l'effet de cette correction hydrophobe sur les énergies libres de
solvatation. Là encore la phénoménologie du changement de comportement
avec l'augmentation en taille des solutés est bien reproduite. L'énergie
libre de solvatation est d'abord proportionnelle au volume du soluté
jusqu'à 6 $\textrm{\AA}$, la pente de la courbe diminue ensuite.
Cependant l'énergie libre n'atteint pas de plateau pour les rayons
considérés et est surestimée par rapport aux simulations. Le doublet
de paramètres utilisé a été choisi pour obtenir le meilleur compromis
entre structures et énergies libres de solvatation et permet d'étudier
l'ensemble des solutés.

On peut vérifier que cette correction, qui contient toujours le bridge
de sphères dures, donne toujours des bons résultats pour les petits
solutés. À titre d'exemple, on présente la fonction de distribution
radiale entre le solvant et le centre de masse du néopentane ainsi
que la carte de densité obtenue autour de ce solvant. Dans la ref.
\cite{jeanmairet_molecular_2013} est également donnée l'énergie libre
de solvatation pour les six premiers alcanes en incluant la correction
hydrophobe, qui est en tout point semblable à la \ref{fig:enercomp_neutre_dipol+HSB}.

\begin{figure}
\noindent \begin{centering}
\begin{tabular}{cc}
\includegraphics[width=0.5\textwidth]{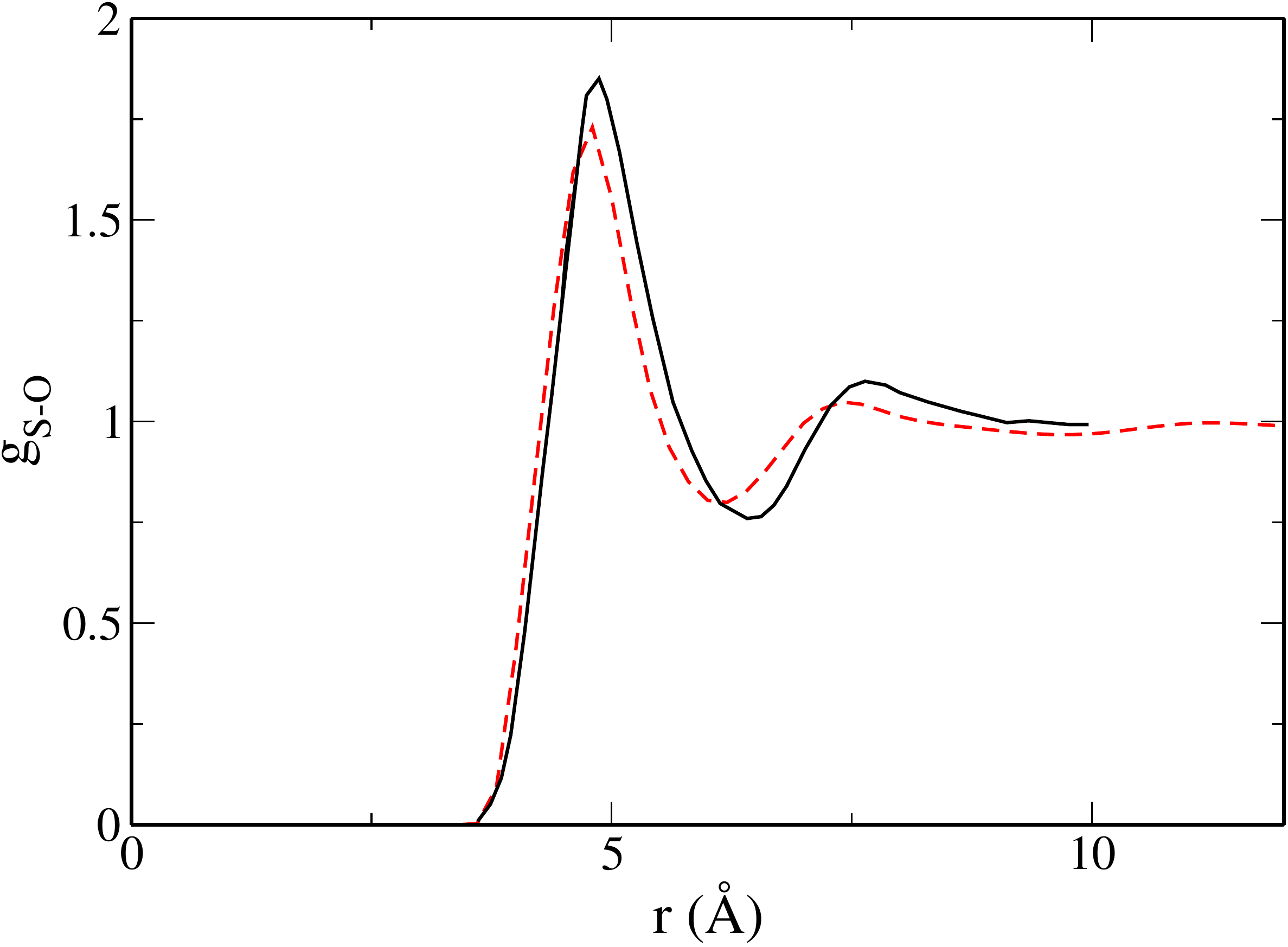} & \includegraphics[width=0.5\textwidth]{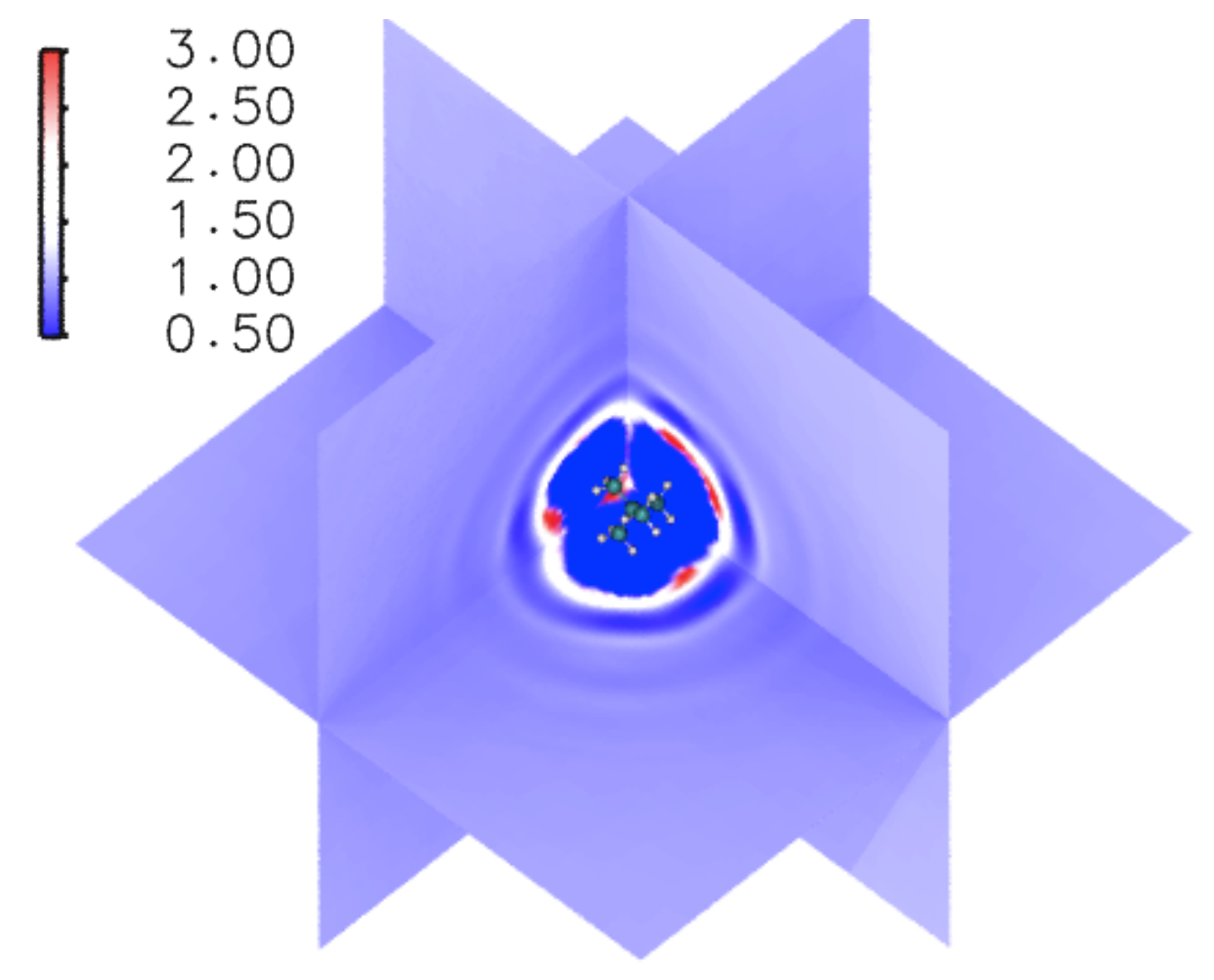}\tabularnewline
\end{tabular}
\par\end{centering}

\noindent \centering{}\protect\caption{À gauche, la fonction de distribution radiale entre le centre de masse
du néopentane et l'eau obtenue par MDFT avec la correction hydrophobe
en tirets rouges, et par MD tirés de la ref. \cite{huang_molecules_2003}
en noir\label{fig:ener_chandler+HANSEN_spce-MDFT-VdW-1}. À droite,
une carte de densité du solvant eau autour du soluté néopentane.}
\end{figure}

\fbox{\begin{minipage}[t]{1\columnwidth}%
On a introduit dans la fonctionnelle une description phénoménologique
qui reproduit correctement la solvatation des solutés hydrophobes
de petite et de grande taille. Cette correction contient des termes
physiques similaires à la théorie de Cahn-Hilliard utilisée pour décrire
les séparations de phases. Il est possible d'utiliser une densité
gros grains pour traiter des phénomènes physiques se passant à des
échelles mésoscopiques. Ceci peut permettre de coupler la MDFT, qui
décrit la solvatation au niveau moléculaire, avec des méthodes permettant
d'étudier des plus grandes échelles comme les méthodes Lattice-Boltzmann.%
\end{minipage}}

\lhead[\chaptername~\thechapter]{\rightmark}

\rhead[\leftmark]{}

\lfoot[\thepage]{}

\cfoot{}

\rfoot[]{\thepage}

\chapter{Implémentation numérique de la MDFT\label{chap:impl=0000E9mentation} }

Dans ce chapitre, on décrit le fonctionnement du code mdft, basé sur
la théorie MDFT présentée dans cette thèse. Ce code vise à trouver
la valeur de la densité de solvant qui minimise la fonctionnelle d'énergie
libre, et la valeur que prend la fonctionnelle pour cette densité
de solvant.

\section{Discrétisation de la densité sur une double grille spatiale et angulaire}

Dans le cadre de la théorie MDFT, la densité est un champ scalaire
$\rho(\bm{r},\bm{\mbox{\ensuremath{\Omega}}})$ qui dépend de six
cordonnées $(\bm{r},\mbox{\ensuremath{\bm{\Omega}}})=(x,y,z,\theta,\phi,\psi)$,
avec $x,y$ et $z$ les coordonnées cartésiennes dans le référentiel
du laboratoire et $\theta,\phi$ et $\psi$ les trois angles d'Euler.
Comme une minimisation analytique n'est pas réalisable, un code, mdft
(molecular density functional theory), est écrit en Fortran moderne.
Celui-ci permet de minimiser numériquement une fonctionnelle de la
densité. Comme on ne peut pas travailler numériquement avec un espace
de variable continu, l'espace et les orientations sont discrétisés
sur deux grilles tridimensionnelles.

\begin{figure}
\noindent \centering{}%
\begin{tabular}{ccc}
\includegraphics[width=0.45\textwidth]{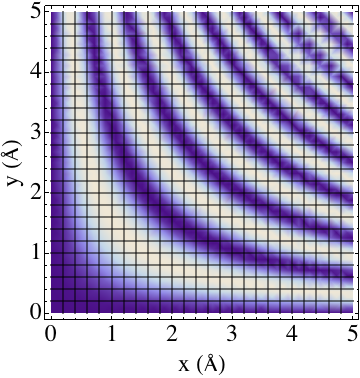} & \includegraphics[width=0.45\textwidth]{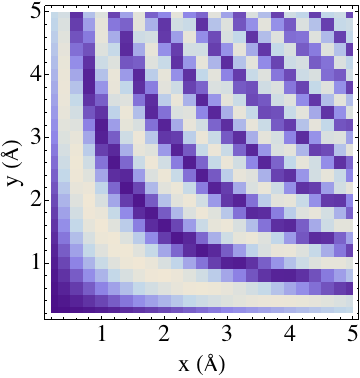} & \includegraphics[width=0.04\textheight]{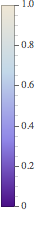}\tabularnewline
\end{tabular}\protect\caption{Illustration de la discrétisation de l'espace. À gauche on présente
une fonction définie sur $\mathbb{R}^{2}$ où l'on a représenté le
grillage utilisé. À droite, la même fonction telle qu'elle est approchée
par la discrétisation dans le code. Elle prend une valeur constante
en chaque point de grille, égale à la valeur moyenne de la fonction
continue sur le pixel contenant le point de grille.\label{fig:2densitywandwoMesh}}
\end{figure}

Cette discrétisation est illustrée sur une densité à deux dimensions
sur la \ref{fig:2densitywandwoMesh}. Au lieu de chercher la valeur
de la densité en tout point $(x,y)$ du plan, on se contente de chercher
cette densité sur les nœuds d'une grille d'un pas régulier de $0.1\ \textrm{\AA}$.
Dans l'implémentation actuelle, la grille spatiale est orthorhombique
(un prisme rectangulaire à base rectangulaire). On utilise typiquement
3 ou 4 nœuds par angström. On appellera $\text{L}_{x},\ \text{L}_{y}\textrm{ et }\text{L}_{z}$
les dimensions de la grille spatiale selon les axes $\text{\ensuremath{\bm{O}_{x}}},$
$\text{\ensuremath{\bm{O}_{y}}}$ et $\text{\ensuremath{\bm{O}_{z}}}$
du repère cartésien. L'espace angulaire est lui aussi discrétisé et
on utilise généralement 6 nœuds pour les angles $(\theta,\phi)$ et
4 nœuds pour l'angle $\psi$, soit environ $1200$ nœuds par angström
cube et un nombre total de l'ordre de $10^{7}$-$10^{8}$ variables
de minimisation $\left\{ \bm{r}_{i},\bm{\Omega}_{j}\right\} $. Pour
que la densité en chaque point ne puisse être que positive, ce qui
est une contrainte physique évidente, la minimisation est en réalité
conduite sur la variable $\Psi\left(\left\{ \bm{r}_{i},\bm{\Omega}_{j}\right\} \right)=\sqrt{\rho\left(\left\{ \bm{r}_{i},\bm{\Omega}_{j}\right\} \right)}$
qui est en quelque sorte un équivalent classique fictif de la fonction
d'onde électronique. Cependant, hormis pour le calcul des gradients,
on travaille toujours avec la densité recalculée à partir de $\Psi$.
Dans la suite du texte, par soucis de simplicité, on continuera donc
à utiliser la densité comme variable de minimisation.

\fbox{\begin{minipage}[t]{1\columnwidth}%
La minimisation réalisée sur la variable $\Psi$ peut être évitée
en réalisant une minimisation sous contrainte de la variable $\rho$,
en imposant comme contrainte que la densité soit positive.%
\end{minipage}}

\section{Fonctionnement du code mdft\label{sec:Algorithme-du-code}}

\begin{figure}
\noindent \centering{}\includegraphics[width=0.8\textwidth]{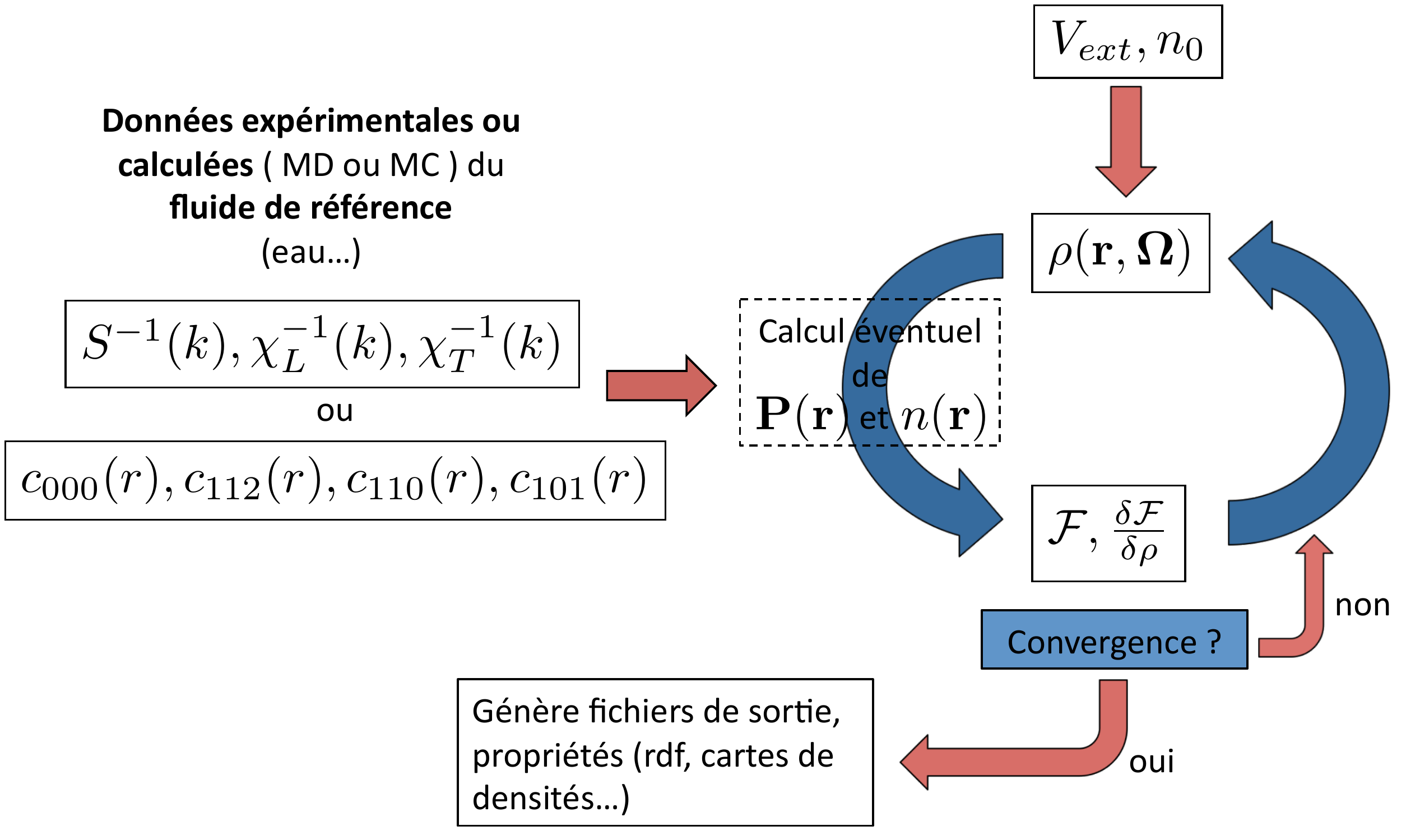}\protect\caption{Représentation schématique de l'algorithme du code mdft.\label{fig:shema_algo}}
\end{figure}

Le but du code mdft est de minimiser une fonctionnelle de la forme
présentée en \ref{eq:F=00003DFid+Fext+Fexc} dans l'approximation
du fluide homogène de référence. Cela consiste à trouver en chaque
point $\left\{ \bm{r}_{i},\bm{\Omega}_{j}\right\} $ des deux grilles
tridimensionnelles spatiale et angulaire, la valeur de la densité
à l'équilibre thermodynamique $\rho_{\mathrm{eq}}(\bm{r},\bm{\Omega})$,
telle que le grand potentiel soit minimal. L'algorithme de ce programme
est schématisé sur la \ref{fig:shema_algo}. 

\subsection{Description moléculaire du soluté et du solvant}

Le soluté est décrit par un fichier ayant un format présenté dans
le \ref{tab:solute.in}. Ici le soluté est décrit par N$^{u}$=12
sites différents. Chacun de ces sites est caractérisé par une charge
et des paramètres Lennard-Jones $\sigma_{i}^{u}$ et $\epsilon_{i}^{u}$.
Les cinquième et sixième colonnes contiennent les paramètres $\lambda^{1\mathrm{S}}$
et $\lambda^{2\mathrm{S}}$ associés à la correction à trois corps
de ce site.

Les trois colonnes suivantes décrivent les cordonnées cartésiennes
des sites du soluté $\bm{r}_{i}^{u}=(x_{i}^{u},y_{i}^{u},z_{i}^{u})$.
Les trois dernières colonnes donnent des informations sur le site
mais ne sont pas utilisées directement dans le code. Pour des raisons
de mémoire, on dit que deux sites sont équivalents s'ils possèdent
les mêmes charges et paramètres Lennard-Jones. Ainsi, dans l'exemple
donné, tous les hydrogènes et tous les carbones sont équivalents entre
eux. 

On étudie un solvant perturbé par un soluté, le fichier d'input du
solvant est similaire à celui du soluté.

Le soluté intervient dans le calcul de la fonctionnelle à travers
le potentiel extérieur, qui est la somme de plusieurs contributions
possibles. Les potentiels extérieurs actuellement utilisables dans
le programme sont celles présentées ci-dessous.
\begin{table}

\begin{tabular}{c|ccccccccccc}
Benzène &  &  &  &  &  &  &  &  &  &  & \tabularnewline
12~2 &  &  &  &  &  &  &  &  &  &  & \tabularnewline
\# & charge  & $\sigma$ & $\epsilon$ & $\lambda^{1S}$ & $\lambda^{2S}$ & $x$ & $y$ & $z$ & Z & atome & surnom\tabularnewline
\hline 
1 & -0.115 & 3.55 & 0.292669 & 0.0 & 0.0 & 1.386 & 0.000 & 0.0 & 6 & C & C\tabularnewline
1 & -0.115 & 3.55 & 0.292669 & 0.0 & 0.0 & 0.693 & -1.200 & 0.0 & 6 & C & C\tabularnewline
1 & -0.115 & 3.55 & 0.292669 & 0.0 & 0.0 & -0.693 & -1.200 & 0.0 & 6 & C & C\tabularnewline
1 & -0.115 & 3.55 & 0.292669 & 0.0 & 0.0 & -1.386 & 0.000 & 0.0 & 6 & C & C\tabularnewline
1 & -0.115 & 3.55 & 0.292669 & 0.0 & 0.0 & -0.693 & 1.200 & 0.0 & 6 & C & C\tabularnewline
1 & -0.115 & 3.55 & 0.292669 & 0.0 & 0.0 & 0.693 & 1.200 & 0.0 & 6 & C & C\tabularnewline
2 & 0.115 & 2.42 & 0.125248 & 0.0 & 0.0 & 2.462 & 0.000 & 0.0 & 1 & H & H\tabularnewline
2 & 0.115 & 2.42 & 0.125248 & 0.0 & 0.0 & 1.231 & -2.132 & 0.0 & 1 & H & H\tabularnewline
2 & 0.115 & 2.42 & 0.125248 & 0.0 & 0.0 & -1.231 & -2.132 & 0.0 & 1 & H & H\tabularnewline
2 & 0.115 & 2.42 & 0.125248 & 0.0 & 0.0 & -2.462 & 0.000 & 0.0 & 1 & H & H\tabularnewline
2 & 0.115 & 2.42 & 0.125248 & 0.0 & 0.0 & -1.231 & 2.132 & 0.0 & 1 & H & H\tabularnewline
2 & 0.115 & 2.42 & 0.125248 & 0.0 & 0.0 & 1.231 & 2.132 & 0.0 & 1 & H & H\tabularnewline
\end{tabular}\protect\caption{Exemple d'un fichier d'input de soluté\label{tab:solute.in}}

\end{table}

\subsection{Calcul du potentiel extérieur}

\paragraph{Calcul du potentiel extérieur électrostatique \protect \\
}

Les valeurs et les positions des charges du soluté définissent une
distribution de charge $\sigma_{c}^{u}(\bm{r})$. Le potentiel électrostatique
généré par cette distribution de charge s'exprime à l'aide de l'équation
de Poisson, en unités SI, \foreignlanguage{english}{
\begin{equation}
\Delta V_{c}=-\frac{\sigma_{c}^{u}}{\epsilon_{0}},\label{eq:poisson}
\end{equation}
}où $\Delta$ est l'opérateur laplacien et $\epsilon_{0}$ est la
permittivité du vide. Dans l'espace de Fourier cette équation de Poisson
se réécrit, 
\begin{equation}
\hat{V_{c}}(\bm{k})=-\frac{\hat{\sigma}_{c}^{u}(\bm{k})}{\epsilon_{0}k^{2}},\label{eq:V_c(k)}
\end{equation}
où $k=\left\Vert \bm{k}\right\Vert $ est la norme du vecteur réciproque
et $\hat{\sigma}_{c}^{u}$ est la transformée de Fourier (voir la
\ref{sec:Transform=0000E9es-de-Fourier}) de la densité de charge
du soluté. Le programme calcule la transformée de Fourier du potentiel
électrostatique à partir de celle de la densité de charge du soluté
et de l'\ref{eq:V_c(k)}. La solution, $V_{c}$, est donnée par la
transformée de Fourier inverse de $\hat{V}_{c}$.

Le potentiel extérieur électrostatique est ensuite calculé à l'aide
de la densité de charge d'une molécule de solvant, $\sigma_{c}^{v}(\bm{r},\bm{\Omega})$,%
\footnote{On rappelle que cette densité de charge moléculaire est utilisée pour
calculer la densité de charge du solvant $\rho_{c}(\bm{r})=\iiint_{\mathbb{R}^{3}}\sigma_{c}^{v}(\bm{r}-\bm{r}^{\prime},\bm{\Omega})\rho(\bm{r}^{\prime},\bm{\Omega})\mathrm{d}\bm{r}^{\prime}$ %
} 
\begin{equation}
v_{c}(\bm{r},\bm{\Omega})=\frac{1}{4\pi}V_{c}(\bm{r})\sigma_{c}^{v}(\bm{r},\bm{\Omega}).
\end{equation}

\paragraph{Calcul du potentiel extérieur Lennard-Jones\protect \\
}

À partir des paramètres de Lennard-Jones des sites du soluté et de
ceux du solvant, le programme calcule des paramètres Lennard-Jones
mixtes \textit{$\sigma_{i}^{uv}$}et \textit{$\epsilon_{i}^{uv}$
via} les règles de Lorentz-Berthelot, qui sont celles utilisées le
plus généralement,
\begin{equation}
\sigma_{i}^{uv}=\frac{\sigma_{i}^{u}+\sigma^{v}}{2},
\end{equation}
\begin{equation}
\epsilon_{i}^{uv}=\sqrt{\epsilon^{v}\epsilon_{i}^{u}}.
\end{equation}
Grâce à ces paramètres mixtes le programme calcule ensuite le potentiel
extérieur Lennard-Jones 
\begin{equation}
v_{\text{LJ}}(\bm{r})=4\sum_{i=1}^{\text{N}^{u}}\epsilon_{i}^{uv}\left[\left(\frac{\sigma_{i}^{uv}}{\left\Vert \bm{r}\right\Vert }\right)^{12}-\left(\frac{\sigma_{i}^{uv}}{\left\Vert \bm{r}\right\Vert }\right)^{6}\right].
\end{equation}

Pour des raisons numériques, on borne la densité d'énergie totale
à une valeur maximale, typiquement $100\ \mathrm{kJ.mol}{}^{-1}$.

\paragraph{Calcul du potentiel extérieur sphères dures\protect \\
}

Il est également possible de rajouter un rayon de sphère dure $\mathrm{R}_{i}^{u}$
sur les différents sites du soluté. On définit également un rayon
de sphère dure pour le solvant R$^{v}$. Le potentiel extérieur sphères
dures vaut alors,
\begin{equation}
v_{SD}(r)=\sum_{i=1}^{\text{N}^{u}}v_{iSD}(r),
\end{equation}
avec
\begin{equation}
\begin{cases}
v_{iSD}(r)=0\  & \text{si }|\bm{r}-\bm{r}_{i}^{u}|\leq\text{R}_{i}^{u}+\text{R}^{v}\\
v_{iSD}(r)=\infty\  & \text{si }|\bm{r}-\bm{r}_{i}^{u}|>\text{R}_{i}^{u}+\text{R}^{v}
\end{cases}.
\end{equation}

\paragraph{Calcul d'un potentiel extérieur sphères molles\protect \\
}

Un potentiel de sphère molle purement répulsive en $r^{-\mathrm{p}}$
où p est un entier positif, peut aussi être rajouté sur les sites
de soluté en précisant la valeur des rayons de sphère molle $r_{iH}$,
\begin{equation}
V_{\mathrm{p}}(r)=\text{k}_{\text{B}}T\sum_{i=1}^{\text{N}^{u}}\left\Vert r-r_{iH}\right\Vert {}^{-\mathrm{p}}\text{ si \ensuremath{\left\Vert r\right\Vert }>}r_{iH},\text{ et \ensuremath{\infty}sinon}\label{eq:VpHansen_gl}
\end{equation}
On note que si p=12, ce potentiel correspond à celui utilisé dans
la référence \cite{DHS_dzubiella_competition_2004}.

\subsection{Initialisation de la densité}

Une fois la densité d'énergie extérieure connue, on peut initier la
densité à 
\begin{equation}
\rho_{\text{init}}(\bm{r},\bm{\Omega})=\rho_{\mathrm{b}}\exp\left[-\beta\left(v_{\mathrm{ext}}(\bm{r})-v_{c}(\bm{r})\right)\right],\label{eq:rho_init}
\end{equation}
où $v_{\mathrm{ext}}$ est la densité d'énergie extérieure totale,
on rappelle que $\rho_{\mathrm{b}}=n_{b}/(8\pi^{2})$. On retire la
contribution due à l'énergie électrostatique, pathologique, qui peut
diverger. Cette initialisation permet d'avoir des densités raisonnables,
c'est-à-dire élevées (respectivement faibles) là où le potentiel extérieur
est faible (respectivement élevé). La plupart du temps on initie donc
la densité à $\rho_{\text{init}}(\bm{r},\bm{\Omega})=\rho_{\mathrm{b}}\exp\left[-\beta v_{\mathrm{LJ}}(\bm{r})\right]$.

\fbox{\begin{minipage}[t]{1\columnwidth}%
On peut remarquer que l'\ref{eq:rho_init} correspond à la solution
exacte pour un fluide idéal (dont le terme d'excès est nul) et un
potentiel extérieur $v_{\mathrm{ext}}(\bm{r})-v_{c}(\bm{r})$.%
\end{minipage}}

\subsection{Fonctions de corrélation directe}

Pour le calcul du terme d'excès, il est nécessaire d'introduire des
fonctions de corrélation obtenues expérimentalement ou calculées par
simulations numériques. Celles-ci dépendent de la forme de la fonctionnelle
que l'on veut utiliser. Ce sont soit les fonctions de corrélation
directe, $c_{000}$, $c_{112}$, $c_{110}$ et $c_{101}$ si on utilise
la forme dipolaire de la fonctionnelle présentée en \ref{sec:Fexcdip},
soit le facteur de structure $S$ et les inverses des constantes diélectriques,
$\chi_{L}^{-1}$ et $\chi_{T}^{-1}$, si on utilise la forme multipolaire
de la \ref{sec:Fexcmulti}. On souligne une fois encore que la force
de la théorie réside dans le fait que ces fonctions de corrélation
ne doivent être déterminées qu'une seule fois pour chaque solvant
bulk de densité donnée, et que l'on peut ensuite étudier la solvatation
de n'importe quel soluté dans ce solvant. Les moments multipolaires
de la molécule d'eau sont également nécessaires et il sont fournis
de manière implicite par le choix du modèle moléculaire.

\subsection{Minimiseur et minimisation}

\paragraph{Description du minimiseur\label{par:Description-du-minimiseur}\protect \\
}

Pour réaliser la minimisation fonctionnelle, le code mdft utilise
le minimiseur L-BFGS (Limited-memory BFGS) \cite{LBFGS_zhu_algorithm_1997}
qui est une méthode quasi-Newton pour trouver les minima d'une fonction.
Ce minimiseur est très efficace pour les problèmes impliquant un grand
nombre de variables. Cette routine trouve les points stationnaires
d'une fonction, c'est-à-dire les points où le gradient est nul. Au
lieu d'utiliser le gradient et la matrice Hessienne (la matrice des
dérivés secondes) de la fonction à minimiser, comme un algorithme
gradient-conjugué classique, le minimiseur utilise les $m$ valeurs
obtenues aux $m$ pas précédents de la densité et du gradient de la
fonctionnelle (dans notre code on utilise généralement $m=4$). Ces
valeurs stockées servent à réaliser implicitement les opérations requierant
normalement l'utilisation de la matrice Hessienne. Ceci permet d'éviter
le calcul de la matrice Hessienne et de limiter la quantité de mémoire
utilisée. Ce minimiseur nécessite donc qu'on lui fournisse, à chaque
pas, les valeurs de l'énergie libre et des gradients en chaque nœud
de la double grille. Il y a deux critères de convergence permettant
de considérer que le minimum de la fonctionnelle est atteint. Le premier
porte sur la norme du gradient, qui doit être inférieure à un critère
arbitraire (généralement $10^{-7}$). Le second porte sur la variation
relative de l'énergie libre entre deux pas successifs ($(F_{n}-F_{n-1})/F$)
qui doit elle aussi être inférieure à un critère arbitraire (généralement
$10^{-6}$).

\paragraph{Processus de minimisation\protect \\
}

À partir de la valeur initiale de la densité, du potentiel extérieur
et des fonctions de corrélation, le code calcule les différentes parties
de la fonctionnelle de l'\ref{eq:F=00003DFid+Fext+Fexc} ainsi que
leur gradient. Le minimiseur calcule alors une nouvelle densité notée
$\rho_{1}$. Le code recalcule les différentes parties de la fonctionnelle
et leur gradient à partir de la densité $\rho_{1}$. Si l'un des deux
critères de convergence décrits en \ref{par:Description-du-minimiseur}
est atteint, la procédure s'arrête et les dernières densités et énergie
libre de solvatation calculées sont considérées comme les grandeurs
physiques à l'équilibre thermodynamique. Si aucun des critères de
convergence n'est vérifié alors le minimiseur trouve une nouvelle
densité $\rho_{2}$ à partir de sa valeur précédente $\rho_{1}$ et
de celles des gradients. À partir de cette densité, les nouvelles
valeurs des différentes composantes de l'énergie libre (de la fonctionnelle)
et des gradients sont calculées. Le procédé est itéré jusqu'à ce qu'un
des critères de convergence soit atteint. Une fois sorti de cette
boucle de minimisation, le code calcule différentes observables d'intérêt
à partir de la densité d'équilibre obtenue : fonctions de distribution
radiale, densité, densité de polarisation à l'équilibre, etc.

\section{Calculs des différentes intégrales rencontrées}

Les différents termes de la fonctionnelle qui doivent être calculés
lors de la minimisation contiennent plusieurs types d'intégrale sur
les différentes variables mises en jeu. On distingue essentiellement
trois types d'intégrale : des intégrales sur les angles, des intégrales
sur le volume et des intégrales doubles sur le volume au carré.

\subsection{Intégrales sur le volume}

Pour les termes qui contiennent une intégrale simple sur les cordonnées
d'espace comme la partie idéale de l'\ref{eq:Fid}, on réalise directement
le calcul dans l'espace direct. L'intégrale est simplement approchée
par une somme finie, 
\begin{equation}
{\cal F}\left[\rho(\bm{r})\right]=\iiint_{\mathbb{R}^{3}}A(\rho(\bm{r}))\text{d}\bm{r}\equiv{\cal F}\left[\left\{ \rho(\bm{r}_{i})\right\} \right]\equiv\sum_{i=1}^{\text{N}}A(\rho(r_{i}))\Delta_{\text{V}},\label{eq:int_simple}
\end{equation}

où $\text{N}=\text{N}_{x}.\text{N}_{y}.\text{N}_{z}$ avec $\text{N}_{x}$
(resp. $\text{N}_{y}$, resp. $\text{N}_{z}$) le nombre de nœuds
de la grille selon la direction $\text{\ensuremath{\bm{O}_{x}}}$
(resp. $\text{\ensuremath{\bm{O}_{y}}}$, resp. $\text{\ensuremath{\bm{O}_{z}}}$)
et $\Delta_{\text{V}}=\text{L}_{x}\cdot\text{L}_{y}\cdot\text{L}_{z}/(\text{N}_{x}\cdot\text{N}_{y}\cdot\text{N}_{z})$
le volume d'un voxel élémentaire. On a fait ici une approximation
implicite, qui consiste en l'utilisation d'une quadrature simple pour
évaluer l'intégrale.

\fbox{\begin{minipage}[t]{1\columnwidth}%
Les méthodes de quadrature sont des approximations pour évaluer la
valeur numérique d'une intégrale. On remplace l'intégrale par une
somme pondérée en un certain nombre de points appartenant au domaine
d'intégration. Par exemple, on a à l'\ref{eq:int_simple} des points
régulièrement espacés avec comme poids le volume du voxel élémentaire. 

Pour évaluer l'intégrale sur le volume d'une fonction, on pourrait
utiliser une quadrature plus adaptée, c'est à dire qui donnerait une
bonne estimation numérique de l'intégrale avec un nombre de point
plus faible. Il est évident, par exemple, que la densité varie peu
loin de la perturbation, on pourrait ainsi utiliser un maillage plus
lâche loin du soluté, plutôt qu'une grille régulière.

Cependant, comme on le verra par la suite, pour calculer efficacement
les produits de convolution, nous faisons un usage intensif de transformées
de Fourier rapides (FFT) et l'algorithme choisi pour faire ces FFT
requiert l'utilisation d'une grille régulière.%
\end{minipage}}

\subsection{Intégrales sur la grille angulaire}

Ce sont des intégrales de la forme,
\begin{equation}
n(\bm{r})=\int_{\theta=0}^{\pi}\int_{\phi=0}^{2\pi}\int_{\psi=0}^{2\pi}\rho(\bm{r},\bm{\Omega})\text{d}\bm{\Omega}.
\end{equation}
On rappelle que $\bm{\Omega}$ est la notation compacte des trois
angles d'Euler.

\paragraph{Quadrature pour les deux premiers angles d'Euler $(\theta,\phi)$\protect \\
}

Les intégrations sur la grille angulaire sont réalisées dans l'espace
direct exactement de la même manière que pour le volume. Cependant,
on n'utilise plus ici de quadrature régulière. À mon arrivée au laboratoire
une quadrature du type Gauss-Legendre\cite{press_numerical_1996}
était utilisée pour les angles $(\theta,\phi)$. Cette quadrature
est conçue pour des intégrales dont le domaine d'intégration est $\left[-1,1\right]$.
Elle donne les $n$ points où la fonction doit être évaluée comme
étant les racines du polynôme de Legendre d'ordre $n$, ainsi que
les poids associés à ces points. Cette quadrature est classique et
donne des résultats exacts pour les fonctions polynomiales de degré
$2n-1$, et donc les harmoniques sphériques d'ordre $n-1$. Pour l'intégration
sur la sphère unité, avec une quadrature de Gauss-Legendre d'ordre
$n$, comme les nœuds $x_{i}$ tels que $i\in\left[1,n\right]$, varie
entre $-1$ et $1$, on définit $\theta_{i}$ tel que $\cos(\theta_{i})=x_{i}$,
de telle sorte que $\theta$ varie entre $0$ et $\pi$. Les angles
$\phi_{i}$ doivent varier entre $0$ et $2\pi$, on les choisit régulièrement
espacés sur cette intervalle. On a donc pour une quadrature de Gauss-Legendre
d'ordre $n$, $2n^{2}$ points pour décrire les deux premiers angles
d'Euler.

\begin{figure}
\noindent \centering{}\includegraphics[width=0.9\textwidth]{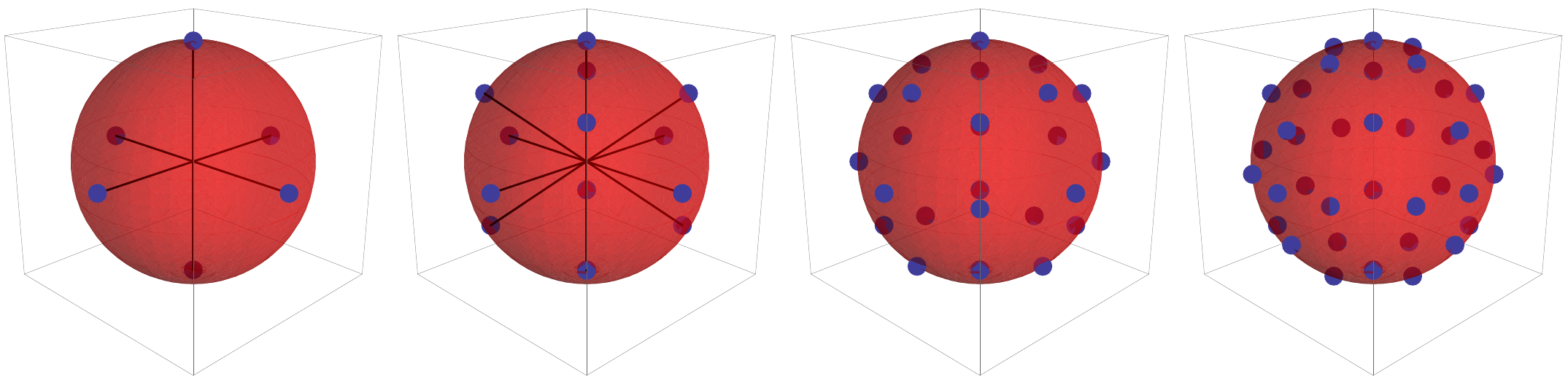}\protect\caption{Visualisation des orientations considérées dans la quadrature de Lebedev,
avec de gauche à droite 6, 14, 26 et 38 points. La sphère unité est
en rouge et les orientations considérées sont représentées par des
points bleus.\label{fig:leb_graph}}
\end{figure}

\begin{figure}
\noindent \centering{}\includegraphics[width=0.8\textwidth]{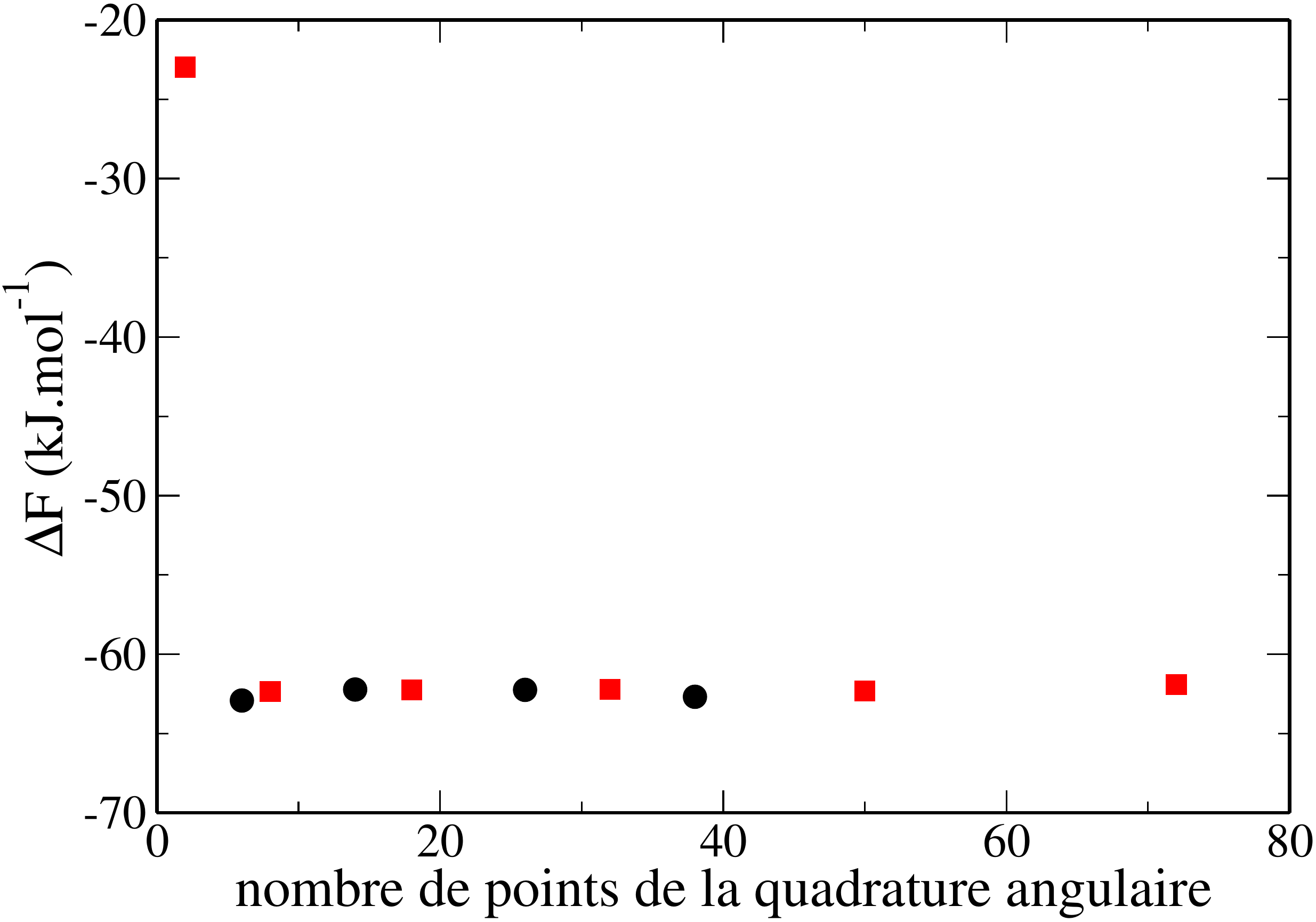}\protect\caption{Valeurs des énergies libres de solvatation obtenues en utilisant une
grille de Lebedev (cercles noirs) ou de Gauss Legendre (carrés rouges)
en fonction du nombre de points de la grille. Le soluté utilisé est
un sodium portant une demi charge élémentaire et le solvant un fluide
de Stockmayer. La boite fait $25\ \textrm{\AA}^{3}$, avec une grille
de $96^{3}$ points.\label{fig:conv_leb}}
\end{figure}

Au cours de ma thèse j'ai implémenté dans le code mdft une autre quadrature
que celle-ci, la quadrature de Lebedev \cite{lebedev_quadrature_1999}.
Cette quadrature est spécialement développée pour réaliser une intégration
sur la sphère unité. Elle est construite sur une symétrie octaédrique.
Ainsi, contrairement à la quadrature de Gauss-Legendre, les contraintes
sur le nombre de points pour calculer numériquement l'intégrale sont
fortes. En conséquence, dans la version actuelle du code, on peut
utiliser 6, 14, 26 ou 38 points pour décrire les orientations sur
la sphère unité. Par exemple, les 6 premières orientations sont dans
les directions des centres des faces du cube unité contenant la sphère,
les 8 points suivants sont dans les directions des sommets, les 12
points suivants sont orientés dans la direction du milieu des arêtes
du cube comme illustré sur la \ref{fig:leb_graph}.
La quadrature de Lebedev est plus efficace qu'une quadrature de Gauss-Legendre
d'un facteur d'environ 2/3. On peut voir sur la \ref{fig:conv_leb}
que les deux quadratures donnent des résultats similaires pour un
soluté et un solvant simple.

\fbox{\begin{minipage}[t]{1\columnwidth}%
L'introduction de cette quadrature a été une bonne façon de \og mettre
les mains \fg{} dans un code existant. De plus, il se peut que pour
étudier des solutés plus complexe, il soit nécessaire d'avoir une
meilleur résolution angulaire, ce qui sera facilité par l'utilisation
de la quadrature de Lebedev.%
\end{minipage}}

\paragraph{Quadrature pour l'angle de rotation propre $\psi$\protect \\
}

Au début de ma thèse, le code ne fonctionnait que pour des solvants
strictement linéaires (acétonitrile, fluide de Stockmayer, etc). Pour
pouvoir l'utiliser pour le solvant eau il a fallu introduire un nouvel
angle d'Euler, celui correspondant à la rotation propre d'une molécule
de solvant sur son axe de symétrie principal $C_{n}$. J'ai donc introduit
l'angle $\psi$ comme une nouvelle variable dont dépendait la densité.
Dans le but d'augmenter l'efficacité numérique du code, cet angle
a été introduit comme variant entre $0$ et $2\pi/n$, où $n$ est
l'ordre de l'axe de symétrie principal (soit 2 pour l'eau). La quadrature
choisie pour cette nouvelle variable est une quadrature régulière,
comparable à celle pour l'angle $\phi$ dans la quadrature de Gauss-Legendre.

\subsection{Intégrales doubles sur le volume}

\begin{figure}
\noindent \centering{}\includegraphics[width=0.4\textwidth]{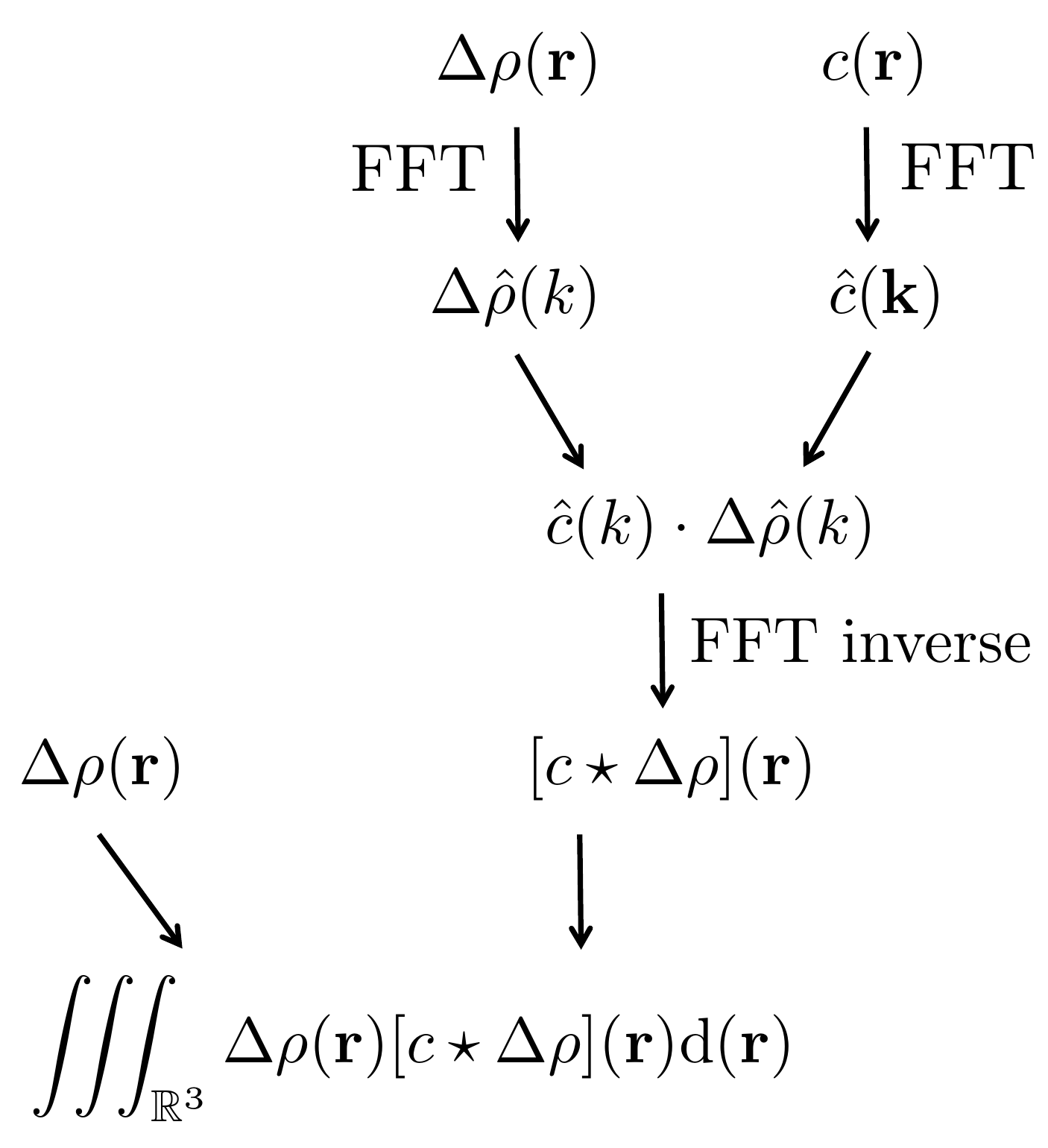}\protect\caption{Représentation schématique de la façon dont sont calculées, dans le
code mdft, les intégrales contenant un produit de convolution comme
dans l'\ref{eq:Fintconvolut}.\label{fig:shema_FFT_Convo}}
\end{figure}

Pour les termes contenant une intégrale double sur le volume, comme
par exemple la partie d'excès de l'\ref{eq:Fexc}, le calcul dans
l'espace direct apparait comme trop coûteux. Cependant, ces intégrales
sont des convolutions de la forme,
\begin{eqnarray}
{\cal F}\left[\rho(\bm{r})\right] & = & \iiint_{\mathbb{R}^{3}}\iiint_{\mathbb{R}^{3}}\Delta\rho(\bm{r}_{1})c(\bm{r}_{12})\Delta\rho(\bm{r}_{2})\text{d}\bm{r}_{1}\text{d}\bm{r}_{2}\label{eq:Fintconvolut}\\
 & = & \iiint_{\mathbb{R}^{3}}\Delta\rho(\bm{r}_{1})\left[\iiint_{\mathbb{R}^{3}}c(\bm{r}_{12})\Delta\rho(\bm{r}_{2})\text{d}\bm{r}_{2}\right]\text{d}\bm{r}_{1}\\
 & = & \iiint_{\mathbb{R}^{3}}\Delta\rho(\bm{r}_{1})\left[c\star\Delta\rho\right](\bm{r}_{1})\text{d}\bm{r}_{1},
\end{eqnarray}
où on a séparé les termes dépendant de $\bm{r}_{2}$, et $\star$
désigne le produit de convolution entre les deux fonctions $\Delta\rho$
et $c$. On calcule ce produit de convolution en se servant de transformées
de Fourier. En effet, une des propriétés intéressantes des transformées
de Fourier (voir la \ref{sec:Transform=0000E9es-de-Fourier}) est
que le produit des transformées de deux fonctions est égal à la transformée
de Fourier du produit de convolution de ces deux fonctions. Par exemple,
la procédure pour calculer l'intégrale de l'\ref{eq:Fintconvolut}
est schématisée sur la \ref{fig:shema_FFT_Convo} :
\begin{itemize}
\item On calcule, si on ne les connait pas déjà, les champs scalaires $\Delta\rho$
et $c$ en chaque point de l'espace, c'est-à-dire pour tous les points
de notre grille. 
\item On calcule leurs transformées de Fourier respectives, que l'on note
$\Delta\hat{\rho}$ et $\hat{c}$. 
\item On multiplie ces deux transformées de Fourier pour calculer la transformée
de Fourier du produit de convolution.
\item On effectue la transformée de Fourier inverse de ce produit pour obtenir
la fonction produit de convolution $c\star\Delta\rho$ dans l'espace
direct. 
\item On peut alors calculer l'intégrale de l'\ref{eq:Fintconvolut} comme
on l'a fait avec l'intégrale simple sur l'espace de l'\ref{eq:int_simple}.
\end{itemize}

\subsection{Calcul des densités de charges et de polarisation microscopique}

La densité de charge microscopique du solvant est calculée dans l'espace
de Fourier,
\begin{equation}
\sigma(\bm{k},\bm{\Omega})=\sum_{m}\mathrm{z}_{m}\mathrm{e}\exp\left(-i\bm{k}\cdot\bm{s}_{m}(\bm{\Omega})\right)
\end{equation}
 où $\mathrm{z}_{m}\mathrm{e}$ désigne la charge partielle et $\bm{s}_{m}(\bm{\Omega})$
la position du m$^{\mathrm{i\grave{e}me}}$ site de solvant d'une
molécule d'eau située à l'origine et d'orientation $\bm{\Omega}$.
La densité de polarisation microscopique est également calculée dans
l'espace réciproque en utilisant l'\ref{eq:mu_k_omega}. Ces deux
densités sont calculées à l'initialisation du programme et stockées
pour être utilisées par la suite. 

Comme les densités de charge et de polarisation sont des produits
de convolution, entre ces densités microscopiques et la densité macroscopique
$\rho(\bm{r},\bm{\Omega})$, on les calcule également dans l'espace
de Fourier :
\begin{equation}
\bm{P}(\bm{k})=\iiint_{8\pi^{2}}\mu(\bm{k},\bm{\Omega})\rho(\bm{k},\bm{\Omega})\mathrm{d}\bm{\Omega}
\end{equation}
et
\begin{equation}
\rho_{c}(\bm{k})=\iiint_{8\pi^{2}}\sigma(\bm{k},\bm{\Omega})\rho(\bm{k},\bm{\Omega})\mathrm{d}\bm{\Omega}.
\end{equation}

Les composantes longitudinale et transverse de la polarisation sont
ensuite calculées directement dans l'espace réciproque grâce à l'\ref{eq:PL/PT(k)}.
On peut ensuite calculer la fonctionnelle multipolaire. 

On souligne qu'il faut donc réaliser un nombre de transformées de
Fourrier directe et inverse égale au nombre total d'angles utilisés.

\section{Transformées de Fourier rapides\label{sec:Transform=0000E9es-de-Fourier}}

Comme on l'a vu dans le calcul de la densité d'énergie électrostatique
et le calcul des intégrales doubles sur le volume, le code mdft requiert
le calcul de transformées de Fourier.

Ceci est fait numériquement par l'utilisation de transformées de Fourier
discrètes efficaces. Celles-ci sont réalisées par des transformées
de Fourier rapides (FFT), qui sont calculées grâce à l'algorithme
FFTW3 (Fastest Fourier Transform in the West)\cite{FFTW_frigo_fftw,FFTW_url,FFTW_frigo_design_2005}.
On ne décrira pas, ici, comment fonctionne cet algorithme en particulier,
ni les principes des transformées de Fourier discrètes et rapides
puisqu'aucun travail n'a été réalisé sur cette partie. On se limite
à l'utilisation de ces techniques numériques. On peut néanmoins souligner
que l'on approxime ce qui est en réalité des intégrales continues
(les transformées de Fourier) par des sommes finies (les transformées
de Fourier discrètes),
\begin{equation}
\hat{f}(k)=\iiint_{\mathbb{R}^{3}}f(\bm{x})e^{-2i\pi\bm{k}\cdot\bm{x}}\text{d}\bm{x}\approx\hat{f}(\bm{k}_{p})=\sum_{n=0}^{\text{N}-1}f(\bm{x}_{n})e^{-2i\pi\bm{k}_{p}\cdot\bm{x}_{n}}\Delta_{V}.\label{eq:FT->DFT}
\end{equation}

La transformée de Fourier discrète transforme alors une suite de N
nombres (possiblement complexes, mais dans notre cas réels) en une
suite $\text{N}$-périodique de nombres (complexes). Notons que la
transformée de Fourrier discrète inverse se construit de manière analogue,
\begin{equation}
f(\bm{x}_{n})=\frac{1}{\left(2\pi\right)^{3}}\sum_{p=0}^{\text{N}-1}\hat{f}(\bm{k}_{p})e^{2i\pi\bm{k}_{p}\cdot\bm{x}_{n}}\Delta_{V_{k}}.\label{eq:FFTW-1}
\end{equation}
Cette équation implique également la N-périodicité des $f(\bm{x}_{n})$
et $\hat{f}(\bm{k}_{p})$ et donc des conditions aux bords périodiques.

On souligne également que les définitions données ici sont celles
de FFTW3 qui ne sont pas normalisées : si on fait une transformée
de Fourier discrète suivi d'une transformée de Fourier discrète inverse
de la même fonction, on voit que le résultat est la série de départ
multipliée par N, il faut donc à posteriori renormaliser ces résultats.

\fbox{\begin{minipage}[t]{1\columnwidth}%
L'utilisation de transformées de Fourier rapides et l'efficacité du
code FFTW3 sont en grande partie à l'origine de la rapidité du code
mdft. On peut par exemple constater, pour la correction à trois corps,
la différence flagrante de temps de calcul entre la routine qui calcule
directement les intégrales doubles dans l'espace direct et celle qui
calcule les produits de convolution à l'aide de transformées de Fourier,
voir la \ref{fig:Comp_OldNew_F3B_temps}.%
\end{minipage}}

\section{Influence du maillage sur la convergence}

Après avoir présenté comment fonctionne le code mdft, on s'intéresse
à l'influence sur les résultats obtenus de la résolution (la finesse
du maillage) des grilles angulaire et spatiale, ainsi qu'à l'efficacité
algorithmique du code.

\begin{figure}
\noindent \centering{}\includegraphics[width=0.6\textwidth]{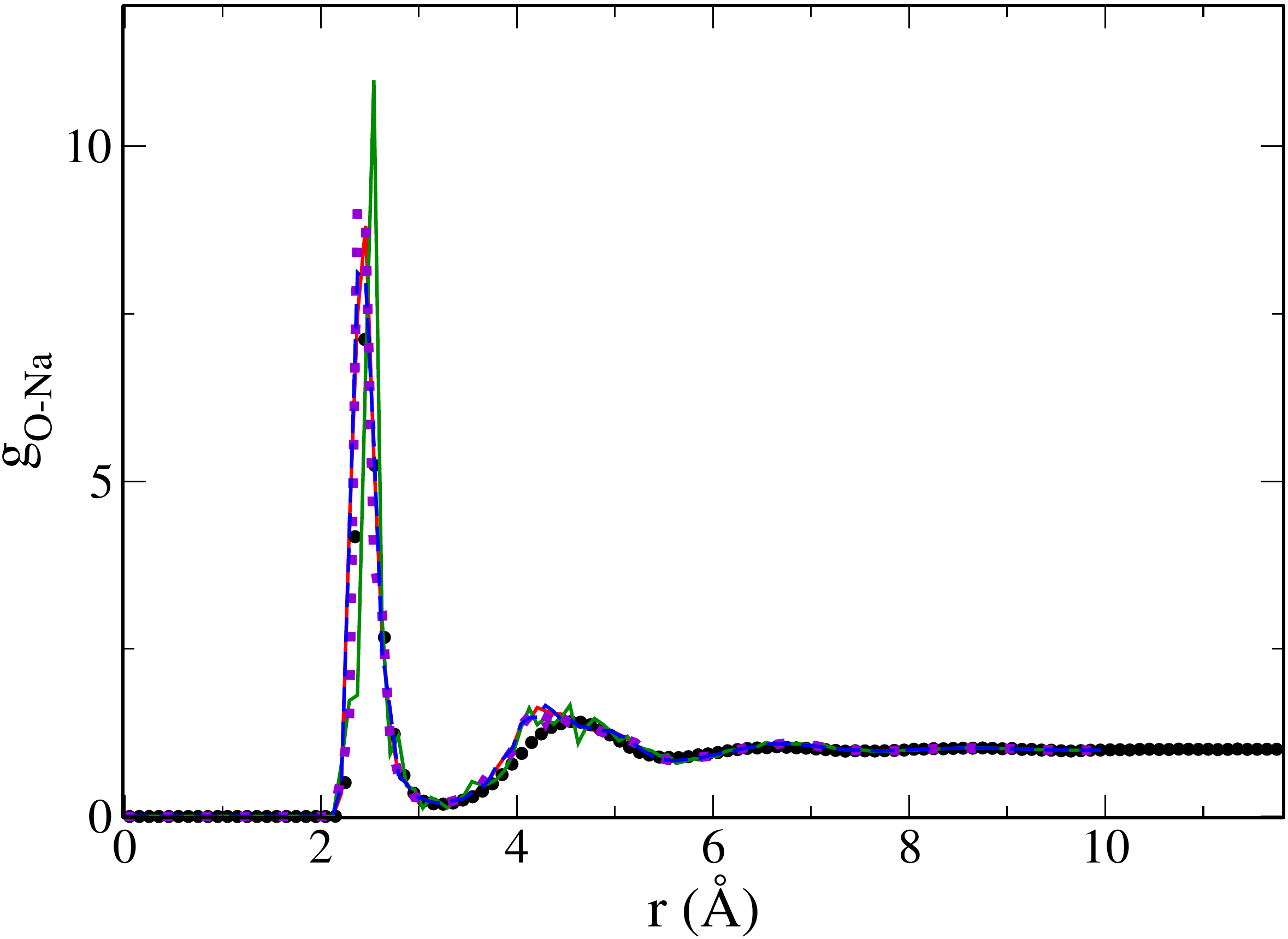}\protect\caption{Fonctions de distribution radiale entre le solvant eau et l'ion sodium.
La référence calculée par dynamique moléculaire est donnée par les
points noirs. Les calculs MDFT on été réalisés avec une boîte de $30\ \textrm{\AA}$,
12 points pour la grille angulaire et 2 (ligne verte), 3 (points violets),
4 (tirets bleus), et 5 (ligne rouge) points par $\textrm{\AA}$ pour
la grille spatiale.\label{fig:rdf_comp_resoluton} }
\end{figure}

\begin{figure}
\noindent \begin{centering}
\hspace{-0.5cm}%
\begin{tabular}{cc}
\includegraphics[width=0.5\textwidth]{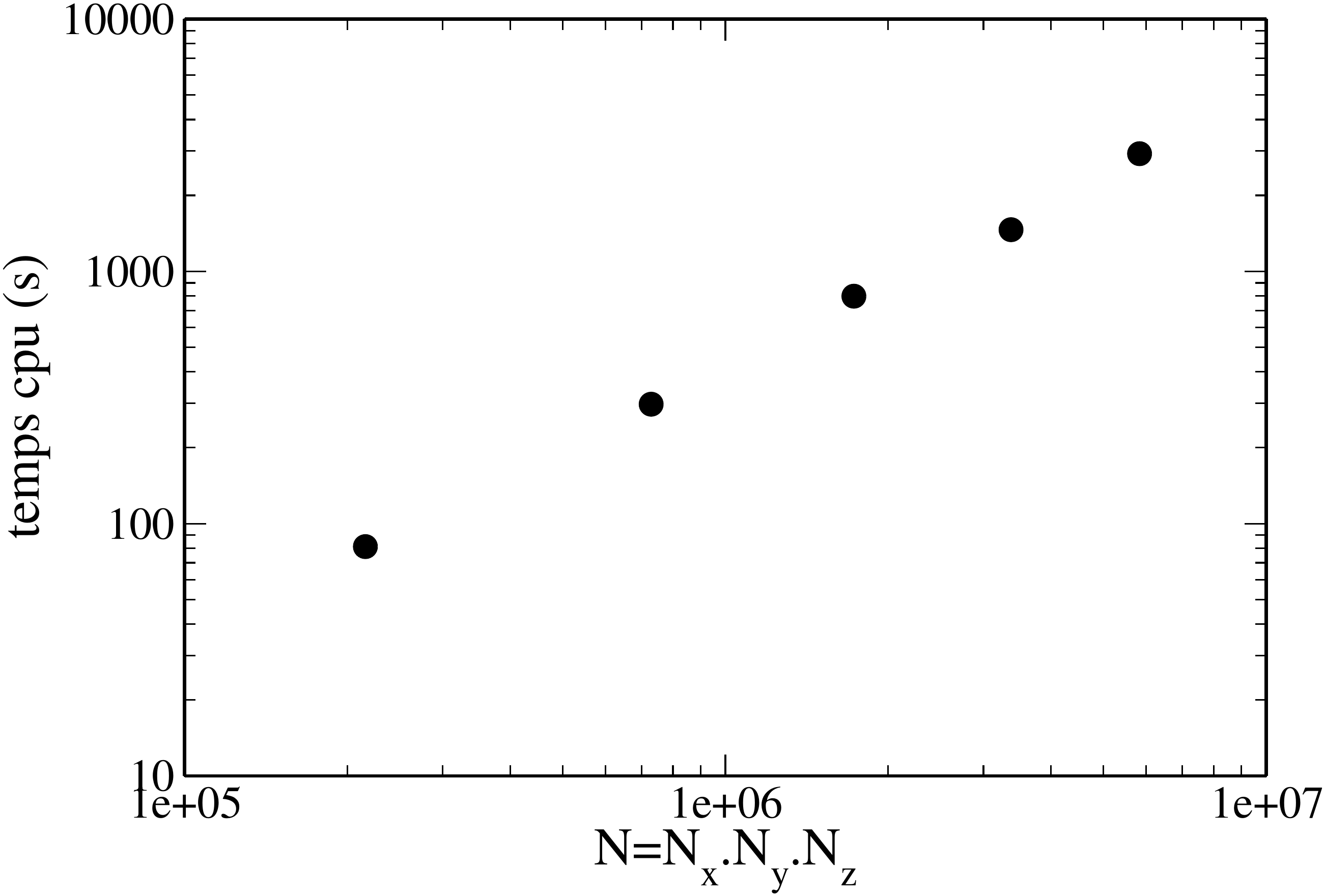} & \includegraphics[width=0.48\textwidth]{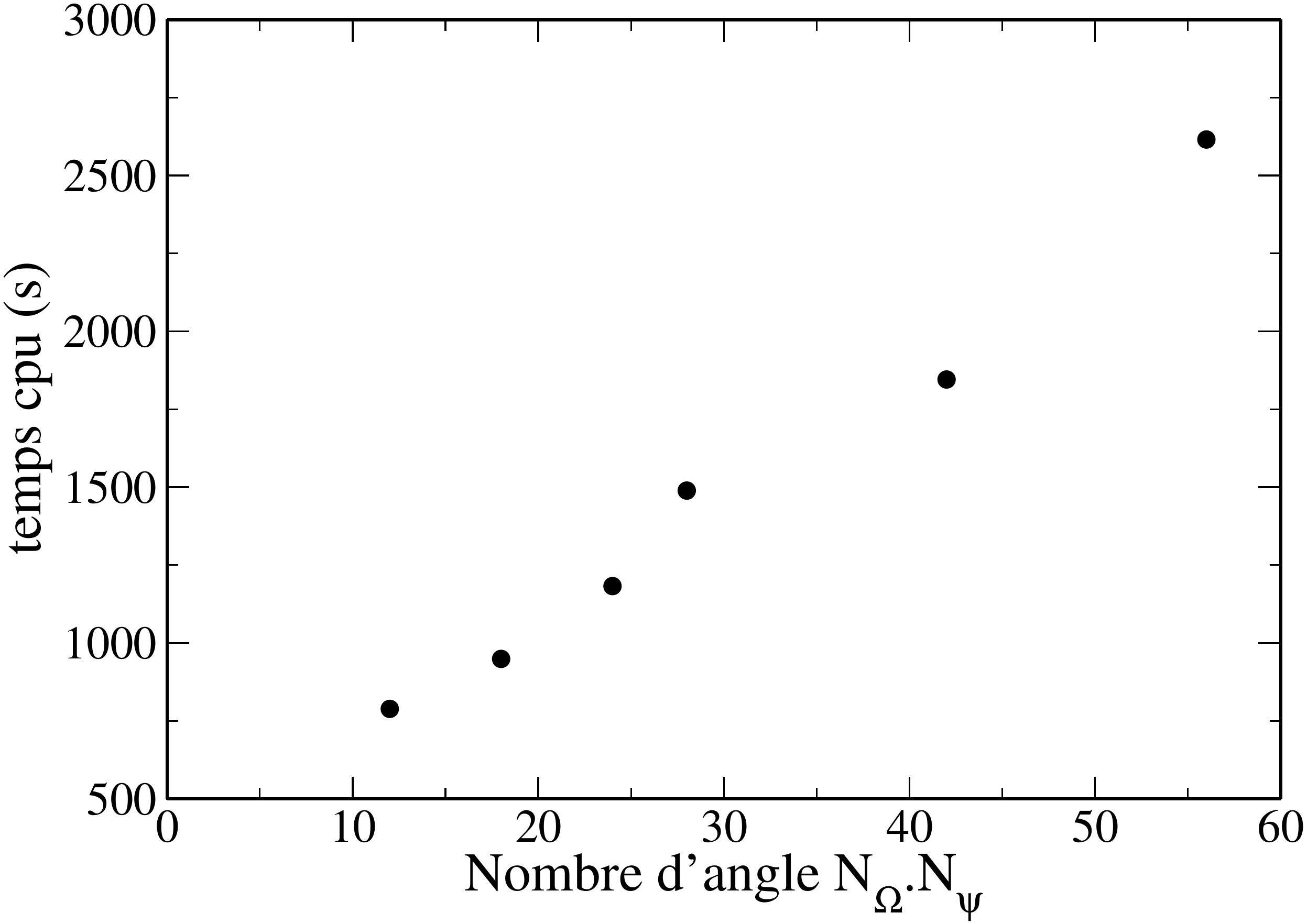}\tabularnewline
\end{tabular}
\par\end{centering}

\noindent \centering{}\protect\caption{Temps CPU pour réaliser la minimisation totale. Le système est le
même que pour la courbe de la \ref{fig:rco,verge_en pas}. À gauche,
ce temps est présenté pour une grille angulaire de 12 points en fonction
du nombre total de points de la grille spatiale. À droite, ce temps
est présenté pour une grille spatiale de $120^{3}$ points en fonction
du nombre de points de la grille angulaire.\label{fig:iterate_time} }
\end{figure}

On montre sur la \ref{fig:rdf_comp_resoluton} l'influence de la résolution
de la grille spatiale sur les fonctions de distribution radiale. On
a choisi comme exemple l'ion sodium, que l'on étudie avec la fonctionnelle
multipolaire et la correction à trois corps décrite au \ref{chap:F3B}.
On obtient un profil correct à partir d'une résolution de 3 points
par $\textrm{\AA}$. Une augmentation de cette résolution n'a pas
de conséquence notable sur la structure. En ce qui concerne l'énergie
libre de solvatation, un profil de convergence typique est présenté
en \ref{fig:rco,verge_en pas} pour un ion sodium, un critère de convergence
de $10^{-4}$ est atteint après une vingtaine de pas. Le temps de
CPU (Central Processing Unit) nécessaire à la minimisation de la fonctionnelle
augmente comme le nombre de points de grille angulaire et spatiale.
Ceci est illustré sur la \ref{fig:iterate_time}. 
\begin{figure}
\noindent \centering{}\includegraphics[width=0.6\textwidth]{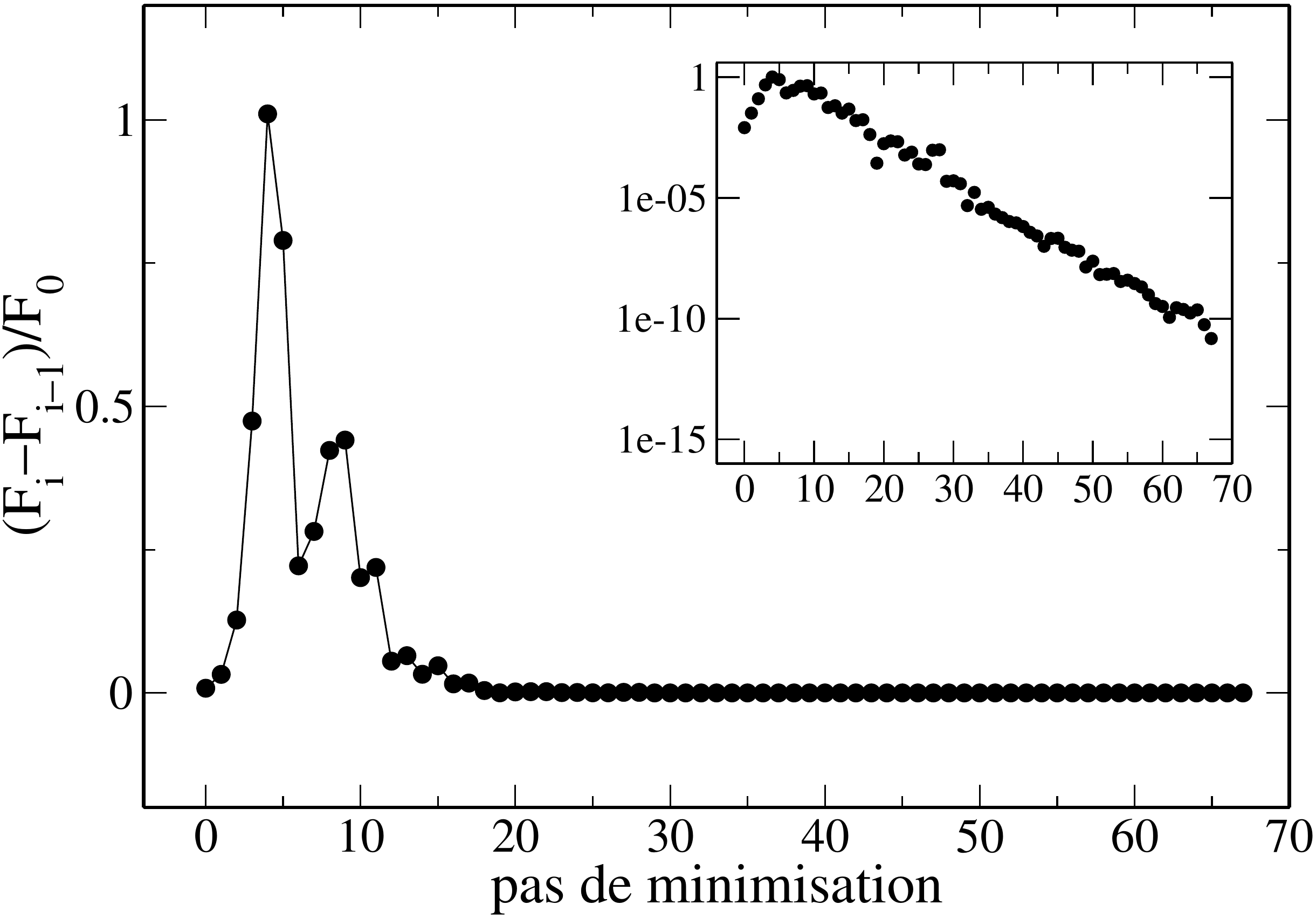}\protect\caption{Différences d'énergies libre de solvatation entre deux pas successifs,
normalisées par l'énergie libre de solvatation initiale, pour un sodium
avec une boîte de $30\ \textrm{\AA}$, 5 points par $\textrm{\AA}$
et 12 points pour la grille angulaire en fonction du pas d'itération.
L'encart représente la même courbe avec une échelle logarithmique
pour l'axe des ordonnées. \label{fig:rco,verge_en pas} }
\end{figure}

Le code est donc linéaire en fonction du nombre de points de grille
utilisé pour la grille angulaire, et pour la grille spatiale. Le temps
de calcul est donc  proportionnel à la taille du système étudié.

\pagebreak{}

\section{Conclusion}

Le code mdft est une implémentation en Fortran moderne de la théorie
MDFT présentée dans ce manuscrit. Il permet la minimisation fonctionnelle
de l'énergie libre sur une double grille angulaire et spatiale (soit
6 dimensions). Une grande partie de l'efficacité numérique de ce programme
provient de l'utilisation de transformées de Fourier rapides\cite{FFTW_frigo_fftw}.
Le calcul de la partie d'excès nécessite l'injection de fonctions
de corrélation obtenues préalablement par simulations numériques ou
par l'expérience. Ce code permet d'obtenir des informations sur la
structure et sur l'énergie libre de solvatation comparables à celles
obtenues par simulations numériques en étant $10^{2}-10^{3}$ fois
plus rapide. Un calcul typique de solvatation d'une petite molécule
prend moins de dix minutes sur un ordinateur portable. Le code n'a
pas été parallélisé pour l'instant et, étant encore en développement,
n'est pas encore mis à disposition de la communauté.

\lhead[\chaptername~\thechapter]{\rightmark}

\rhead[\leftmark]{}

\lfoot[\thepage]{}

\cfoot{}

\rfoot[]{\thepage}

\part{Applications}

\chapter{Applications}

Après avoir introduit la théorie de la fonctionnelle de la densité
classique moléculaire et validé son utilisation sur un ensemble de
petites molécules tests, nous allons montrer dans ce chapitre qu'elle
est un outil adapté à l'étude de systèmes complexes.

Nous présenterons d'abord des calculs de potentiels de force moyenne.
Nous nous intéressons ensuite à l'étude d'une argile, la pyrophyllite
puis à celle d'un système biologique, la protéine lysozyme du blanc
d'œuf de poule.\newpage{}

\section{Calculs de potentiels de force moyenne (PMF)}

Nous avons utilisé la théorie MDFT pour calculer des potentiels de
force moyenne entre des petites molécules hydrophobes dans l'eau.
Un potentiel de force moyenne (PMF) représente l'évolution de l'énergie
libre d'un système en fonction d'une coordonnée de réaction (CR),
tous les autres degrés de liberté étant moyennés thermodynamiquement.
C'est une grandeur couramment utilisée en chimie numérique, par exemple
pour étudier les interactions entre médicaments et récepteurs, protéines
et membranes cellulaires. Dans un solvant, l'interaction entre solutés
se fait à travers les molécules de solvant et le PMF contient l'effet
du solvant sur cette l'interaction. Les PMF sont aussi utilisés pour
la construction de modèles gros-grains pour les simulations moléculaires\cite{marrink_martini_2007}. 

Pour ces raisons, des méthodes basées sur des simulations MD ou MC
ont été développées pour calculer des PMF. C'est un problème particulièrement
complexe car il nécessite un échantillonnage des configurations du
solvant pour chaque valeur de la coordonnée réactionnelle. Certaines
valeurs de la CR sont particulièrement difficiles à échantillonner
car elles possèdent une énergie d'activation élevée. Il faut donc
biaiser les simulations moléculaires pour forcer l'exploration le
long de la coordonnée réactionnelle\cite{torrie_nonphysical_1977}.
Ces méthodes fonctionnent mais nécessitent des simulations longues
pour avoir une statistique suffisante. 

L'utilisation d'une méthode implicite peut faciliter l'accès aux PMF.
Par exemple, 3D-RISM avec une fermeture HNC ou Kovalenko-Hirata a
été utilisé pour calculer les PMF d'ions polyatomiques dans des solvants
polaires\cite{kovalenko_potential_1999} et d'ions simples dans l'eau\cite{kovalenko_potentials_2000}.
On se propose d'utiliser la MDFT pour calculer les PMF\cite{Stojanovic_2014}
entre deux solutés sphériques. On considérera deux petits solutés
(méthanes), deux gros solutés (néopentanes), deux solutés de tailles
différentes (un méthane et un néopentane). On s'intéresse aussi à
un PMF de torsion dans une molécule (butane). Pour le méthane et le
néopentane, on utilise les modèles unifiés de OPLS/AA. Ces molécules
sont donc représentées par des sphères Lennard-Jones.

\subsection{Fonctionnelle}

Comme les solutés étudiés ici ne sont pas chargés, on utilise une
fonctionnelle où le terme d'excès se limite au terme fonction de $c_{000}(r)$
de l'\ref{eq:Fexc_exacte} et au terme correctif. On a donc une fonctionnelle
de la forme,
\begin{equation}
{\cal F}[n(\bm{r})]={\cal F}_{\mathrm{id}}[n(\bm{r})]+{\cal F}_{\mathrm{ext}}[n(\bm{r})]-\frac{1}{2}\text{k}_{\text{B}}\text{T}\iiint_{\mathbb{R}^{3}}\iiint_{\mathbb{R}^{3}}\left[c_{000}(\bm{r-r^{\prime}})\Delta n(\bm{r})\Delta n(\bm{r}^{\prime})\right]\text{d}\bm{r}\text{d}\bm{r}^{\prime}+{\cal F}_{\mathrm{cor}}[n(\bm{r})].
\end{equation}
Le terme correctif choisi est le bridge de sphères dures décrit à
l'\ref{eq:HSB} car il permet d'obtenir des valeurs des énergies libres
de solvatation correctes pour les solutés neutres. Le terme extérieur
est dû à la présence des deux solutés à une distance fixée.

\subsection{Résultats}

\paragraph{Méthane-méthane:}

\begin{figure}
\noindent \begin{centering}
\includegraphics[width=0.6\textwidth]{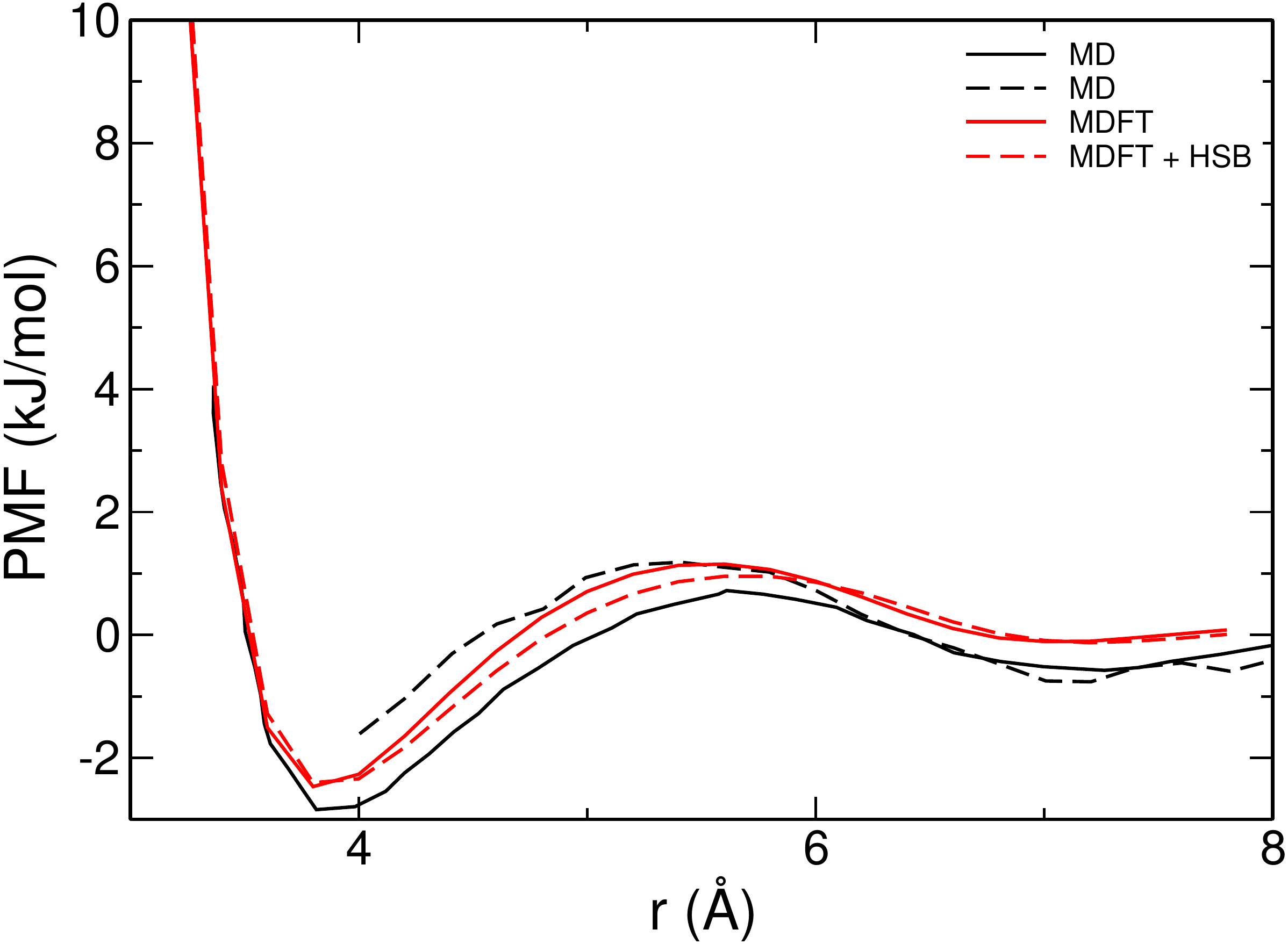}
\par\end{centering}

\protect\caption{PMF méthane-méthane calculé par MDFT en rouge, sans bridge de sphères
dures (traits pleins) et avec bridge de sphères dures (tirets) et
par MD en noir (les résultats de Asthagiri\cite{asthagiri_role_2008}
sont en traits pleins et ceux de Shimizu en tirets\cite{shimizu_temperature_2000}).\label{fig:PMF CH4_CH4}}
\end{figure}

Le modèle de méthane utilisé est celui de Asthagiri et collaborateurs\cite{asthagiri_role_2008}.
On utilise le même modèle d'eau, SPC. Sur la \ref{fig:PMF CH4_CH4},
on présente le PMF obtenu entre deux molécules de méthane par MDFT
et par deux dynamiques moléculaires différentes\cite{asthagiri_role_2008,shimizu_temperature_2000}.
Le PMF obtenu est en accord quantitatif avec ceux obtenus par les
deux simulations. L'ajout du bridge de sphères dures ne modifie pas
sensiblement le résultat, ce qui est attendu pour les petits solutés. 

Les potentiels de force moyenne dans le cas de particules sphériques
présentent généralement deux minima séparés par un maximum. C'est
le cas ici. Le premier minimum, appelé \textit{minimum de contact}
(CM), est généralement plus faible que le second minimum, appelé\textit{
minimum séparé par le solvant} (SSM). Le CM correspond au cas où les
deux solutés sont en contact et ne sont donc pas séparés par des molécules
de solvant. Le SSM correspond à un état métastable où les deux solutés
sont séparés par une couche unique de solvant. Dans le cas du PMF
présenté en \ref{fig:PMF CH4_CH4}, le premier minimum se situe à
$\approx\ 3.9\ \textrm{\AA}$ et correspond à la configuration la
plus stable de deux molécules de méthane dans l'eau. Dans le cas de
particules hydrophobes, le CM est stabilisé pour des raisons entropiques
tandis que le SSM l'est pour des raisons enthalpiques\cite{smith_entropy_1992,huang_molecules_2003}.

Comme l'obtention des PMF entre petits solutés sphériques est possible,
on s'intéresse maintenant à des solutés plus gros.

\paragraph{Néopentane-néopentane:}

Le comportement hydrophobe des molécules de néopentane a fait l'objet
de plusieurs études théoriques\cite{huang_molecules_2003}. Huang
et collaborateurs ont étudié l'argon, le méthane et le néopentane
dans l'eau et ont trouvé que, contrairement aux deux autres solutés,
le néopentane à un comportement similaire aux solutés hydrophobes
de grande taille. Une analyse orientationnelle de la molécule montre
notamment que les molécules d'eau de la première couche de solvatation
ne peuvent former quatre liaisons hydrogènes avec les molécules d'eau
voisines ce qui est typique des gros hydrophobes. 
\begin{figure}
\noindent \begin{centering}
\includegraphics[width=0.6\textwidth]{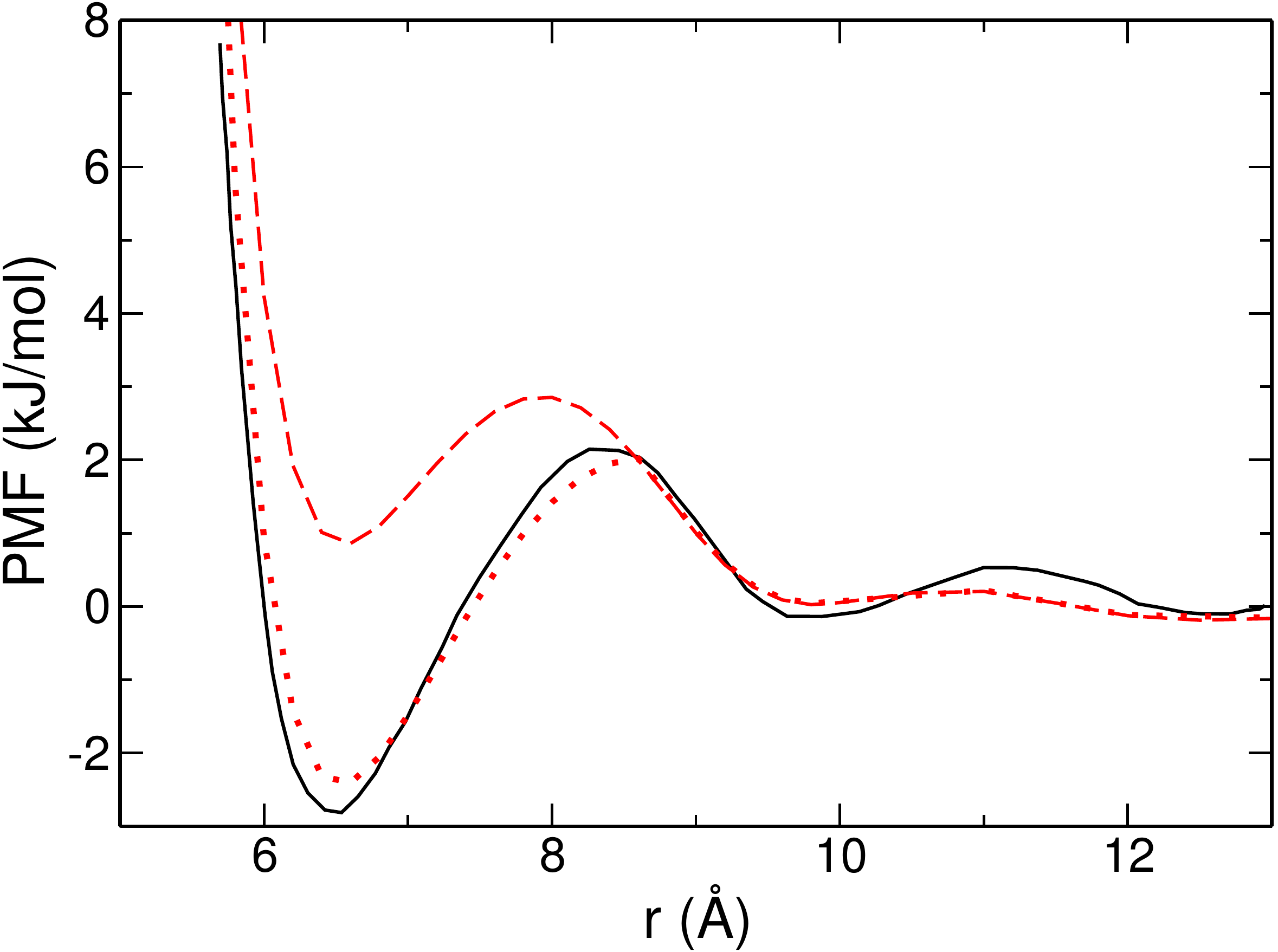}
\par\end{centering}

\protect\caption{PMF néopentane-néopentane calculé par MDFT en rouge, avec bridge de
sphères dures (tirets) et avec la correction hydrophobe à longue distance
(pointillés) et par MD\cite{huang_molecules_2003} en noir.\label{fig:PMF neo_neo}}
\end{figure}
Les potentiels de force moyenne entre deux molécules de néopentane,
représentées par un modèle unifié (une seule sphère Lennard-Jones),
sont présentés sur la \ref{fig:PMF neo_neo}. On voit que les résultats
obtenus par minimisation fonctionnelle avec un bridge de sphères dures
sont très différents des résultats de MD. En effet, les positions
des CM et SSM sont légèrement décalées vers les petites distances
entre néopentane et les valeurs des deux extrema sont surestimées.
Ces résultats mitigés ne sont pas surprenants car la solvatation du
néopentane, qui est un gros soluté, est gouvernée par des effets hydrophobes
à longue portée. Pour améliorer la description du néopentane, on ajoute
la correction hydrophobe à longue portée du \ref{chap:hydro}, ce
qui améliore l'accord entre MD et MDFT. Il est donc possible d'obtenir
les PMF entre deux petits solutés et entre deux gros solutés. Qu'en
est-il des PMF entre solutés de tailles différentes où ces deux échelles
sont mélangées?

\paragraph{Méthane-néopentane :}

\begin{figure}
\noindent \begin{centering}
\includegraphics[width=0.6\textwidth]{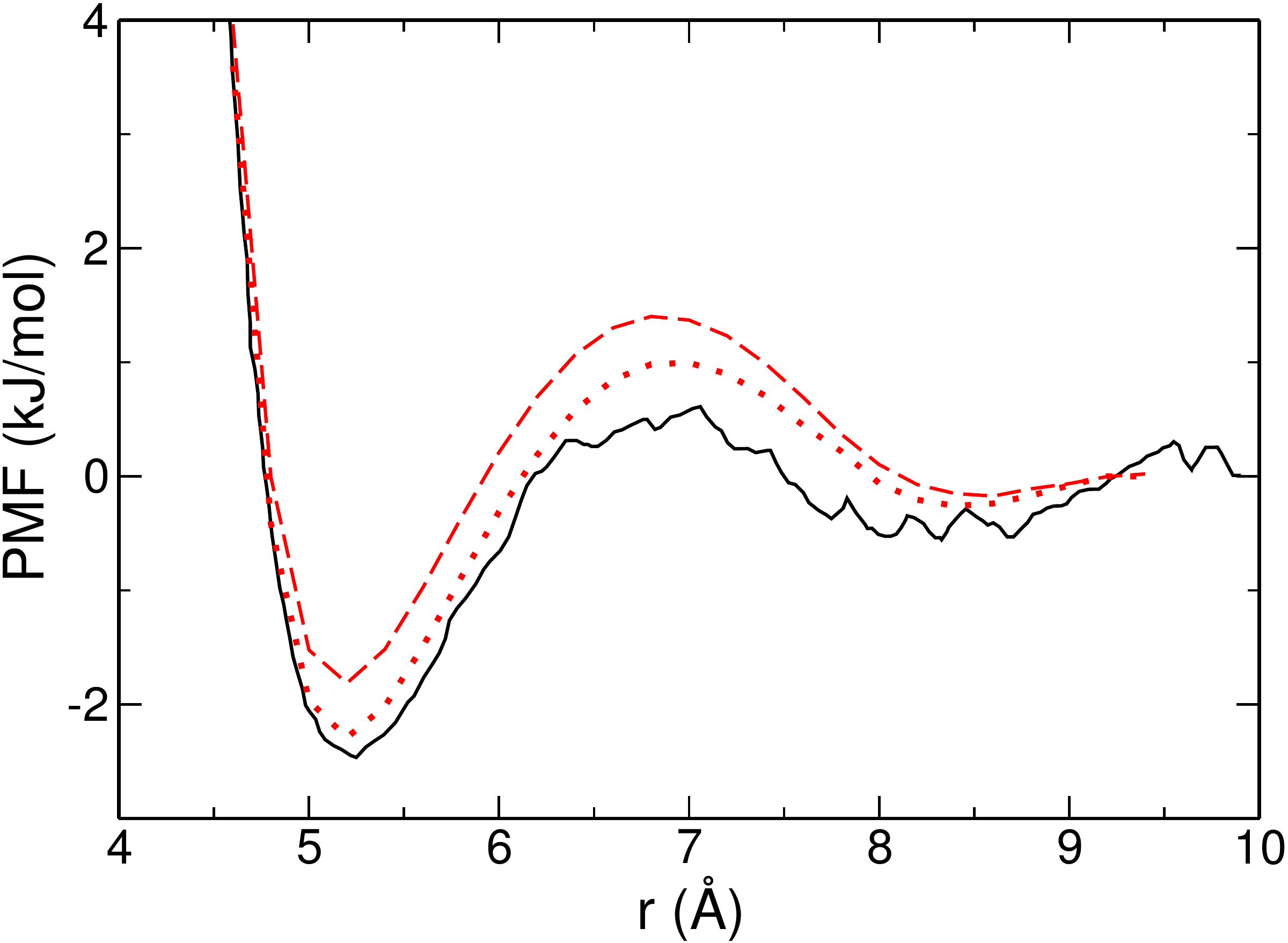}
\par\end{centering}

\protect\caption{PMF méthane-néopentane calculé par MDFT en rouge, avec bridge de sphères
dures (tirets) et avec la correction hydrophobe à longue distance
(pointillés) et par MD\cite{shimizu_origins_2002} en noir .\label{fig:PMF met_neo}}
\end{figure}

La comparaison des PMF entre méthane et néopentane obtenus avec et
sans la correction hydrophobe, présentée en \ref{fig:PMF met_neo},
prouve que l'inclusion de la correction hydrophobe à longue portée
est nécessaire pour reproduire correctement les PMF impliquant des
gros hydrophobes. L'accord entre les résultats obtenus par dynamique
moléculaire\cite{shimizu_origins_2002} et par minimisation fonctionnelle
est encore une fois très bon.

La MDFT permet donc d'obtenir des PMF corrects entre solutés sphériques,
le prochain paragraphe s'intéresse à un PMF intramoléculaire.

\paragraph{Torsion du butane:}

\begin{figure}
\noindent \begin{centering}
\includegraphics[width=0.6\textwidth]{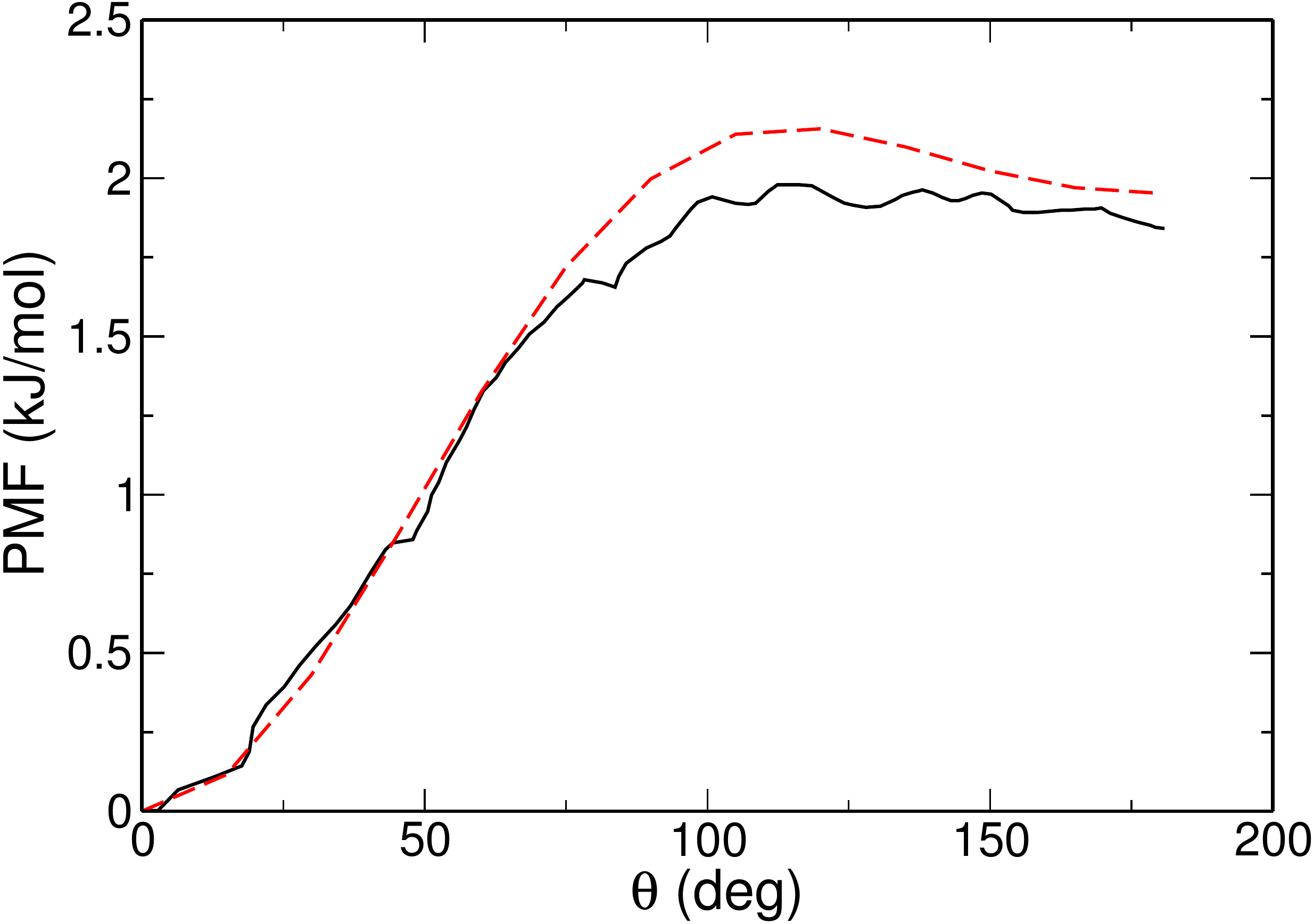}
\par\end{centering}

\protect\caption{PMF selon la torsion de l'angle dièdre du butane, calculé par MDFT
avec bridge sphères dures en tirets rouges et par MD\cite{beglov_finite_1994}
en traits noirs continus.\label{fig:PMF butane}}
\end{figure}

Nous avons calculé le potentiel de force moyenne dû à la torsion d'une
molécule de butane selon son angle dièdre. Cette molécule est représentée
par un modèle à quatre sites. Les paramètres Lennard-Jones utilisés
sont les mêmes que dans une étude de MD calculant ce PMF\cite{beglov_finite_1994}.
Ce sont les paramètres OPLS présentés dans le \ref{tab:ParamOPLSalcane}.
Les longueurs de liaisons ainsi que les angles entre sites Lennard-Jones
sont fixés à leurs valeurs idéales 1.54$\ \textrm{\AA}$ et 111\textdegree .
Les résultats obtenus par minimisation fonctionnelle avec la correction
bridge de sphères dures et par MD \cite{beglov_finite_1994} sont
présentés en \ref{fig:PMF butane}. L'accord entre les deux méthodes
est une fois de plus très bon.

\fbox{\begin{minipage}[t]{1\columnwidth}%
Le calcul de PMF en utilisant la théorie MDFT est possible pour des
molécules neutres. Nous avons pu reproduire très rapidement des PMF
intermoléculaires pour des solutés sphériques et un PMF intramoléculaire.

La rapidité de la méthode permet de réaliser des études systématiques
qui sont hors de portée des techniques de simulations moléculaires.

Ces résultats sont encourageants, et ouvrent la voie à l'utilisation
de la MDFT pour calculer des PMF pour des solutés plus complexes,
notamment des solutés chargés.%
\end{minipage}}

Dans les prochaines parties nous présentons l'étude de solutés complexes
par MDFT.

\clearpage{}

\section{Étude de la solvatation d'une argile}

Les phénomènes aux interfaces jouent un rôle important dans de nombreuses
applications, notamment en catalyse hétérogène et dans les propriétés
d'absorption des matériaux poreux. Ces phénomènes peuvent être étudiés
expérimentalement, par spectroscopie ou diffraction\cite{richmond_molecular_2002,marry_water_2011}.
Ces expériences donnent des renseignements sur les quelques couches
moléculaires de liquide adsorbées à la surface d'un matériaux. L'utilisation
de méthodes numériques peut permettre de clarifier ou de rationaliser
des observations expérimentales en donnant une explication moléculaire
à ces phénomènes. Les méthodes théoriques couramment utilisées reposent
généralement sur des simulations moléculaires (MD ou MC)\cite{rotenberg_molecular_2011}.
L'étude de l'adsorption dans les matériaux poreux est un problème
multi-échelle puisqu'on s'intéresse à l'adsorption de molécules nanométriques
dans des pores de taille mésoscopique. Il est nécessaire de réaliser
des simulations comportant des milliers ou des millions d'atomes pour
étudier ces systèmes, ce qui requiert l'utilisation de supercalculateurs.
Il est donc difficile de réaliser des études systématiques de ces
systèmes ou de systèmes de plus grande taille étant donné les ressources
computationelles nécessaires. L'utilisation de méthodes implicites
est une alternative pour éviter ce coût numérique et énergétique.
La théorie MDFT a le double avantage d'être bien plus rapide et moins
coûteuse numériquement que les simulations moléculaires, tout en conservant
une description moléculaire du solvant. 

On se propose d'utiliser la MDFT dans l'approximation du fluide homogène
de référence pour étudier la solvatation d'une argile, la pyrophyllite.
Les argiles sont des matériaux d'intérêt pour diverses applications.
Par exemple, l'Andra envisage de stocker sur des longues durées des
déchets nucléaires dans les couches géologiques profondes constituées
d'argile du site de Bure. En effet, les argiles sont imperméables,
ce qui empêcherait la diffusion des radionucléotides dans les couches
inférieures si les matériaux d'enrobage des déchets radioactifs venaient
à se fissurer. Les argiles sont des matériaux multi-échelle constitués
d'un empilement de feuillets atomiques séparés par des pores nanométriques.
L'étude de ces pores de petite taille est tout à fait possible par
MD. Les matrices d'argile sont également constituées de pores à l'échelle
mésoscopique dont l'étude n'est pas réalisable par MD. On peut néanmoins
les étudier par des simulations n'ayant pas une description moléculaire,
comme la méthode de Lattice-Boltzmann. Pour avoir une description
complète du matériau, il faudrait pouvoir étudier la solvatation dans
ces pores d'échelles différentes, tout en conservant une description
moléculaire. La MDFT peut permettre d'étudier ces matériaux à l'échelle
mésoscopique, avec une description moléculaire. De plus, il pourrait
être possible de coupler la MDFT avec des modélisations macroscopiques,
par exemple de dynamique des fluides, pour avoir un traitement multi-échelle
complet de ces matériaux.

On utilisera tout d'abord un solvant simple, le fluide de Stockmayer,
pour tester la faisabilité de l'approche\cite{levesque_solvation_2012},
avant d'étudier la solvatation par l'eau\cite{jeanmairet_hydration_2014}
qui est un solvant complexe nécessitant l'utilisation d'une fonctionnelle
adaptée. Les résultats obtenus par MDFT seront dans les deux cas comparés
à des résultats de dynamique moléculaire.

\subsection{Solvatation par le fluide de Stockmayer}

On utilise comme solvant le fluide de Stockmayer constitué d'une sphère
Lennard-Jones centrale et de deux sites portant une charge de $\pm1.91\ \mathrm{e}$
situés à une distance de $0.1\ \textrm{\AA}$ de part et d'autre de
ce site, résultant en un dipôle de 1.85 D. Les paramètres de Lennard-Jones
donnés en \ref{tab:param-pyro} et le dipôle ont été ajustés pour
que le fluide mime certaines propriétés de l'eau (rayon moléculaire,
dipôle, constante diélectrique). Cependant, ce solvant ne permet pas
de reproduire toutes les propriétés de l'eau, comme la polarisation
au-delà de l'ordre dipolaire, la création de liaisons hydrogènes intermoléculaires
et par conséquent la structure locale tétraédrique de l'eau.

\subsubsection{Description du système}

\begin{figure}
\noindent \begin{centering}
\includegraphics[width=0.8\textwidth]{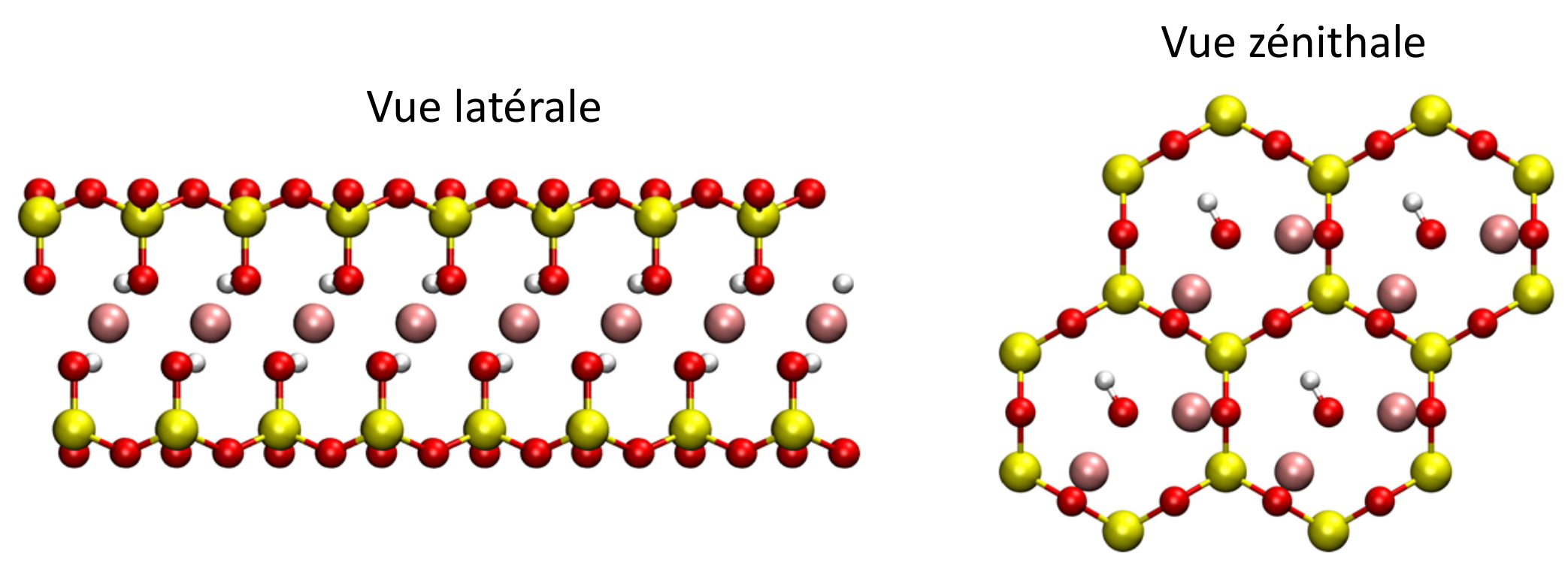}
\par\end{centering}

\protect\caption{Vues latérale et zénithale des feuillets de pyrophyllite. Les atomes
de silicium sont en jaune, les atomes d'aluminium en rose, les oxygènes
en rouge et les hydrogènes en blanc.\label{fig:Vues-lat=0000E9rales-et}}
\end{figure}

La pyrophyllite est une argile neutre qui cristallise dans le groupe
monoclinique 2/m. Elle est clivée selon le plan \{011\}. Elle est
constituée d'un empilement de neufs couches atomiques. La couche supérieure
est constituée d'atomes d'oxygène (O) qui se trouvent juste au-dessus
d'une couche d'atomes de silicium (Si). Ces deux couches forment une
structure hexagonale en nids d'abeilles. La couche centrale est composée
d'atomes d'aluminium (Al). Entre cette couche d'aluminium centrale
et les deux couches Si-O se trouvent deux couches constituées de groupements
hydroxyles, dont l'oxygène se trouve au centre des hexagones formés
par les deux couches supérieures. Les vues latérale et zénithale des
feuillets d'argiles sont données en \ref{fig:Vues-lat=0000E9rales-et}.
Les coordonnées des couches atomiques selon l'axe $\mathrm{O_{z}}$
perpendiculaire aux feuillets sont données dans le \ref{tab:coord pyro},
la couche d'aluminium centrale a été choisie comme l'origine de cet
axe. 

\begin{table}
\noindent \centering{}%
\begin{tabular}{|c|c|c|c|c|c|c|c|c|c|}
\hline 
Atome & O & Si & O & H & Al & H & O & Si & O\tabularnewline
\hline 
\hline 
z ($\textrm{\AA}$) & 39.03 & 39.62 & 41.21 & 41.30 & 0.00 & 1.00 & 1.09 & 2.68 & 3.27\tabularnewline
\hline 
\end{tabular}\protect\caption{Coordonnées des couches atomiques de l'argile pyrophyllite selon l'axe
$\mathrm{O_{z}}$ perpendiculaire au plan de l'argile.\label{tab:coord pyro}}
\end{table}

\subsubsection{Aspects numériques}

\paragraph{Champ de force:}

À la fois dans les simulations de dynamique moléculaire et dans les
calculs MDFT, les atomes de la surface interagissent avec le solvant
au travers du champ de force CLAYFF\cite{CLAYFF_cygan_molecular_2004},
constitué de sites Lennard-Jones et de charges ponctuelles. C'est
un champ de force générique pour les systèmes minéraux et leurs interfaces
avec les solutions. Les paramètres de champ de force pour l'argile
et le solvant sont donnés dans le \ref{tab:param-pyro}. La boîte
de simulation contient deux demi-couches d'argile de surface $\mathrm{L_{x}\times L_{y}}=41.44\times35.88\ \textrm{\AA}^{2}$.
La distance entre deux surfaces est de $\mathrm{L_{z}}=45.57\ \textrm{\AA}$.

\begin{table}
\noindent \centering{}%
\begin{tabular}{|c|c|c|c|c|}
\hline 
Molécule & Atome & $\epsilon\ \mathrm{(kJ.mol^{-1})}$ & $\sigma\ $($\textrm{\AA}$) & q (e)\tabularnewline
\hline 
Pyrophyllite & Al & 5.56388e-6 & 4.27120 & 1.575\tabularnewline
\hline 
 & Si & 7.7005e-6 & 3.30203 & 2.100\tabularnewline
\hline 
 & O$\mathrm{_{G}}$ & 0.650190 & 3.16554 & -1.050\tabularnewline
\hline 
 & O$\mathrm{_{H}}$ & 0.650190 & 3.16554 & -0.950\tabularnewline
\hline 
 & H$\mathrm{_{G}}$ & 0.0 & 0.0 & 0.425\tabularnewline
\hline 
Stockmayer & Central & 1.847 & 3.024 & 0.0\tabularnewline
\hline 
 & Site 1 & 0.0 & 0.0 & 1.91\tabularnewline
\hline 
 & Site 2 & 0.0 & 0.0 & -1.91\tabularnewline
\hline 
\end{tabular}\protect\caption{Champ de force utilisé pour le soluté pyrophyllite et le fluide de
Stockmayer dans les simulations de dynamique moléculaire et les calculs
MDFT. Les paramètres de la pyrophyllite sont tirés du champ de force
CLAYFF.\label{tab:param-pyro}}
\end{table}

\paragraph{Dynamique moléculaire :}

Les simulations de dynamique moléculaire ont été réalisées par Virginie
Marry au laboratoire Phenix de l'UPMC. Les simulations sont menées
dans l'ensemble canonique avec un thermostat de Nosé-Hoover. Après
une étape d'équilibration, les simulations ont été menées sur une
trajectoire de 5~ns. La densité de solvant locale $n(\bm{r})$ est
calculée par moyenne temporelle dans des voxels de volume $0.1^{3}\ \textrm{\AA}^{3}$.
Les densités dans la direction perpendiculaire au plan des feuillets
sont calculées à partir de la densité moyenne précédente que l'on
moyenne dans les plans $(x,y)$ parallèles aux feuillets. Les simulations
de dynamique moléculaire ont été réalisées avec le logiciel DLPOLY\cite{DL_POLY_todorov_dl_poly_3,DL_POLY_url}.

\paragraph{MDFT :}

Comme le solvant est strictement dipolaire, on utilisera la MDFT avec
une fonctionnelle d'excès dipolaire, telle qu'elle a été présentée
dans la \ref{sec:Fexcdip}. La MDFT est environ mille fois plus rapide
que la MD.

\subsubsection{Résultats}

On compare d'abord les profils de densité et d'orientation obtenus
par MDFT aux résultats de MD. On analyse ensuite l'influence des interactions
électrostatiques sur ces grandeurs.

\paragraph{Profil de densité :}

\begin{figure}
\noindent \begin{centering}
\includegraphics[width=0.6\textwidth]{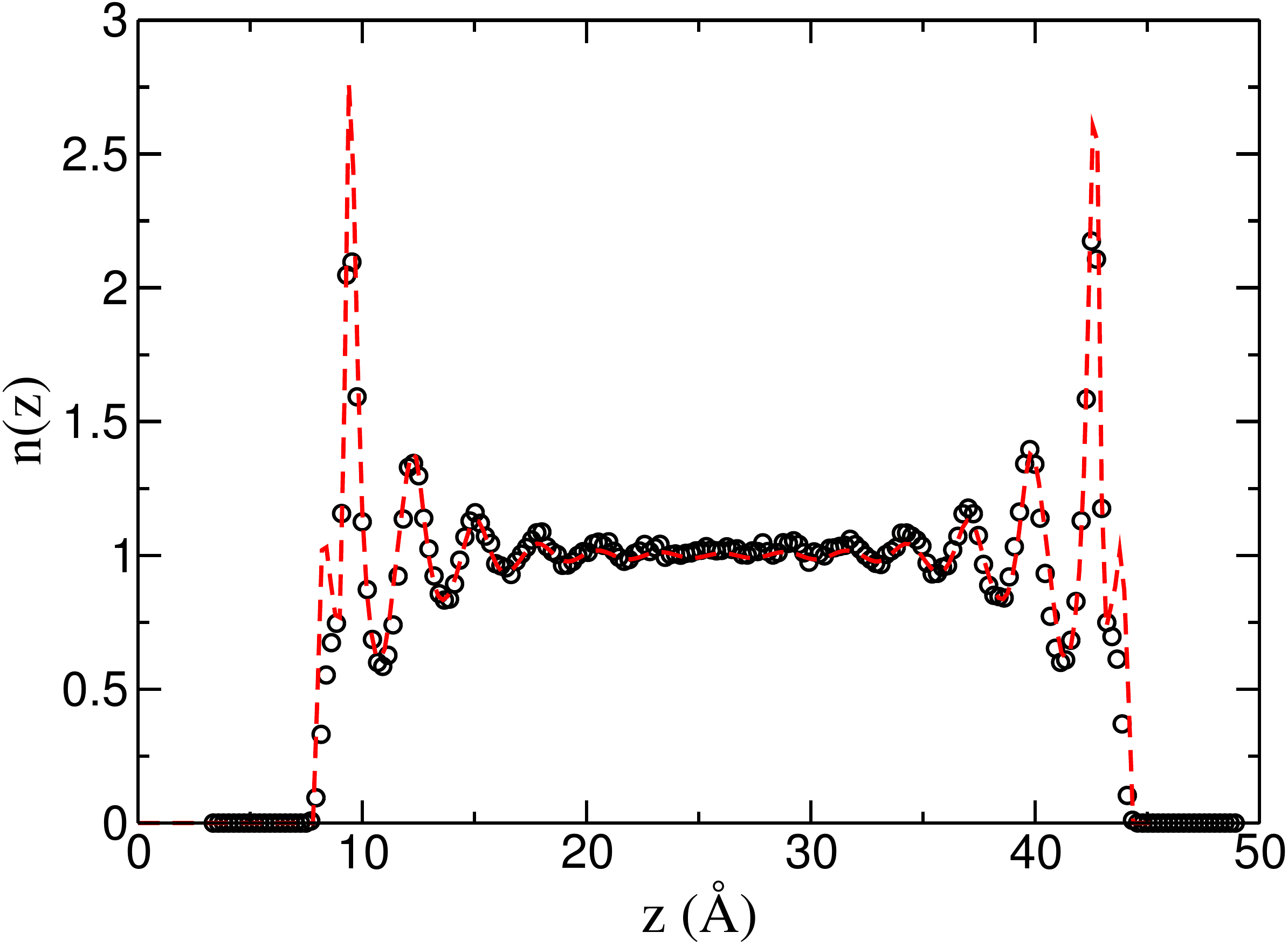}
\par\end{centering}

\protect\caption{Profils de la densité planaire selon l'axe perpendiculaire au plan
de l'argile. Les résultats calculés par dynamique moléculaire sont
les cercles noirs. Ceux calculés par MDFT sont en tirets rouges.\label{fig:densit=0000E9_plan=0000B0_stock}}
\end{figure}

On analyse l'interaction du solvant avec l'interface grâce à la densité
planaire normalisée selon l'axe z, définie comme,
\begin{equation}
n(\mathrm{z})=\int_{x=0}^{\mathrm{L_{x}}}\int_{y=0}^{\mathrm{L_{y}}}\frac{n\text{(\ensuremath{\bm{r}})}}{n_{b}}\mathrm{d}x\mathrm{d}y.\label{eq:densit=0000E9_planaire}
\end{equation}
Les profils de densité obtenus par dynamique moléculaire et par minimisation
fonctionnelle sont donnés en \ref{fig:densit=0000E9_plan=0000B0_stock}.
 On trouve un bon accord global entre les deux méthodes. En MD, le
pic principal observé présente un épaulement alors qu'en MDFT on observe
un \og prépic \fg{} au lieu de cet épaulement. Le premier pic est
suivi par un second pic plus faible et par des oscillations à longue
distance. L'épaulement ou \og prépic \fg{} se situe à z=$8.3\ \textrm{\AA}$
c'est-à-dire à seulement $1.76\ \textrm{\AA}$ de la couche de surface,
son intensité est de $\approx$0.7 en MD et de $\approx$1 en MDFT.
Le pic principal, qu'on appelle premier pic, se situe à z=$9.5\ \textrm{\AA}$
soit à $2.96\ \textrm{\AA}$ de la surface. Le second pic se trouve
à $12.3\ \textrm{\AA}$, soit à $5.76\ \textrm{\AA}$ de la surface.
Au centre de la cellule on retrouve une densité égale à celle de l'eau
bulk, $n=n_{b}$. Le solvant se trouve alors dans les mêmes conditions
que dans la phase homogène. 

\begin{figure}
\noindent \begin{centering}
\includegraphics[width=0.8\textwidth]{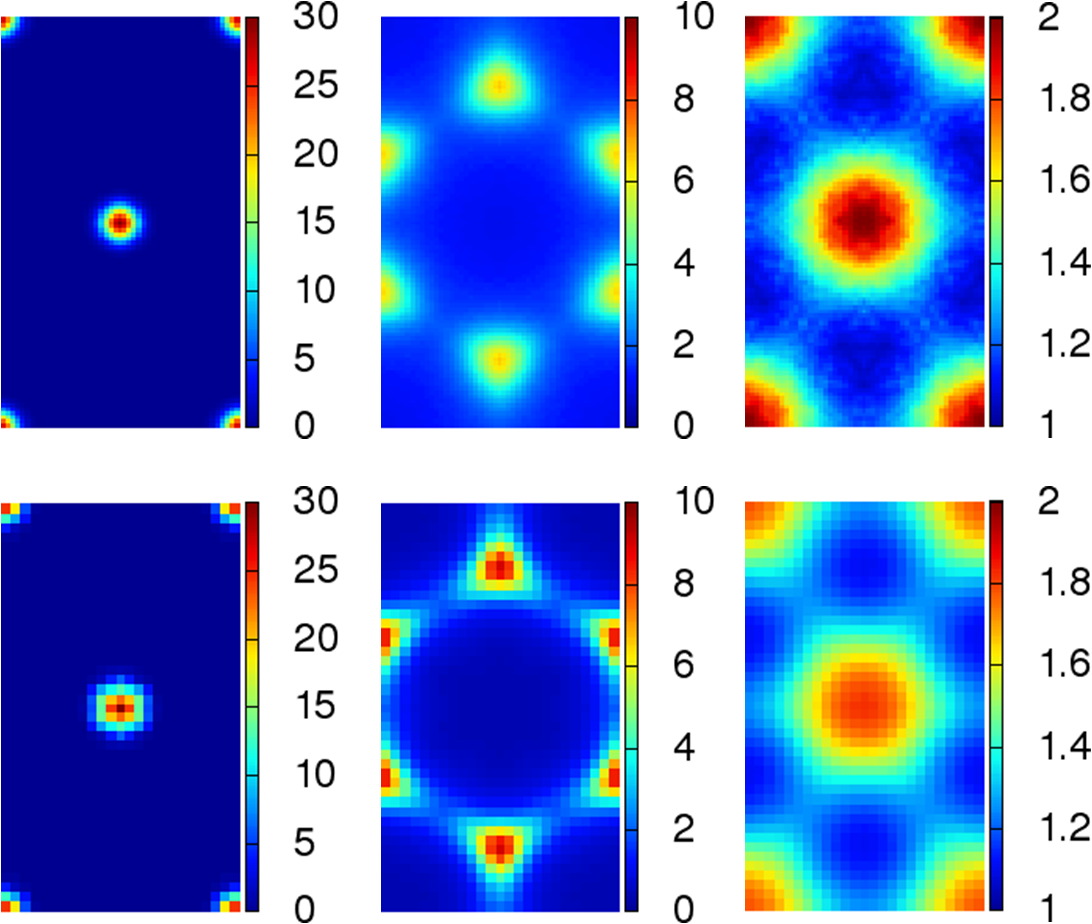}
\par\end{centering}

\protect\caption{Cartes de densité normalisée, $n(\bm{r})/n_{b}$, dans les plans du
prépic (gauche), du pic principal (centre) et du second pic (droite),
calculées par MD en haut et par MDFT en bas.\label{fig:densit=0000E9_maps_plan_stock}}
\end{figure}

La seule différence dans les résultats obtenus par les deux méthodes
réside donc dans l'intensité du prépic, qui est plus marquée pour
la MDFT. Ce prépic est très localisé. Cette observation est confirmée
par la \ref{fig:densit=0000E9_maps_plan_stock} où sont présentées
les cartes de densité, calculées par MD et par MDFT. Ces cartes présentent
les densités dans les plans parallèles au plan de l'argile situés
à une distance correspondant aux maxima de densité planaire de la
\ref{fig:densit=0000E9_plan=0000B0_stock}. 

On obtient des cartes globalement similaires avec les deux méthodes.
Pour MDFT, dans le plan du prépic, le pic de densité est légèrement
plus large, avec une intensité similaire. Ceci explique la plus grande
valeur lorsque l'on fait la moyenne dans le plan. Le prépic se trouve
au centre des hexagones Si-O. On peut souligner la valeur élevée de
la densité à cet endroit ($n/n_{b}\approx3\text{0}$). L'intégrale
de ce pic correspond à la présence d'une molécule unique dans cette
zone. 

Sur la \ref{fig:densit=0000E9_maps_plan_stock} sont également présentées
les cartes de densité dans les plans du premier et du second pic.
Là encore, les cartes obtenues par les deux méthodes sont très similaires.
Le premier pic est localisé essentiellement au-dessus des atomes de
silicium alors que l'on observe une déplétion au-dessus des atomes
d'oxygène. Le second pic est globalement localisé au centre des hexagones.

On a désormais une idée claire de la structure tridimensionnelle du
solvant à la surface de la pyrophyllite. Quelques molécules de solvant
sont adsorbées très près de la surface, au centre des hexagones. Au-dessus
de ces molécules se trouve une couche très structurée de molécules
de solvant localisées au-dessus des atomes de Si. Enfin, une troisième
couche de molécules se trouve au centre des hexagones avec une structure
moins bien définie. La distance relativement courte entre les couches
est signe d'une interaction et donc d'une cohésion importante entre
ces couches. Cette conclusion est en accord avec Rotenberg et collaborateurs
qui ont montré récemment que la compétition entre adhésion et cohésion
détermine l'hydrophobicité de ces surfaces\cite{rotenberg_molecular_2011}.

\paragraph{Propriétés orientationnelles: }

On s'intéresse désormais aux propriétés orientationnelles des molécules
solvatées à la surface de la pyrophylitte. Même si cela est fait de
manière routinière, on peut souligner que les propriétés orientationnelles
sont particulièrement lentes à sonder par des simulations moléculaires,
MD ou MC. En effet, chaque élément de volume doit être échantillonné
pour un grand nombre d'angles. La MDFT donne directement accès à ces
propriétés puisque la densité spatialement et angulairement dépendante
$\rho(\bm{r},\bm{\Omega})$ et la densité de polarisation $\bm{P}(\bm{r})$
définie dans l'\ref{eq:Pola_dip} sont les variables sur lesquelles
est conduite la minimisation.

On constate, avec les deux méthodes utilisées, que la densité de polarisation
est alignée selon l'axe Oz.
\begin{figure}
\noindent \begin{centering}
\includegraphics[width=0.6\textwidth]{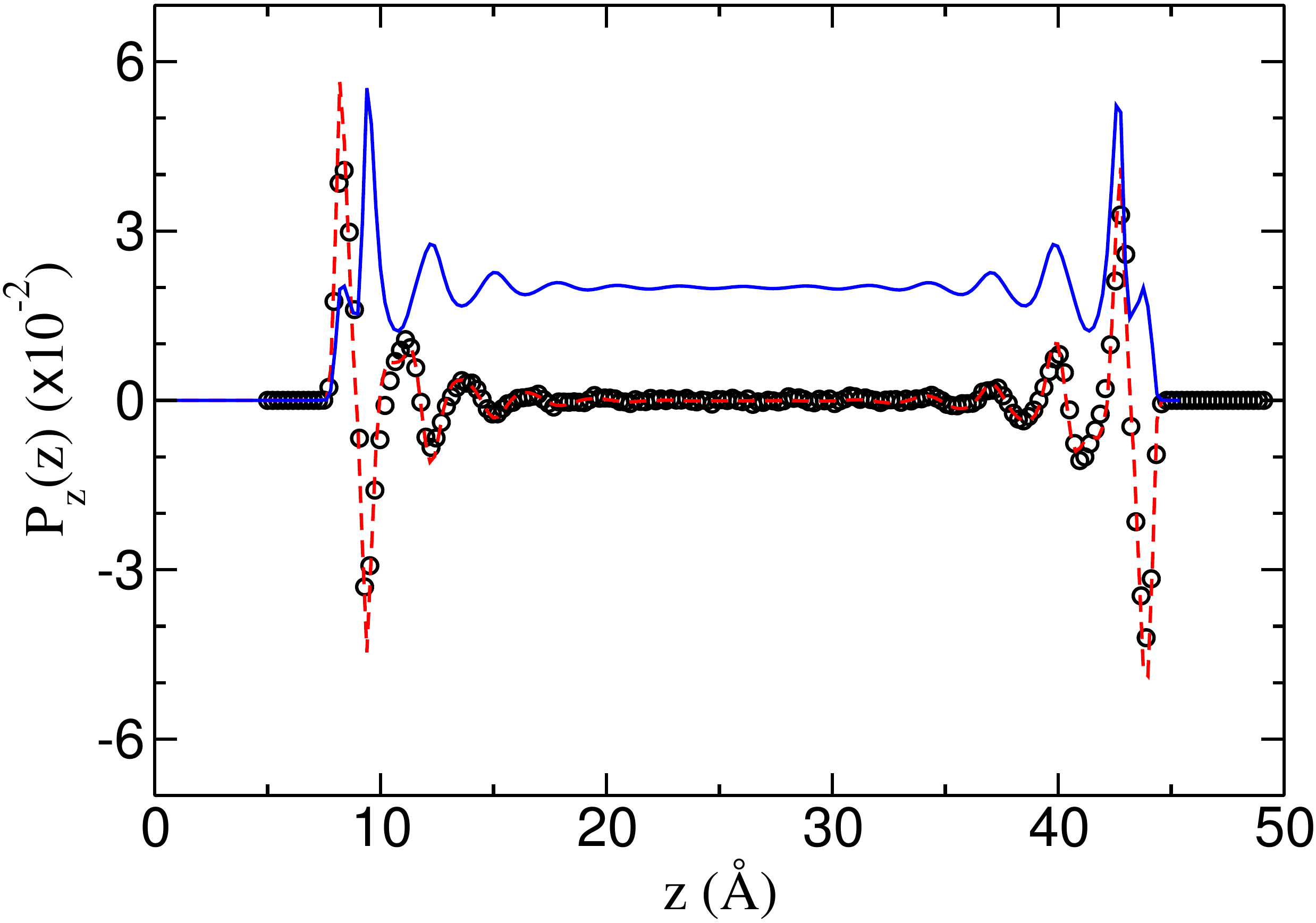}
\par\end{centering}

\protect\caption{Composante normale au plan de l'argile de la polarisation adimensionnée,
entre deux couches de pyrophyllite calculée par MD (cercles noirs)
et MDFT (tirets rouges). On rappelle le profil de densité en trait
bleu pour repère. Les maxima de polarisation correspondent à des maxima
de densité.\label{fig:pola_plan=0000B0_stock}}
\end{figure}
Sur la \ref{fig:pola_plan=0000B0_stock} est présentée la projection
$P_{\mathrm{z}}$ de la polarisation $\bm{P}$ selon l'axe Oz en fonction
de la cordonnée z (c'est-à-dire moyennée dans le plan ($x,y$)). Là
encore, on trouve un accord quantitatif entre MD et MDFT. Un maxima
est trouvé là où se trouve le prépic de densité, un second là où se
trouve le premier pic. Entre ces deux maxima, le signe de $P_{z}$
change.
\begin{figure}
\noindent \begin{centering}
\includegraphics[width=0.8\textwidth]{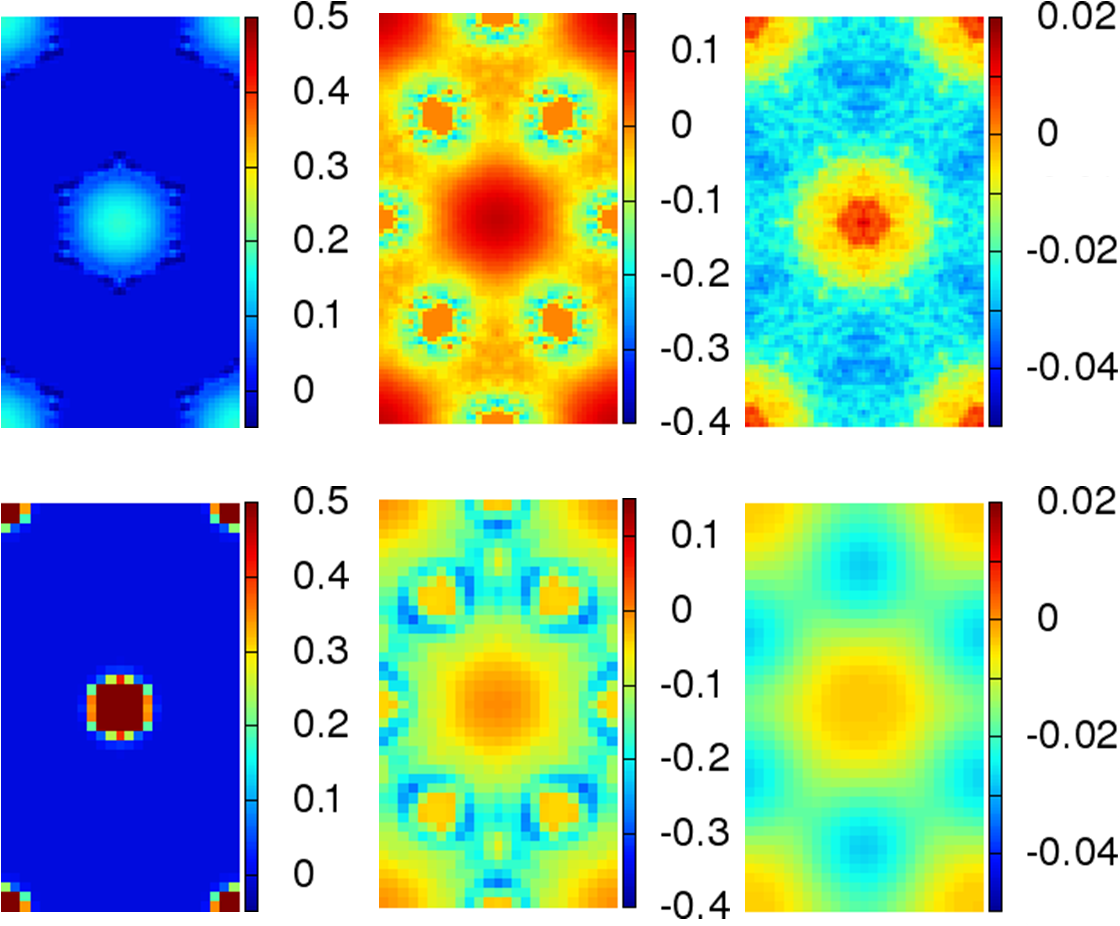}
\par\end{centering}

\protect\caption{Cartes de la polarisation $P_{z}$ dans les plans du prépic (gauche),
du pic principal (centre) et du second pic (droite), calculées par
MD en haut et par MDFT en bas.\label{fig:pola_pmap_stock-1}}
\end{figure}

\begin{figure}
\noindent \begin{centering}
\includegraphics[width=0.6\textwidth]{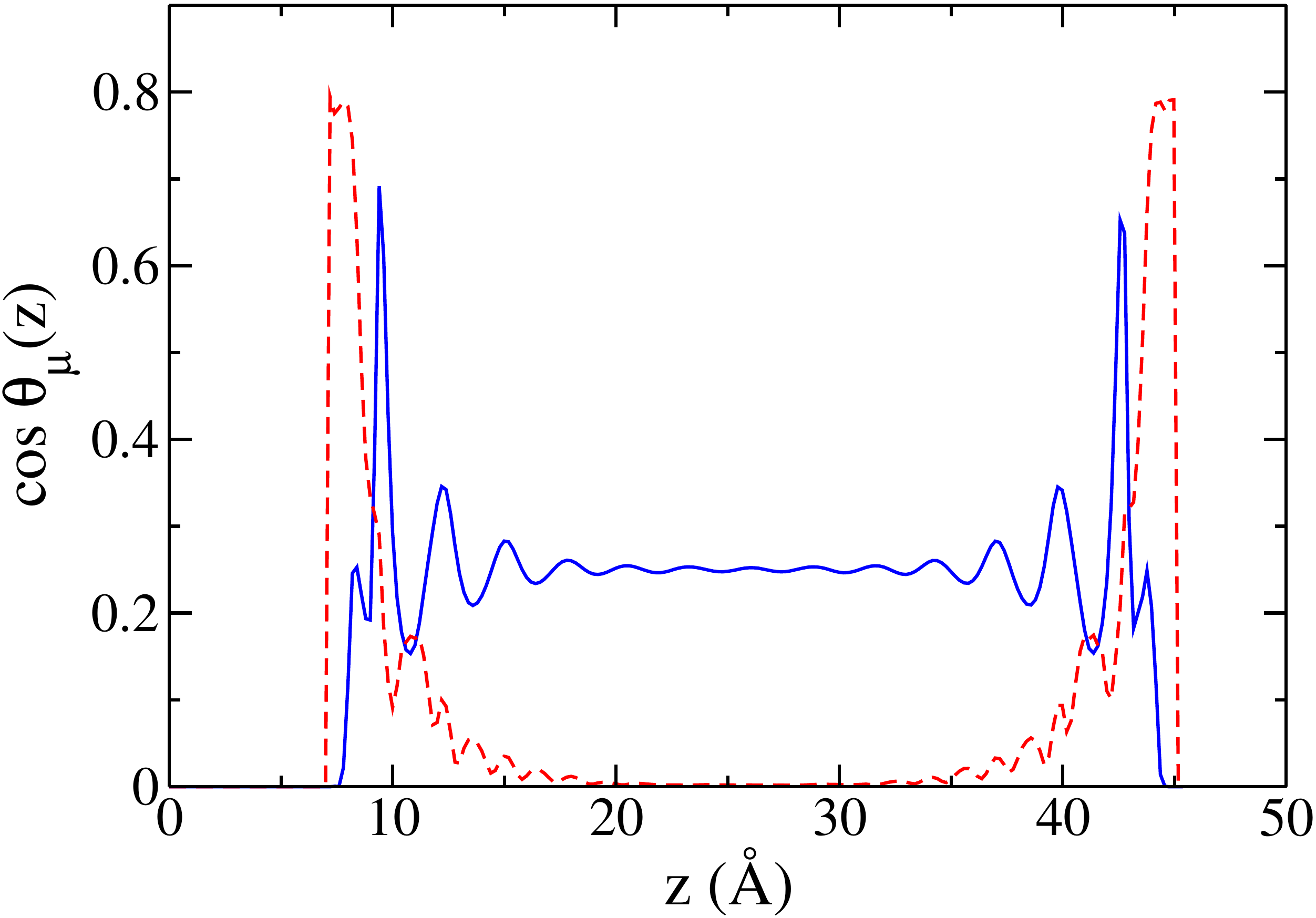}
\par\end{centering}

\protect\caption{Profil de l'orientation moléculaire moyenne $\cos(\theta_{\mu}(z))$
selon la normale au plan de l'argile obtenu par MDFT en tirets rouges.
Comme sur la \ref{fig:pola_pmap_stock-1}, on a rappelé le profil
de densité en bleu. \label{fig:profil_orientat_stoc}}
\end{figure}

\begin{figure}
\noindent \begin{centering}
\includegraphics[width=0.8\textwidth]{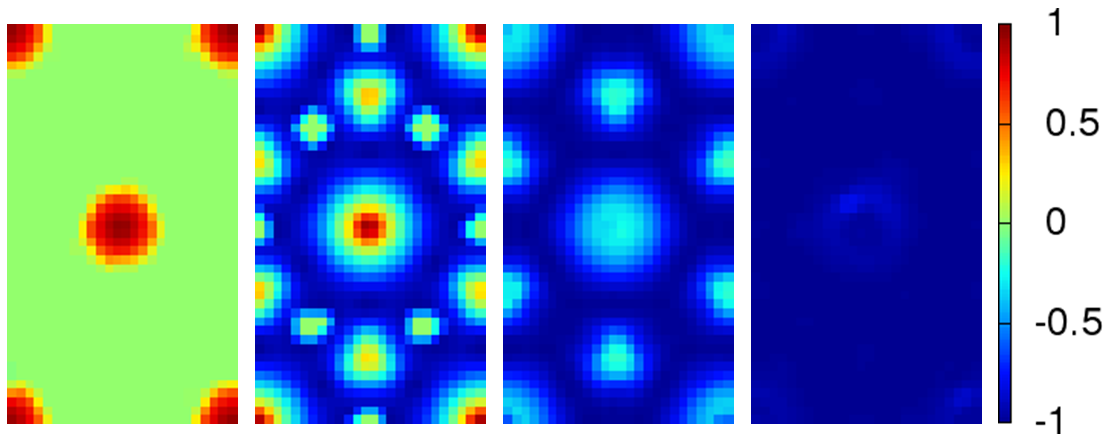}
\par\end{centering}

\protect\caption{De gauche à droite, paramètre d'ordre $\cos(\theta_{P}(\bm{r})$ dans
le plan du prépic, dans le plan médian entre prépic et premier pic,
dans le plan du premier pic et celui du second pic. \label{fig:map_orienta_stock}}
\end{figure}

On présente sur la \ref{fig:pola_pmap_stock-1} les cartes de la composante
selon z de la polarisation dans les plans des trois premiers pics
de densité. La densité de polarisation est surestimée par MDFT dans
le prépic. Comme l'axe $\mathrm{O_{z}}$ est toujours orienté vers
l'extérieur de la surface, la projection $P_{z}$ est positive quand
le dipôle de l'eau, qui est orienté de l'oxygène vers le centre de
masse des hydrogènes, pointe vers la surface. Dans le premier et le
second pic, la polarisation est trouvée légèrement plus grande par
MD. Pour essayer d'obtenir plus d'informations sur les propriétés
orientationnelles des molécules, on définit deux paramètres d'ordre.
Le premier est l'orientation locale de la polarisation définie par
le cosinus de l'angle entre le champ de polarisation et la normale
à la surface,
\begin{equation}
\cos(\theta_{P}(\bm{r}))=\frac{\bm{P}(\bm{r})\cdot\bm{z}}{\left\Vert \bm{P}(\bm{r})\right\Vert }=\frac{P_{\mathrm{z}}(\bm{r})}{\left\Vert \bm{P}(\bm{r})\right\Vert }.\label{eq:CosThetaP}
\end{equation}
 Le second est l'orientation moléculaire moyenne, 
\begin{equation}
\cos(\theta_{\mu}(\bm{r}))=\frac{\left\Vert P_{z}(\bm{r})\right\Vert }{n(\bm{r})}.\label{eq:costheta_mu}
\end{equation}
Quand $\cos(\theta_{P}(\bm{r}))$ est positif (resp. négatif) le dipôle
pointe vers l'extérieur (resp. l'intérieur) de la surface. On représente
sur la \ref{fig:profil_orientat_stoc} la valeur moyenne dans le plan
$(x,y)$ de l'orientation moléculaire moyenne $\cos(\theta_{\mu}(z))=\left\langle \cos(\theta_{\mu}(\bm{r}))\right\rangle _{x,y}$,
obtenue par MDFT.
 Les molécules de solvant sont fortement orientées dans le plan du
prépic, avec une valeur élevée de l'orientation moyenne. Cette orientation
préférentielle est moins marquée dans le plan du premier pic de solvatation.
Dans le plan du second, le pic de polarisation correspond à un grand
nombre de molécules ayant une faible orientation préférentielle selon
l'axe $\mathrm{Oz}$. On peut confirmer cette analyse en regardant
des cartes d'orientation locale obtenue par MDFT dans le plan des
trois pics, définie par l'\ref{eq:CosThetaP}. Ces cartes sont présentées
sur la \ref{fig:map_orienta_stock}. 
On observe que les molécules dans le plan du prépic sont très orientées,
et les dipôles, orientés du site négatif vers le site positif, pointent
vers l'extérieur de la surface. L'orientation est plus faible dans
le pic principal et les molécules situées au sommet des atomes de
silicium sont orientées le dipôle pointant vers l'intérieur de la
surface. Dans le plan du second pic, on trouve une orientation préférentielle
faible avec un dipôle pointant vers l'extérieur de la surface.

On a maintenant une vision claire des propriétés de solvatation de
l'argile par le fluide de Stockmayer. Dans le plan du prépic, une
molécule de solvant se trouve au centre des hexagones, avec un dipôle
figé pointant vers l'extérieur de la surface. Un peu au-dessus, dans
le plan du premier pic, les molécules de solvant sont essentiellement
situées au dessus des atomes de silicium, avec une orientation préférentielle
moins marquée des dipôles vers l'extérieur de la surface. Enfin, la
seconde couche de solvatation est globalement située au centre des
hexagones avec une faible orientation des dipôles vers l'intérieur
de la surface.

\paragraph{Rôle de l'électrostatique:}

\begin{figure}
\noindent \begin{centering}
\includegraphics[width=0.6\textwidth]{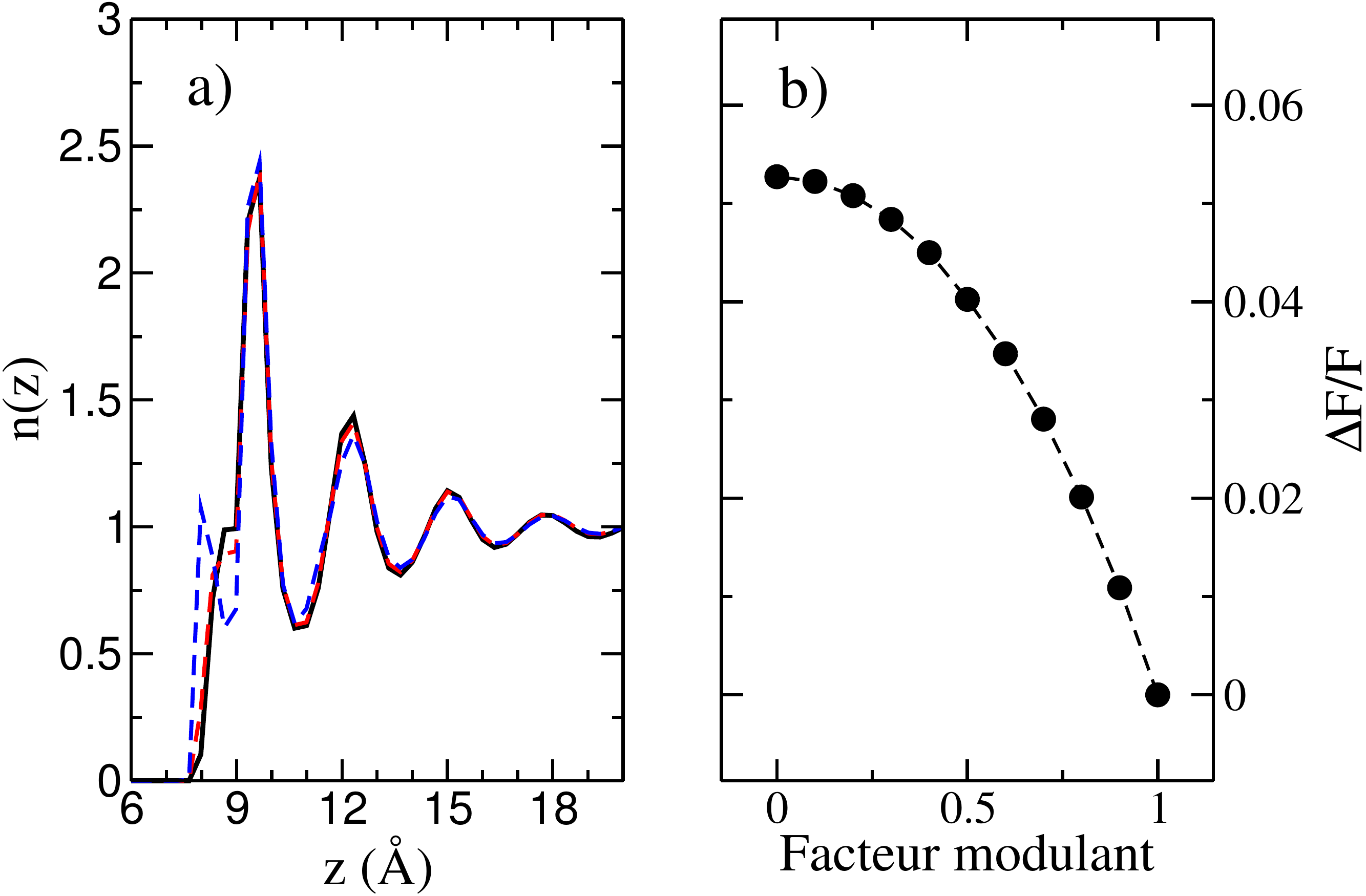}
\par\end{centering}

\protect\caption{À gauche: densités moyennes le long de l'axe perpendiculaire pour
des charges correspondant à celles du champ de force CLAYFF (bleu),
à moitié éteintes (rouge) et complètement éteintes (noir). À droite,
le changement relatif de l'énergie libre de solvatation en fonction
du facteur modulant les charges, à 0 les charges sont éteintes. Ce
changement relatif reste inférieur à 6\%.\label{fig:electsotsta_Stoc}}
\end{figure}

L'efficacité numérique de la MDFT permet une étude systématique de
la structure du solvant et du rôle des interactions de Van-der-Waals
et électrostatique dans le champ de force CLAYFF, qui serait trop
coûteuse par MD. Ces interactions sont modélisées par des interactions
Lennard-Jones et électrostatique dont les paramètres sont données
dans le \ref{tab:param-pyro}. On reporte les profils de densité moyenne
le long l'axe perpendiculaire à la surface sur la \ref{fig:electsotsta_Stoc},
pour des charges modulées par un facteur variant de 0 (charges nulles)
à 1 (charges habituelles du champ de force). 
Sans surprise, seule la forme du prépic, où les molécules sont les
plus fortement polarisées, est modifiée quand on supprime les charges.
Sur la \ref{fig:electsotsta_Stoc} on présente les changements relatifs
de l'énergie libre de solvatation en fonction du facteur modulant
les charges. L'énergie libre de solvatation n'est affectée que de
6\% quand on éteint les charges. La contribution principale à l'énergie
libre de solvatation provient donc des forces de Van-der-Waals. Ce
résultat n'est pas évident à prévoir, même pour une surface globalement
non chargée.

\fbox{\begin{minipage}[t]{1\columnwidth}%
On a montré que la MDFT est une méthode viable pour étudier des systèmes
complexes constitués de plusieurs milliers d'atomes comme les argiles.
Elle permet de reproduire quantitativement les prédictions MD concernant
les propriétés structurelles et orientationnelles du solvant, tout
en réduisant le temps CPU de trois ordres de grandeur. La MDFT permet
une étude systématique de ces systèmes. En particulier, il devient
possible d'étudier l'influence des paramètres de champ de force sur
les propriétés de solvatation. Les profils de densité et les énergies
libres de solvatation sont ici peu sensibles à la valeur des charges
de l'argile. Le développement de champ de force pourrait ainsi être
grandement facilité par l'étude rapide de l'influence des paramètres. %
\end{minipage}}

\subsection{Solvatation par l'eau}

Nous avons étudié la solvatation de la même argile, la pyrophyllite,
avec le solvant eau SPC/E. On utilise donc la fonctionnelle dont la
partie d'excès contient un traitement multipolaire de la polarisation
décrite dans la \ref{sec:Fexcmulti}. Pour une grille spatiale avec
un maillage comportant 64 points par $\textrm{\AA}^{3}$, et 30 points
pour décrire les orientations, la convergence est atteinte en une
quinzaine d'itérations. Le calcul dure environ 25 minutes. C'est environ
trois ordres de grandeur plus rapide que la dynamique moléculaire.

\subsubsection{Résultats}

\paragraph{Profils de densité :}

\begin{figure}
\noindent \begin{centering}
\includegraphics[width=0.6\textwidth]{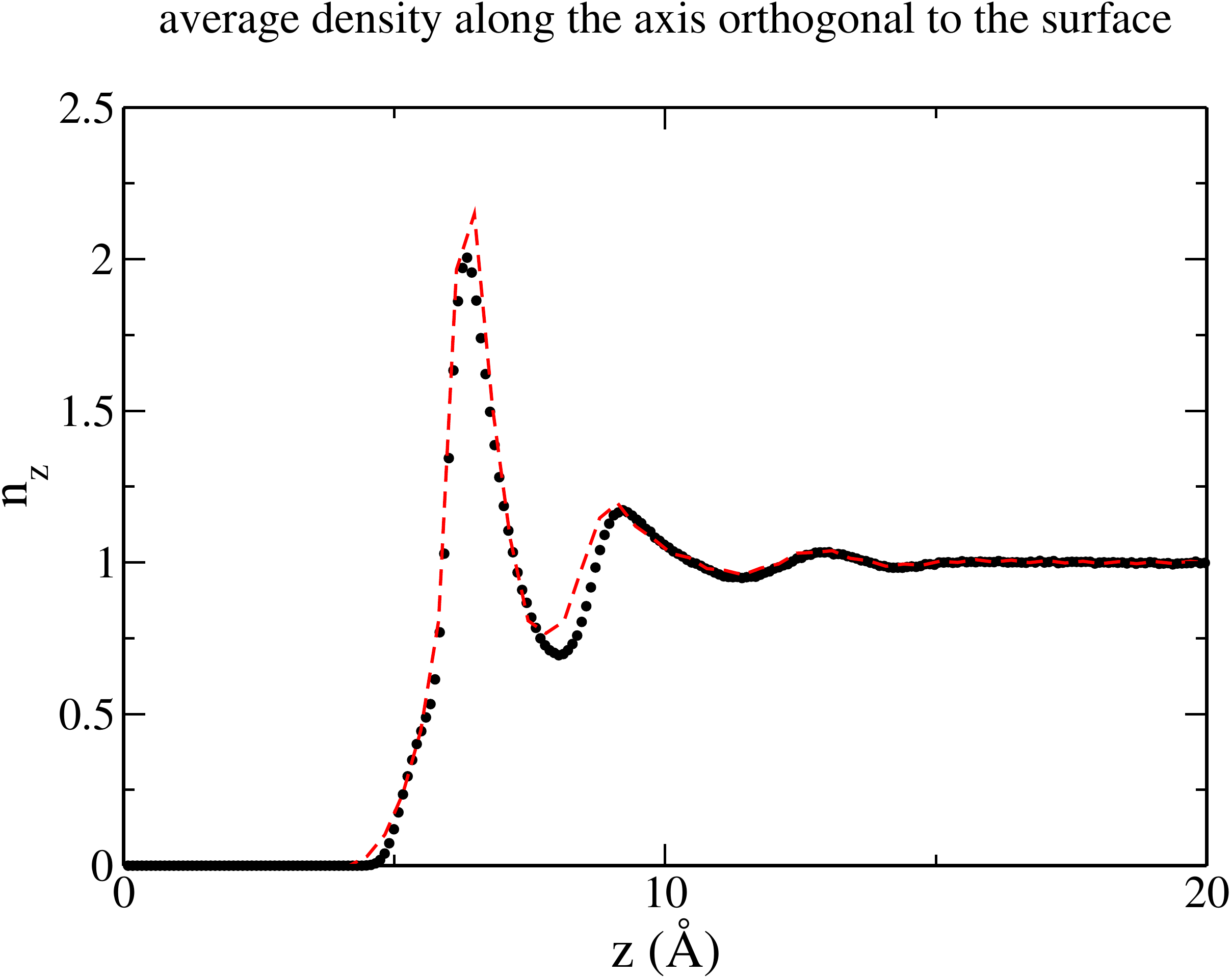}
\par\end{centering}

\protect\caption{Profil de la densité planaire selon l'axe perpendiculaire aux feuillets,
définie à l'\ref{eq:densit=0000E9_planaire}. Les résultats calculés
par dynamique moléculaire sont les cercles noirs. Ceux calculés par
MDFT sont en tirets rouges.\label{fig:densit=0000E9_plan=0000B0_water}}
\end{figure}

Sur la \ref{fig:densit=0000E9_plan=0000B0_water} est présentée la
densité planaire. On observe un accord quantitatif entre MD et MDFT.
Ce profil de densité comporte trois pics. Le premier, qui correspond
à la première couche de solvatation de l'argile, se trouve à $6.3\ \textrm{\AA}$
de la couche centrale de l'argile et comporte un épaulement à $5.7\ \textrm{\AA}$.
Il est intéressant de remarquer qu'avec une description de l'eau dipolaire,
se limitant aux premiers invariants, on trouve, comme dans le cas
du fluide de Stockmayer, que l'on surestime le premier épaulement.
Une meilleure description du solvant corrige cet effet. Le second
pic, dû à la seconde couche de solvatation, se trouve à $2.9\ \textrm{\AA}$
du premier. Il y a une déplétion entre ces deux premiers pics. Enfin,
un troisième pic se situe à $z=13.0\thinspace\textrm{\AA}$. Son l'intensité
est pratiquement négligeable mais on le distingue néanmoins en raison
des deux zones de déplétion qui l'encadrent. À une distance plus grande
de la surface, la densité tend vers celle du solvant homogène $n_{b}=0.033\ \textrm{\AA}^{-3}$. 

\begin{figure}
\noindent \begin{centering}
\includegraphics[width=0.6\textwidth]{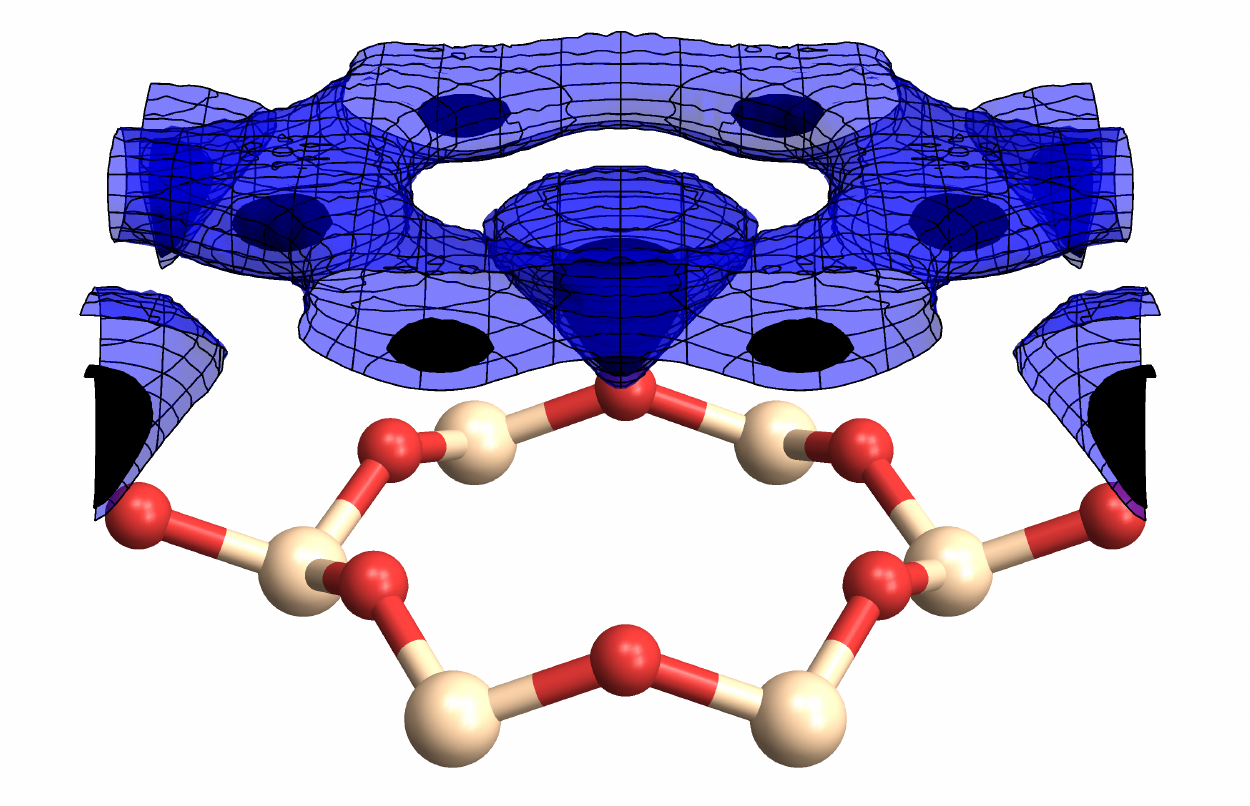}
\par\end{centering}

\protect\caption{Surfaces d'isodensité d'eau, avec $n(\bm{r})$=2 (en bleu) et 4 (en
noir). On représente la structure en hexagone de la pyrophyllite pour
aider à la visualisation, les atomes d'oxygène sont en rouge, les
atomes de silicium en blanc.\label{fig:isodens_water_pyro}}
\end{figure}

Le calcul de la densité locale $n(\bm{r})$ ou de la densité moyenne
$n(z)$ par MD n'est pas fait de manière exacte. On mesure plutôt
$n(z)\star\Delta\mathrm{z}$ et $n(\bm{r})\star\Delta\mathrm{V}$,
où $\star$ désigne une convolution et $\Delta\mathrm{z}$ et $\Delta\mathrm{V}$
sont des dimensions caractéristiques choisies arbitrairement. Ce sont
respectivement la distance et le volume dans lesquels est réalisée
la moyenne autour du point d'intérêt. Plus ces dimensions caractéristiques
sont petites plus le temps nécessaire pour obtenir une bonne statistique
augmente. Ce problème n'existe pas en MDFT puisque l'on n'effectue
pas d'échantillonnage. Il est aisé de tracer des cartes d'isodensité,
comme celles montrées en \ref{fig:isodens_water_pyro}, car la densité
est la variable naturelle. Cette carte présente deux surfaces d'isodensité
($n(\bm{r})$=2, 4) au dessus d'un hexagone Si-O de la surface. Les
zones de plus haute densité se trouvent au centre des hexagones et
des atomes de silicium (ces zones sont respectivement responsables
de l'épaulement et du premier pic dans le profil de densité de la
\ref{fig:densit=0000E9_plan=0000B0_water}). Une zone de densité moins
forte se trouve au dessus des hexagones, en forme d'anneau (cette
zone est responsable du second pic dans le profil de densité).

\begin{figure}
\noindent \begin{centering}
\includegraphics[width=0.8\textwidth]{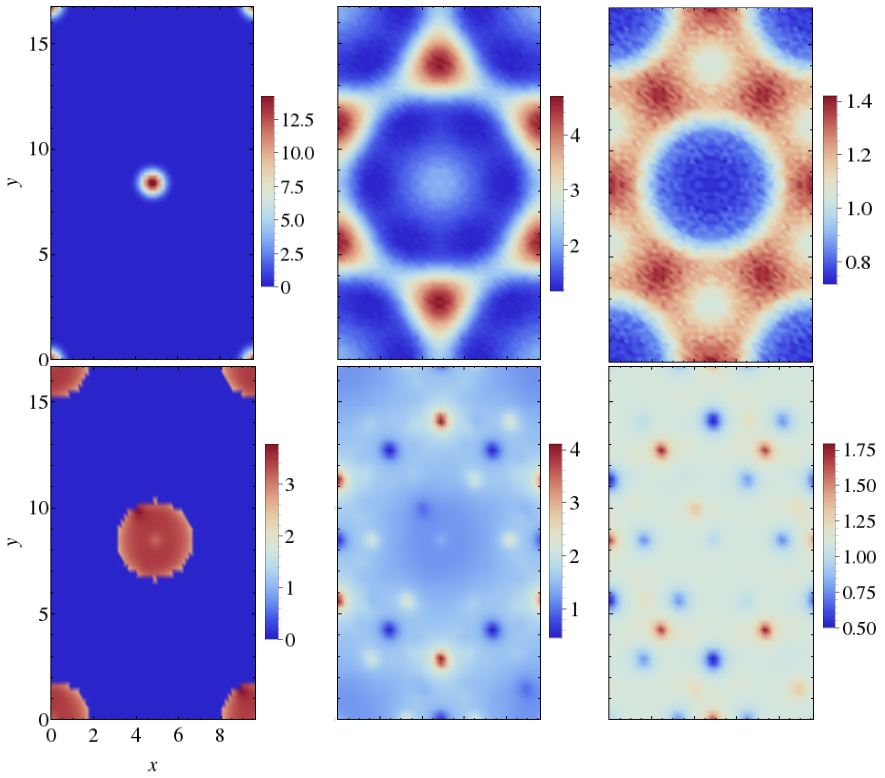}
\par\end{centering}

\protect\caption{Cartes de densité normalisée $n(\bm{r})/n_{b}$ de l'eau dans les
plans de l'épaulement (gauche), du pic principal (centre) et du second
pic (droite). Les résultats calculés par MD sont en haut, ceux calculés
par MDFT en bas.\label{fig:densit=0000E9_plan_maps_water}}
\end{figure}

On représente des cartes de densité dans les plans de l'épaulement,
du pic principal et du second pic sur la \ref{fig:densit=0000E9_plan_maps_water}.
Dans le plan du prépic, la densité est fortement localisée au centre
des hexagones, très proche de la surface hydrophobe. À cette distance
l'eau est repoussée partout ailleurs que dans cette petite zone, ce
qui explique que lorsqu'on moyenne la densité dans le plan, comme
sur la \ref{fig:densit=0000E9_plan=0000B0_water}, on observe qu'un
épaulement. Dans le plan du premier pic de la \ref{fig:densit=0000E9_plan=0000B0_water},
c'est-à-dire dans la première couche de solvatation, on trouve une
densité très localisée au dessus des atomes de silicium. Pour la seconde
couche de solvatation, qui correspond au deuxième pic du profil de
densité, on observe une densité importante au sommet des atomes d'oxygène
et une déplétion au sommet des atomes de silicium. On constate sur
la \ref{fig:densit=0000E9_plan_maps_water} que l'accord entre MD
et MDFT est seulement qualitatif : on observe les mêmes tendances
mais les cartes obtenues par MD sont plus diffuses. On peut faire
l'hypothèse que les convolutions évoquées plus tôt jouent un rôle
ici. On a vérifié que convoluer les profils de densité obtenus par
MDFT avec un volume $\Delta$V de $1\ \textrm{\AA}^{3}$ permettait
d'obtenir des cartes en bien meilleur accord avec celles obtenues
par MD. Cette différence illustre l'avantage de la méthode MDFT qui
évite un échantillonnage d'autant plus coûteux qu'on veut des cartes
précises.

\paragraph{Propriétés orientationnelles :}

\begin{figure}
\noindent \begin{centering}
\includegraphics[width=0.6\textwidth]{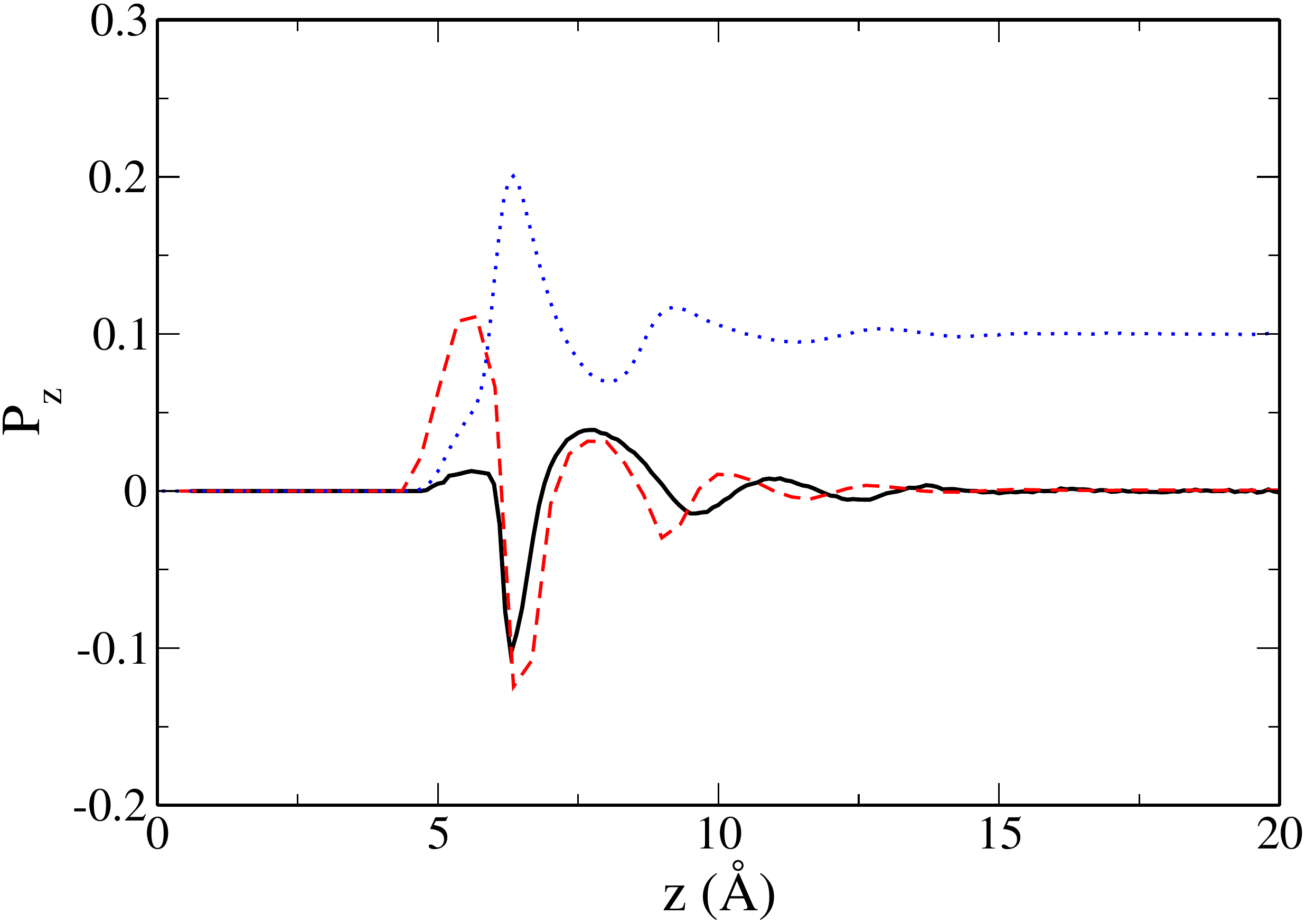}
\par\end{centering}

\protect\caption{Projections de la polarisation selon l'axe z, moyennée dans chaque
plan $(x,y)$. Le profil obtenu par MD est en trait noir plein, celui
obtenu par MDFT en tirets rouges. On rappelle le profil de la densité
$n_{\mathrm{z}}$ en pointillés bleu, en unité arbitraire.\label{fig:pola_plan=0000B0_water}}
\end{figure}

On s'intéresse maintenant aux propriétés orientationnelles. On présente
sur la \ref{fig:pola_plan=0000B0_water} la projection $P_{\mathrm{z}}$
de la polarisation sur l'axe perpendiculaire à la surface de l'argile,
moyennée dans chaque plan. Les composantes transverses sont trouvées
négligeables ($P_{\mathrm{x}}/P_{\mathrm{z}}$ et $P_{\mathrm{y}}/P_{\mathrm{z}}$
de l'ordre de $10^{-3}$), la polarisation est donc alignée dans la
direction perpendiculaire à la surface. Les extrema dans le profil
de polarisation correspondent à des extrema dans le profil de densité.
Un premier maximum de polarisation se trouve à la même abscisse que
l'épaulement. Celui-ci est beaucoup plus marqué dans les résultats
obtenus par MDFT. Le profil obtenu par MDFT est très semblable à celui
obtenu pour le fluide de Stockmayer. Un second extremum, négatif,
se trouve dans le plan du premier pic de densité. Entre ces deux extrema
le signe de $P_{\mathrm{z}}$ change. L'intensité de la polarisation
macroscopique est liée d'une part à la densité de molécule et d'autre
part à leur orientation préférentielle dans la région de l'espace
considérée. Ainsi, dans le plan du prépic, la polarisation est positive,
c'est donc les atomes d'oxygène des molécules d'eau qui sont globalement
les plus proches de la surface. Dans le plan du pic principal, les
molécules d'eau sont essentiellement situées au-dessus des silicium.
Leurs hydrogènes sont les plus proches de la surface, puisque le pic
de polarisation correspondant est négatif. Enfin le second pic de
densité est pratiquement non polarisé, il n'y presque pas d'orientation
préférentielle dans cette région de l'espace.

\begin{figure}
\noindent \begin{centering}
\includegraphics[width=0.6\textwidth]{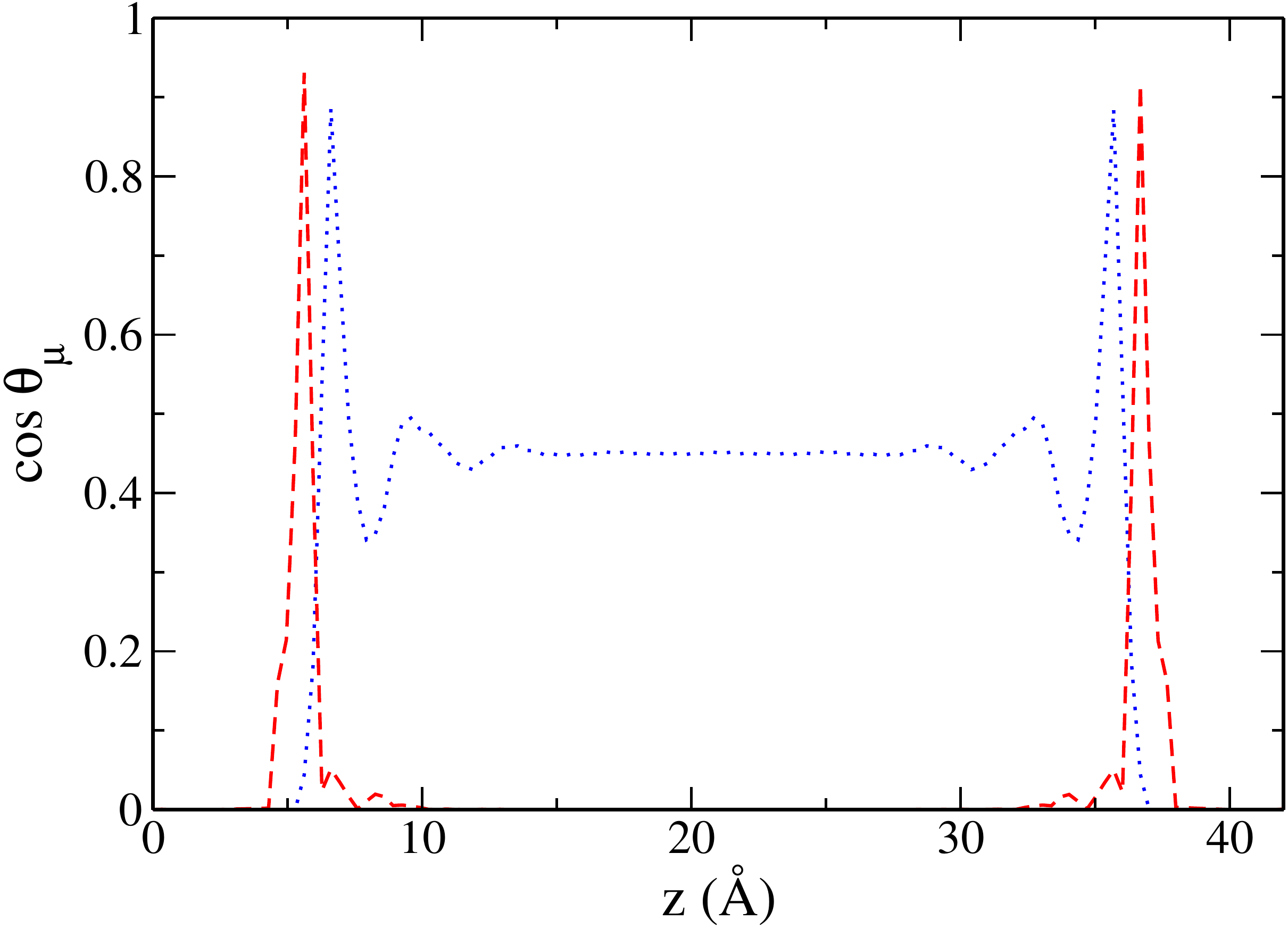}
\par\end{centering}

\protect\caption{Valeur moyenne de l'orientation moléculaire locale $\left\langle \cos\theta_{\mu}(\bm{r})\right\rangle _{xy}$
obtenue par MDFT en tirets rouges. Pour faciliter la lecture on rappelle
le profil de densité en pointillés bleus en unités arbitraires.\label{fig:costetha_mu_wat_pyro}}
\end{figure}

L'orientation moléculaire locale $\cos\theta_{\mu}(\bm{r})$, définie
à l'\ref{eq:costheta_mu} obtenue par MDFT, moyennée dans chaque plan
$(x,y)$ est donnée en \ref{fig:costetha_mu_wat_pyro}. Dans le plan
de l'épaulement, les molécules d'eau sont particulièrement orientées.
L'atome d'oxygène de la molécule d'eau est le plus proche possible
de la surface et de la charge positive des atomes de silicium. Au
contraire, dans le plan du pic principal, les molécules d'eau sont
globalement orientées dans la direction opposée. La faible valeur
de $\cos\theta_{\mu}(z)$ correspondant à ce pic principal suggère
une orientation préférentielle faible. La seconde couche de solvatation
ne possède pas d'orientation préférentielle.

\paragraph{Propriétés énergétiques:}

On peut analyser l'influence des différentes composantes énergétiques
en regardant la valeur de chacun des termes de la fonctionnelle utilisée.
La partie d'excès est décomposée en une partie radiale et une partie
de polarisation. L'énergie libre de solvatation par unité de surface
de la pyrophyllite $\gamma_{\mathrm{solv}}$ est donnée au \ref{tab:-composantes_ener_pyro_wat}.
Elle est de $332.4\ \mathrm{mJ.m^{-2}}$. La valeur positive confirme
la nature hydrophobe du matériau. La contribution énergétique principale
provient de la partie radiale des interactions solvant-solvant. La
partie d'interaction soluté-solvant et la partie de polarisation ont
une contribution très limitée à l'énergie totale. Ceci ne signifie
pas pour autant que la polarisation et la perturbation créées par
le soluté n'ont qu'une influence limitée dans le processus de solvatation.

\fbox{\begin{minipage}[t]{1\columnwidth}%
En effet, on peut imaginer la solvatation d'un mur dur dans un gaz
parfait. Le potentiel extérieur infini créé par le mur va évidemment
entrainer une densité nulle à l'intérieur du mur. Il en résulte un
terme extérieur ${\cal F}_{\mathrm{ext}}$ nul pour la densité d'équilibre.
Cependant, c'est la minimisation de ce terme extérieur qui impose
un profil de densité différent de celui obtenu en l'absence du mur
dur.%
\end{minipage}}
\begin{table}
\noindent \begin{centering}
\begin{tabular}{|c|c|c|c|c|c|}
\hline 
 & ${\cal F}$ & ${\cal F}_{\mathrm{ext}}$ & ${\cal F}_{\mathrm{id}}$ & ${\cal F}_{\mathrm{exc(rad)}}$ & ${\cal F}_{\mathrm{exc(ori\text{\textdegree})}}$\tabularnewline
\hline 
\hline 
$\gamma_{\mathrm{solv}}\ (\mathrm{mJ.m^{-2}})$ & 332.4 & -12.8 & 100.4 & 244.5 & 0.3\tabularnewline
\hline 
\end{tabular}\protect\caption{Les différentes composantes de l'énergie libre de solvatation par
unité de surface de l'argile.\label{tab:-composantes_ener_pyro_wat} }

\par\end{centering}

\end{table}

\paragraph{Le rôle de l'électrostatique :}

\begin{figure}
\noindent \begin{centering}
\includegraphics[width=0.6\textwidth]{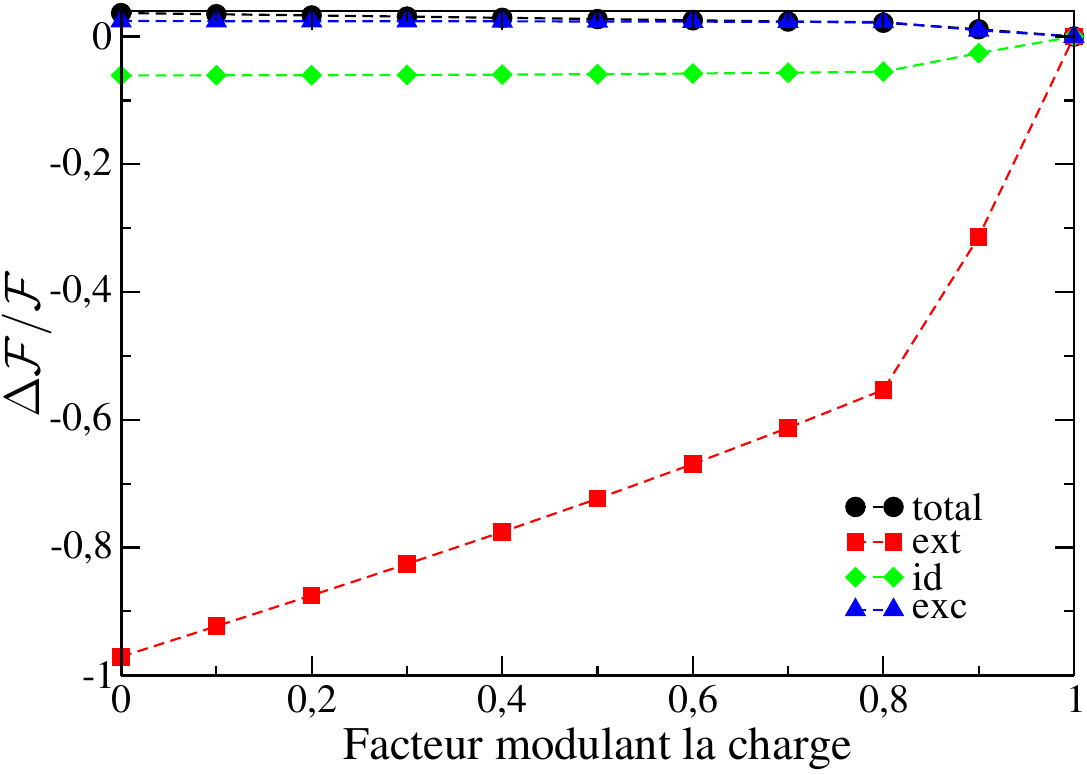}
\par\end{centering}

\protect\caption{Évolution relative de l'énergie libre par unité de surface et de ses
composantes, en fonction du facteur modulant la charge. Ces grandeurs
sont normalisées par rapport aux charges non modifiées. Le changement
relatif de l'énergie de solvatation est de 3\%. \label{fig:evol_ener_wat_pyro}}
\end{figure}

L'évolution de l'énergie libre et de ses composantes avec le facteur
modulant les charges de la surface de la pyrophyllite est donnée en
\ref{fig:evol_ener_wat_pyro}. L'énergie libre de solvatation totale
ne change pas de plus de 3\% quand on éteint totalement les charges.
De plus, les profils de densité ne sont pas modifiés par ce procédé.
Quand on éteint les charges, les énergies libres de solvatation totale,
idéale et la partie radiale du terme d'excès sont peu modifiées en
valeur absolue. La partie extérieure change beaucoup relativement
mais sa valeur reste faible. Ceci confirme que l'électrostatique a
peu d'importance pour étudier la solvatation d'argiles globalement
neutres. 

\fbox{\begin{minipage}[t]{1\columnwidth}%
Nous avons montré que la MDFT est adaptée à l'étude d'un matériaux
constitué de plus de mille atomes, même dans un solvant complexe comme
l'eau. Ceci peut être réalisé à un coût numérique mille fois plus
faible que celui de la MD.%
\end{minipage}}

\newpage{}

\section{Étude d'une molécule d'intérêt biologique: le lysozyme}

Les molécules d'eau confinées dans des cavités jouent un rôle important
dans la compréhension de la structure, de la stabilité et des fonctions
des biomolécules, notamment des protéines\cite{eriksson_response_1992,zhou_chemistry_2001}.
Ces molécules d'eau internes peuvent être observées par cristallographie\cite{nakasako_waterprotein_2004}.
Il est difficile de sonder la présence de molécule d'eau dans des
cavités internes de protéines par dynamique moléculaire. Elles sont
généralement piégées grâce à des processus impliquant de larges fluctuations
de la structure des protéines (repliement et dépliement) sur des échelles
de temps trop longues (ms) pour être facilement simulées par MD. Des
simulations sur des trajectoires de l'ordre de la microseconde permettant
de voir des échanges de molécules d'eau entre l'intérieur de la protéine
et le solvant bulk ont néanmoins été publiées récemment\cite{grossfield_internal_2008,shaw_atomic-level_2010}.
Cependant, ces simulations nécessitent l'utilisation de supercalculateurs
et restent très coûteuses numériquement à mettre en œuvre. Hirata
et collaborateurs ont montré que l'utilisation de la méthode de solvant
implicite 3D-RISM permet de détecter ces molécules piégées dans les
protéines. Ces simulations ont un coût numérique beaucoup plus faible
puisqu'elles évitent un traitement dynamique de la protéine\cite{imai_water_2005,3D_RISM_yoshida_molecular_2009,imai_locating_2007}.
Nous avons donc testé la théorie MDFT sur le même exemple que ces
auteurs : le lysozyme du blanc d'œuf de poule. La structure 3D de
cette protéine (code 1HEL) a été extraite de la protein data bank\cite{wilson_structural_1992}.
Cette protéine est connue pour avoir trois cavités (notées W1, W2
et W3) dans lesquelles sont respectivement détectées par cristallographie
quatre, une et aucune molécules d'eau. Il existe d'autres structures
cristallographiques\cite{kundrot_crystal_1987} de la même protéine
qui détectent une molécule d'eau dans la cavité W3 mais avec un facteur
B très grand, signe d'une précision très faible quant à la position
de cette molécule.

Toutes les molécules d'eau présentes dans la structure cristallographique
ont été retirées pour ne conserver comme soluté que la protéine en
elle-même. Le potentiel d'interaction de ce soluté a été généré en
utilisant les paramètres de champ de force Amber (parm99)\cite{wang_how_2000}
en utilisant le programme GROMACS\cite{berendsen_gromacs:_1995} pour
ajouter les atomes d'hydrogène qui ne sont pas visibles en cristallographie
et absents dans le fichier PDB.

\begin{figure}
\noindent \begin{centering}
\includegraphics[width=0.8\textwidth]{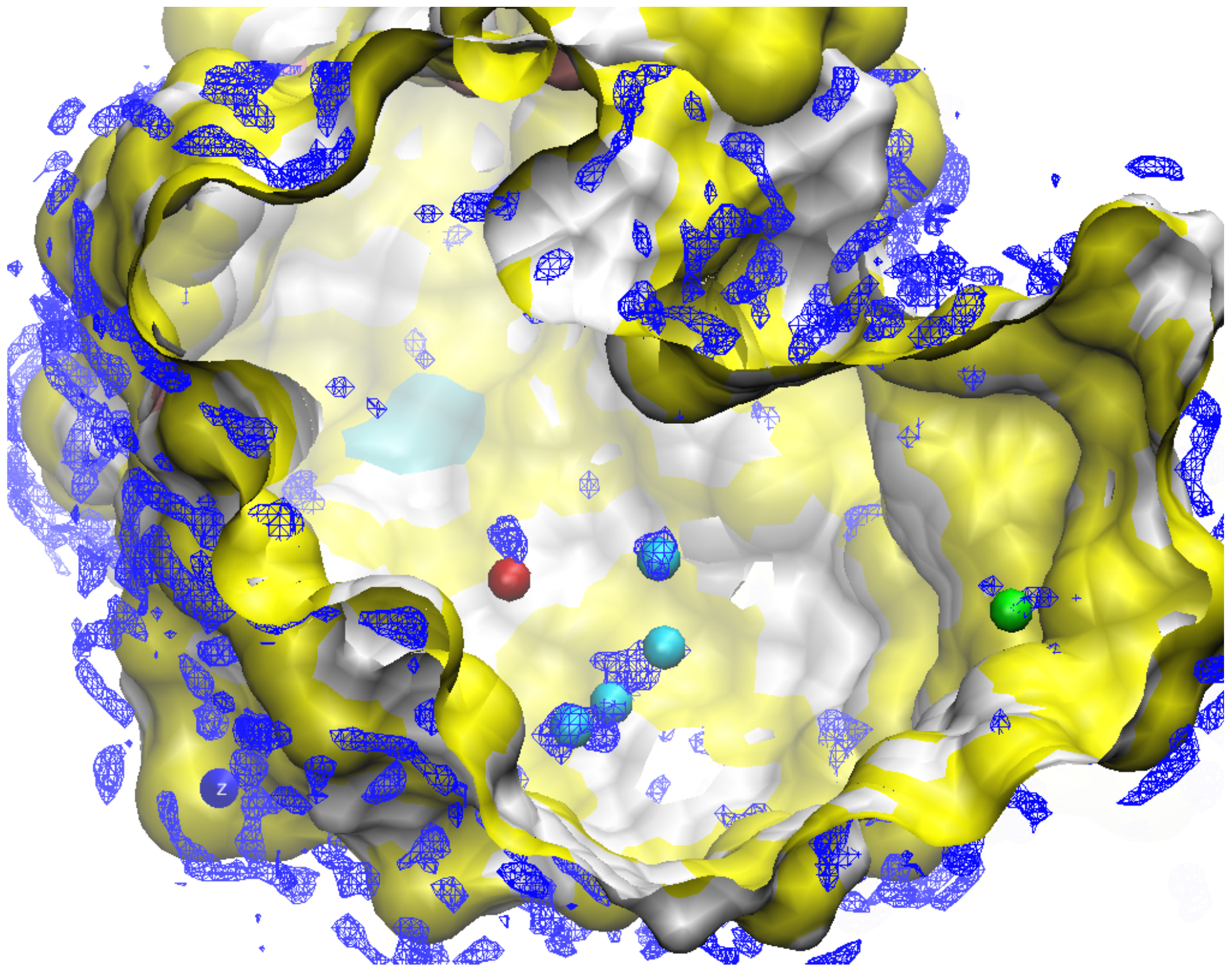}
\par\end{centering}

\protect\caption{Vue de l'intérieur de la protéine lysozyme dans la représentation
SURF. Les molécules d'eau visibles par cristallographie sont  représentées
par des sphères cyans pour la cavité W1, vertes pour la cavité W2
et rouges pour la cavité W3. Les surfaces de haute isodensité $n(\bm{r})/n_{b}=10$
obtenues par MDFT sont représentées avec un maillage bleu. \label{fig:lyzo_cav_toute}}
\end{figure}

Nous avons utilisé la fonctionnelle multipolaire, sans terme à trois
corps, avec le modèle d'eau SPC/E. La minimisation s'est faite avec
une boîte de $64\ \textrm{\AA}^{3}$ pour une grille spatiale de $192^{3}$
points et une grille angulaire de 24 angles. La minimisation se fait
en 38 pas et 3270 secondes sur un seul CPU. On obtient la carte de
densité du solvant à l'équilibre. Sur la \ref{fig:lyzo_cav_toute}
est présentée une vue de l'intérieure de la protéine en représentation
SURF où les positions des cavités sont représentées par des sphères
colorées situées aux coordonnées des molécules d'eau détectées par
cristallographie. On a également représenté des isosurfaces de haute
densité. On observe que la densité n'est pas uniforme selon que les
groupements situés à la surface de la protéine sont plutôt hydrophobes
ou hydrophiles. À l'intérieur de la cavité, les seules zones où on
rencontre une densité notable sont les zones où on détecte des molécules
d'eau internes par cristallographie. Ces densités se trouvent exactement
à l'endroit où les molécules d'eau sont détectées pour les cavités
W1 et W2. La densité est légèrement décalée pour la cavité W3. Le
facteur B pour cette molécule étant très faible, sa position cristallographique
est incertaine. La carte de densité tend donc à confirmer la présence
d'une molécule d'eau interne dans cette cavité.

\fbox{\begin{minipage}[t]{1\columnwidth}%
La MDFT est donc une méthode de choix pour vérifier la présence de
molécule de solvant à l'intérieur de cavité de molécules biologiques.
Ceci permet de compléter les résultats expérimentaux de cristallographie,
à un coût numérique très raisonnable, alors que ces systèmes sont
pratiquement hors de portée pour les simulations numériques.%
\end{minipage}}

\newpage{}

\lhead[\chaptername~\thechapter]{\rightmark}

\rhead[\leftmark]{}

\lfoot[\thepage]{}

\cfoot{}

\rfoot[]{\thepage}

\chapter{Conclusions et perspectives}

\section{Conclusions}

L'objectif de cette thèse est le développement d'une méthode de solvant
implicite, la théorie fonctionnelle de la densité moléculaire, MDFT,
pour son application à l'étude de la solvatation en milieu aqueux.
C'est un problème physico-chimique important, généralement appréhendé
théoriquement par des simulations moléculaires de type Monte Carlo
ou dynamique moléculaire, couplées à des techniques d'intégration
thermodynamique. Si ces méthodes donnent en principe des résultats
avec une \og précision chimique \fg{}, leur utilisation pour l'étude
de solutés complexes présente un coût numérique important. Ceci limite
leur utilisation systématique, les systèmes les plus complexes restant
à ce jours hors de portée. L'utilisation de méthodes de solvant implicite,
basées sur des théories des liquides, comme la MDFT, permet de réduire
considérablement ce coût. La méthode MDFT est exacte théoriquement
et conserve un niveau de description moléculaire du solvant.

La théorie de la fonctionnelle de la densité classique est basée sur
la preuve que le grand potentiel d'un solvant, en présence d'une perturbation
créant un champ extérieur, peut s'écrire comme une fonctionnelle unique
de la densité de solvant. Cette fonctionnelle étant inconnue, la clé
de l'utilisation de cette méthode réside dans la formulation d'une
approximation correcte de la fonctionnelle, et plus particulièrement
du terme d'excès relatif aux interactions solvant-solvant. Dans cette
thèse, nous nous sommes placés dans l'approximation du fluide homogène
de référence qui postule que la fonctionnelle d'excès peut s'écrire
en fonction de la fonction de corrélation directe du solvant pur.

Pendant cette thèse, j'ai développé la théorie de la fonctionnelle
de la densité dans l'approximation du fluide de référence pour étudier
la solvatation d'un soluté quelconque dans l'eau. Cette théorie existait
déjà pour des solvants plus simples, tels que l'acétonitrile ou le
fluide de Stockmayer, mais son extension à l'eau a posé des problèmes
particuliers, notamment dus au fait que l'eau (représentée par le
modèle à trois sites SPC/E) ne possède pas une distribution de charge
strictement dipolaire. Ceci a nécessité la formulation d'une fonctionnelle
d'excès dépendant de la densité de particule et de la densité de polarisation.
L'écriture de cette fonctionnelle fait intervenir le facteur de structure
et les composantes transverse et longitudinale des susceptibilités
diélectriques. L'utilisation de cette fonctionnelle permet de calculer
la structure du solvant à l'équilibre thermodynamique autour d'un
soluté quelconque ainsi que l'énergie libre de solvatation du soluté.
Ces fonctions de corrélation peuvent être obtenues expérimentalement
ou calculées par des simulations de dynamique moléculaire. On insiste
sur le fait qu'il est nécessaire de déterminer ces fonctions de corrélation
pour le solvant pur, à une densité donnée. Il suffit donc de connaître
un jeu de trois fonctions pour chaque solvant pour pouvoir étudier
la solvatation de n'importe quel soluté dans ce solvant.

Cette théorie a néanmoins montré ses limites, notamment pour l'étude
des solutés chargés. La cause de cette limitation provient des approximations
réalisées lors du développement de la théorie. La fonctionnelle développée
est incapable de reproduire l'ordre tétraédrique local de l'eau dû
aux liaisons hydrogènes. Nous avons donc introduit un terme de correction
à trois corps qui renforce l'ordre tétraédrique entre le solvant et
les solutés. Une implémentation numériquement efficace de cette correction
a aussi été réalisée. En incorporant cette correction, la structure
du solvant autour des solutés chargés obtenue est très proche des
résultats obtenus par des simulations numériques.

Une autre limite de la théorie développée est son incapacité à reproduire
correctement le changement de comportement à petite et grande échelle
de la structure et de l'énergie de solvatation de solutés hydrophobes.
Cet effet peut être corrigé par la séparation de la fonctionnelle
en une partie courte et longue échelle, et par l'introduction d'une
correction longue échelle de la fonctionnelle. Le développement de
cette correction longue portée et son implémentation ont constitué
une autre partie importante des travaux effectués au cours de cette
thèse. Cette formulation prouve qu'il est possible de traiter des
problèmes multi-échelles avec la théorie de la fonctionnelle de la
densité dans le cadre de l'approximation du fluide homogène de référence. 

Une autre partie importante de ce travail de thèse a consisté en l'implémentation
de cette théorie dans un code de minimisation numérique de la fonctionnelle.
Ce code, nommé mdft, utilise une discrétisation de l'espace et des
orientations sur une double grille tridimensionnelle. Il est écrit
en Fortran moderne.

Nous avons appliqué notre théorie sur un certain nombre de systèmes.
Les résultats obtenus sur des solutés simples, atomes, ions et petites
molécules sont en accord avec des résultats de référence obtenus par
MD qui utilisent les mêmes champs de force. Nous avons également étudié
des systèmes complexes. L'étude de la solvatation d'une argile hydrophobe
par le solvant de Stockmayer et l'eau par la théorie MDFT donne accès
aux propriétés structurales et orientationnelles du solvant, en bon
accord avec les simulations moléculaires menées sur le même système,
tout en étant plus rapide d'environ trois ordres de grandeurs. La
MDFT est donc une bonne alternative pour l'étude de ces systèmes.
Le gain numérique vis-à-vis de la MD permet de plus une étude énergétique
systématique en fonction de paramètres des champ de forces, notamment
les charges utilisées pour décrire ces systèmes. La MDFT peut donc
aussi être un outil permettant d'optimiser des champs de force moléculaires
avec un coût raisonnable.

Nous nous sommes également servi de notre théorie pour étudier un
soluté d'intérêt biologique, le lysozyme, et nous avons pu confirmer
l'existence de cavités contenant des molécules d'eau à l'intérieur
de la protéine. Une telle étude est impossible par des méthodes de
simulation car l'entrée et la sortie des molécules d'eau à l'intérieur
de la protéine requiert des dépliements et des repliements de la protéine
qui ne sont pas observables par MD. L'échelle de temps de ces mouvements
étant à ce jour encore hors de portée pour ces techniques. La force
de la méthode MDFT réside ici dans le fait que l'équilibre thermodynamique
est atteint par minimisation fonctionnelle et non par échantillonnage.

Enfin, la théorie a été utilisée pour calculer des potentiels de force
moyenne. La précision des résultats obtenus est comparable à celle
obtenue par MD. L'important gain de temps de calcul par rapport à
la MD laisse à penser que la MDFT pourrait devenir un outil standard
pour calculer des PMF, qui sont des observables très utiles notamment
pour la paramétrisation de champs de force gros-grains.

La construction de la fonctionnelle proposée est également intéressante
pour la compréhension des phénomènes physiques mis en jeu dans la
solvatation. En effet, on peut rajouter différentes corrections et
voir leurs effets sur la solvatation. Ceci permet de voir quels phénomènes
physiques, modélisés par ces corrections gouvernent les propriétés
de solvatation des systèmes étudiés. Il est plus compliqué de rationaliser
ce lien entre théorie et résultats pour les simulations moléculaires
puisque les seules choses que l'on peut modifier sont les paramètres
des champs de force utilisés dont l'effet sur les propriétés d'intérêt
est très sensible.

\section{Perspectives}

Si les résultats obtenus pendant cette thèse sont encourageants, il
reste néanmoins de nombreuses tâches à accomplir avant de faire de
la MDFT, et du code correspondant, une méthode standard. Tout d'abord,
il est prévu d'utiliser une fonction de corrélation du solvant utilisant
l'ensemble des projections sur les invariants rotationnels. Pour cela,
une collaboration avec Luc Bellonni du CEA, qui est un expert du calcul
de ces projections, a commencé en octobre 2013, sous forme de la thèse
de Lu Ding. Ceci permettra d'avoir une bien meilleure description
des interactions entre molécules de solvant et permettra peut-être
de s'affranchir des différentes corrections: terme à trois corps,
bridge de sphères dures et hydrophobicité à longue portée. Un point
théoriquement capital, soulevé par Bob Evans, est de vérifier en quel
endroit du diagramme de phase de l'eau se trouve le solvant dans notre
théorie. Ceci peut se faire en vérifiant un certain nombre de règles
thermodynamiques, telle que la règle d'adsorption de Gibbs et les
règles de transition de phase. Un bon point de départ serait la construction
d'un modèle d'eau spécifique à la MDFT, au lieu de prendre ceux couramment
utilisés en MD. On pense, par exemple, partir d'un fluide de sphères
dures, auquel serait ajouté une terme de polarisation, un terme à
trois corps et une perturbation Lennard-Jones. Il est également prévu
d'introduire un terme de polarisabilité électrostatique du solvant.

Enfin, l'objectif ultime, qui était à l'origine le but de cette thèse
et qui s'est révélé irréalisable pendant ces trois ans faute de temps,
est un couplage entre DFT électronique et MDFT. Ceci permettrait d'utiliser
la théorie MDFT pour traiter le solvant tout en ayant une description
quantique du soluté à un coût bien plus faible que les méthodes QM/MM
et avec une précision supérieure aux méthodes utilisant un continuum
comme PCM. La partie la plus technique résidera dans l'écriture d'un
potentiel de couplage entre la partie classique et la partie quantique.
On se propose comme point de départ de réaliser des minimisations
successives avec un couplage nul. D'abord d'eDFT pour obtenir une
densité électronique, puis de se servir de cette densité électronique
comme potentiel extérieur pour trouver la densité de solvant d'équilibre
par MDFT. On pourra se servir du champ électrique créé par cette densité
de solvant comme potentiel extérieur pour mener une nouvelle minimisation
eDFT, et itérer le processus jusqu'à atteindre une convergence des
densités moléculaire et électronique calculées par les deux méthodes. 

Enfin, une perspective primordiale est la mise à disposition à moyen
terme du programme mdft à l'ensemble de la communauté.
\cleardoublepage{}

\lhead[Publications issues de ce travail]{}

\rhead[]{}

Ci-dessous, l'ensemble des articles publiés ou acceptés sur le sujet
de ce manuscrit avant sa rédaction, le premier juin 2014.\\

M. Levesque, V. Marry, B. Rotenberg, G. Jeanmairet, R. Vuilleumier,
D. Borgis. (2012) Solvation of complex surfaces via molecular density
functional theory. \textit{J. Chem. Phys.} \textbf{137}, 224107.

G. Jeanmairet, M. Levesque, R. Vuilleumier, D. Borgis. (2013) Molecular
Density Functional Theory of Water, \textit{J. Phys. Chem. Lett.}
\textbf{4}, 619-624 .

G. Jeanmairet, M. Levesque, D. Borgis. (2013) Molecular Density Functional
Theory of Water describing Hydrophobicity at Short and Long Length
Scales, \textit{J. Chem. Phys.} \textbf{139}, 154101. 

G. Jeanmairet, V. Marry, M. Levesque, B. Rotenberg, D. Borgis. (2014)
Hydration of Clays at the Molecular Scale: The Promising Perspective
of Classical Density Functional Theory, \textit{Mol. Phys.} \textbf{112},
1320-1329. 

V. P. Sergiievskyi, G. Jeanmairet, M. Levesque, D. Borgis. (2014)
Fast Computation of Solvation Free Energies with Molecular Density
Functional Theory: Thermodynamic-Ensemble Partial Molar Volume Corrections,\textit{
J. Phys. Chem. Lett. }\textbf{5}, 1935-1942. 

G. Jeanmairet, N. Levy, M. Levesque, D. Borgis. (2014) Introduction
to Classical Density Functional Theory by Computational Experiment,
\textit{arXiv preprint} arXiv:1401.1679. \\

Plusieurs autres articles sont soumis ou en cours de rédaction.

\appendix

\lhead[\chaptername~\thechapter]{\rightmark}

\rhead[\leftmark]{}

\lfoot[\thepage]{}

\cfoot{}

\rfoot[]{\thepage}

\part{Appendices}

\chapter{\label{chap:D=0000E9riv=0000E9-fonctionelle}Notions de fonctionnelle
et de dérivation fonctionnelle}

\section{Fonctionnelle}

On appellera fonctionnelle ${\cal F}$ une fonction prenant comme
argument une autre fonction $f$. La fonctionnelle renvoie un scalaire
(on se limitera aux scalaires réels) en fonction de la valeur prise
par la fonction $f$, sur un intervalle $I$. Par exemple,
\begin{equation}
{\cal F}_{\mathrm{id}}\left[\rho(\bm{r})\right]=\mathrm{k_{B}}\mathrm{T}\iiint_{\mathbb{R}^{3}}\left(\rho(\bm{r})\ln\left(\frac{\rho(\bm{r})}{\rho_{\mathrm{b}}}\right)-\rho(\bm{r})+\rho_{\mathrm{b}}\right)\mathrm{d}\bm{r},\label{eq:Fid_appendix}
\end{equation}
où ${\cal F}_{\mathrm{id}}$ est une fonctionnelle de la densité $\rho$,
elle-même fonction des coordonnées cartésiennes. La valeur de la fonctionnelle
est liée à celle de la densité sur l'ensemble de son domaine de définition,
ici $\mathbb{R}^{3}$.

\section{Dérivée fonctionnelle}

Soit la fonction $f(\bm{x})$ où $\bm{x}=x_{1},x_{2},...,x_{\mathrm{n}}$
est un vecteur à $\mathrm{n}$ dimensions. La variation de $f$ due
à une variation infinitésimale de $\bm{x}$ s'écrit
\begin{equation}
\mathrm{d}f=f(\bm{x}+d\bm{x})-f(\bm{x})=\sum_{i=1}^{\mathrm{n}}f_{i}^{\prime}(\bm{x})\mathrm{d}x_{i},
\end{equation}
avec,
\begin{equation}
f_{i}^{\prime}(\bm{x})=\frac{\partial f}{\partial x_{i}}.
\end{equation}
De même, si ${\cal A}$ est une fonctionnelle de $f(\bm{x})$, c'est-à-dire
qui dépend de la valeur de $f$ sur un intervalle $\left[a;b\right]$,
la variation de la fonctionnelle s'exprime comme
\begin{equation}
\delta{\cal A}={\cal A}\left[f(\bm{x})+\delta f(\bm{x})\right]-{\cal A}\left[f(\bm{x})\right]=\intop_{a}^{b}{\cal A}^{\prime}\left[f(\bm{x})\right]\delta f(\bm{x})\mathrm{d}\bm{x}
\end{equation}
avec la dérivée fonctionnelle
\begin{equation}
{\cal A}^{\prime}\left[f(\bm{x})\right]=\frac{\delta{\cal A}}{\delta f(\bm{x})}.\label{eq:derivefonctionnelle}
\end{equation}
Illustrons ces notions sur un cas utile en DFT classique, par exemple
la contribution idéale à la fonctionnelle d'énergie libre de l'\ref{eq:Fid_appendix}.
\begin{eqnarray*}
\beta\delta{\cal F}_{\mathrm{id}} & = & \beta{\cal F}_{\mathrm{id}}\left[\rho(\bm{r})+\delta\rho(\bm{r})\right]-\beta{\cal F}_{\mathrm{id}}\left[\rho(\bm{r})\right]\\
 & = & \iiint_{\mathbb{R}^{3}}\left[\left(\rho(\bm{r})+\delta\rho(\bm{r})\right)\ln\left(\frac{\rho(\bm{r})+\delta\rho(\bm{r})}{\rho_{\mathrm{b}}}\right)-\rho(\bm{r})+\delta\rho(\bm{r})+\rho_{\mathrm{b}}\right]\mathrm{d}\bm{r}\\
 &  & -\iiint_{\mathbb{R}^{3}}\left(\rho(\bm{r})\ln\left(\frac{\rho(\bm{r})}{\rho_{b}}\right)-\rho(\bm{r})+\rho_{b}\right)\mathrm{d}\bm{r}\\
 & = & \iiint_{\mathbb{R}^{3}}\left[\left(\rho(\bm{r})+\delta\rho(\bm{r})\right)\left[\ln\left(\frac{\rho(\bm{r})}{\rho_{b}}\right)+\frac{\delta\rho(\bm{r})}{\rho(\bm{r})}+{\cal O}(\delta\rho(\bm{r})^{2})\right]-\rho(\bm{r})-\delta\rho(\bm{r})+\rho_{b}\right]\mathrm{d}\bm{r}\\
 &  & -\iiint_{\mathbb{R}^{3}}\left(\rho(\bm{r})\ln\left(\frac{\rho(\bm{r})}{\rho_{b}}\right)-\rho(\bm{r})+\rho_{b}\right)\mathrm{d}\bm{r}\\
 & = & \iiint_{\mathbb{R}^{3}}\left(\delta\rho(\bm{r})\ln\left(\frac{\rho(\bm{r})}{\rho_{b}}\right)+\delta\rho(\bm{r})+{\cal O}(\delta\rho(\bm{r})^{2})-\delta\rho(\bm{r})\right)\mathrm{d}\bm{r}\\
 & = & \iiint_{\mathbb{R}^{3}}\left(\delta\rho(\bm{r})\ln\left(\frac{\rho(\bm{r})}{\rho_{b}}\right)+{\cal O}(\delta\rho(\bm{r})^{2})\right)\mathrm{d}\bm{r}
\end{eqnarray*}
qui, par identification avec l'\ref{eq:derivefonctionnelle}, donne
la dérivée fonctionnelle de la partie idéale,

\begin{equation}
\beta\frac{\delta{\cal F}_{\mathrm{id}}}{\delta\rho(\bm{r})}=\ln\left(\frac{\rho(\bm{r})}{\rho_{b}}\right).
\end{equation}

\lhead[\chaptername~\thechapter]{\rightmark}

\rhead[\leftmark]{}

\lfoot[\thepage]{}

\cfoot{}

\rfoot[]{\thepage}

\chapter{Démonstration : Le grand potentiel est une fonctionnelle des densités
de particule et de polarisation \label{sec:gdPotmini}}

\section{La densité qui minimise le grand potentiel est la densité d'équilibre}

Cette démonstration a déjà été réalisée en \ref{sub:Principe-Variationel-pour}
et ne supposait pas d'écriture particulière du hamiltonien. Les résultats
obtenus sont toujours vrais pour le hamiltonien de l'\ref{eq:Hwater_multi}.
On a donc : %
\footnote{On rappelle l'expression de la trace classique : \\
\[
\text{Tr}_{\text{cl}}=\sum_{\text{N}=0}^{\infty}\frac{1}{\mathrm{h}\text{\ensuremath{^{3\mathrm{N}}}N}!}\idotsint_{\mathbb{R}^{3\text{N}}}\idotsint_{\mathbb{R}^{3\text{N}}}\text{d}\bm{r}...\text{d}\bm{r}_{\text{N}}\text{d}\bm{p}_{1}...\text{d}\bm{p}_{\text{N}}
\]
}
\begin{equation}
\Theta\left[f\right]=\mathrm{Tr_{cl}}\left[f({\cal H}_{N}-\mu N+\beta^{-1}\ln f)\right],\label{eq:DefThetaTRace}
\end{equation}
\begin{equation}
\Theta\left[f_{0}\right]=-\mathrm{k_{B}}T\ln\Xi=\Theta,\label{eq:Theta0=00003DTheta(f0)}
\end{equation}
\begin{equation}
\Theta\left[f\right]>\Theta\left[f_{0}\right],\text{ si \ensuremath{f\neq f_{0}}}.\label{eq:Thetaf>theta}
\end{equation}

\section{L'énergie libre intrinsèque est une fonctionnelle unique des densités
de particule et de polarisation \label{sec:L'=0000E9nergie-libre-intrins=0000E8que_unique_fonctioelle}}

On veut introduire les deux nouvelles fonctionnelles,
\begin{equation}
{\cal F}[n(\bm{r}),\bm{P}(\bm{r})]=\mathrm{Tr_{cl}}\left[f_{0}\left({\cal K}+{\cal U}+\beta^{-1}\ln f_{0}\right)\right],\label{eq:F(n,P)}
\end{equation}
et
\begin{equation}
\Theta_{\mathrm{{\cal V}}}[n(\bm{r}),\bm{P}(\bm{r})]={\cal F}[n(\bm{r}),\bm{P}(\bm{r})]+\iiint_{\mathbb{R}^{3}}\Psi_{n}(\bm{r})n(\bm{r})\mathrm{d}\bm{r}+-\iiint_{\mathbb{R}^{3}}\bm{P}(\bm{r})\bm{E}(\bm{r})\mathrm{d}\bm{r},\label{eq:Theta_V}
\end{equation}
où $\Psi_{n}(\bm{r})=\Phi_{n}(\bm{r})-\mu$. L'indice $\mathrm{{\cal V}}$
dans l'\ref{eq:Theta_V} rappelle que cette fonctionnelle dépend de
la forme choisie pour le potentiel extérieur, c'est donc une notation
compacte du couple $(\phi_{n},\bm{E})$. Nous allons montrer qu'il
est possible d'écrire ces deux fonctionnelles.

\section{Démonstration : ${\cal F}$ est une fonctionnelle unique de $n(\boldsymbol{r})$
et $\boldsymbol{P}(\boldsymbol{r})$}

La densité de probabilité d'équilibre est une fonction des champs
extérieurs $\Phi_{n}$ et $\bm{E}$. Puisque
\begin{equation}
n_{0}(\bm{r})=\mathrm{Tr_{cl}}\left[f_{0}\tilde{n}(\bm{r})\right]
\end{equation}
et
\begin{equation}
\bm{P}_{0}(\bm{r})=\mathrm{Tr_{cl}}\left[f_{0}\tilde{\bm{P}}(\bm{r})\right],
\end{equation}
$n_{0}(\bm{r})$ et $\bm{P}_{0}(\bm{r})$ à l'équilibre sont eux aussi
des fonctionnelles de ces champs extérieurs%
\footnote{On rappelle que les \textasciitilde{} désignent les fonctions microscopiques. %
}. Supposons qu'il existe deux couples de champs extérieurs différents
qui engendrent les mêmes densités de particule et de polarisation
à l'équilibre $(n(\bm{r}),\bm{P}(\bm{r}))$. Au hamiltonien ${\cal H}_{N}$
on associe la distribution de probabilité à l'équilibre $f_{0}$ et
le grand potentiel $\Theta$. Au hamiltonien ${\cal H}_{N}^{\prime}$
on associe la distribution de probabilité à l'équilibre $f_{0}^{\prime}$
et le grand potentiel $\Theta^{\prime}$. D'après l'\ref{eq:Thetaf>theta},
on a
\begin{equation}
\Theta^{\prime}=\mathrm{Tr_{cl}}f_{0}^{\prime}\Bigl[\left({\cal H}_{N}^{\prime}-\mu N+\beta^{-1}\ln f_{0}^{\prime}\right)\Bigr]<\Theta\left[f_{0}\right],
\end{equation}
\begin{equation}
\Theta^{\prime}=\mathrm{Tr_{cl}}\Bigl[f_{0}^{\prime}\left({\cal H}_{N}^{\prime}-\mu N+\beta^{-1}\ln f_{0}^{\prime}\right)\Bigr]<\mathrm{Tr_{cl}}\Bigl[f_{0}\left({\cal H}_{N}^{\prime}-\mu N+\beta^{-1}\ln f_{0}\right)\Bigr],
\end{equation}
\begin{equation}
\Theta^{\prime}<\Theta+\iiint_{\mathbb{R}^{3}}n_{0}(\bm{r})(\phi_{N}^{\prime}(\bm{r})-\phi_{N}(\bm{r}))\mathrm{d}\bm{r}-\iiint_{\mathbb{R}^{3}}\bm{P}_{0}(\bm{r})\cdot(\bm{E}^{\prime}(\bm{r})-\bm{E}(\bm{r}))\mathrm{d}\bm{r}.\label{eq:theta<thetaprime+...}
\end{equation}
En échangeant les rôles de $\Theta$ et $\Theta^{\prime}$, on trouve
\begin{equation}
\Theta<\Theta^{\prime}+\iiint_{\mathbb{R}^{3}}n_{0}(\bm{r})(\phi_{N}(\bm{r})-\phi_{N}^{\prime}(\bm{r}))\mathrm{d}\bm{r}-\iiint_{\mathbb{R}^{3}}\bm{P}_{0}(\bm{r})\cdot(\bm{E}(\bm{r})-\bm{E}^{\prime}(\bm{r}))\mathrm{d}\bm{r}.\label{eq:thetaprime<tehta+...}
\end{equation}
En sommant l'\ref{eq:theta<thetaprime+...} et l'\ref{eq:thetaprime<tehta+...}
on obtient $\Theta+\Theta^{\prime}<\Theta^{\prime}+\Theta$, ce qui
est absurde. À un couple de densité de particule et de polarisation
n'est donc associé qu'un unique potentiel extérieur. Le potentiel
extérieur est déterminé de manière unique par les densités d'équilibre
et réciproquement. Il en suit que la distribution de probabilité est
une fonctionnelle unique des densités d'équilibre. La fonctionnelle
donnée à l'\ref{eq:F(n,P)} est donc une unique fonctionnelle des
densités d'équilibre $n$ et $\bm{P}$. Elle est dite universelle
car elle possède la même expression pour tous les potentiels extérieurs.

Montrons que la fonctionnelle introduite en \ref{eq:Theta_V} est
égale au grand potentiel à son minimum.

Pour les densités d'équilibre $n_{0}(\bm{r})$ et $\bm{P}_{0}(\bm{r})$,
cette fonctionnelle est égale au grand potentiel,
\begin{equation}
\Theta_{{\cal V}}[n_{0}(\bm{r}),\bm{P}(\bm{r})]=\Theta.
\end{equation}
Supposons qu'il existe des densités de particule et polarisation $n^{\prime}(\bm{r})$
et $\bm{P}^{\prime}(\bm{r})$ associées à une distribution de probabilité
$f^{\prime}$.
\begin{eqnarray*}
\Theta\left[f^{\prime}\right] & = & \mathrm{Tr_{cl}}\left[f^{\prime}({\cal H}_{N}-\mu N+\beta^{-1}\ln f^{\prime})\right]\\
 & = & {\cal F}\left[n^{\prime}(\bm{r}),\bm{P}^{\prime}(\bm{r})\right]+\iiint_{\mathbb{R}^{3}}\Psi_{n}(\bm{r})n^{\prime}(\bm{r})\mathrm{d}\bm{r}+\iiint_{\mathbb{R}^{3}}\bm{P}^{\prime}(\bm{r})\bm{E}(\bm{r})\mathrm{d}\bm{r}\\
 & = & \Theta_{{\cal V}}\left[n^{\prime}(\bm{r}),\bm{P}^{\prime}(\bm{r})\right].
\end{eqnarray*}
En utilisant l'\ref{eq:Thetaf>theta}, on a alors
\begin{equation}
\Theta_{{\cal V}}\left[n^{\prime}(\bm{r}),\bm{P}^{\prime}(\bm{r})\right]>\Theta_{{\cal V}}\left[n_{0}(\bm{r}),\bm{P}_{0}(\bm{r})\right].
\end{equation}
Ainsi, les densités d'équilibre $n_{0}(\bm{r})$ et $\bm{P}_{0}(\bm{r})$
minimisent la fonctionnelle $\Theta_{V}\left[n(\bm{r}),\bm{P}(\bm{r})\right]$,
ce que l'on peut noter :

\begin{equation}
\left.\frac{\delta\Theta_{{\cal V}}\left[n(\bm{r}),\bm{P}(\bm{r})\right]}{\delta n(\bm{r})}\right|_{n_{0}}=0\text{ et }\left.\frac{\delta\Theta_{{\cal V}}\left[n(\bm{r}),\bm{P}(\bm{r})\right]}{\delta\bm{P}(\bm{r})}\right|_{\bm{P}_{0}}=0.
\end{equation}

\chapter{Fonctionnelle de la densité du fluide de sphères dures\label{sec:FMT}}

Nous avons implémenté une fonctionnelle pour le fluide de sphères
dures. La théorie de la mesure fondamentale (FMT) a été introduite
par Rosenfeld, nous la notons R-FMT\cite{rosenfeld_free-energy_1989}.
Cette fonctionnelle est une excellente approximation de la fonctionnelle
d'excès pour un fluide composé de sphères dures de rayon $\mathrm{R_{0}}$.
Elle nécessite le calcul de densités convoluées avec des fonctions
poids : 
\begin{equation}
{\cal F}_{\mathrm{exc}}\left[\rho(\bm{r})\right]=\mathrm{k_{B}T}\iiint_{\mathbb{R}^{3}}\Phi(\{n_{\alpha}(\bm{r})\})\mathrm{d}\bm{r},
\end{equation}

avec,
\begin{equation}
n_{\alpha}(\bm{r})=\iiint_{\mathbb{R}^{3}}\rho(\bm{r})\omega_{\alpha}(\bm{r}-\bm{r}^{\prime})\mathrm{d}\bm{r}^{\prime}=\rho(\bm{r})\star\omega_{\alpha}(\bm{r}),
\end{equation}

où les $\omega_{\alpha}$ désignent des fonctions poids géométriques
que l'on définit plus bas, $\star$ désigne le produit de convolution
et $\Phi$ est une densité d'énergie libre.

On donne la dérivée de cette énergie libre par rapport à la densité
:
\begin{eqnarray}
\frac{\delta{\cal F}_{\mathrm{exc}}}{\delta\rho(\bm{r})} & = & \mathrm{k_{B}T}\sum_{\alpha=1}^{\mathrm{N_{w}}}\iiint_{\mathbb{R}^{3}}\frac{\delta\Phi}{\delta n_{\alpha}(\bm{r}^{\prime})}\frac{\delta n_{\alpha}(\bm{r}^{\prime})}{\delta\rho(\bm{r})}\mathrm{d}\bm{r}^{\prime}=\mathrm{\mathrm{k_{B}}T}\sum_{\alpha=1}^{\mathrm{N_{w}}}\iiint_{\mathbb{R}^{3}}\frac{\delta\Phi}{\delta n_{\alpha}(\bm{r}^{\prime})}\omega_{\alpha}(\bm{r}-\bm{r}^{\prime})\mathrm{d}\bm{r}^{\prime}\nonumber \\
 & = & \mathrm{k_{B}T}\sum_{\alpha=1}^{\mathrm{N_{w}}}\frac{\delta\Phi}{\delta n_{\alpha}(\bm{r}^{\prime})}\star\omega_{\alpha}(\bm{r})
\end{eqnarray}

qui s'exprime comme une somme des dérivés partielles de $\Phi$ par
rapport aux fonctions poids.

Kierlik et Rosinberg ont dérivé une version alternative plus simple
de FMT, que nous notons KR-FMT. Phan\cite{phan_equivalence_1993}
et collaborateurs ont démontré que la formulation KR-FMT est équivalente
à la formulation vectorielle R-FMT. Cette formulation fait intervenir
moins de fonctions poids que la formulation de Rosenfeld. On calcule
ces fonctions poids, qui sont des produits de convolution, par FFT.
La formulation de KR diminue donc le nombre de FFT à réaliser. C'est
donc la version que nous avons implémenté dans le code mdft\cite{MDFT_levesque_krfmt}.

Elle nécessite quatre fonctions poids scalaires $\omega_{\alpha}$
avec $\alpha=0,1,2,3$. 
\begin{gather}
\omega_{0}(r)=-\frac{1}{8\pi}\delta^{\prime\prime}(\mathrm{R_{0}}-r)+\frac{1}{2\pi r}\delta^{\prime}(\mathrm{R_{0}}-r),\\
\omega_{1}(r)=\frac{1}{8\pi}\delta^{\prime}(\mathrm{R_{0}}-r),\\
\omega_{2}(\bm{r})=4\pi\mathrm{R_{0}}\omega_{1}(\bm{r})=4\pi\mathrm{R_{0}^{2}}\omega_{0}(\bm{r})=\delta(R_{0}-r),\\
\omega_{3}(\bm{r})=\Theta(\mathrm{R_{0}}-r),
\end{gather}

où $\Theta$ désigne la fonction de Heaviside et $\delta$ la distribution
de Dirac. Cette formulation peut être utilisée avec une densité d'énergie
libre $\Phi$ dérivant de la théorie de Percus-Yevick ou de Carnahan-Starling.
Les deux densités d'énergie libre, celle de Percus-Yevick (PY) 
\begin{equation}
\Phi^{\mathrm{PY}}\left[n_{\alpha}\right]=-n_{0}\ln\left(1-n_{3}\right)+\frac{n_{1}n_{2}}{1-n_{3}}+\frac{1}{24\pi}\frac{n_{2}^{3}}{(1-n_{3})^{2}}
\end{equation}
 et celle de Carnahan-Starling (CS) 
\begin{equation}
\Phi^{\mathrm{CS}}\left[n_{\alpha}\right]=\left(\frac{1}{36\pi}\frac{n_{2}^{3}}{n_{3}^{2}}-n_{0}\right)\ln\left(1-n_{3}\right)+\frac{n_{1}n_{2}}{1-n_{3}}+\frac{1}{36\pi}\frac{n_{2}^{3}}{(1-n_{3})^{2}n_{3}},
\end{equation}

sont implémentées dans le code. Kierlik et Rosinberg ont montré que
CS donne des résultats plus précis mais que son utilisation sans modification
des fonctions poids cause une inconsistance thermodynamique. Dans
la pratique, nous utilisons toujours la relation de CS. 

Le potentiel chimique du fluide de sphères dures et la fonction de
corrélation directe peuvent être calculés par dérivation fonctionnelle,
\begin{equation}
\mu_{\mathrm{exc}}^{\mathrm{HS}}=\left.\frac{\delta{\cal F}_{\mathrm{exc}}\left[\rho(\bm{r})\right]}{\delta\rho(\bm{r})}\right|_{\rho(\bm{r})=\rho_{\mathrm{b}}},
\end{equation}
 et
\begin{equation}
c_{000}^{\mathrm{HS}}(r;\rho_{b})=-\left.\frac{\delta^{2}{\cal F}_{\mathrm{exc}}\left[\rho(\bm{r})\right]}{\delta\rho(\bm{r})\delta\rho(\bm{r}^{\prime})}\right|_{\rho(\bm{r})=\rho_{\mathrm{b}}}.
\end{equation}
Ces fonctions sont à calculer si on veut utiliser le bridge de sphères
dures décrit au \ref{chap:MDFT_dup_multu}. En pratique, la fonction
de corrélation directe est calculée efficacement dans l'espace de
Fourier, 
\begin{equation}
c_{000}^{\mathrm{HS}}(k;\rho_{\mathrm{b}})=-\sum_{\alpha,\beta}\frac{\partial^{2}\Phi}{\partial n_{\alpha}\partial n_{\beta}}(\left\{ n_{\gamma}^{b}\right\} )\hat{\omega}_{\alpha}(k)\hat{\omega}_{\beta}(k),
\end{equation}
où $\left\{ n_{\gamma}^{b}\right\} $ représente l'ensemble des fonctions
pondérées pour le fluide de sphères dures homogène de densité $\rho_{\mathrm{b}}$;
et $\hat{\omega}_{\alpha}(k)$ les transformées de Fourier des fonctions
poids.

\lhead[\chaptername~\thechapter]{\rightmark}

\rhead[\leftmark]{}

\lfoot[\thepage]{}

\cfoot{}

\chapter{Dérivées des fonctionnelles\label{sec:D=0000E9riv=0000E9es-des-fonctionnelles}}

Les dérivées des différentes parties des fonctionnelles utilisées
dans ce manuscrit sont rassemblées ici.

La dérivée de la partie idéale de l'\ref{eq:FidRho(r,Omega)} est
\begin{equation}
\frac{\delta{\cal F}_{\text{id}}\left[\rho(\bm{r},\bm{\Omega})\right]}{\delta\rho(\bm{r},\bm{\Omega})}=\text{k}_{\text{B}}\mathrm{T}\ln\left(\frac{\rho(\bm{r},\bm{\Omega})}{\rho_{b}}\right).
\end{equation}
Celle de la partie extérieure donnée en \ref{eq:Fext(rho(r,Omega))}
est
\begin{equation}
\frac{\delta{\cal F}_{\text{ext}}\left[\rho(\bm{r},\bm{\Omega})\right]}{\delta\rho(\bm{r},\bm{\Omega})}=v_{\mathrm{ext}}(\bm{r},\bm{\Omega}).
\end{equation}

Ces deux termes ne posent aucun problème particulier pour être calculés.
La dérivée de la partie d'excès dipolaire de l'équation \ref{eq:Fexc_exacte}
est 
\begin{gather}
\frac{\delta{\cal F}_{\text{exc}}\left[\rho(\bm{r},\bm{\Omega})\right]}{\delta\rho(\bm{r}_{2},\bm{\Omega})}=\text{k}_{\text{B}}\mathrm{T}\iiint_{\mathbb{R}^{3}}\left[c_{000}(\left\Vert \bm{r}_{12}\right\Vert )\Delta n(\bm{r}_{1})\right]\text{d}\bm{r}_{1}\nonumber \\
-\frac{1}{2\mu}\text{k}_{\text{B}}T\iiint_{\mathbb{R}^{3}}\left[c_{101}(\left\Vert \bm{r}_{12}\right\Vert )\Delta n(\bm{r}_{1})\right]\text{d}\bm{r}_{1}\nonumber \\
-\frac{1}{2}\text{k}_{\text{B}}T\iiint_{\mathbb{R}^{3}}\left[c_{101}((\left\Vert \bm{r}_{12}\right\Vert )\bm{P}(\bm{r}_{2})\cdot\tilde{\bm{u}}_{12}\right]\text{d}\bm{r}_{1}\nonumber \\
+\frac{1}{2\mu}\text{k}_{\text{B}}T\iiint_{\mathbb{R}^{3}}\left[c_{011}(\left\Vert \bm{r}_{12}\right\Vert )\Delta n(\bm{r}_{1})\right]\text{d}\bm{r}_{1}\nonumber \\
+\frac{1}{2}\text{k}_{\text{B}}T\iiint_{\mathbb{R}^{3}}\left[c_{011}(\left\Vert \bm{r}_{12}\right\Vert )\bm{P}(\bm{r}_{2})\cdot\tilde{\bm{u}}_{12}\right]\text{d}\bm{r}_{1}\\
-\frac{1}{\mu}\text{k}_{\text{B}}T\iiint_{\mathbb{R}^{3}}\left[c_{112}(\left\Vert \bm{r}_{12}\right\Vert )3(\bm{P}(\bm{r}_{1})\cdot\tilde{\bm{u}}_{12})-\bm{P}(\bm{r}_{1})\right]\text{d}\bm{r}_{1}\nonumber \\
+\frac{\delta{\cal F}_{\mathrm{cor}}\left[\rho(\bm{r},\bm{\Omega})\right]}{\delta\rho(\bm{r},\bm{\Omega})}
\end{gather}

Ces gradients sont des produits de convolution faciles à calculer
par FFT, méthode employée de manière systématique dans le code.

La dérivée de la partie d'excès multipolaire de l'\ref{eq:Fexcmulti}
peut se décomposer en la somme de sa dérivée par rapport à la densité
et celle par rapport à la polarisation,
\begin{equation}
\beta\frac{\delta{\cal F}_{\mathrm{exc}}[\Delta n(\bm{r}),\bm{P}(\bm{r})]}{\delta\Delta n(\bm{r}_{2})}=\frac{1}{2}\iiint_{\mathbb{R}^{3}}S^{-1}(\left\Vert \bm{r}_{12}\right\Vert )\Delta n(\bm{r}_{2})\mathrm{d}\bm{r}_{2}-\iiint_{\mathbb{R}^{3}}\frac{\Delta n(\bm{r}_{2})}{n_{0}}\mathrm{d}\bm{r}_{2}+\frac{\delta\beta{\cal F}_{\mathrm{corr}}[\Delta n(\bm{r}),\bm{P}(\bm{r})]}{\delta n(\bm{r}_{2})}.
\end{equation}

La dérivée par rapport à la polarisation est beaucoup plus facile
à exprimer directement dans l'espace de Fourier,
\begin{gather}
\beta\frac{\delta{\cal F}_{\mathrm{exc}}[\Delta\hat{n}(\bm{k}),\hat{\bm{P}}(\bm{k})]}{\delta\hat{\bm{P}}(\bm{k})}=\frac{3}{\mu_{0}n_{0}^{2}}\hat{\bm{P}}\left(\bm{k}\right)\hat{\bm{\mu}}(\bm{k},\boldsymbol{\Omega})+\frac{1}{8\pi\epsilon_{0}}\hat{\chi}_{L}^{-1}(k)\hat{\boldsymbol{P}}_{\mathrm{L}}(\bm{\bm{k}})\left(\hat{\bm{P}}(\bm{k})\cdot\bm{k}\right)\bm{k}\nonumber \\
+\frac{1}{8\pi\epsilon_{0}}\hat{\chi}_{T}^{-1}(k)\hat{\bm{P}}_{\mathrm{T}}(\bm{k})\left[\hat{\bm{\mu}}(\bm{k},\boldsymbol{\Omega})-\left(\hat{\bm{P}}(\bm{k})\cdot\bm{k}\right)\bm{k}\right].
\end{gather}

Il faut ensuite calculer les composantes de cette dérivée dans l'espace
direct, donc en effectuer la transformée de Fourier inverse.

La dérivée du terme de correction à trois corps de l'\ref{eq:F3B}
peut se faire composante par composante:
\begin{equation}
\left.\frac{\delta{\cal F}_{\mathrm{cor}}^{\mathrm{3B-1S}}[n(\bm{r})]}{\delta n(\bm{r})}\right|_{\bm{r}=\bm{r}_{2}}=\sum_{m}\lambda_{m}^{1\mathrm{S}}\iiint_{\mathbb{R}^{3}}f_{m}(r_{m2})f_{m}(r_{m3})\left(\frac{\bm{r}_{m2}\cdot\bm{r}_{m3}}{r_{m2}r_{m3}}-\cos\theta_{0}\right)^{2}n(\bm{r}_{3})d\bm{r}_{3}
\end{equation}
\begin{gather}
\left.\frac{\delta{\cal F}_{\mathrm{cor}}^{\mathrm{3B-2S}}[n(\bm{r})]}{\delta n(\bm{r})}\right|_{\bm{r}=\bm{r}_{2}}=\frac{1}{2}\sum_{m}\lambda_{m}^{2\mathrm{S}}\iiint_{\mathbb{R}^{3}}f_{m}(r_{m3})f_{w}(r_{23})\left(\frac{\bm{r}_{m3}\cdot\bm{r}_{23}}{r_{m3}r_{23}}-\cos\theta_{0}\right)^{2}n(\bm{r}_{3})d\bm{r}_{3}\nonumber \\
+\frac{1}{2}\sum_{m}\lambda_{m}^{2\mathrm{S}}\iiint_{\mathbb{R}^{3}}f_{m}(r_{m2})f_{w}(r_{23})\left(\frac{\bm{r}_{m2}\cdot\bm{r}_{23}}{r_{m2}r_{23}}-\cos\theta_{0}\right)^{2}n(\bm{r}_{3})d\bm{r}_{3}
\end{gather}
\begin{gather}
\left.\frac{\delta{\cal F}_{\mathrm{cor}}^{\mathrm{3B-ww}}[n(\bm{r})]}{\delta n(\bm{r})}\right|_{\bm{r}=\bm{r}_{1}}=\nonumber \\
\frac{1}{2}\lambda_{w}\iiint_{\mathbb{R}^{3}}\iiint_{\mathbb{R}^{3}}f_{w}(r_{12})f_{w}(r_{13})\left(\frac{\bm{r}_{12}\cdot\bm{r}_{13}}{r_{12}r_{13}}-\cos\theta_{0}\right)^{2}n(\bm{r}_{2})n(\bm{r}_{3})d\bm{r}_{2}d\bm{r}_{3}\nonumber \\
+\lambda_{w}\iiint_{\mathbb{R}^{3}}n(\bm{r}_{2})\left[\iiint_{\mathbb{R}^{3}}f_{w}(r_{12})f_{w}(r_{23})\left(\frac{\bm{r}_{12}\cdot\bm{r}_{23}}{r_{12}r_{23}}-\cos\theta_{0}\right)^{2}n(\bm{r}_{3})d\bm{r}_{3}\right]d\bm{r}_{2}\nonumber \\
\end{gather}

\chapter{Couplage densité-polarisation dans la fonctionnelle d'excès multipolaire\label{sec:Couplage-densit=0000E9-polarisation}}

Si, contrairement à la \ref{sec:Fexcmulti}, on ne néglige pas les
couplages entre densité et polarisation, on peut écrire la fonctionnelle
d'excès: \foreignlanguage{english}{
\begin{gather}
\beta{\cal F}_{\mathrm{exc}}[\Delta n(\bm{r}),\bm{P}_{\mathrm{L}}(\bm{r}),\bm{P}_{\mathrm{T}}(\bm{r})]=\beta{\cal F}_{\mathrm{int}}[\Delta n(\bm{r}),\bm{P}_{\mathrm{L}}(\bm{r}),\bm{P}_{\mathrm{T}}(\bm{r})]-\beta{\cal F}_{\mathrm{id}}[\Delta n(\bm{r}),\bm{P}_{\mathrm{L}}(\bm{r}),\bm{P}_{\mathrm{T}}(\bm{r})]\nonumber \\
=\frac{1}{2}\iiint_{\mathbb{R}^{3}}\iiint_{\mathbb{R}^{3}}C_{\mathrm{nn}}(r_{12})\Delta n(\bm{r}_{1})\Delta n(\bm{r}_{2})\mathrm{d}\bm{r}_{1}\mathrm{d}\bm{r}_{2}+\iiint_{\mathbb{R}^{3}}\iiint_{\mathbb{R}^{3}}C_{\mathrm{nP}}(r_{12})\Delta n(\bm{r}_{1})P_{\mathrm{L}}(\bm{r}_{2})\mathrm{d}\bm{r}_{1}\mathrm{d}\bm{r}_{2}\nonumber \\
+\frac{1}{2}\iiint_{\mathbb{R}^{3}}\iiint_{\mathbb{R}^{3}}C_{\mathrm{PP}}(r_{12})P_{\mathrm{L}}(\bm{r}_{1})P_{\mathrm{L}}(\bm{r}_{2})\mathrm{d}\bm{r}_{1}\mathrm{d}\bm{r}_{2}\nonumber \\
+\frac{\beta}{8\pi\epsilon_{0}}\iiint_{\mathbb{R}^{3}}\iiint_{\mathbb{R}^{3}}\chi_{\mathrm{T}}^{-1}(r_{12})\bm{P}_{\mathrm{T}}(\bm{r}_{1})\bm{P}_{\mathrm{T}}(\bm{r}_{2})\mathrm{d}\bm{r}_{1}\mathrm{d}\bm{r}_{2},\label{eq:Fexmulti_coupl}
\end{gather}
}où on a définit un champ de polarisation longitudinal, qui est la
transformée inverse de $\hat{P}_{\mathrm{L}}(\bm{k})$, défini par
\begin{equation}
\hat{\bm{P}}_{\mathrm{L}}(\bm{k})=\hat{P}_{\mathrm{L}}(\bm{k})\tilde{\bm{k}}
\end{equation}
 avec
\begin{equation}
\hat{P}_{\mathrm{L}}(\bm{k})=\tilde{\bm{k}}\cdot\hat{\bm{P}}(\bm{k}),
\end{equation}
où $\tilde{\bm{k}}$ désigne le vecteur unitaire de l'espace réciproque.

L'\ref{eq:Fexmulti_coupl} se réecrit dans l'espace de Fourier :\foreignlanguage{english}{
\begin{gather}
\beta{\cal F}_{\mathrm{exc}}[\Delta\hat{n}(\bm{k}),\hat{\bm{P}}_{\mathrm{L}}(\bm{k}),\hat{\bm{P}}_{\mathrm{T}}(\bm{k})]=\frac{1}{2}\iiint_{\mathbb{R}^{3}}\hat{C}_{\mathrm{nn}}(k)\Delta\hat{n}(\bm{k})\Delta\hat{n}(-\bm{k})\mathrm{d}\bm{k}\nonumber \\
+\frac{1}{2}\iiint_{\mathbb{R}^{3}}\hat{C}_{\mathrm{PP}}(k)\hat{P}_{\mathrm{L}}(\bm{k})\hat{P}_{\mathrm{L}}(-\bm{k})\mathrm{d}\bm{k}+\iiint_{\mathbb{R}^{3}}\hat{C}_{\mathrm{nP}}(k)\Delta\hat{n}(\bm{k})\hat{P}_{\mathrm{L}}(\bm{-k})\mathrm{d}\bm{k}\nonumber \\
\frac{1}{8\pi\epsilon_{0}}\iiint_{\mathbb{R}^{3}}\hat{\chi}_{\mathrm{T}}^{-1}(k)\hat{\bm{P}}_{\mathrm{T}}(\bm{k})\cdot\hat{\bm{P}}_{\mathrm{T}}(-\bm{k})\mathrm{d}\bm{k}\label{eq:Fexc_multi_coupl_k}
\end{gather}
}

On reconnait des expressions couplant les densités et les polarisations
similaires à celles de l'\ref{eq:Fexcmulti}. Cependant, les fonctions
de corrélation intervenant dans ces expressions sont modifiées. Il
y a un nouveau terme qui couple densité et polarisation.

On définit les facteurs de structure densité-densité, densité-polarisation
longitudinale et polarisation longitudinale-polarisation longitudinale:
\begin{equation}
\hat{S}_{\mathrm{nn}}(k)=\left\langle \Delta\hat{n}(\bm{k})\Delta\hat{n}(-\bm{k})\right\rangle =\hat{S}(\bm{k})
\end{equation}
\begin{equation}
\hat{S}_{\mathrm{nP}}(k)=\left\langle \Delta\hat{n}(\bm{k})\hat{P}_{\mathrm{L}}(-\bm{k})\right\rangle =-ik\left\langle \Delta\hat{n}(\bm{k})\hat{\rho}_{\mathrm{c}}(-\bm{k})\right\rangle =-ikS_{\mathrm{nc}}(k)
\end{equation}
\begin{eqnarray}
\hat{S}_{\mathrm{PP}}(k) & = & \left\langle \hat{P}_{\mathrm{L}}(\bm{k})\hat{P}_{\mathrm{L}}(-\bm{k})\right\rangle =-4\pi\epsilon_{0}\mathrm{k_{B}T}\hat{\chi}_{L}(k)\\
 & = & -\left\langle \hat{\rho}_{\mathrm{c}}(\bm{k})\hat{\rho}_{\mathrm{c}}(-\bm{k})\right\rangle /k^{2}=-S_{cc}(k)/k^{2}
\end{eqnarray}

Les expressions des fonctions de corrélation intervenant dans l'\ref{eq:Fexc_multi_coupl_k}
sont les suivantes :\foreignlanguage{english}{
\begin{gather}
\hat{C}_{\mathrm{nn}}(k)=\frac{\hat{S}_{\mathrm{PP}}(k)}{\hat{S}_{\mathrm{PP}}(k)\hat{S}_{\mathrm{nn}}(k)-\hat{S}_{\mathrm{nP}}(k)^{2}},\\
\hat{C}_{\mathrm{nP}}(k)=\frac{\hat{S}_{\mathrm{nP}}(k)}{\hat{S}_{\mathrm{PP}}(k)\hat{S}_{\mathrm{nn}}(k)-\hat{S}_{\mathrm{nP}}(k)^{2}},\\
\hat{C}_{\mathrm{cc}}(k)=\frac{\hat{S}_{\mathrm{nn}}(k)}{\hat{S}_{\mathrm{PP}}(k)\hat{S}_{\mathrm{nn}}(k)-\hat{S}_{\mathrm{nP}}(k)^{2}},
\end{gather}
}

Ainsi, si on néglige le couplage densité-polarisation, on retrouve
l'\ref{eq:Fexcmulti} avec, $\hat{C}_{\mathrm{cc}}=-\frac{1}{4\pi\epsilon_{0}}\hat{\chi}_{\mathrm{L}}^{-1}$,
et $\hat{C}_{\mathrm{nn}}=\hat{S}^{-1}$.

\begin{figure}[h]
\noindent \centering{}\includegraphics[width=0.6\textwidth]{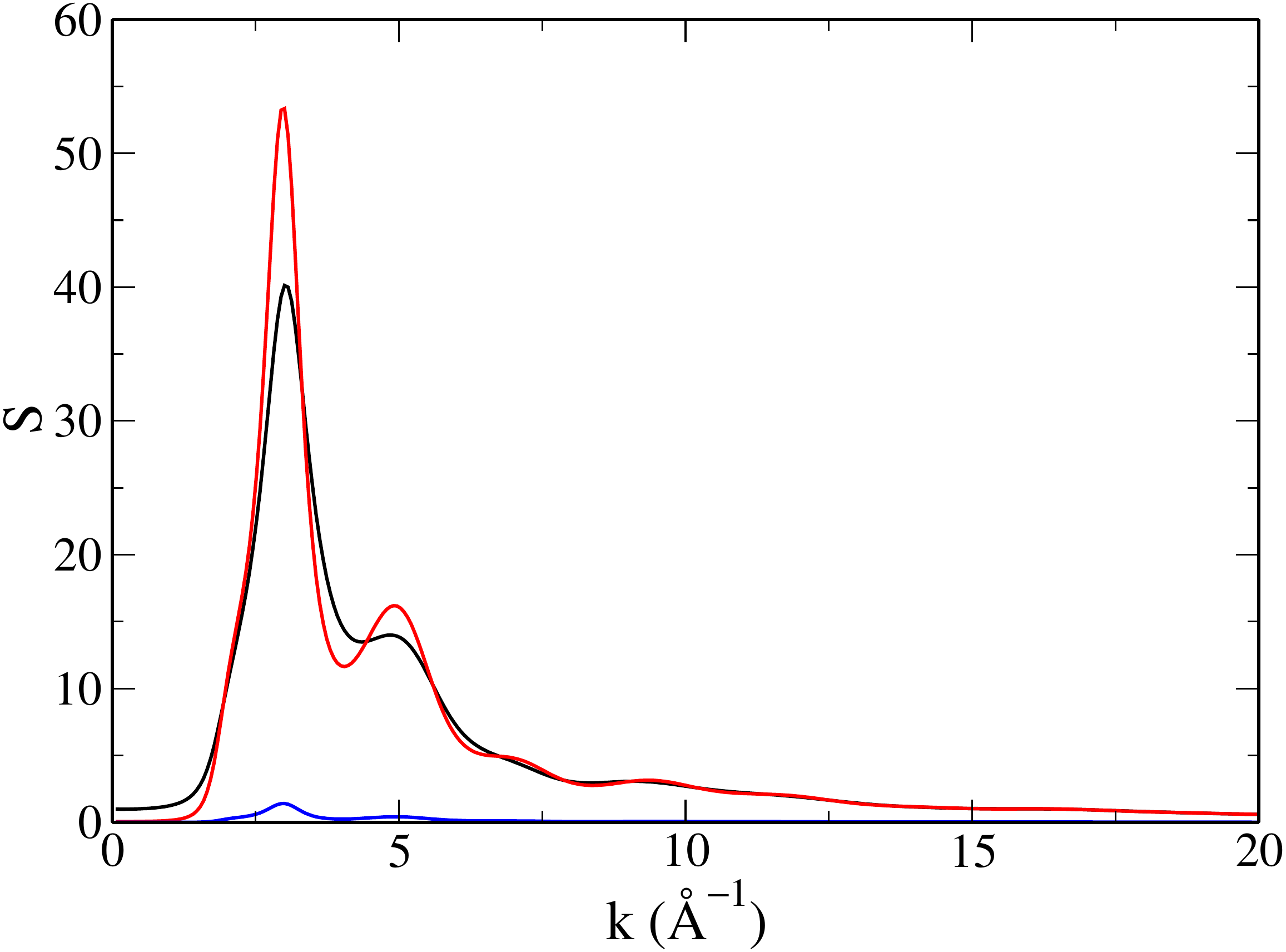}\protect\caption{Comparaison des facteurs de structure $\hat{S}_{\mathrm{PP}}$ (noir),
$\left(\hat{S}_{\mathrm{nP}}\right)^{2}$ (bleu) et du produit $\hat{S}_{\mathrm{PP}}\hat{S}_{\mathrm{nn}}$
(rouge) adimensionnés. \label{fig:ChiLcompwithandwithoutcoupling}}
\end{figure}

L'inclusion d'un terme de couplage entre densité et polarisation modifie
donc aussi les fonctions de corrélation couplant les densités et les
polarisations longitudinales. Néanmoins, l'effet sur ces fonctions
est faible, comme on peut le voir sur la \ref{fig:ChiLcompwithandwithoutcoupling},
où on voit que le terme $\left(\hat{S}_{\mathrm{nP}}\right)^{2}$
au dénominateur des fonctions $\hat{C}$ est négligeable par rapport
au terme $\hat{S}_{\mathrm{PP}}\hat{S}_{\mathrm{nn}}$ et qu'il est
petit devant le terme $\hat{S}_{\mathrm{PP}}$.

Cette fonctionnelle a été implémentée de manière à être calculée dans
l'espace de Fourier.

Sa dérivée est également calculée dans l'espace de Fourier,\foreignlanguage{english}{
\begin{gather*}
\frac{\delta\beta{\cal F}_{\mathrm{exc}}[\Delta\hat{n}(\bm{k}),\hat{\bm{P}}_{\mathrm{L}}(\bm{k}),\hat{\bm{P}}_{\mathrm{T}}(\bm{k})]}{\delta\Delta\hat{n}(\bm{k})}=\hat{C}_{\mathrm{nn}}(k)\Delta\hat{n}(\bm{k})+\hat{C}_{\mathrm{nP}}(k)\hat{P}_{\mathrm{L}}(\bm{-k}),
\end{gather*}
\begin{equation}
\frac{\delta\beta{\cal F}_{\mathrm{exc}}[\Delta\hat{n}(\bm{k}),\hat{\bm{P}}_{\mathrm{L}}(\bm{k}),\hat{\bm{P}}_{\mathrm{T}}(\bm{k})]}{\delta\Delta\hat{P}_{\mathrm{L}}(\bm{k})}=\hat{C}_{\mathrm{PP}}(k)\hat{P_{\mathrm{L}}}(\bm{k})+\hat{C}_{\mathrm{nP}}(k)\Delta\hat{n}(\bm{k}),
\end{equation}
\begin{equation}
\frac{\delta\beta{\cal F}_{\mathrm{exc}}[\Delta\hat{n}(\bm{k}),\hat{\bm{P}}_{\mathrm{L}}(\bm{k}),\hat{\bm{P}}_{\mathrm{T}}(\bm{k})]}{\delta\Delta\hat{\bm{P}}_{\mathrm{T}}(\bm{k})}=\frac{1}{4\pi\epsilon_{0}}\hat{\chi}_{\mathrm{T}}^{-1}(k)\hat{\bm{P}}_{\mathrm{T}}(\bm{k})\tilde{\bm{k}},
\end{equation}
}

\begin{figure}[h]
\noindent \centering{}\includegraphics[width=0.8\textwidth]{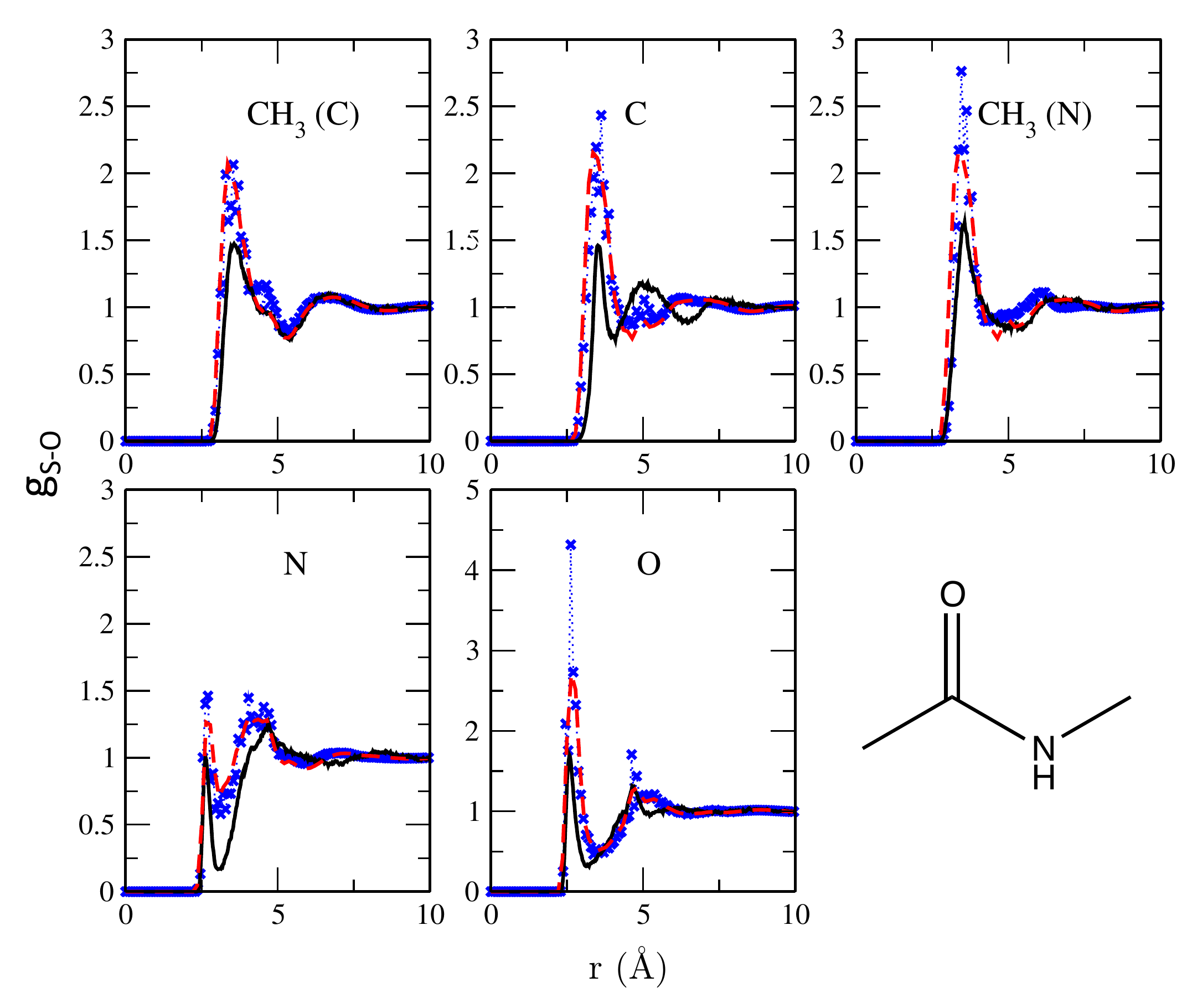}\protect\caption{Fonctions de distribution radiale entre le solvant et les différents
sites de la N-méthylacétamide obtenus par minimisation fonctionnelle,
sans terme de couplage (en croix bleues) et avec (en tirets rouges).
En noir, les références de MD.\label{fig:rdfNMAcoupl}}
\end{figure}

\begin{figure}[h]
\noindent \centering{}\includegraphics[width=0.8\textwidth]{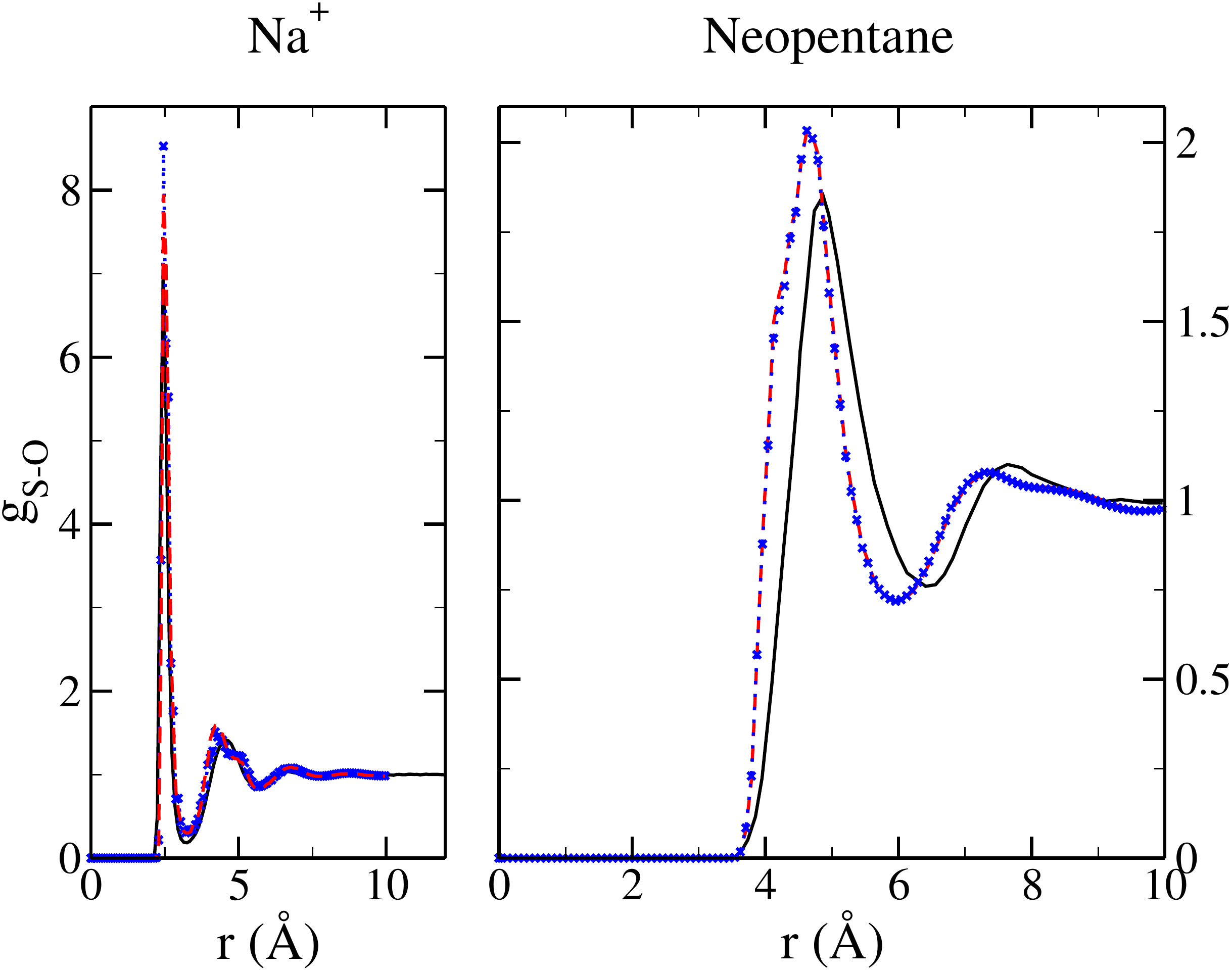}\protect\caption{Fonctions de distribution radiale entre le solvant et l'ion sodium
(à gauche) et le solvant et le centre de masse du néopentane (à droite).
La légende est la même que celle de la \ref{fig:rdfNMAcoupl}.\label{fig:rdfcoupl_Na_neop}}
\end{figure}

L'inclusion de ce terme de couplage n'a aucune influence visible sur
les fonctions de distribution radiale obtenues par minimisation fonctionnelle
pour les molécules hydrophobes comme le néopentane présenté en \ref{fig:rdfcoupl_Na_neop}.
Pour les molécules hydrophiles comme la N-méthylacétamide présentée
en \ref{fig:rdfNMAcoupl} ou les ions comme le sodium présenté en
\ref{fig:rdfcoupl_Na_neop}, l'ajout du couplage améliore légèrement
les fonctions de distribution radiale : la hauteur des pics dus à
la première couche de solvatation diminue mais reste surestimée. Le
second pic est peu affecté.

En ce qui concerne l'énergie libre de solvatation, l'ajout du terme
de couplage modifie l'énergie libre de solvatation d'environ 5\%.
Parce que ce terme augmente le temps de calcul et que son influence
sur les structures de solvatation et les énergies libres de solvatation
obtenues est limitée, nous l'avons négligé dans le manuscrit.

\cleardoublepage{}

\lhead[]{\rightmark}

\rhead[\leftmark]{}

\bibliographystyle{pnas2009}
\bibliography{text}

\cleardoublepage{}

\lhead[]{Nomenclature}

\rhead[Nomenclature]{}

\printnomenclature[2.5cm]{}
\end{document}